\pdfoutput=1
\newif\ifproofcopy
\proofcopytrue
\ifproofcopy
  \documentclass[12pt,twoside,openright]{memoir}
\else
  \documentclass[11pt,twoside,openright]{memoir}
\fi

\usepackage[T1]{fontenc}
\usepackage[utf8]{inputenc}
\usepackage{newtxtext}
\usepackage{newtxmath}
\usepackage[scaled=0.85]{beramono}
\usepackage{microtype}
\UseMicrotypeSet[protrusion]{alltext-nott}

\usepackage{amsmath}

\usepackage{amssymb}

\usepackage{amsthm}
\usepackage{booktabs}
\usepackage{longtable}
\usepackage{tabularx}
\usepackage{multirow}
\usepackage{enumitem}
\setlist{nosep,leftmargin=1.4em}
\setlist[enumerate]{labelindent=0.4em,leftmargin=*}

\usepackage{graphicx}
\usepackage{tikz}
\usetikzlibrary{positioning,arrows.meta,calc,decorations.pathreplacing,fit}

\usepackage[most]{tcolorbox}
\usepackage{seqsplit}
\usepackage{listings}

\usepackage[round,authoryear]{natbib}
\setcitestyle{aysep={},yysep={,}}

\usepackage[hyphens]{url}
\usepackage[colorlinks=true,linkcolor=black,citecolor=black,urlcolor=black,
            pdfusetitle]{hyperref}
\usepackage{memhfixc}   

\ifproofcopy
  \setstocksize{9.818in}{7.636in}
  \settrimmedsize{9.818in}{7.636in}{*}
  \setlrmarginsandblock{0.955in}{0.955in}{*}
  \setulmarginsandblock{0.982in}{1.036in}{*}
\else
  \setstocksize{9in}{7in}
  \settrimmedsize{9in}{7in}{*}
  \setlrmarginsandblock{0.875in}{0.875in}{*}
  \setulmarginsandblock{0.9in}{0.95in}{*}
\fi
\setheadfoot{\onelineskip}{2\onelineskip}
\setheaderspaces{*}{1.5\onelineskip}{*}
\settrims{0pt}{0pt}
\checkandfixthelayout

\makechapterstyle{mitpress}{%

  }
\chapterstyle{mitpress}

\setsecheadstyle{\Large\bfseries\raggedright}
\setsubsecheadstyle{\large\bfseries\raggedright}
\setsubsubsecheadstyle{\normalsize\bfseries\raggedright}

\makepagestyle{mitpress}
\makeevenhead{mitpress}{\thepage}{\small\itshape\leftmark}{}
\makeoddhead{mitpress}{}{\small\itshape\rightmark}{\thepage}
\makepsmarks{mitpress}{%
  \createmark{chapter}{left}{shownumber}{}{\quad}
  \createmark{section}{right}{shownumber}{}{\quad}
  \createplainmark{toc}{both}{\contentsname}
  \createplainmark{bib}{both}{\bibname}}
\aliaspagestyle{chapter}{plain}
\nouppercaseheads
\setsecnumdepth{subsection}
\settocdepth{section}
\nonzeroparskip

\DeclareRobustCommand{\code}[1]{\texttt{\small #1}}

\DeclareRobustCommand{\codebrk}[1]{\texttt{\small\seqsplit{#1}}}

\newcommand{\ev}[1]{}

\newcommand{\Sign}{\mathbf{Sign}}
\newcommand{\Sen}{\mathrm{Sen}}
\newcommand{\Mod}{\mathrm{Mod}}

\newcommand{\Prop}{\mathrm{Prop}}
\newcommand{\Srt}{\mathrm{Sort}}

\newtcolorbox{aside}[1]{
  colback=gray!8, colframe=gray!50, arc=2mm, boxrule=0.4pt,
  left=8pt, right=8pt, top=6pt, bottom=6pt,
  before skip=10pt, after skip=10pt,
  fonttitle=\bfseries, title={#1}, fontupper=\small}

\theoremstyle{plain}
\newtheorem{theorem}{Theorem}[chapter]
\newtheorem{lemma}[theorem]{Lemma}
\newtheorem{proposition}[theorem]{Proposition}
\newtheorem{corollary}[theorem]{Corollary}
\newtheorem{conjecture}[theorem]{Conjecture}
\newtheorem{hypothesis}[theorem]{Hypothesis}

\theoremstyle{definition}
\newtheorem{definition}[theorem]{Definition}
\newtheorem{example}[theorem]{Example}

\theoremstyle{remark}
\newtheorem{remark}[theorem]{Remark}

\usepackage{mathpartir}

\newcommand{\Setu}{\mathrm{Set}}            
\newcommand{\Typeu}{\mathrm{Type}}          
\newcommand{\lzero}{\mathsf{0}}
\newcommand{\lsucc}{\mathsf{s}}
\newcommand{\lmax}{\mathsf{max}}
\newcommand{\limax}{\mathsf{imax}}
\newcommand{\cumle}{\le_{\mathrm{c}}}       
\newcommand{\Id}{\mathsf{Id}}
\newcommand{\refl}{\mathsf{refl}}
\newcommand{\Jelim}{\mathsf{J}}
\newcommand{\irrel}{\mathsf{irrel}}
\newcommand{\fst}{\mathsf{fst}}
\newcommand{\snd}{\mathsf{snd}}
\newcommand{\rec}{\mathsf{rec}}
\newcommand{\Nat}{\mathsf{Nat}}
\newcommand{\List}{\mathsf{List}}
\newcommand{\Vect}{\mathsf{Vec}}
\newcommand{\Unit}{\mathbf{1}}
\newcommand{\Empty}{\mathbf{0}}
\newcommand{\defeq}{\equiv}                 
\newcommand{\subty}{\leq}                   
\newcommand{\chk}{\Leftarrow}               
\newcommand{\infr}{\Rightarrow}             
\newcommand{\ctx}{\Gamma}
\providecommand{\tttEnv}{\mathcal{S}}                 
\providecommand{\tttlam}[2]{\lambda #1{:}#2.\,}      
\providecommand{\tttpair}[2]{\langle #1, #2\rangle}  
\providecommand{\tttann}[2]{(#1 : #2)}               
\providecommand{\tttinst}[2]{#1.\{#2\}}              
\providecommand{\tttwf}{\ \mathsf{wf}}               
\providecommand{\tttctx}{\ \mathsf{ctx}}             
\providecommand{\tttstep}{\longrightarrow}           
\providecommand{\tttBool}{\mathsf{Bool}}
\providecommand{\ttttt}{\mathsf{tt}}
\providecommand{\tttzero}{\mathsf{zero}}
\providecommand{\tttsucc}{\mathsf{succ}}
\providecommand{\tttnil}{\mathsf{nil}}
\providecommand{\tttcons}{\mathsf{cons}}
\providecommand{\tttmatch}{\mathsf{match}}
\providecommand{\tttExists}{\mathsf{Exists}}
\providecommand{\tttOr}{\mathsf{Or}}
\providecommand{\tttFalse}{\mathsf{False}}
\providecommand{\tttAcc}{\mathsf{Acc}}
\providecommand{\tttFV}{\mathrm{FV}}
\providecommand{\tttlev}[1]{[\![#1]\!]}              
\providecommand{\tttsim}{\simeq}                     
\providecommand{\tttrule}[1]{\textsc{#1}}

\newcommand{\JTerm}{\mathsf{JTerm}}         
\newcommand{\JB}{\mathsf{JustifiedBy}}      
\newcommand{\jdecl}{\mathsf{declared}}
\newcommand{\jobs}{\mathsf{observed}}
\newcommand{\jver}{\mathsf{verified}}
\newcommand{\japp}{\mathsf{app}}
\newcommand{\jsum}{\mathsf{sum}}
\newcommand{\jinst}{\mathsf{inst}}
\newcommand{\jpair}{\mathsf{pair}}
\newcommand{\Evid}{\mathcal{E}}             
\newcommand{\supp}{\mathrm{supp}}           

\newcommand{\Th}{\mathbf{Th}}
\newcommand{\Thz}{\mathbf{Th}_0}
\newcommand{\proofs}{\mathit{proofs}}
\newcommand{\Ins}{\mathbf{Ins}}
\newcommand{\piIns}{\pi\mathbf{Ins}}
\newcommand{\Ent}{\mathbf{Ent}}
\newcommand{\Log}{\mathbf{Log}}
\newcommand{\Cat}{\mathbf{Cat}}
\newcommand{\SetC}{\mathbf{Set}}           

\newcommand{\Entity}{\mathsf{Entity}}
\newcommand{\Class}{\mathsf{Class}}
\newcommand{\isa}{\mathsf{is\_a}}
\newcommand{\subclassof}{\mathsf{subclass\_of}}
\newcommand{\sqsubclass}{\sqsubseteq}       
\newcommand{\rtype}[1]{\{#1\}}              
\newcommand{\RC}[1]{R_{#1}}                 
\newcommand{\ttrmeet}{\wedge}
\newcommand{\ttrjoin}{\vee}
\newcommand{\sing}[1]{T_{#1}}               

\newif\ifdraftmark
\draftmarktrue
\ifdraftmark
  \newcommand{\draftmarktext}{\footnotesize\scshape Draft --- 10.5281/zenodo.23884679}
  \makeevenfoot{mitpress}{}{\draftmarktext}{}
  \makeoddfoot{mitpress}{}{\draftmarktext}{}
  \makeevenfoot{plain}{\draftmarktext}{\thepage}{}
  \makeoddfoot{plain}{}{\thepage}{\draftmarktext}
  \makeevenfoot{empty}{}{\draftmarktext}{}
  \makeoddfoot{empty}{}{\draftmarktext}{}
\fi

\title{Typed Knowledge over Constructive Institutions\\[1ex]
  \large Eigenius, Volume~I: Theory}
\author{Hans-Martin Will}
\date{Draft --- \today}

\begin{document}

\frontmatter
\maketitle
\thispagestyle{empty}
\vspace*{\fill}
\noindent Copyright \copyright\ 2026 Hans-Martin Will.

\medskip
\noindent This work is licensed under the Creative Commons
Attribution-NonCommercial-NoDerivatives 4.0 International License (CC BY-NC-ND 4.0). To view a
copy of this license, visit \url{https://creativecommons.org/licenses/by-nc-nd/4.0/}.

\medskip
\noindent DOI: \href{https://doi.org/10.5281/zenodo.23884679}{10.5281/zenodo.23884679}

\medskip
\noindent The Lean~4 formalization that accompanies this volume is published at
\url{https://github.com/eigenius/theory}.
\clearpage
\begin{abstract}
Type theories designed as foundations for mathematics or for program verification need only that a
proposition holds; a theory with proof irrelevance in conversion erases, at the kernel level, how
it was established. A persistent knowledge store, by contrast, must retain that record: who
established a claim, by which derivation or external tool, and under which assumptions. This book,
the first volume of a series on Eigenius, presents the theory of Eigenius---a store of typed
knowledge over constructive institutions---and develops its metatheory from that requirement.

The kernel theory is a dependent type theory with a cumulative universe hierarchy, an impredicative
universe of propositions, inductive families with recursors, and an intensional identity type. It
departs from the theories of current proof assistants in two ways. First, proof irrelevance is a
provable fact with an explicit witness rather than a rule of conversion, so proofs of a single
proposition are provably equal yet remain distinct terms. Second, definitional equality is
joinability under an orthogonal rewriting system---making it confluent, injective on type formers,
and decidable by a type-free test---while a separate subsumption judgment covers universe
cumulativity, function types covariant in their codomains, and declared class inclusions.

The metatheory is developed in five parts. First, a set-theoretic model that interprets terms
directly proves consistency, relative to a single obligation on the global environment, for two
bidirectional fragments that the kernel implements and, by translation, for the declarative theory.
Second, a justification logic over the kernel's propositions records evidence together with its
support. Third, the theory of an environment is organized as an institution, so that extending the
store is a signature morphism and a verdict obtained in an external system enters along a
comorphism. Fourth, a typed knowledge-graph theory with record types and width subtyping gives the
store its classes, resources, and refinements. Finally, the procedures that decide the
bidirectional fragments---conversion by reduction to weak-head normal form and normalization by
evaluation---are stated with their correctness relative to the termination of reduction. A Lean~4
formalization accompanies the volume, and an appendix maps it to the definitions and results in the
text. Later volumes of the series treat the implementation, the language pipeline, and the system.
\end{abstract}
\tableofcontents

\mainmatter

\chapter{Introduction}
\label{ch:introduction}

Eigenius~\citep{will-brown-fuchs2026} is a store for claims in which the standing of every claim
is a property the store computes. A claim enters as a typed resource; a kernel decides whether it
is well formed; the grounds it rests on are recorded as terms the kernel evaluates; and results
produced by external reasoners enter through declared interfaces whose obligations the kernel
checks. This book, the first volume of a series, states the theory behind that design. It presents
the four foundational theories---a dependent type theory (Chapter~\ref{ch:mltt}), justification
logic (Chapter~\ref{ch:justification-logic}), institution theory over constructive type theories
(Chapter~\ref{ch:institutions}), and a type-theoretic semantics of knowledge graphs
(Chapter~\ref{ch:knowledge-graphs})---together with the procedures that decide the kernel type
theory (Chapter~\ref{ch:checking}). Later volumes treat its realization: the implementation
(Volume~II), the language pipeline (Volume~III), and the system (Volume~IV).

\section{Claims and their standing}
\label{sec:intro-standing}

Software agents now author a growing share of the scientific and engineering record: analyses in
drug-discovery pipelines, autonomy stacks, climate models, and materials screens. Today's record
leaves a reader with two options: trust the author or do not use the result. This trade-off is
tractable at the rate at which human authors produce results, but intractable at the rate at
which language models produce them.

Consider the following sentence from a typical results section:
\begin{quote}
EIG\_0291 was identified as a strong CDK2 inhibitor ($\mathrm{IC}_{50} = 85 \pm 5$\,nM, $n = 6$).
\end{quote}
The sentence relies on five implicit assertions:
\begin{enumerate}[label=(\arabic*)]
\item \emph{measurement}: six $\mathrm{IC}_{50}$ readings were taken under a stated assay protocol;
\item \emph{statistics}: $85 \pm 5$\,nM is the summary those six readings produce under a stated
  procedure;
\item \emph{vocabulary translation}: a low $\mathrm{IC}_{50}$ at this target is what the domain
  term ``strong inhibitor'' denotes;
\item \emph{decision rule}: a stated threshold and test decide that $85$\,nM is low enough;
\item \emph{sample attribution}: the six readings are readings of EIG\_0291 and of no other
  compound or batch.
\end{enumerate}
The sentence states none of these five assertions, yet its conclusion depends on all of them. A
hallucination or an unjustified assumption at any of the five yields a sentence of the exact same
form as a sound one, whether a human or an agent drafted it.

Moreover, the five assertions rest on different kinds of support. Assertions (1) and (5) rest on
measurement, assertion (2) on computation, and assertions (3) and (4) on decisions---conventions
adopted by someone with the authority to adopt them. The conclusion that EIG\_0291 is a strong
inhibitor composes all five, and that composition is a proof. These four kinds of support fail in
different ways: a measurement fails through its instrument or protocol, a computation through its
recipe or inputs, a decision through the authority behind it, and a proof through a logical gap. A
reader who acts on a result needs to know the support behind each part of it, but the published
record does not state it. Instead, the reader must reconstruct it from the methods section, the
deposited code, and the cited sources, and that reconstruction remains incomplete wherever a step
goes undocumented.

Such reconstructions are incomplete even when authors deposit their analysis code. For example, the
report identifying \emph{WRN} helicase as a synthetic-lethal target in microsatellite-unstable
cancers \citep{chan2019} includes its analysis code, yet the statistical model behind one of its
panels is recoverable only by reading that code: the figure caption names the test, but omits the
model terms, the rows entering the model, and the normalization.

Four existing kinds of systems each address only part of this problem:
\begin{itemize}
\item A \emph{knowledge graph} with a provenance vocabulary \citep{prov,moreau2010} records the
  assertion that a result was derived from a dataset. However, that provenance assertion has the
  same author as the result itself, and the graph has no means of distinguishing a derivation that
  actually took place from one that was merely asserted.
\item A \emph{database with integrity constraints} checks properties of stored values. A
  constraint can verify that a $p$-value lies in $[0,1]$, but it cannot verify that the $p$-value
  is the one produced by the stated recipe, because the recipe itself is not checked against the
  value.
\item A \emph{proof assistant} checks whether an argument composes. However, the leaves of a
  mathematical proof come from a short, fixed list of axioms, whereas the leaves of an empirical
  argument are instrument readings, thresholds, and conventions---an open set that grows with every
  experiment.
\item A \emph{language-model pipeline} produces text describing a derivation, but provides no
  procedure to decide whether the derivation holds. A flawed inference and a valid one have the
  same surface form.
\end{itemize}

In software development, a coding agent converges because a compiler and a type checker reject its
ill-formed output with a location and a reason, prompting the agent to revise. No analogous
procedure rejects the output of an agent that reasons in prose. Eigenius provides such a procedure
by representing claims as typed resources and making admission to the store a type-checking
problem. The store computes its verdict at commit time from the data it holds; an author cannot
simply assert the verdict.

Pharos, presented in 2008 as a platform for enforcing the scientific method in the life sciences
\citep{brown2008pharos}, is a direct predecessor of Eigenius; it combined a typed graph store with
a port of a higher-order-logic proof assistant. However, when independent components (a graph
store, a proof assistant, a reasoning engine) each maintain their own type discipline, they require
pairwise translation code that lies outside the verified record. In Eigenius, every component
beyond the kernel's type theory is either expressed directly within that type theory or connected
to it through a declared, typed interface.

\section{Five requirements}
\label{sec:intro-requirements}

This design imposes five core requirements. The type theory of Chapter~\ref{ch:mltt} addresses the
first two, and Chapters~\ref{ch:justification-logic} through~\ref{ch:knowledge-graphs} address the
remaining three.

\paragraph{Decidable claims.}
A claim is stated as a type, its evidence is a term of that type, and the store admits a claim
only if the kernel accepts the term. Under this propositions-as-types reading of dependent type
theory, type checking serves as the decision procedure required by the agent loop. Because claims
about empirical data refer to specific objects---a claim about a measurement names that measurement,
and its type depends on it---the theory requires dependent function and pair types, inductive
families to define the data domains, and an identity type to state equations. Two further features
support the rest of the architecture: a cumulative hierarchy of universes with subsumption gives
class hierarchies (Chapter~\ref{ch:knowledge-graphs}) a direct typing, and a kernel theory close to
that of Lean~4 \citep{lean4} allows results verified in Lean to enter the store with their meaning
preserved. Chapter~\ref{ch:mltt} presents this system: Martin-L\"of's intuitionistic type theory
\citep{martinlof1984}, extended with the universe structure of Luo's Extended Calculus of
Constructions and the inductive families and impredicative universe of propositions of the Calculus
of Inductive Constructions. Section~\ref{sec:mltt-metatheory} details the departures from
Martin-L\"of's theory and establishes consistency and key metatheoretic properties of the
combination, while Chapter~\ref{ch:checking} presents the procedures that decide its bidirectional
fragments.

\paragraph{Retained proofs.}
Whereas a proof assistant needs only to establish that a proposition holds, a knowledge store must
also record who established a claim, by which route, under which assumptions, and whether that
evidence still stands. Consequently, the proof of a claim is not discarded once the claim is
admitted: it remains an explicit term that the store retains, inspects, and compares with other
proofs of the same claim, while keeping the empirical or declared grounds of a claim distinct from
its deductive proofs (Chapter~\ref{ch:justification-logic}). At this point, the theory of
Chapter~\ref{ch:mltt} departs from Lean~4 and from Coq's universe of strict propositions. Those
kernels build proof irrelevance into conversion, identifying all proofs of a proposition during
type checking so that the specific proof supplied is erased at the kernel level. In the theory of
Chapter~\ref{ch:mltt}, by contrast, proof irrelevance is an axiom rather than a conversion rule:
two proofs of a proposition are provably equal via an explicit witness, yet they remain distinct
terms under reduction. Because conversion no longer identifies inhabitants of propositions, the
equality the kernel computes does not need to inspect the types of terms; accordingly, conversion
is untyped joinability without extensionality rules, and $\lambda$-abstractions carry their domain
types. Sections~\ref{sec:mltt-prop} and~\ref{sec:mltt-metatheory} develop this design, on which the
justification terms of Chapter~\ref{ch:justification-logic}, the witness transport of
Chapter~\ref{ch:institutions}, and the refined records of Chapter~\ref{ch:knowledge-graphs} all
rely.

\paragraph{Computed standing.}
A deductive proof shows how a claim follows from its premises, but the record must also track the
ultimate grounds of those premises: which observations were recorded, which declarations were made
by which agents, and which steps were verified by an external checker. Grounds and proofs are
distinct objects: under proof irrelevance, any two deductive proofs of the same proposition are
provably equal, whereas two sets of grounds for it are not---one may rest on a single measurement
and another on three independent assays---and the set of possible grounds is open-ended, growing
with every new observation. Justification logic \citep{artemov2008justification} provides the
corresponding formal object: the justification term, built from atomic justifications by
application and sum, paired with an evidence function that determines which formulas a term
justifies. Chapter~\ref{ch:justification-logic} adapts this framework to the store. Justification
terms form an inductive algebra in the lowest universe whose leaves represent the informal support
kinds of Section~\ref{sec:intro-standing}---\emph{declared} grounds (decisions and conventions),
\emph{observed} grounds (measurements), and \emph{verified} grounds (checked computations and
proofs)---and whose compound terms record how derivations combine them. Whether a justification term
supports a given claim is decided by a computable evidence function
(Section~\ref{sec:jl-evidence}), and the \emph{support} of a term relative to a proposition
(Section~\ref{sec:jl-support}) answers the counterfactual question of whether the claim remains
justified when a set of resources is withdrawn. Every coarse grade of standing---such as an ordering
that ranks \emph{declared} below \emph{verified}---arises as a homomorphic image of support
(Theorem~\ref{thm:jl-lattice-hom}), so the store records fine-grained justification terms and
computes summary grades on demand.

\paragraph{Several logics.}
No single logic covers every mode of reasoning in a scientific record. A statistical test, a
theorem proved in Lean, an inference by the store's own reasoner, and a typing derivation in the
kernel belong to different logical systems, and the store must host their results without
re-implementing each system inside the kernel. Institution theory
\citep{goguen-burstall1984,goguen1992} provides a general framework for relating logical systems:
each logic is presented by its signatures, sentences, and models, and a comorphism translates
sentences from one logic into another while preserving semantic satisfaction. Because Eigenius
retains evidence, a translation between logics must also transport proof witnesses. When a source
theory relies on an axiom that has no proof witness to transport, the translation must explicitly
adopt the axiom in the target, discharge it by supplying a proof, or retain it as an explicit
hypothesis. Chapter~\ref{ch:institutions} develops this framework across three levels of
abstraction, each obtained from the one above by forgetting structure: the \emph{witness} level,
where each logic is a proof calculus and translations transport proof terms; the \emph{provability}
level, where each logic is a $\pi$-institution (a syntactic consequence relation over translated
sentences, without models); and the \emph{model} level, where a canonical construction yields an
institution whose satisfaction condition holds automatically. The chapter concludes with a summary
table of every institutional obligation and its current status (proved, cited, or open).

\paragraph{Typed data.}
The resources of the store form a knowledge graph: resources carry properties with typed values,
and classes declare parent classes alongside required and recommended properties. Unlike closed
database schemas, a resource may carry additional properties not mentioned by its classes, and two
classes with the same properties remain distinct unless declared otherwise.
Chapter~\ref{ch:knowledge-graphs} gives this data model a formal semantics based on Cooper's Type
Theory with Records \citep{cooper2023perception}, embedded within the kernel type theory of
Chapter~\ref{ch:mltt}. In this semantics, a class denotes a named record type and a resource
denotes a record, while class membership is captured at three levels
(Section~\ref{sec:kg-membership}): a declared subclass judgment from the environment, a structural
typing judgment over record fields, and an internal proposition within the logic. Width subtyping
of records accommodates undeclared properties, joins of singleton types represent enumerations, and
dependent fields express conditional property requirements. Because the grammatical framework of
Volume~III builds on the same record types, Chapter~\ref{ch:knowledge-graphs} also develops the
general record theory used across the series (Section~\ref{sec:kg-ttr}) and connects it to the
typed feature structures of computational linguistics. An appendix to the chapter maps Atomic Data
\citep{atomicdata}---the knowledge-graph format from which Eigenius derives its data model---into
this type theory (Appendix~\ref{sec:kg-atomic-data}).

\section{The structure of this volume}
\label{sec:intro-structure}

The chapters of this volume build on one another sequentially. Chapter~\ref{ch:mltt} defines the
kernel type theory in which all subsequent constructions are formulated.
Chapter~\ref{ch:justification-logic} defines justification terms as inductive data within that type
theory and formulates the justification relation as an internal proposition.
Chapter~\ref{ch:institutions} organizes both the kernel type theory and justification logic as
hosted logics within the institutional framework, and connects them to external systems such as
Lean and statistical inference engines. Chapter~\ref{ch:knowledge-graphs} uses the subsumption
judgment of Chapter~\ref{ch:mltt} (Section~\ref{sec:mltt-subsumption}) to model class hierarchies,
providing the data signatures over which the logics of Chapter~\ref{ch:institutions} state claims
about stored resources. Chapter~\ref{ch:checking} presents the algorithms that decide the
bidirectional fragments of Chapter~\ref{ch:mltt}---weak-head reduction, the conversion test,
normalization by evaluation, and bidirectional type checking---and proves their correctness relative
to the termination of reduction. Finally, Appendix~\ref{app:formalization} maps every definition
and theorem in the volume to its counterpart in the accompanying Lean~4 formalization.

This volume focuses exclusively on theory. Each chapter presents the theory in the target form the
system is designed to realize; points where the current implementation still diverges from this
target theory are documented in Volume~II. Every formal claim in the text is either proved, cited
with its source, or explicitly marked as a conjecture or hypothesis with the remaining obligation
identified. Notation is introduced at its first use, and each symbol retains a single meaning
throughout the volume.

\section{Readers}
\label{sec:intro-readers}

This volume is written primarily for researchers in type theory and logic. Five additional groups
of readers can enter through specific chapters:
\begin{itemize}
\item \emph{Researchers in formal verification} will find a small trusted kernel that gates every
  admission to the store, external provers whose outputs enter as kernel-checked certificates, and
  proof-witness transport across logical systems (Chapters~\ref{ch:mltt}, \ref{ch:institutions},
  and~\ref{ch:checking}, together with Volume~II).
\item \emph{Researchers in knowledge representation} will find a type-theoretic semantics for
  open-world knowledge graphs and a formal mapping of the Atomic Data model
  (Chapter~\ref{ch:knowledge-graphs}).
\item \emph{Computational linguists} will find the dependent record theory that underpins the
  type-logical grammar of Volume~III (Chapter~\ref{ch:knowledge-graphs}).
\item \emph{Researchers in databases and provenance} will find epistemic standing formulated as a
  computed property over fine-grained justification terms, supporting counterfactual invalidation
  and algebraic provenance grading (Chapter~\ref{ch:justification-logic}).
\item \emph{Researchers in trustworthy AI and scientific computing} will find an architecture that
  turns claim admission into a decidable type-checking problem for autonomous agents
  (Section~\ref{sec:intro-standing}).
\end{itemize}

The text assumes familiarity with graduate-level logic, the typed lambda calculus, and basic
category theory (categories, functors, and natural transformations). No prior background in
justification logic, institution theory, or Type Theory with Records is assumed;
Chapters~\ref{ch:justification-logic}--\ref{ch:knowledge-graphs} introduce the definitions they
require.

\chapter{Dependent Type Theory: from Martin-L\"of to Inductive Constructions}
\label{ch:mltt}

\providecommand{\tttEnv}{\mathcal{S}}                 
\providecommand{\tttlam}[2]{\lambda #1{:}#2.\,}      
\providecommand{\tttpair}[2]{\langle #1, #2\rangle}  
\providecommand{\tttann}[2]{(#1 : #2)}               
\providecommand{\tttinst}[2]{#1.\{#2\}}              
\providecommand{\tttwf}{\ \mathsf{wf}}               
\providecommand{\tttctx}{\ \mathsf{ctx}}             
\providecommand{\tttstep}{\longrightarrow}           
\providecommand{\tttCat}{\mathsf{Cat}}
\providecommand{\tttAtom}{\mathsf{Atom}}
\providecommand{\tttMode}{\mathsf{Mode}}
\providecommand{\tttNum}{\mathsf{Num}}
\providecommand{\tttBool}{\mathsf{Bool}}
\providecommand{\ttttt}{\mathsf{tt}}
\providecommand{\tttzero}{\mathsf{zero}}
\providecommand{\tttsucc}{\mathsf{succ}}
\providecommand{\tttnil}{\mathsf{nil}}
\providecommand{\tttcons}{\mathsf{cons}}
\providecommand{\tttmatch}{\mathsf{match}}
\providecommand{\tttExists}{\mathsf{Exists}}
\providecommand{\tttOr}{\mathsf{Or}}
\providecommand{\tttFalse}{\mathsf{False}}
\providecommand{\tttAcc}{\mathsf{Acc}}
\providecommand{\tttFV}{\mathrm{FV}}
\providecommand{\tttlev}[1]{[\![#1]\!]}              
\providecommand{\tttsim}{\simeq}                     
\providecommand{\tttrule}[1]{\textsc{#1}}

This chapter states the dependent type theory that the Eigenius kernel decides,
restricted to its logical core: dependent functions and pairs, a cumulative
hierarchy of universes with an impredicative universe of propositions at the
bottom, an intensional identity type, inductive families with derived
recursors, a global environment of declarations, and a proof-irrelevant
universe of propositions. Program constructs, knowledge-graph types, and
institution-facing references are excluded; Section~\ref{sec:mltt-scope} lists
them together with the chapters and volumes that treat them. The theory is a
descendant of Martin-L\"of's intuitionistic type
theory~\citep{martinlof1984}. Its universe structure follows Luo's Extended
Calculus of Constructions (ECC)~\citep{luo1990,luo1994}, while its inductive
families and universe polymorphism follow the Calculus of Inductive
Constructions (CIC) as realized by the Lean~4
kernel~\citep{coquand-paulin1990,lean4,carneiro2024}. Proof irrelevance is the
central point where the theory departs from Lean: because Eigenius stores
evidence rather than truth alone (Chapter~\ref{ch:introduction}), a proof must
remain a distinct term that the kernel can inspect and compare. Consequently,
proof irrelevance is a provable fact with an explicit witness
(Section~\ref{sec:mltt-prop}) rather than a rule of conversion, and
Section~\ref{sec:mltt-metatheory} traces the consequences of this choice for
definitional equality. Table~\ref{tab:mltt-comparison} in
Section~\ref{sec:mltt-metatheory} summarizes the differences between this
theory and Martin-L\"of type theory (MLTT), ECC, CIC, and the Lean~4 kernel.

The primary specification in Sections~\ref{sec:mltt-judgements}--\ref{sec:mltt-prop}
is declarative. Alongside it, Section~\ref{sec:mltt-pi-sigma} introduces the
syntax-directed bidirectional judgments (inference and checking) that explain
where type annotations and subsumption act; Section~\ref{sec:mltt-consistency}
uses these bidirectional fragments to establish consistency via a set-theoretic
model, and Chapter~\ref{ch:checking} gives the procedures that decide them
(weak-head reduction, the conversion test, and normalization by evaluation).
Points where the current kernel implementation still diverges from this target
theory, together with the adjustments that align it, are recorded in
Volume~II. Section~\ref{sec:mltt-metatheory} highlights the one syntactic
adjustment on which the consistency proof of this chapter relies:
$\lambda$-abstractions carry their domain types.

\section{Judgments, contexts, and propositions as types}
\label{sec:mltt-judgements}

\paragraph{Syntax.} Terms and types belong to a single syntactic category, as in
any pure type system with Russell-style universes.

\begin{definition}[Terms]\label{def:mltt-terms}
Let $x, y, z$ range over a countable set of variables, $c$ over constants of a
global environment (Section~\ref{sec:mltt-environment}), and $\ell$ over level
expressions (Definition~\ref{def:mltt-levels}). The terms are generated by the
grammar
\[
\begin{array}{r@{\ }c@{\ }l}
a, b, A, B, C, P &::=& x \mid \tttinst{c}{\bar\ell} \mid \Srt\,\ell
  \mid \Pi x{:}A.\,B \mid \tttlam{x}{A} b \mid b\,a \\[2pt]
&\mid& \Sigma x{:}A.\,B \mid \tttpair{a}{b} \mid \fst\,a \mid \snd\,a \\[2pt]
&\mid& \Id\,A\,a\,b \mid \refl\,a \mid \Jelim\,a\,C\,d\,b\,p \mid \tttann{a}{A}.
\end{array}
\]
Inductive types, their constructors, and their recursors enter as constants
introduced by inductive declarations (Section~\ref{sec:mltt-inductive}).
The binders $\Pi$, $\Sigma$, and $\lambda$ bind $x$ in $B$ and in $b$, but not
in the domain $A$. Each abstraction carries its domain type $A$, as in ECC and
Lean's kernel theory; Section~\ref{sec:mltt-metatheory} explains why the domain
is part of the term syntax. Terms are identified up to $\alpha$-equivalence.
\end{definition}

\paragraph{Reading the term shapes.} Aside from variables, constants, and
annotations, each term shape is a type former, an introduction form, or an
elimination form, and under the propositions-as-types reading each has a
logical interpretation. The metavariable letters are a reading aid only: $a, b$
stand for general terms, $A, B$ for types, $C$ for a type family (a
\emph{motive}), and $P$ for a proposition.
\begin{itemize}
\item $x$ is a variable, representing a local hypothesis or parameter declared
  in the context.
\item $\tttinst{c}{\bar\ell}$ is a constant from the global environment (a
  definition, axiom, inductive type, constructor, or recursor), instantiated at
  the universe levels $\bar\ell$.
\item $\Srt\,\ell$ is the universe at level $\ell$, a type whose elements are
  types. In particular, $\Srt\,0$ is $\Prop$, the universe of propositions, and
  $\Srt\,1$ is $\Setu$, the universe of small data types.
\item $\Pi x{:}A.\,B$ is the dependent function type mapping each $x : A$ to an
  element of $B$, where $B$ may depend on $x$. It is introduced by the
  abstraction $\tttlam{x}{A} b$ and eliminated by application $b\,a$. Read
  logically, it represents universal quantification $\forall x{:}A.\,B$, and
  reduces to implication $A \to B$ when $x$ does not occur free in $B$.
\item $\Sigma x{:}A.\,B$ is the dependent pair type in which the type $B$ of the
  second component depends on the first component $x : A$. It is introduced by
  the pair $\tttpair{a}{b}$ and eliminated by the projections $\fst\,a$ and
  $\snd\,a$. Read logically, it represents constructive existential
  quantification with an explicit witness, and reduces to conjunction
  $A \wedge B$ when $x$ does not occur free in $B$.
\item $\Id\,A\,a\,b$ is the proposition that $a$ and $b$ are equal elements of
  $A$. It is introduced by $\refl\,a$, the proof that $a$ equals itself, and
  eliminated by $\Jelim\,a\,C\,d\,b\,p$: given the base point $a$, a motive $C$
  parameterized by an endpoint and an equality proof, a base case $d$ for the
  endpoint $a$ and the proof $\refl\,a$, a target endpoint $b$, and a proof
  $p : \Id\,A\,a\,b$, it produces an element of $C\,b\,p$. This expresses the
  indiscernibility of identicals.
\item $\tttann{a}{A}$ is a term annotated with an expected type $A$. It has no
  logical content of its own; instead, it fixes the type against which the
  bidirectional rules of Figure~\ref{fig:mltt-bidir} check $a$.
\end{itemize}

\begin{definition}[Substitution]\label{def:mltt-substitution}
$b[a/x]$ denotes the capture-avoiding substitution of $a$ for the free
occurrences of $x$ in $b$. For a list $\bar a = a_1, \ldots, a_n$ and distinct
variables $\bar x$, $b[\bar a/\bar x]$ denotes simultaneous substitution. Level
substitution $b[\bar\ell/\bar u]$ replaces the level parameters $\bar u$ by
$\bar\ell$ inside every level expression occurring in $b$.
\end{definition}

\begin{definition}[Context]\label{def:mltt-context}
A \emph{context} is a finite sequence $\ctx ::= \cdot \mid \ctx, x{:}A$ of
variable declarations with pairwise distinct variables. A \emph{telescope}
$\Delta = (x_1{:}A_1, \ldots, x_n{:}A_n)$ is a context fragment in which each
$A_i$ may depend on $x_1, \ldots, x_{i-1}$. For a telescope $\Delta$,
$\Pi\Delta.\,B$ abbreviates $\Pi x_1{:}A_1.\cdots\Pi x_n{:}A_n.\,B$,
$\lambda\Delta.\,b$ abbreviates $\lambda x_1{:}A_1.\cdots\lambda x_n{:}A_n.\,b$,
and $b\,\bar a$ abbreviates the iterated application $b\,a_1 \cdots a_n$.
\end{definition}

\begin{definition}[Judgment forms]\label{def:mltt-judgements}
Relative to a well-formed global environment $\tttEnv$ (left implicit when
fixed), the declarative system has three primary judgment forms:
\begin{enumerate}
\item $\ctx \tttctx$: $\ctx$ is a well-formed context;
\item $\ctx \vdash A : \Srt\,\ell$: $A$ is a well-formed type in universe
  $\Srt\,\ell$ (in Russell style, ``$A$ is a type'' is simply a typing judgment
  whose type is a universe);
\item $\ctx \vdash a : A$: $a$ is a well-typed term of type $A$.
\end{enumerate}
Definitional equality $a \defeq b$ is an untyped relation on terms rather than a
typing judgment: Definition~\ref{def:mltt-defeq} generates it from the
computation rules introduced in Sections~\ref{sec:mltt-pi-sigma}--\ref{sec:mltt-environment},
and the typing rules consult it via Rule~\tttrule{Conv}. Three auxiliary
judgment forms complete the presentation: the directional subsumption judgment
$\ctx \vdash A \subty B$ (Section~\ref{sec:mltt-subsumption}), which enters
typing through Rule~\tttrule{Sub}; the bidirectional judgments
$\ctx \vdash a \chk A$ (checking) and $\ctx \vdash a \infr A$ (inference),
introduced in Section~\ref{sec:mltt-pi-sigma} to make the placement of
subsumption and type annotations syntax-directed; and the well-formedness
judgment $\tttEnv \tttwf$ for global environments
(Section~\ref{sec:mltt-environment}).
\end{definition}

\begin{figure}[t]
\begin{mathpar}
\inferrule*[right=\tttrule{C-Empty}]{ }{\cdot \tttctx}
\and
\inferrule*[right=\tttrule{C-Ext}]
  {\ctx \vdash A : \Srt\,\ell \\ x \notin \mathrm{dom}(\ctx)}
  {\ctx, x{:}A \tttctx}
\and
\inferrule*[right=\tttrule{Var}]
  {\ctx \tttctx \\ (x{:}A) \in \ctx}
  {\ctx \vdash x : A}
\and
\inferrule*[right=\tttrule{Conv}]
  {\ctx \vdash a : A \\ A \defeq B \\ \ctx \vdash B : \Srt\,\ell}
  {\ctx \vdash a : B}
\and
\inferrule*[right=\tttrule{Sub}]
  {\ctx \vdash a : A \\ \ctx \vdash A \subty B \\ \ctx \vdash B : \Srt\,\ell}
  {\ctx \vdash a : B}
\end{mathpar}
\caption{Contexts, variables, conversion, and subsumption.}
\label{fig:mltt-structural}
\end{figure}

\paragraph{Structural rules.} Figure~\ref{fig:mltt-structural} presents the
rules for context formation and variable lookup, together with the two rules
(\tttrule{Conv} and \tttrule{Sub}) that change the type of a term without
modifying the term itself.

\begin{definition}[Definitional equality]\label{def:mltt-defeq}
One-step reduction $\tttstep$ is the compatible closure of the computation
rules of the calculus: $\beta$-reduction and pair projections
(Figure~\ref{fig:mltt-pi-sigma}), annotation erasure
(Rule~\tttrule{Ann-$\tttstep$}), $\delta$-unfolding of transparent definitions
(Section~\ref{sec:mltt-environment}), $\iota$-reduction for recursors
(Section~\ref{sec:mltt-inductive}), and \tttrule{Id-$\iota$} for the identity
eliminator (Section~\ref{sec:mltt-identity}). Two terms are \emph{joinable} if
they reduce in zero or more steps to terms that are identical up to replacing
level expressions by equivalent ones (Definition~\ref{def:mltt-levels},
Rule~\tttrule{Sort-$\defeq$}). \emph{Definitional equality} $a \defeq b$ is
joinability. Because $\tttstep$ is defined purely on term syntax, no reduction
rule carries a typing premise, every rule fires on any term matching its
left-hand side, and whether $a \defeq b$ holds does not depend on a type at
which $a$ and $b$ are compared. Moreover, no reduction rule is headed by a
$\lambda$-abstraction on its left-hand side; consequently, two abstractions are
joinable only if both their domains and their bodies are joinable (although
$\beta$-reduction discards the domain of the abstraction it contracts,
standalone abstractions are compared in both components).
Lemma~\ref{lem:mltt-confluence} proves that $\tttstep$ is confluent, which
makes $\defeq$ an equivalence relation closed under every term former and under
substitution.
\end{definition}

This definition mirrors the kernel's conversion test, which reduces both terms
and compares their normal forms (Chapter~\ref{ch:checking}), up to one syntactic
difference: the current kernel's $\lambda$-abstractions omit domain annotations,
so its conversion test compares function bodies only and identifies
$\tttlam{x}{\Nat} x$ with $\tttlam{x}{\Setu} x$, whereas
Definition~\ref{def:mltt-defeq} distinguishes them. Volume~II records the
adjustment that aligns the implementation with Definition~\ref{def:mltt-defeq};
modulo that adjustment and the termination of reduction
(Section~\ref{sec:mltt-metatheory}), the kernel decides this relation. Type
theories whose definitional equality is instead a typed judgment are compared
with this design in Section~\ref{sec:mltt-metatheory}.

The complete set of typing and subsumption rules is given across
Sections~\ref{sec:mltt-judgements}--\ref{sec:mltt-prop}. We state the
structural weakening and substitution properties for the full system here, since
the intervening sections invoke them.

\begin{lemma}[Weakening and substitution]\label{lem:mltt-substitution}
For each judgment form $\mathcal{J}$ of Definition~\ref{def:mltt-judgements}
(including $\ctx \vdash A \subty B$):
\begin{enumerate}
\item if $\ctx, \ctx' \vdash \mathcal{J}$ and $\ctx \vdash B : \Srt\,\ell$ with
  $y$ fresh, then $\ctx, y{:}B, \ctx' \vdash \mathcal{J}$;
\item if $\ctx, x{:}A, \ctx' \vdash \mathcal{J}$ and $\ctx \vdash a : A$, then
  $\ctx, \ctx'[a/x] \vdash \mathcal{J}[a/x]$;
\item if $a \defeq a'$, then $b[a/x] \defeq b[a'/x]$, and if $b \defeq b'$,
  then $b[a/x] \defeq b'[a/x]$.
\end{enumerate}
\end{lemma}

\begin{proof}
As in ECC~\citep{luo1994}, by simultaneous induction on derivations across the
rules of Sections~\ref{sec:mltt-judgements}--\ref{sec:mltt-prop}. Every
inference rule is closed under weakening and term substitution: no rule depends
on specific variable names, and the side conditions on universe levels
(Definition~\ref{def:mltt-levels}) do not mention term variables. The case for
Rule~\tttrule{Sub} uses the induction hypothesis for $\ctx \vdash A \subty B$,
and the case for Rule~\tttrule{Conv} uses part~(3), which holds because
one-step reduction is closed under substitution and commutes with it.
\end{proof}

\paragraph{Propositions as types.} A \emph{proposition} is a type $P$ with
$\ctx \vdash P : \Prop$, where $\Prop = \Srt\,\lzero$
(Section~\ref{sec:mltt-universes}), and a \emph{proof} of $P$ is a term $p$
with $\ctx \vdash p : P$. Under this Brouwer--Heyting--Kolmogorov (BHK)
interpretation, a proof of an implication is a function transforming proofs, a
proof of a universal statement is a dependent function, and a proof of a
conjunction is a pair. Table~\ref{tab:mltt-bhk} summarizes the encoding of the
logical connectives. Note that existential quantification and disjunction are
defined as inductive families in $\Prop$ rather than as $\Sigma$-types or
disjoint sums: by Rule~\tttrule{$\Sigma$-F} below, a $\Sigma$-type over a data
domain $A : \Setu$ lives in $\Setu$, making it a data type rather than a
proposition.

\begin{table}[t]
\centering\small
\begin{tabular}{@{}lll@{}}
\toprule
Connective & Type & Universe of the type \\
\midrule
$P \to Q$ & $\Pi x{:}P.\,Q$, $x \notin \tttFV(Q)$ & $\Prop$ \\
$\forall x{:}A.\,P$ & $\Pi x{:}A.\,P$ & $\Prop$, for every $A$ (impredicativity) \\
$P \wedge Q$ & $\Sigma x{:}P.\,Q$, $x \notin \tttFV(Q)$ & $\Prop$ \\
$\exists x{:}A.\,P$ & $\tttExists\,A\,(\tttlam{x}{A} P)$, inductive & $\Prop$ \\
$P \vee Q$ & $\tttOr\,P\,Q$, inductive & $\Prop$ \\
$\bot$ & $\tttFalse$, inductive without constructors & $\Prop$ \\
$a =_A b$ & $\Id\,A\,a\,b$ & $\Prop$ \\
\bottomrule
\end{tabular}
\caption{The propositional connectives as types.}
\label{tab:mltt-bhk}
\end{table}

A proof term is thus a self-contained record of how a proposition was derived.
Chapter~\ref{ch:justification-logic} relates this term algebra to the
justification terms of justification logic that Eigenius stores alongside
claims.

\section{Dependent functions and dependent pairs}
\label{sec:mltt-pi-sigma}

\paragraph{Rules.} Figure~\ref{fig:mltt-pi-sigma} states the formation,
introduction, elimination, and computation rules for $\Pi$- and $\Sigma$-types.
In the formation rules, the universe level of a composite type is computed from
the levels $\ell_1$ and $\ell_2$ of its components by two operators defined in
Section~\ref{sec:mltt-universes} (Definition~\ref{def:mltt-levels}): the
ordinary maximum $\lmax\,\ell_1\,\ell_2$ for $\Sigma$-types, and the
impredicative maximum $\limax\,\ell_1\,\ell_2$ for $\Pi$-types (which evaluates
to $0$ whenever $\ell_2$ evaluates to $0$, keeping $\Pi x{:}A.\,P$ in $\Prop$,
and otherwise coincides with $\lmax\,\ell_1\,\ell_2$). Each abstraction carries
its domain: Rule~\tttrule{$\Pi$-I} types $\tttlam{x}{A} b$ at a $\Pi$-type with
domain $A$, and in the bidirectional system below an abstraction infers its
type whenever its body infers a type.

\begin{figure}[t]
\begin{mathpar}
\inferrule*[right=\tttrule{$\Pi$-F}]
  {\ctx \vdash A : \Srt\,\ell_1 \\ \ctx, x{:}A \vdash B : \Srt\,\ell_2}
  {\ctx \vdash \Pi x{:}A.\,B : \Srt\,(\limax\,\ell_1\,\ell_2)}
\and
\inferrule*[right=\tttrule{$\Pi$-I}]
  {\ctx, x{:}A \vdash b : B}
  {\ctx \vdash \tttlam{x}{A} b : \Pi x{:}A.\,B}
\and
\inferrule*[right=\tttrule{$\Pi$-E}]
  {\ctx \vdash f : \Pi x{:}A.\,B \\ \ctx \vdash a : A}
  {\ctx \vdash f\,a : B[a/x]}
\\
\inferrule*[right=\tttrule{$\Pi$-$\beta$}]
  { }
  {(\tttlam{x}{A} b)\,a \tttstep b[a/x]}
\\
\inferrule*[right=\tttrule{$\Sigma$-F}]
  {\ctx \vdash A : \Srt\,\ell_1 \\ \ctx, x{:}A \vdash B : \Srt\,\ell_2}
  {\ctx \vdash \Sigma x{:}A.\,B : \Srt\,(\lmax\,\ell_1\,\ell_2)}
\and
\inferrule*[right=\tttrule{$\Sigma$-I}]
  {\ctx \vdash a : A \\ \ctx \vdash b : B[a/x] \\ \ctx, x{:}A \vdash B : \Srt\,\ell}
  {\ctx \vdash \tttpair{a}{b} : \Sigma x{:}A.\,B}
\\
\inferrule*[right=\tttrule{$\Sigma$-E$_1$}]
  {\ctx \vdash p : \Sigma x{:}A.\,B}
  {\ctx \vdash \fst\,p : A}
\and
\inferrule*[right=\tttrule{$\Sigma$-E$_2$}]
  {\ctx \vdash p : \Sigma x{:}A.\,B}
  {\ctx \vdash \snd\,p : B[\fst\,p/x]}
\\
\inferrule*[right=\tttrule{$\Sigma$-$\beta$}]
  { }
  {\fst\,\tttpair{a}{b} \tttstep a \\ \snd\,\tttpair{a}{b} \tttstep b}
\end{mathpar}
\caption{Dependent functions and dependent pairs. The $\lmax$ in
\tttrule{$\Sigma$-F} makes $\Sigma$ predicative; the $\limax$ in
\tttrule{$\Pi$-F} makes $\Pi$ impredicative at $\Prop$.}
\label{fig:mltt-pi-sigma}
\end{figure}

\paragraph{Derived forms.} The non-dependent function type
$A \to B := \Pi x{:}A.\,B$ and the Cartesian product
$A \times B := \Sigma x{:}A.\,B$, in both cases with $x \notin \tttFV(B)$, are
the special cases of Figure~\ref{fig:mltt-pi-sigma} in which the substitutions
$B[a/x]$ and $B[\fst\,p/x]$ are vacuous.

\paragraph{Annotation and bidirectional typing.} The annotated term
$\tttann{a}{A}$ records an expected type $A$ inside a term. In the declarative
system it acts as the identity and erases under reduction:
\begin{mathpar}
\inferrule*[right=\tttrule{Ann}]
  {\ctx \vdash A : \Srt\,\ell \\ \ctx \vdash a : A}
  {\ctx \vdash \tttann{a}{A} : A}
\and
\inferrule*[right=\tttrule{Ann-$\tttstep$}]
  { }
  {\tttann{a}{A} \tttstep a}
\end{mathpar}
Its nontrivial role is in the \emph{bidirectional system}, which splits typing
into two syntax-directed modes: \emph{inference} $\ctx \vdash a \infr A$, where
the type $A$ is synthesized as an output from $a$, and \emph{checking}
$\ctx \vdash a \chk A$, where the expected type $A$ is supplied as an input.
Figure~\ref{fig:mltt-bidir} shows the rules that govern mode switching and the
introduction and elimination forms whose typing depends on mode. Variables,
applications, projections, constants, universes, and type formers synthesize
their types, as does a $\lambda$-abstraction whose body synthesizes a type; by
contrast, dependent pairs, $\refl$, and abstractions whose bodies only check
require an expected type in checking mode. Type annotation (\tttrule{Inf-Ann})
switches from checking to inference, whereas Rule~\tttrule{Chk-Sub} switches
from inference to checking and is the sole point at which subsumption
(Section~\ref{sec:mltt-subsumption}) is applied. Consequently, a $\beta$-redex
$(\tttlam{x}{A} b)\,a$ is directly typable in the bidirectional system when $b$
infers a type; when $b$ only checks (for instance, when $b$ is a pair), the
abstraction in head position carries an annotation, as in
$\tttann{\tttlam{x}{A} \tttpair{c}{d}}{\Pi x{:}A.\,B}\,a$. Every bidirectional
derivation erases to a declarative one; conversely, every declarative
derivation can be annotated into a bidirectional one
(Proposition~\ref{prop:mltt-k1-complete}). Section~\ref{sec:mltt-consistency}
makes the bidirectional fragments ($\mathcal{K}_0$ and $\mathcal{K}_1$) precise
and uses them to construct the set-theoretic consistency model, while
Chapter~\ref{ch:checking} gives the algorithms that decide both modes.

\begin{figure}[t]
\begin{mathpar}
\inferrule*[right=\tttrule{Inf-Var}]
  {(x{:}A) \in \ctx}
  {\ctx \vdash x \infr A}
\and
\inferrule*[right=\tttrule{Inf-App}]
  {\ctx \vdash f \infr C \\ \ctx \vdash C \defeq \Pi x{:}A.\,B : \Srt\,\ell \\
   \ctx \vdash a \chk A}
  {\ctx \vdash f\,a \infr B[a/x]}
\\
\inferrule*[right=\tttrule{Inf-Lam}]
  {\ctx \vdash A \infr \Srt\,\ell \\ \ctx, x{:}A \vdash b \infr B}
  {\ctx \vdash \tttlam{x}{A} b \infr \Pi x{:}A.\,B}
\and
\inferrule*[right=\tttrule{Chk-Lam}]
  {\ctx \vdash C \defeq \Pi x{:}A'.\,B : \Srt\,\ell \\
   \ctx \vdash A \infr \Srt\,\ell' \\ A \defeq A' \\
   \ctx, x{:}A \vdash b \chk B}
  {\ctx \vdash \tttlam{x}{A} b \chk C}
\\
\inferrule*[right=\tttrule{Chk-Pair}]
  {\ctx \vdash C \defeq \Sigma x{:}A.\,B : \Srt\,\ell \\
   \ctx \vdash a \chk A \\ \ctx \vdash b \chk B[a/x]}
  {\ctx \vdash \tttpair{a}{b} \chk C}
\\
\inferrule*[right=\tttrule{Inf-Ann}]
  {\ctx \vdash A \infr \Srt\,\ell \\ \ctx \vdash a \chk A}
  {\ctx \vdash \tttann{a}{A} \infr A}
\and
\inferrule*[right=\tttrule{Chk-Sub}]
  {\ctx \vdash a \infr A \\ \ctx \vdash A \subty B}
  {\ctx \vdash a \chk B}
\end{mathpar}
\caption{Mode-relevant bidirectional rules. \tttrule{Chk-Lam} checks that the
domain of the abstraction is definitionally equal to the domain of the expected
$\Pi$-type. The remaining term formers infer their types, placing premises in
checking mode wherever the corresponding rule of Figure~\ref{fig:mltt-pi-sigma}
specifies an expected type. A premise written
$\ctx \vdash C \defeq D : \Srt\,\ell$ abbreviates the conjunction of
$C \defeq D$ (Definition~\ref{def:mltt-defeq}) and $\ctx \vdash D \infr \Srt\,\ell$;
it is not a typed equality judgment.}
\label{fig:mltt-bidir}
\end{figure}

\paragraph{No extensionality rules.} Definitional equality includes no
$\eta$-rules: $f$ and $\tttlam{x}{A} f\,x$ are not definitionally equal, nor
are $p$ and $\tttpair{\fst\,p}{\snd\,p}$; two inhabitants of a unit type are
not definitionally equal; and two functions that agree when applied to a fresh
variable are not definitionally equal. Function extensionality,
\[
\Pi f{:}(\Pi x{:}A.\,B).\,\Pi g{:}(\Pi x{:}A.\,B).\,
  (\Pi x{:}A.\,\Id\,B\,(f\,x)\,(g\,x)) \to \Id\,(\Pi x{:}A.\,B)\,f\,g,
\]
is not derivable in the core calculus, but it may be declared as an axiom in
the global environment; it satisfies the semantic obligation of
Definition~\ref{def:mltt-obligation} because data functions denote their
set-theoretic graphs in the model of Section~\ref{sec:mltt-metatheory}. Omitting
$\eta$-conversion matches the kernel's type-free conversion test
(Chapter~\ref{ch:checking}). Table~\ref{tab:mltt-comparison} compares the
$\eta$-rules of related type theories; because all of them hold in the
set-theoretic model, adding $\eta$-conversion would modify the kernel's
conversion algorithm without requiring a new model. Record types, the labeled
generalization of $\Sigma$-types treated in Chapter~\ref{ch:knowledge-graphs},
likewise have no $\eta$-rule.

\section{Universes}
\label{sec:mltt-universes}

\paragraph{Russell-style hierarchy.} A universe is a type whose elements are
themselves types, and any term inhabiting a universe is used directly as a type
without a decoding function. The universes are $\Srt\,\ell$ for level
expressions $\ell$:
\[
\Prop := \Srt\,\lzero, \qquad
\Setu := \Srt\,(\lsucc\,\lzero), \qquad
\Typeu_k := \Srt\,(\lsucc^{k+1}\,\lzero) \quad (k \ge 1).
\]
Here $\Prop$ is the universe of propositions, $\Setu$ is the universe of small
data types such as $\Nat$, $\Typeu_1$ contains $\Setu$, $\Typeu_2$ contains
$\Typeu_1$, and so on.

\begin{definition}[Level expressions and their order]\label{def:mltt-levels}
Let $u, v$ range over level parameters. Level expressions are generated by
\[
\ell ::= \lzero \mid \lsucc\,\ell \mid \lmax\,\ell\,\ell \mid \limax\,\ell\,\ell
  \mid u .
\]
A \emph{level assignment} $\rho$ maps level parameters to natural numbers. The
value $\tttlev{\ell}_\rho \in \mathbb{N}$ is defined by
$\tttlev{\lzero}_\rho = 0$, $\tttlev{\lsucc\,\ell}_\rho = \tttlev{\ell}_\rho + 1$,
$\tttlev{\lmax\,\ell_1\,\ell_2}_\rho = \max(\tttlev{\ell_1}_\rho, \tttlev{\ell_2}_\rho)$,
$\tttlev{u}_\rho = \rho(u)$, and
\[
\tttlev{\limax\,\ell_1\,\ell_2}_\rho =
\begin{cases}
0 & \text{if } \tttlev{\ell_2}_\rho = 0,\\
\max(\tttlev{\ell_1}_\rho, \tttlev{\ell_2}_\rho) & \text{otherwise.}
\end{cases}
\]
The \emph{order} on levels is $\ell \le \ell'$ iff
$\tttlev{\ell}_\rho \le \tttlev{\ell'}_\rho$ for every assignment $\rho$, and
\emph{level equivalence} is $\ell \tttsim \ell'$ iff $\ell \le \ell'$ and
$\ell' \le \ell$. The \emph{order of cumulativity} is $\ell \cumle \ell'$ iff
$\ell \le \ell'$ and $\tttlev{\ell'}_\rho = 0$ whenever
$\tttlev{\ell}_\rho = 0$, for every assignment $\rho$.
\end{definition}

The operator $\limax$, the \emph{impredicative maximum} (following Lean's
terminology), computes the universe level of a dependent function type: if
$A : \Srt\,\ell_1$ and $B : \Srt\,\ell_2$, then
$\Pi x{:}A.\,B : \Srt\,(\limax\,\ell_1\,\ell_2)$
(Rule~\tttrule{$\Pi$-F}, Figure~\ref{fig:mltt-pi-sigma}). Whenever the codomain
$B$ is a proposition, the function type remains a proposition regardless of the
universe level of $A$, making $\Prop$ impredicative; otherwise its level is the
ordinary maximum $\lmax\,\ell_1\,\ell_2$. In the presence of level parameters,
$\limax\,\ell\,u$ cannot be expressed using $\lmax$ and $\lsucc$ alone, because
its value depends on whether $u$ is instantiated to $0$; it is therefore
included as a primitive constructor of level expressions.

\begin{proposition}\label{prop:mltt-levels}
The relation $\le$ is a preorder on level expressions, compatible with $\lsucc$,
$\lmax$, and level substitution, and $\tttsim$ is a congruence. For all $\ell$
and $m$: $\limax\,\ell\,\lzero \tttsim \lzero$ and
$\limax\,\ell\,(\lsucc\,m) \tttsim \lmax\,\ell\,(\lsucc\,m)$. The relation
$\cumle$ is a preorder contained in $\le$ and compatible with level
substitution; it coincides with $\le$ on levels whose value is never $0$, and
$\lzero \cumle \ell'$ holds iff $\ell' \tttsim \lzero$. The relations $\le$,
$\tttsim$, and $\cumle$ are decidable.
\end{proposition}

\begin{proof}
Reflexivity, transitivity, and compatibility are inherited pointwise from the
standard order on $\mathbb{N}$; compatibility with level substitution holds
because $\tttlev{\ell[\bar\ell'/\bar u]}_\rho = \tttlev{\ell}_{\rho'}$ for the
assignment $\rho'(u_i) = \tttlev{\ell'_i}_\rho$. The two $\limax$ equations
follow immediately from the definition since $\tttlev{\lsucc\,m}_\rho \ge 1$.
To decide $\ell \le \ell'$, let $u_1, \ldots, u_n$ be the level parameters
occurring in $\ell$ and $\ell'$. Case-split on each $u_i$ being either $\lzero$
or $\lsucc\,u_i'$ for a fresh parameter $u_i'$ ranging over $\mathbb{N}$; the
resulting $2^n$ cases partition all assignments. After this substitution,
whether an expression evaluates to $0$ is determined syntactically ($\lzero$
evaluates to $0$, $\lsucc\,e$ never does, $\lmax\,e_1\,e_2$ evaluates to $0$
iff both arguments do, and $\limax\,e_1\,e_2$ iff $e_2$ does), so every
occurrence of $\limax$ rewrites to $\lzero$ or to $\lmax$ via the two equations
above. Each side then simplifies to a maximum of terms of the form $u + k$ and
a constant, and $\max_i(u_i + k_i, c) \le \max_j(v_j + m_j, d)$ holds under all
assignments iff $c \le \max_j(m_j, d)$ and for each $i$ there is some $j$ with
$v_j = u_i$ and $k_i \le m_j$: the first condition checks the zero assignment,
while if the second fails for some $i$, choosing a sufficiently large value for
$u_i$ violates the inequality. Each of the $2^n$ cases is therefore a finite
check. For $\cumle$, the additional zero-preservation condition is decided in
each case by the same syntactic zero test.
\end{proof}

\begin{figure}[t]
\begin{mathpar}
\inferrule*[right=\tttrule{Sort}]
  {\ctx \tttctx \\ \tttFV_{\mathrm{lvl}}(\ell) \subseteq \bar u}
  {\ctx \vdash \Srt\,\ell : \Srt\,(\lsucc\,\ell)}
\and
\inferrule*[right=\tttrule{Sort-$\defeq$}]
  {\ell \tttsim \ell'}
  {\Srt\,\ell \defeq \Srt\,\ell'}
\and
\inferrule*[right=\tttrule{$\Pi$-Prop}]
  {\ctx \vdash A : \Srt\,\ell \\ \ctx, x{:}A \vdash P : \Prop}
  {\ctx \vdash \Pi x{:}A.\,P : \Prop}
\end{mathpar}
\caption{Universe rules. $\bar u$ are the level parameters in scope
(Section~\ref{sec:mltt-environment}). \tttrule{$\Pi$-Prop} is the special case
$\ell_2 = \lzero$ of \tttrule{$\Pi$-F}, displayed separately for reference.}
\label{fig:mltt-universes}
\end{figure}

\paragraph{Impredicativity.} A type is \emph{predicative} if it resides in a
universe strictly above every universe it quantifies over, and
\emph{impredicative} otherwise. Rule~\tttrule{$\Pi$-F} assigns
$\Pi x{:}A.\,B$ the universe level $\limax\,\ell_1\,\ell_2$. When $B$ is a
proposition ($\ell_2 = \lzero$), the dependent product is also a proposition
regardless of how large $A$ is (Rule~\tttrule{$\Pi$-Prop}); in particular,
$\Pi X{:}\Prop.\,X : \Prop$ quantifies over $\Prop$ itself. When $\ell_2 \ge 1$,
the product resides at $\lmax\,\ell_1\,\ell_2$ and is predicative. By contrast,
Rule~\tttrule{$\Sigma$-F} uses $\lmax$ unconditionally, so $\Sigma$-types are
predicative at every level, including $\Prop$: for example,
$\Sigma X{:}\Prop.\,X$ lives in $\Setu$. Allowing a $\Sigma$-type with both
projections to live in $\Prop$ over a domain from a higher universe would embed
that universe into $\Prop$ and reproduce Girard's paradox
(Theorem~\ref{thm:mltt-girard})~\citep{coquand1986}. The operator $\limax$
allows a single polymorphic declaration to cover both regimes when $\ell_2$ is
a level parameter: $\Pi x{:}A.\,B$ with $B : \Srt\,u$ has level
$\limax\,\ell_1\,u$, which specializes to $\lzero$ under assignments with
$\rho(u) = 0$ and to $\lmax\,\ell_1\,u$ otherwise
(Proposition~\ref{prop:mltt-levels}). Without $\limax$, a level algebra with
$\lmax$ alone would have to commit to one of the two cases at declaration time.

\paragraph{Stratification.} Rule~\tttrule{Sort} places each universe inside the
next higher universe, so no universe is a member of itself.

\begin{theorem}[Girard; Hurkens]\label{thm:mltt-girard}
Extend the theory with a universe $\Typeu_\star$ satisfying
$\Typeu_\star : \Typeu_\star$ and closed under $\Pi$. Then there is a closed
term of type $\Pi X{:}\Typeu_\star.\,X$; the extended theory is inconsistent.
\end{theorem}

This is Girard's paradox for System~U$^-$, simplified by
Hurkens~\citep{girard1972,hurkens1995}. The paradox requires two impredicative
universes with one inside the other; having a single impredicative universe
$\Prop$ at the bottom of a predicative hierarchy is consistent
(Theorem~\ref{thm:mltt-consistency-cic}).

\paragraph{Universe polymorphism.} Any declaration in the global environment may
abstract over a list of level parameters $\bar u$. A constant expression
$\tttinst{c}{\bar\ell}$ instantiates the declaration $c$ at the level
expressions $\bar\ell$: its type---and, for a transparent definition, its
body---is obtained from the declared one by the level substitution
$[\bar\ell/\bar u]$ (Section~\ref{sec:mltt-environment}). For instance, the
list type $\tttinst{\List}{u} : \Srt\,(\lsucc\,u) \to \Srt\,(\lsucc\,u)$ is a
single declaration that can be instantiated at every data universe at or above
$\Setu$. Because level parameters are bound only at top-level declarations and
never inside term formers, universe polymorphism is
prenex~\citep{harper-pollack1991}.

\paragraph{Cumulativity.} The data universes above $\Prop$ are cumulative:
every type in a data universe is also a type in every higher universe,
\[
\Setu \subty \Typeu_1 \subty \Typeu_2 \subty \cdots,
\qquad\text{and in general}\quad \Srt\,m \subty \Srt\,n \iff m \cumle n .
\]
Here $\subty$ is the subsumption judgment of Section~\ref{sec:mltt-subsumption},
and universe cumulativity is its base rule \tttrule{S-Cum}. Crucially, $\Prop$
is not cumulative into $\Setu$: by Proposition~\ref{prop:mltt-levels},
$\lzero \cumle n$ holds only when $n \tttsim \lzero$, so a proposition is not a
type in $\Setu$. This gives the universe hierarchy of
ECC~\citep{luo1990,luo1994} minus its inclusion $\Prop \preceq \Typeu_0$: an
impredicative universe of propositions sitting below a cumulative, predicative
tower of data universes. Within the data tower, any type may be used wherever a
type of a higher universe is expected without an explicit lifting operator; for
example, $\Nat : \Setu$ yields $\Nat : \Typeu_1$ directly by Rule~\tttrule{Sub}.

\paragraph{Propositions as data.} In the presence of impredicativity and proof
irrelevance, including $\Prop$ in $\Setu$ would make the property of being a
proof unstable under substitution: $\tttlam{x}{X} x$ is a data function on the
elements of a variable type $X : \Setu$, yet it would become a proof when $X$
is instantiated by a proposition. Excluding $\Prop \subty \Setu$ ensures that
whether a typed term is a proof is determined by the universe of its type and
is preserved under substitution. When a proposition must be packaged as a type
in $\Setu$, one uses a one-field inductive wrapper, as in Lean:
\[
\mathsf{PLift}\,(P : \Prop) : \Setu, \qquad
\mathsf{up} : P \to \mathsf{PLift}\,P,
\]
whose recursor defines the projection $\mathsf{down} : \mathsf{PLift}\,P \to P$.
A wrapped term $\mathsf{up}\,p$ is a data value carrying the proof $p$ as a
subterm; two such wrapped proofs are propositionally equal by proof irrelevance
while remaining distinct terms. Similarly, a proposition paired with a proof of
it is an element of $\Sigma P{:}\Prop.\,P$, which lives in $\Setu$ because
$\Prop : \Setu$.

\paragraph{Comparison with non-cumulative universes.} The Lean~4 kernel uses
the same level algebra without cumulativity~\citep{lean4,type_checking_in_lean4}.
In Lean, every well-typed term has a type unique up to definitional equality,
the kernel has conversion but no subtyping relation, and elaboration solves
level \emph{equations} rather than level inequalities. Moving a type from a
lower universe to a higher one requires explicit lifting through the one-field
inductive wrappers $\mathsf{ULift}$ (for data types) and $\mathsf{PLift}$ (for
propositions), while universe polymorphism handles definitions that operate
uniformly across levels. Consequently, the translation from Eigenius into Lean
(Volume~II) replaces each use of universe cumulativity by an explicit
$\mathsf{ULift}$, whereas $\mathsf{PLift}$ already has a direct counterpart in
both systems. Among other proof assistants, Coq is cumulative and extends
cumulativity to inductive types covariant in their universe
parameters~\citep{timany-sozeau2018}, while Agda is non-cumulative by default
and provides cumulativity as an option~\citep{agda}.

\section{Subsumption}
\label{sec:mltt-subsumption}

\paragraph{Uses of subsumption.} Both the knowledge-graph theory of
Chapter~\ref{ch:knowledge-graphs} and the type-logical grammar of Volume~III
rely on a directional subsumption judgment. In the knowledge-graph theory,
classes are types and each resource is a single record term: an instance of a
class $C$ must also be an instance of every superclass of $C$ without altering
the term, since inserting a coercion function would create a distinct term for
the same resource. Declared class inclusion therefore enters as a base case of
a subsumption judgment whose coercions are the identity. In the grammar, a
predicate typed at a superclass applies to arguments typed at its subclasses,
and a function category is subsumed by function categories with wider result
types; Rule~\tttrule{S-$\Pi$} below captures this codomain covariance, while
$\eta$-expansion (Proposition~\ref{prop:mltt-spi-eta}) adapts a function to a
narrower argument domain. Once subsumption is present in the kernel theory,
universe cumulativity (\tttrule{S-Cum}) is simply another base case of the same
judgment. Because a term can have multiple types related by subsumption, types
in this chapter are not unique up to definitional equality; translating
derivations into Lean's non-cumulative kernel therefore requires the coherence
property discussed after Theorem~\ref{thm:mltt-declarative-consistency}.

This design adapts the treatment of natural-language semantics in Modern Type
Theories~\citep{luo2012cnt,luo2012coercive,chatzikyriakidis-luo-2020}, where
common nouns denote types, subtyping relates hyponyms to hypernyms, and
predicates typed at a supertype apply to subtype arguments. Here knowledge-graph
classes play the role of common nouns and declared class inclusions provide the
subtyping relation, restricted at the kernel level to the fragment in which
every coercion is the identity.

\paragraph{The judgment.} Subsumption $\ctx \vdash A \subty B$ is a directional
judgment on types, kept separate from definitional equality. It enters typing
solely through Rule~\tttrule{Sub}: from $\ctx \vdash a : A$ and
$\ctx \vdash A \subty B$ one derives $\ctx \vdash a : B$ with the term $a$
unchanged. Subtyping in the kernel is thus \emph{subsumptive}, and every
coercion it induces on terms is the identity.

\begin{definition}[Subsumption]\label{def:mltt-subsumption}
$\ctx \vdash A \subty B$ is the least relation closed under the rules of
Figure~\ref{fig:mltt-subsumption}. It relates well-formed types: the typing
premises of the rules ensure that $\ctx \vdash A : \Srt\,\ell$ and
$\ctx \vdash B : \Srt\,\ell'$ hold for some levels whenever
$\ctx \vdash A \subty B$ is derivable, and the intermediate term $B$ in
\tttrule{S-Trans} is likewise required to be a type.
\end{definition}

\begin{figure}[t]
\begin{mathpar}
\inferrule*[right=\tttrule{S-Conv}]
  {A \defeq B \\ \ctx \vdash A : \Srt\,\ell \\ \ctx \vdash B : \Srt\,\ell'}
  {\ctx \vdash A \subty B}
\and
\inferrule*[right=\tttrule{S-Cum}]
    {\ctx \tttctx \\ m \cumle n \\ \tttFV_{\mathrm{lvl}}(m, n) \subseteq \bar u}
  {\ctx \vdash \Srt\,m \subty \Srt\,n}
\\
\inferrule*[right=\tttrule{S-Trans}]
  {\ctx \vdash A \subty B \\ \ctx \vdash B : \Srt\,\ell \\
   \ctx \vdash B \subty C}
  {\ctx \vdash A \subty C}
\\
\inferrule*[right=\tttrule{S-$\Pi$}]
  {A \defeq A' \\ \ctx, x{:}A' \vdash B \subty B' \\\\
   \ctx \vdash \Pi x{:}A.\,B : \Srt\,\ell \\
   \ctx \vdash \Pi x{:}A'.\,B' : \Srt\,\ell'}
  {\ctx \vdash \Pi x{:}A.\,B \subty \Pi x{:}A'.\,B'}
\end{mathpar}
\caption{The subsumption judgment. \tttrule{S-$\Pi$} requires definitionally
equal domains and is covariant in the codomain. The typing premises ensure that
both sides of every subsumption, as well as the intermediate type in
\tttrule{S-Trans}, are well-formed types.}
\label{fig:mltt-subsumption}
\end{figure}

\paragraph{The structural $\Pi$ rule.} Rule~\tttrule{S-$\Pi$} states that a
function returning values in $B$ also returns values in any supertype $B'$,
provided the two function types have definitionally equal domains. The codomain
premise $\ctx, x{:}A' \vdash B \subty B'$ is checked in the context extended by
$x{:}A'$, where $B$ remains well-formed via Rule~\tttrule{Sub} on $x$ with
\tttrule{S-Conv}. Universe cumulativity therefore propagates through the
codomains of function types; for example, $\Setu \subty \Typeu_1$ yields
\[
\Nat \to \Setu \;\subty\; \Nat \to \Typeu_1
\qquad\text{and}\qquad
\Pi X{:}\Setu.\,X \to \Setu \;\subty\; \Pi X{:}\Setu.\,X \to \Typeu_1 .
\]
Because the domains are compared by definitional equality rather than
contravariant subsumption, as in ECC and Coq~\citep{luo1994},
$\Pi X{:}\Typeu_1.\,X \to X$ is not a subtype of $\Pi X{:}\Setu.\,X \to X$.
Requiring convertible domains has two key consequences: in the set-theoretic
model of Section~\ref{sec:mltt-metatheory}, subsumption between $\Pi$-types is
a plain set inclusion (so a function denotes the exact same graph at both
types), and in the syntax-directed system $\mathcal{K}_1$, transitivity of
subsumption holds directly by structural induction
(Lemma~\ref{lem:mltt-k1-trans}). When a function typed on a larger domain $A$
is needed on a smaller domain $A' \subty A$, elaboration either inserts an
explicit $\eta$-expansion (Proposition~\ref{prop:mltt-spi-eta}) or instantiates
a universe-polymorphic definition at the smaller level.

\begin{proposition}[Contravariant subsumption by $\eta$-expansion]
\label{prop:mltt-spi-eta}
Let $\ctx \vdash f : \Pi x{:}A.\,B$, $\ctx \vdash A' \subty A$, and
$\ctx, x{:}A' \vdash B \subty B'$, with $\ctx, x{:}A' \vdash B' : \Srt\,\ell$.
Then $\ctx \vdash \tttlam{x}{A'} f\,x : \Pi x{:}A'.\,B'$. The original term $f$
and its $\eta$-expansion $\tttlam{x}{A'} f\,x$ are not definitionally equal,
since the calculus has no $\eta$-rule; in the model of
Section~\ref{sec:mltt-metatheory}, the $\eta$-expansion denotes the restriction
of $f$ to the smaller domain $A'$.
\end{proposition}

\begin{proof}
In the extended context $\ctx, x{:}A'$, Rule~\tttrule{Var} gives $x : A'$, and
Rule~\tttrule{Sub} with $A' \subty A$ (weakened by
Lemma~\ref{lem:mltt-substitution}) gives $x : A$. Rule~\tttrule{$\Pi$-E} then
gives $f\,x : B$, Rule~\tttrule{Sub} with $B \subty B'$ gives $f\,x : B'$, and
Rule~\tttrule{$\Pi$-I} yields $\tttlam{x}{A'} f\,x : \Pi x{:}A'.\,B'$. The
semantic statement follows immediately from the interpretation clause for
$\lambda$-abstractions in Section~\ref{sec:mltt-metatheory}.
\end{proof}

\paragraph{Where subsumption applies.} In the bidirectional system, subsumption
is invoked exclusively by Rule~\tttrule{Chk-Sub} when a term with an inferred
type $A$ is checked against an expected type $B$. Definitional equality is
unaffected: \tttrule{S-Conv} embeds $\defeq$ into $\subty$, but no rule embeds
$\subty$ back into $\defeq$. Thus two $\Pi$-types that differ only in their
codomain universes, such as $\Nat \to \Setu$ and $\Nat \to \Typeu_1$, are
related one-way by $\subty$ without being definitionally equal. Keeping
subsumption out of definitional equality preserves the symmetry and type-freedom
of $\defeq$, which the conversion algorithm of Chapter~\ref{ch:checking} relies
on.

In particular, because definitional equality is an untyped relation on terms
(Definition~\ref{def:mltt-defeq}), an equation cannot hold at a supertype while
failing at a subtype. By contrast, in a calculus combining $\eta$-conversion
with domain-contravariant function subtyping, two functions
$f, g : \Typeu_1 \to \Nat$ satisfying $f\,X \defeq g\,X$ for all $X : \Setu$
would be $\eta$-equal at the supertype $\Setu \to \Nat$ without being equal at
$\Typeu_1 \to \Nat$, forcing conversion to depend on the type at which terms
are compared. Omitting $\eta$-conversion keeps definitional equality
type-independent.

\paragraph{Relation to ECC.} Luo's cumulativity relation $\preceq$ in
ECC~\citep{luo1994} includes conversion, the universe chain
$\Prop \preceq \Typeu_0 \preceq \Typeu_1 \preceq \cdots$, the codomain-covariant
$\Pi$-rule with convertible domains, and a $\Sigma$-rule covariant in both
components. Rule~\tttrule{S-$\Pi$} is identical to ECC's $\Pi$-rule, whereas
our system omits the inclusion $\Prop \preceq \Typeu_0$ and includes no
structural rule for $\Sigma$.\footnote{ECC's covariant $\Sigma$-rule---inferring
$\Sigma x{:}A.\,B \subty \Sigma x{:}A'.\,B'$ from $A \subty A'$ and
$x{:}A \vdash B \subty B'$---is omitted because later chapters do not require it.
In the set-theoretic model of Section~\ref{sec:mltt-metatheory} it would also
hold as a plain set inclusion, since pairs denote untagged Kuratowski pairs, so
adding it would require no change to the model.}

\paragraph{Coercive subtyping.} Luo's later framework of \emph{coercive
subtyping}~\citep{luo1999} witnesses $A \le_\kappa B$ by an explicit function
$\kappa : A \to B$ inserted during elaboration, so that a term $a : A$ used in
a context expecting $B$ elaborates to $\kappa\,a$. In that setting, coercions
may compute, change data representations, or project components of a dot-type
to model copredication in natural-language
semantics~\citep{chatzikyriakidis-luo-2020}. The subsumption judgment of this
chapter corresponds to the kernel fragment in which every coercion is the
identity. Coercive subtyping with non-trivial coercions belongs to elaboration:
because the inserted coercions are ordinary function terms of the calculus,
supporting them in the elaborator (Volume~III) leaves the kernel theory of this
chapter unchanged. Chapter~\ref{ch:knowledge-graphs} extends the kernel's
identity-coercion subsumption with width and depth rules for records, declared
class inclusions, and rules for joins and singletons.

\begin{remark}[Tarski-style universes]\label{rem:mltt-tarski}
In a Tarski-style presentation, each universe $U_n$ is a type of codes equipped
with a decoding family $T_n : U_n \to \Srt$, and cumulativity is mediated by an
explicit lifting operator $\uparrow : U_n \to U_{n+1}$ satisfying
$T_{n+1}(\uparrow a) = T_n(a)$; a Russell-style cumulative hierarchy elides
both the decoding families and the lifts~\citep{luo2012universes}. Because the
institutional models constructed in Chapter~\ref{ch:institutions} are syntactic
closed theories, this volume uses the Russell-style presentation throughout.
\end{remark}

\section{Identity types}
\label{sec:mltt-identity}

\paragraph{Rules.} Propositional equality is Martin-L\"of's intensional
identity type with the based eliminator $\Jelim$, placed in $\Prop$
(Figure~\ref{fig:mltt-identity}). Because the motive $C$ of $\Jelim$ may target
an arbitrary universe $\Srt\,v$, $\Jelim$ provides large elimination from
$\Prop$.

\begin{figure}[t]
\begin{mathpar}
\inferrule*[right=\tttrule{Id-F}]
  {\ctx \vdash A : \Srt\,\ell \\ \ctx \vdash a : A \\ \ctx \vdash b : A}
  {\ctx \vdash \Id\,A\,a\,b : \Prop}
\and
\inferrule*[right=\tttrule{Id-I}]
  {\ctx \vdash a : A}
  {\ctx \vdash \refl\,a : \Id\,A\,a\,a}
\\
\inferrule*[right=\tttrule{Id-E}]
  {\ctx \vdash C : \Pi y{:}A.\,\Id\,A\,a\,y \to \Srt\,v \\
   \ctx \vdash d : C\,a\,(\refl\,a) \\
   \ctx \vdash b : A \\ \ctx \vdash p : \Id\,A\,a\,b}
  {\ctx \vdash \Jelim\,a\,C\,d\,b\,p : C\,b\,p}
\\
\inferrule*[right=\tttrule{Id-$\iota$}]
  { }
  {\Jelim\,a\,C\,d\,b\,(\refl\,a') \tttstep d}
\end{mathpar}
\caption{The identity type. The base point $a$ is an explicit argument of
$\Jelim$, as in the recursor for equality in Lean's kernel theory, so a type
checker reads $a$ directly from the term rather than extracting it from the
type of $C$. Rule~\tttrule{Id-$\iota$} matches the base point $a$, the target
endpoint $b$, and the argument $a'$ of $\refl$ as independent pattern variables
and is therefore left-linear; when $\Jelim\,a\,C\,d\,b\,(\refl\,a')$ is well
typed, Rule~\tttrule{Id-I} ensures $a \defeq a' \defeq b$, so the reduct $d$
has the type of the redex up to $\defeq$.}
\label{fig:mltt-identity}
\end{figure}

\paragraph{Uniqueness of identity proofs.} Because $\Id\,A\,a\,b$ lives in
$\Prop$, the proof-irrelevance axiom $\irrel$ (Section~\ref{sec:mltt-prop})
proves any two identity proofs propositionally equal.

\begin{proposition}[UIP]\label{prop:mltt-uip}
Let the global environment declare $\irrel$ (Section~\ref{sec:mltt-prop}). For
all $\ctx \vdash p : \Id\,A\,a\,b$ and $\ctx \vdash q : \Id\,A\,a\,b$,
\[
\ctx \vdash \irrel\,(\Id\,A\,a\,b)\,p\,q : \Id\,(\Id\,A\,a\,b)\,p\,q .
\]
Consequently, Streicher's axiom~K \citep[Section~1.2]{streicher1993},
\[
\Pi C{:}\Id\,A\,a\,a \to \Srt\,v.\;C\,(\refl\,a) \to \Pi p{:}\Id\,A\,a\,a.\;C\,p,
\]
is inhabited by the term
\[
\begin{gathered}
\tttlam{C}{\Id\,A\,a\,a \to \Srt\,v}\tttlam{d}{C\,(\refl\,a)}\tttlam{p}{\Id\,A\,a\,a} \\
\Jelim\,(\refl\,a)\,M\,d\,p\,(\irrel\,(\Id\,A\,a\,a)\,(\refl\,a)\,p), \\
\text{where}\quad
M := \tttlam{y}{\Id\,A\,a\,a}\tttlam{\_}{\Id\,(\Id\,A\,a\,a)\,(\refl\,a)\,y}\,C\,y .
\end{gathered}
\]
\end{proposition}

\begin{proof}
Since $\Id\,A\,a\,b : \Prop$ by \tttrule{Id-F}, Rule~\tttrule{PI}
(Figure~\ref{fig:mltt-prop}) types $\irrel\,(\Id\,A\,a\,b)\,p\,q$ at
$\Id\,(\Id\,A\,a\,b)\,p\,q$. For axiom~K, in the context extended by $C, d, p$,
the motive $M$ has type
$\Pi y{:}\Id\,A\,a\,a.\,\Id\,(\Id\,A\,a\,a)\,(\refl\,a)\,y \to \Srt\,v$, the
base case $d$ has type $C\,(\refl\,a)$, and
$\irrel\,(\Id\,A\,a\,a)\,(\refl\,a)\,p$ has type
$\Id\,(\Id\,A\,a\,a)\,(\refl\,a)\,p$ by \tttrule{PI}; applying \tttrule{Id-E}
at base point $\refl\,a$ yields the conclusion $C\,p$.
\end{proof}

Notice that this inhabitant of axiom~K does not compute on $\refl\,a$: when
applied to $C$, $d$, and $\refl\,a$, it $\beta$-reduces to
\[
\Jelim\,(\refl\,a)\,M\,d\,(\refl\,a)\,
  \bigl(\irrel\,(\Id\,A\,a\,a)\,(\refl\,a)\,(\refl\,a)\bigr),
\]
whose major premise is an application of the axiom $\irrel$ rather than
$\refl$, so neither \tttrule{Id-$\iota$} nor any other reduction rule fires.

In pure Martin-L\"of type theory with $\Jelim$ alone, uniqueness of identity
proofs is independent: the groupoid model validates $\Jelim$ while refuting
UIP~\citep[Section~5.1]{hofmann-streicher1998}, and homotopy type theory treats
identity types as proof-relevant path spaces in which univalence contradicts
UIP~\citep{hottbook2013}. Placing $\Id$ in $\Prop$ with the axiom $\irrel$ makes
UIP hold propositionally, whereas in Lean~4, where proof irrelevance is built
into conversion, UIP holds definitionally~\citep{lean4}.

\paragraph{Primitive versus inductive presentation.} Although $\Id$ is included
as a primitive type former in Figure~\ref{fig:mltt-identity}, defining equality
as an inductive family in $\Prop$ yields the exact same typing and computation
rules.

\begin{proposition}\label{prop:mltt-id-inductive}
Let $\mathsf{Eq}$ be the inductive family (Definition~\ref{def:mltt-inductive})
with parameters $(A : \Srt\,\ell)(a : A)$, a single index of type $A$, result
sort $\Prop$, and a single constructor $\mathsf{rfl} : \mathsf{Eq}\,A\,a\,a$
taking no arguments. Then $\mathsf{Eq}$ is a subsingleton
(Definition~\ref{def:mltt-subsingleton}), its derived recursor
(Definition~\ref{def:mltt-recursor}) has the type of $\Jelim$ in
\tttrule{Id-E}, and its $\iota$-rule coincides with \tttrule{Id-$\iota$}.
Accordingly, the translation $\Id \mapsto \mathsf{Eq}$,
$\refl \mapsto \mathsf{rfl}$, $\Jelim \mapsto \rec_{\mathsf{Eq}}$ preserves and
reflects all judgments.
\end{proposition}

\begin{proof}
Because $\mathsf{Eq}$ has a single constructor with no arguments, the
subsingleton condition holds vacuously, permitting large elimination into any
universe $\Srt\,v$. The motive abstracts over the index telescope $(y : A)$ and
the major premise, giving $C : \Pi y{:}A.\,\mathsf{Eq}\,A\,a\,y \to \Srt\,v$.
The single minor premise takes no constructor arguments and no induction
hypotheses, so its type is $C\,a\,\mathsf{rfl}$ (since $a$ is the result index
of $\mathsf{rfl}$), and the recursor's conclusion is
$\Pi y{:}A.\,\Pi p{:}\mathsf{Eq}\,A\,a\,y.\,C\,y\,p$. This matches the premises
and conclusion of \tttrule{Id-E} with arguments $a, C, d, b, p$ in the same
order (the type parameter $A$ being determined by the domain of $C$). Its
$\iota$-rule applies the minor premise $d$ to the empty argument list, which is
\tttrule{Id-$\iota$}. Because the translation is a rule-for-rule bijection, it
preserves and reflects derivations.
\end{proof}

Making $\Id$ primitive allows us to state its rules and use propositional
equality in Section~\ref{sec:mltt-prop} without first developing the general
inductive schema of Section~\ref{sec:mltt-inductive};
Proposition~\ref{prop:mltt-id-inductive} ensures that the two presentations are
interchangeable.

\section{Inductive families}
\label{sec:mltt-inductive}

Inductive families are the sole mechanism for introducing data types in the
calculus. Each family is introduced by a declaration in the global environment,
which adds a type former, its constructors, and a derived recursor as constants.
The declaration schema follows Dybjer~\citep{dybjer1994}, with the
parameter/index distinction and elimination restrictions of
CIC~\citep{coquand-paulin1990}. Because the theory includes neither primitive
labeled sums nor general fixpoint recursion, every function over an inductive
family is defined via its recursor and is guaranteed to be total.

\begin{definition}[Inductive declaration]\label{def:mltt-inductive}
An \emph{inductive declaration} has the form
\[
\mathsf{inductive}\ \tttinst{D}{\bar u}\ (\Delta_P) : \Pi\Delta_I.\,\Srt\,\ell
\quad\text{with}\quad
c_k : \Pi\Delta_k.\, D\,\bar p\,\bar t_k \qquad (1 \le k \le n),
\]
where $\bar u$ are level parameters; $\Delta_P$ is the \emph{parameter}
telescope with variables $\bar p$; $\Delta_I$ is the \emph{index} telescope;
$\Srt\,\ell$ is the \emph{result sort}; and for each constructor $c_k$,
$\Delta_k = (y_1{:}B_1, \ldots, y_m{:}B_m)$ is its \emph{argument} telescope
and $\bar t_k$ are its result indices. Parameters are uniform across the
declaration: every constructor return type and every recursive argument applies
$D$ to the exact parameter variables $\bar p$. Indices, by contrast, may vary
between constructors and between recursive occurrences. The declaration
introduces the constants
\[
D : \Pi\Delta_P.\,\Pi\Delta_I.\,\Srt\,\ell, \qquad
c_k : \Pi\Delta_P.\,\Pi\Delta_k.\,D\,\bar p\,\bar t_k,
\]
together with the recursor $\rec_{D}$ of Definition~\ref{def:mltt-recursor},
all universe-polymorphic in $\bar u$.
\end{definition}

\begin{definition}[Strict positivity]\label{def:mltt-positivity}
An argument type $B_j$ of a constructor of $D$ is
\begin{itemize}
\item \emph{non-recursive} if $D$ does not occur in $B_j$;
\item \emph{recursive} if $B_j = \Pi\Theta_j.\,D\,\bar p\,\bar s_j$, where $D$
  occurs neither in the telescope $\Theta_j$ nor in the indices $\bar s_j$.
\end{itemize}
The family name $D$ occurs \emph{strictly positively} in the declaration if
every argument type of every constructor is either non-recursive or recursive,
and $D$ does not occur in any result index $\bar t_k$. A recursive argument
with a non-empty telescope $\Theta_j$ is called \emph{higher-order}. Throughout
this check, a type annotation $\tttann{a}{A}$ is treated as transparent,
counting as $a$: occurrences of $D$ and the recursive shape
$\Pi\Theta_j.\,D\,\bar p\,\bar s_j$ are inspected after erasing all annotations.
\end{definition}

Strict positivity permits $D$ to appear in the codomain of a $\Pi$-typed
constructor argument while rejecting three problematic shapes: negative
occurrences in the domain of a $\Pi$-type (such as $(D \to A) \to D$);
\emph{nested} occurrences inside an argument of another type former (such as
$\List\,D \to D$); and occurrences inside the index terms of a recursive
argument (such as $D\,\bar p\,(f\,(D\,\bar p\,\bar s))$). Although nested
occurrences are positive in the broader sense of CIC with nested inductive
types, excluding them here ensures that every recursive argument is identified
purely by its syntactic shape. Treating type annotations as transparent matches
the erasure operation of Section~\ref{sec:mltt-consistency}, since neither
Rule~\tttrule{$\iota$} nor the semantic interpretation inspects annotations; in
particular, annotating a strictly positive declaration
(Lemma~\ref{lem:mltt-env-annotation}) preserves strict positivity.

\begin{definition}[Admissible declaration]\label{def:mltt-admissible}
An inductive declaration is \emph{admissible} in a well-formed environment
$\tttEnv$ if:
\begin{enumerate}
\item $\tttEnv; \cdot \vdash \Pi\Delta_P.\,\Pi\Delta_I.\,\Srt\,\ell : \Srt\,\ell'$
  for some $\ell'$, with level parameters $\bar u$ in scope;
\item for each $k$, $\tttEnv'; \Delta_P \vdash \Pi\Delta_k.\,D\,\bar p\,\bar t_k :
  \Srt\,\ell_k$, where $\tttEnv'$ extends $\tttEnv$ with $D$ alone;
\item $D$ occurs strictly positively (Definition~\ref{def:mltt-positivity});
\item (universe constraint) for each argument $y_j{:}B_j$ of each constructor,
  with
  \[\Delta_P, y_1{:}B_1, \ldots, y_{j-1}{:}B_{j-1} \vdash B_j : \Srt\,\ell_j,\]
  either $\ell_j \le \ell$ or $\ell \tttsim \lzero$;
\item (non-dependence) for each recursive argument $y_i$ of a constructor and
  each subsequent argument $y_j{:}B_j$, the variable $y_i$ occurs neither in
  $B_j$ (if $B_j$ is non-recursive) nor in $\Theta_j$ (if
  $B_j = \Pi\Theta_j.\,D\,\bar p\,\bar s_j$), where annotations are again erased
  before checking variable occurrences.
\end{enumerate}
\end{definition}

The universe constraint (4) prevents a data family from packaging types or
values from a strictly higher universe. Without it, declaring
$\mathsf{inductive}\ U : \Setu$ with $\mathsf{code} : \Setu \to U$ would embed
$\Setu$ into one of its own inhabitants, reproducing Girard's paradox
(Theorem~\ref{thm:mltt-girard}) at $\Setu$. Inductive families in $\Prop$ are
exempt from this constraint: their inhabitants are proofs, and proof
irrelevance together with the elimination restriction on recursors prevents
extracting a stored type. Because uniform parameters precede the family's result
sort rather than being stored as constructor fields in a smaller sort, they are
likewise unconstrained; for example,
$\tttinst{\List}{u}\,(A : \Srt\,(\lsucc\,u)) : \Srt\,(\lsucc\,u)$ is admissible.

The non-dependence condition (5) matches the inductive schemas of
\citet{dybjer1994} and \citet[Section~5]{carneiro2019}, which write recursive
premises as non-dependent function arrows, and every inductive family used in
this book satisfies it. Note that strict positivity alone does not imply
non-dependence in the presence of type annotations, because an annotation can
mention a recursive argument inside a type whose erasure does not mention $D$
(for example,
$\fst\,\tttann{\tttpair{\Nat}{y}}{\Sigma z{:}\Setu.\,D\,\bar p\,\bar s}$ erases
to $\fst\,\tttpair{\Nat}{y}$, which mentions $y$ but not $D$). The stratified
model of Section~\ref{sec:mltt-consistency} uses non-dependence to bound the
branching factor of the least fixed point that interprets an inductive family.

\begin{figure}[t]
\begin{mathpar}
\inferrule*[right=\tttrule{Ind-F}]
  {\ctx \vdash \bar a : \Delta_P[\bar\ell/\bar u] \\
   \ctx \vdash \bar i : \Delta_I[\bar\ell/\bar u][\bar a/\bar p]}
  {\ctx \vdash \tttinst{D}{\bar\ell}\,\bar a\,\bar i :
     \Srt\,(\ell[\bar\ell/\bar u])}
\and
\inferrule*[right=\tttrule{Ind-I}]
  {\ctx \vdash \bar a : \Delta_P[\bar\ell/\bar u] \\
   \ctx \vdash \bar b : \Delta_k[\bar\ell/\bar u][\bar a/\bar p]}
  {\ctx \vdash \tttinst{c_k}{\bar\ell}\,\bar a\,\bar b :
     \tttinst{D}{\bar\ell}\,\bar a\,\bar t_k[\bar a/\bar p, \bar b/\bar y]}
\end{mathpar}
\caption{Formation and introduction for a declared family $D$. Elimination is
given by the typing of the constant $\rec_{D}$ (Definition~\ref{def:mltt-recursor})
and computation by Rule~\tttrule{$\iota$}. The notation
$\ctx \vdash \bar a : \Delta$ abbreviates typing each term in the list against
the corresponding entry of the telescope instantiated at the preceding terms.}
\label{fig:mltt-inductive}
\end{figure}

\paragraph{Recursors.} The eliminator of an inductive family is derived
uniformly from its declaration: it takes a \emph{motive} abstracting over the
index telescope and the major premise, one \emph{minor premise} for each
constructor, and one \emph{induction hypothesis} for each recursive argument.

\begin{definition}[Recursor]\label{def:mltt-recursor}
For an admissible declaration of $D$ and a fresh level parameter $v$, the
recursor $\tttinst{\rec_{D}}{v, \bar u}$ has type
\[
\begin{array}{l}
\Pi\Delta_P.\;
\Pi\big(C : \Pi\Delta_I.\,D\,\bar p\,\bar\imath \to \Srt\,v\big).\;
\Pi(m_1 : M_1) \cdots \Pi(m_n : M_n).\; \\[2pt]
\qquad \Pi\Delta_I.\;\Pi(x : D\,\bar p\,\bar\imath).\; C\,\bar\imath\,x,
\end{array}
\]
where $\bar\imath$ are the variables of $\Delta_I$ and the minor premise type
$M_k$ for constructor $c_k$ is
\[
M_k := \Pi\Delta_k.\;\Pi\big(h_j : \Pi\Theta_j.\,C\,\bar s_j\,(y_j\,\bar\theta_j)\big)_{j \in
R_k}.\;
C\,\bar t_k\,(c_k\,\bar p\,\bar y),
\]
with $R_k$ the set of positions of the recursive arguments of $c_k$,
$B_j = \Pi\Theta_j.\,D\,\bar p\,\bar s_j$ for $j \in R_k$, and $\bar\theta_j$ the
variables of $\Theta_j$. The \emph{elimination level} $v$ is an unrestricted
level parameter whenever $\lsucc\,\lzero \le \ell$ or $D$ is a subsingleton
(Definition~\ref{def:mltt-subsingleton}); otherwise $v$ is fixed to $\lzero$, so
the recursor eliminates only into $\Prop$.
\end{definition}

\paragraph{Computation.} Applying the recursor to a constructor term reduces to
the corresponding minor premise, supplying each induction hypothesis via a
recursive call to $\rec_{D}$:
\begin{mathpar}
\inferrule*[right=\tttrule{$\iota$}]
  { }
  {\rec_{D}\,\bar p\,C\,\bar m\,\bar t_k[\bar a/\bar y]\,(c_k\,\bar p\,\bar a)
   \;\tttstep\; \\
   m_k\,\bar a\,\big(\lambda\Theta_j[\bar a/\bar y].\,
     \rec_{D}\,\bar p\,C\,\bar m\,\bar s_j[\bar a/\bar y]\,(a_j\,\bar\theta_j)\big)_{j \in R_k}}
\end{mathpar}
Because each recursive call is applied to a subterm $a_j\,\bar\theta_j$ coming
from a field $a_j$ of the constructor term, $\iota$-reduction is structurally
decreasing.

Two details of how Rule~\tttrule{$\iota$} matches its left-hand side and forms
its right-hand side are essential for the metatheory:
\begin{enumerate}
\item \emph{Left-linearity.} Rule~\tttrule{$\iota$} fires on any untyped term matching
  the pattern in which $\rec_{D}$ is applied to the required number of arguments
  and the major premise is headed by $c_k$ applied to its parameters and fields.
  Although the rule is displayed above with the expected index instances
  $\bar t_k[\bar a/\bar y]$ and shared parameters $\bar p$ to show how typing
  aligns the two, the rewrite pattern itself matches the recursor's index
  arguments $\bar\imath$, the constructor's parameter arguments, and the
  constructor's level instance as unconstrained, distinct pattern variables. As
  a result, no variable appears twice on the left-hand side, making the rewrite
  rule left-linear---a property required by Klop's confluence theorem
  (Lemma~\ref{lem:mltt-confluence}). On any well-typed redex, the constructor
  application has both its declared return type
  $D\,\bar p\,\bar t_k[\bar a/\bar y]$ and the type $D\,\bar p\,\bar\imath$
  expected by the recursor, so injectivity of applied inductive families up to
  definitional equality (Lemma~\ref{lem:mltt-confluence}) guarantees that the
  matched variables are definitionally equal to the displayed instances, and both
  sides of the rule have type $C\,\bar t_k[\bar a/\bar y]\,(c_k\,\bar p\,\bar a)$.
\item \emph{Index forcing.} Whenever a constructor argument $y_j$ appears directly
  as one of the constructor's result indices at the exact type of that index---that
  is, $t_{k,i} = y_j$ and $B_j$ equals the $i$-th type of $\Delta_I$ instantiated
  at $t_{k,1}, \ldots, t_{k,i-1}$---Rule~\tttrule{$\iota$} reads that field off
  the recursor's $i$-th index argument $\imath_i$ rather than off the constructor's
  field $a_j$, replacing $a_j$ by $\imath_i$ throughout the right-hand side. On
  the canonical instance written above, $\imath_i$ is $t_{k,i}[\bar a/\bar y] = a_j$,
  so the two coincide; on an arbitrary well-typed redex, $\imath_i \defeq a_j$.
  Reading such forced fields from the recursor's index arguments ensures that the
  reduct depends only on arguments fixed by the recursor call itself. This
  distinction matters in the stratified model of
  Section~\ref{sec:mltt-consistency} for subsingletons
  (Definition~\ref{def:mltt-subsingleton}), such as $\tttAcc$, whose major premise
  is a proof (interpreted as $\emptyset$) and whose data fields are fixed by the
  indices (whereas Lean's kernel reads the fields from the constructor term).
\end{enumerate}

\begin{definition}[Subsingleton]\label{def:mltt-subsingleton}
An admissible declaration of $D$ with result sort $\ell \tttsim \lzero$ is a
\emph{subsingleton} if it has at most one constructor and every argument
$y_j{:}B_j$ of that constructor either is a proof
($\Delta_P, y_1{:}B_1, \ldots, y_{j-1}{:}B_{j-1} \vdash B_j : \Prop$) or occurs
as one of the result indices of the constructor at the type of that index
($t_{k,i} = y_j$, with $B_j$ equal to the $i$-th type of $\Delta_I$ at
$t_{k,1}, \ldots, t_{k,i-1}$, after erasing annotations). A constructor argument
that merely appears inside a complex result index without being the index itself
does not qualify.
\end{definition}

This criterion admits $\tttFalse$ (zero constructors), the unit proposition
$\top$, conjunction ($\mathsf{And}\,P\,Q$, whose sole constructor takes proofs
of $P$ and $Q$), the inductive equality family $\mathsf{Eq}$ of
Proposition~\ref{prop:mltt-id-inductive}, and the accessibility predicate
\[
\mathsf{inductive}\ \tttAcc\ (A : \Srt\,\ell)(R : A \to A \to \Prop) : A \to \Prop
\]
with sole constructor
\[
\mathsf{intro} : \Pi x{:}A.\,(\Pi y{:}A.\,R\,y\,x \to \tttAcc\,A\,R\,y) \to
  \tttAcc\,A\,R\,x.
\]
In $\tttAcc$, the element $x : A$ must be an index rather than a uniform
parameter because the recursive occurrence $\tttAcc\,A\,R\,y$ varies it.
Although $x : A$ is a data argument in $\Srt\,\ell$, it equals the sole result
index at its exact type $A$ and is therefore admitted by the second clause of
Definition~\ref{def:mltt-subsingleton}. Requiring the field's type $B_j$ to
match the index's declared type keeps index-forcing $\iota$-reduction well-typed
in the presence of universe cumulativity: if one declared
$D : \Typeu_1 \to \Prop$ with $\mathsf{mk} : \Pi X{:}\Setu.\,D\,X$, an index
argument checked only against $\Typeu_1$ could not be substituted for $X : \Setu$,
so such a family is restricted to eliminating into $\Prop$. Whenever a data
argument is a result index, its value is uniquely fixed by the type of the
constructor term; any two constructor terms of the same type therefore agree on
all data arguments, and proof irrelevance (\tttrule{PI}) equates their proof
arguments. Conversely, the subsingleton test rejects disjunction $\tttOr$ (two
constructors) and existential quantification $\tttExists$ (whose witness
$x : A$ does not appear as a result index). Those families eliminate only into
$\Prop$: one cannot project a data witness $x : A$ out of a proof of
$\exists x{:}A.\,P$.

\begin{proposition}[Strict positivity is necessary]\label{prop:mltt-negative}
Extend the theory with a family $\mathsf{Bad} : \Setu$ having a single
constructor $\mathsf{mk} : (\mathsf{Bad} \to \tttFalse) \to \mathsf{Bad}$ and
the recursor of Definition~\ref{def:mltt-recursor}, treating the argument as
non-recursive. Then $\tttFalse$ has a closed inhabitant.
\end{proposition}

\begin{proof}
Define $g := \rec_{\mathsf{Bad}}\,(\tttlam{x}{\mathsf{Bad}}\tttFalse)\,
(\tttlam{h}{\mathsf{Bad} \to \tttFalse} h\,(\mathsf{mk}\,h))
: \mathsf{Bad} \to \tttFalse$. The minor premise is well typed: for
$h : \mathsf{Bad} \to \tttFalse$, we have $\mathsf{mk}\,h : \mathsf{Bad}$ and
hence $h\,(\mathsf{mk}\,h) : \tttFalse$. The application
$g\,(\mathsf{mk}\,g) : \tttFalse$ is then a closed inhabitant of $\tttFalse$;
moreover, its $\iota$-reduct is $g\,(\mathsf{mk}\,g)$ itself, so the term has no
normal form.
\end{proof}

While the occurrence of $\mathsf{Bad}$ in Proposition~\ref{prop:mltt-negative}
is negative, non-strict positivity is also inconsistent in the presence of an
impredicative universe: Coquand and Paulin-Mohring show that a family $D$ with a
constructor of type $((D \to \Prop) \to \Prop) \to D$---where $D$ occurs
positively (to the left of two arrows) but not strictly positively---likewise
yields a contradiction~\citep{coquand-paulin1990}.

\paragraph{Pattern matching.} Non-dependent case analysis is a derived form.
Given a scrutinee $e : D\,\bar p\,\bar\imath$ and a result type $T$ in which
neither the indices $\bar\imath$ nor $e$ occur free, we define
\[
\begin{array}{l}
\tttmatch\ e : T\ \{\,c_k\,\bar y_k \Rightarrow b_k\,\}_{k} \;:=\; \\[2pt]
\qquad \rec_{D}\,\bar p\,(\lambda\Delta_I.\,\tttlam{x}{D\,\bar p\,\bar\imath} T)\,
  \big(\lambda\Delta_k.\,\lambda(h_j : \Pi\Theta_j.\,T)_{j \in R_k}.\,b_k\big)_{k}\,\bar\imath\,e ,
\end{array}
\]
where each branch $b_k : T$ is checked in the context extended by the pattern
variables $\bar y_k$, and the induction hypotheses $\bar h_k$ are ignored.
Because this encoding uses a constant motive and does not refine the index
variables $\bar\imath$ to $\bar t_k$ inside the branches, dependent case analysis
in which the result type depends on the scrutinee or its indices is written
directly with $\rec_{D}$ and an explicit motive.

\begin{example}[Natural numbers]\label{ex:mltt-nat}
$\mathsf{inductive}\ \Nat : \Setu$ with constructors $\tttzero : \Nat$ and
$\tttsucc : \Nat \to \Nat$. The sole argument of $\tttsucc$ is recursive with an
empty telescope $\Theta$. The derived recursor is
\[
\begin{array}{l}
\tttinst{\rec_{\Nat}}{v} : \Pi(C : \Nat \to \Srt\,v).\;C\,\tttzero \to \\[2pt]
\qquad \big(\Pi n{:}\Nat.\,C\,n \to C\,(\tttsucc\,n)\big) \to \Pi n{:}\Nat.\,C\,n,
\end{array}
\]
with computation rules
\[
\rec_{\Nat}\,C\,z\,s\,\tttzero \defeq z, \qquad
\rec_{\Nat}\,C\,z\,s\,(\tttsucc\,n) \defeq s\,n\,(\rec_{\Nat}\,C\,z\,s\,n).
\]
Choosing $C := \tttlam{n}{\Nat}\Prop$ defines predicates by structural recursion
on $\Nat$, while choosing a family $C : \Nat \to \Prop$ yields mathematical
induction.
\end{example}

\begin{example}[Lists]\label{ex:mltt-list}
$\mathsf{inductive}\ \tttinst{\List}{u}\ (A : \Srt\,(\lsucc\,u)) :
\Srt\,(\lsucc\,u)$ with constructors $\tttnil : \List\,A$ and
$\tttcons : A \to \List\,A \to \List\,A$. Here $A$ is a uniform parameter, and
the head argument of $\tttcons$ has type $A : \Srt\,(\lsucc\,u)$, satisfying the
universe constraint with equality.
\end{example}

\begin{example}[Vectors]\label{ex:mltt-vec}
$\mathsf{inductive}\ \Vect\ (A : \Setu) : \Nat \to \Setu$ with constructors
$\tttnil : \Vect\,A\,\tttzero$ and
$\tttcons : \Pi n{:}\Nat.\,A \to \Vect\,A\,n \to \Vect\,A\,(\tttsucc\,n)$. Here
the length index varies: $\tttnil$ and $\tttcons$ target distinct result
indices, and the recursive argument of $\tttcons$ has index $n$. The recursor's
motive has type $C : \Pi n{:}\Nat.\,\Vect\,A\,n \to \Srt\,v$, and the minor
premise for $\tttcons$ has type
$\Pi n.\,\Pi a.\,\Pi w{:}\Vect\,A\,n.\,C\,n\,w \to C\,(\tttsucc\,n)\,(\tttcons\,n\,a\,w)$.
Defining a total head function on non-empty vectors $\Vect\,A\,(\tttsucc\,n)$
requires an index-dependent motive such as
\[
C := \tttlam{m}{\Nat}\tttlam{w}{\Vect\,A\,m}
  \rec_{\Nat}\,(\tttlam{\_}{\Nat}\Setu)\,\Unit\,(\tttlam{\_}{\Nat}\tttlam{\_}{\Setu} A)\,m,
\]
which evaluates to $\Unit$ at index $\tttzero$ and to $A$ at successor indices;
the non-dependent $\tttmatch$ construct cannot express this.
\end{example}

\begin{example}[Syntactic categories]\label{ex:mltt-cat}
The categorial grammar of Volume~III types each lexical item by a syntactic
category drawn from an inductive family. A simplified version of that family is
\[
\begin{array}{l}
\mathsf{inductive}\ \tttCat : \Typeu_1 \quad\text{with}\\[2pt]
\quad \mathsf{atom} : \tttAtom \to \tttCat, \qquad
\mathsf{fwd},\ \mathsf{bwd} : \tttMode \to \tttCat \to \tttCat \to \tttCat, \\[2pt]
\quad \mathsf{all} : \tttNum \to (\Setu \to \tttCat) \to \tttCat,
\end{array}
\]
for small inductive types $\tttAtom, \tttMode, \tttNum : \Setu$. Here
$\mathsf{fwd}\,m\,A\,B$ is a functor category that consumes an argument of
category $B$ on the right to yield $A$, and $\mathsf{bwd}$ consumes its argument
on the left. The second argument of $\mathsf{all}$ is a higher-order recursive
argument with telescope $\Theta = (X : \Setu)$: it represents a category
parameterized by a small type $X : \Setu$, which is how determiners polymorphic
in the class of their head noun receive a syntactic category. Because $\tttCat$
appears only in the codomain of $\Setu \to \tttCat$, the declaration is strictly
positive. However, the argument type $\Setu \to \tttCat$ has universe level
$\limax\,2\,2 = 2$ (corresponding to $\Typeu_1$), so the universe constraint of
Definition~\ref{def:mltt-admissible} requires $\tttCat$ itself to reside in
$\Typeu_1$ or higher. The induction hypothesis generated for
$\mathsf{all}\,n\,f$ has type $\Pi X{:}\Setu.\,C\,(f\,X)$.
\end{example}

\section{Definitions and the global environment}
\label{sec:mltt-environment}

\paragraph{Declarations.} Every constant of the calculus is introduced by a
declaration in a global environment $\tttEnv$, and every typing judgment is
parameterized by $\tttEnv$.

\begin{definition}[Global environment]\label{def:mltt-environment}
A \emph{global environment} is a finite sequence $\tttEnv$ of declarations,
each binding a fresh constant name (or family of names) and a list $\bar u$ of
level parameters:
\begin{enumerate}
\item an \emph{axiom} $\mathsf{axiom}\ \tttinst{c}{\bar u} : A$, introducing an
  opaque constant with declared type $A$ and no body;
\item a \emph{definition} $\mathsf{def}\ \tttinst{c}{\bar u} : A := t$, which is
  either \emph{transparent} (so that $c$ unfolds to $t$ under $\delta$-reduction)
  or \emph{opaque} (so that $t$ is type-checked against $A$ at declaration time,
  but $c$ does not unfold);
\item an \emph{inductive declaration} (Definition~\ref{def:mltt-inductive}),
  which simultaneously binds the inductive type former, its constructors, and
  its recursor.
\end{enumerate}
\end{definition}

\begin{figure}[t]
\begin{mathpar}
\inferrule*[right=\tttrule{E-Empty}]{ }{\cdot \tttwf}
\and
\inferrule*[right=\tttrule{E-Ax}]
  {\tttEnv \tttwf \\ c \notin \tttEnv \\ \tttEnv; \cdot \vdash_{\bar u} A : \Srt\,\ell}
  {\tttEnv, \mathsf{axiom}\ \tttinst{c}{\bar u} : A \tttwf}
\and
\inferrule*[right=\tttrule{E-Def}]
  {\tttEnv \tttwf \\ c \notin \tttEnv \\ \tttEnv; \cdot \vdash_{\bar u} A : \Srt\,\ell \\
   \tttEnv; \cdot \vdash_{\bar u} t : A}
  {\tttEnv, \mathsf{def}\ \tttinst{c}{\bar u} : A := t \tttwf}
\and
\inferrule*[right=\tttrule{E-Ind}]
  {\tttEnv \tttwf \\ D, \bar c, \rec_{D} \notin \tttEnv \\
   \text{the declaration is admissible in } \tttEnv}
  {\tttEnv, \mathsf{inductive}\ \tttinst{D}{\bar u}\ldots \tttwf}
\\
\inferrule*[right=\tttrule{Const}]
  {\ctx \tttctx \\ (\tttinst{c}{\bar u} : A) \in \tttEnv \\ |\bar\ell| = |\bar u| \\
   \tttFV_{\mathrm{lvl}}(\bar\ell) \subseteq \bar v}
  {\ctx \vdash_{\bar v} \tttinst{c}{\bar\ell} : A[\bar\ell/\bar u]}
\and
\inferrule*[right=\tttrule{$\delta$}]
  {(\mathsf{def}\ \tttinst{c}{\bar u} : A := t) \in \tttEnv
   \text{ transparent}}
  {\tttinst{c}{\bar\ell} \tttstep t[\bar\ell/\bar u]}
\end{mathpar}
\caption{Well-formed environments, constants, and $\delta$-unfolding. The
subscript $\vdash_{\bar u}$ records the level parameters in scope; in
\tttrule{Const}, $\bar u$ are the level parameters of the declaration and
$\bar v$ are those in scope at the call site. Rule~\tttrule{E-Def} applies to
both transparent and opaque definitions.}
\label{fig:mltt-environment}
\end{figure}

\paragraph{Well-formedness.} Figure~\ref{fig:mltt-environment} defines the
well-formedness judgment $\tttEnv \tttwf$. Because each declaration is checked
in the prefix of declarations preceding it, every well-formed environment is
acyclic: no constant's type or body can refer to itself or to a later
declaration, and the only recursion in the calculus is the structural recursion
provided by inductive recursors. Adding an axiom extends the theory without
supplying a witness, whereas adding a definition is conservative: every
occurrence of a transparent constant can be unfolded to its body via $\delta$,
and an opaque definition differs from an axiom in that its type has been proved
inhabited in the preceding environment. Propositional axioms act as global
hypotheses available to every derivation over $\tttEnv$.

\paragraph{Constants and levels.} Rule~\tttrule{Const} instantiates a
universe-polymorphic constant $c$ at level expressions $\bar\ell$, substituting
$\bar\ell$ for the declared parameters $\bar u$ in its type $A$ (and, via
Rule~\tttrule{$\delta$}, in its body $t$). Requiring the free level parameters of
$\bar\ell$ to lie in the currently scoped parameters $\bar v$ ensures that the
instantiated type $A[\bar\ell/\bar u]$ remains well-formed over $\bar v$. Aside
from matching the parameter count ($|\bar\ell| = |\bar u|$), no further
constraint is needed on $\bar\ell$ because every level expression denotes a
valid universe level. The constants introduced by an inductive declaration
($D$, $c_k$, and $\rec_{D}$) are typed by the same Rule~\tttrule{Const} using the
declared types of Definitions~\ref{def:mltt-inductive}
and~\ref{def:mltt-recursor}.

\begin{lemma}[Environment weakening]\label{lem:mltt-env-weakening}
If $\tttEnv; \ctx \vdash \mathcal{J}$ and $\tttEnv, \tttEnv' \tttwf$, then
$\tttEnv, \tttEnv'; \ctx \vdash \mathcal{J}$.
\end{lemma}

\begin{proof}
By induction on the derivation of $\mathcal{J}$. The only rules that consult the
global environment are \tttrule{Const}, \tttrule{$\delta$}, and the rules for
inductive families, each of which depends only on declaration membership in
$\tttEnv$. Because $\tttEnv, \tttEnv' \tttwf$ requires $\tttEnv'$ to bind fresh
constant names, all prior declarations remain in force.
\end{proof}

In Chapter~\ref{ch:institutions}, a well-formed environment $\tttEnv$ serves as
a signature: its constants provide the signature's vocabulary, its propositional
axioms form the axioms of a theory over that signature, and an environment
extension $\tttEnv \subseteq \tttEnv, \tttEnv'$ defines a signature inclusion
along which Lemma~\ref{lem:mltt-env-weakening} translates derivations.

\section{\texorpdfstring{$\Prop$}{Prop} and proof irrelevance}
\label{sec:mltt-prop}

\paragraph{The axiom.} Because Eigenius retains the proof of a claim as an
explicit term that can be inspected and compared with other proofs of the same
claim (Chapter~\ref{ch:introduction}), proof irrelevance is formulated as a
provable propositional equality with an explicit witness rather than as a rule of
definitional conversion. Specifically, any two proofs of the same proposition are
equated by the global axiom
\[
\mathsf{axiom}\ \irrel : \Pi P{:}\Prop.\,\Pi p{:}P.\,\Pi q{:}P.\,\Id\,P\,p\,q ,
\]
which enters the environment via Rule~\tttrule{E-Ax} without any $\delta$- or
$\iota$-reduction rule (Figure~\ref{fig:mltt-environment}). By impredicativity of
$\Prop$, the type of $\irrel$ is itself a proposition, and applying
Rule~\tttrule{Const} followed by three uses of \tttrule{$\Pi$-E} yields the
derived rule \tttrule{PI} of Figure~\ref{fig:mltt-prop}. All results in this
chapter that invoke proof irrelevance assume an environment declaring $\irrel$.

\begin{figure}[h]
\begin{mathpar}
\inferrule*[right=\tttrule{PI}]
  {\ctx \vdash P : \Prop \\ \ctx \vdash p : P \\ \ctx \vdash q : P}
  {\ctx \vdash \irrel\,P\,p\,q : \Id\,P\,p\,q}
\end{mathpar}
\caption{Proof irrelevance, derived from the axiom $\irrel$.}
\label{fig:mltt-prop}
\end{figure}

Crucially, Rule~\tttrule{PI} introduces a term of the identity type $\Id\,P\,p\,q$
rather than a definitional equation $p \defeq q$: the proof terms $p$ and $q$
remain syntactically distinct, $p \defeq q$ does not hold in general, and no
reduction rule unfolds $\irrel\,P\,p\,q$. Instead, $\irrel\,P\,p\,q$ provides a
propositional equality proof along which $\Jelim$ can transport any predicate on
$p$ to $q$. Because \tttrule{Id-$\iota$} fires only when the equality proof is
of the form $\refl\,a$, the transport term
$\Jelim\,p\,C\,d\,q\,(\irrel\,P\,p\,q)$ is a neutral term that does not reduce.
Note also that the premise $P : \Prop$ applies to the type of the proofs $p$ and
$q$~\citep{pfenning2001irrelevance,abel2012irrelevance}: Rule~\tttrule{PI}
equates inhabitants of a single proposition $P$, not propositions themselves. Two
propositions $P, Q : \Prop$ are inhabitants of $\Prop : \Setu$, which is not a
proposition, so $\Id\,\Prop\,P\,Q$ is not inhabited by $\irrel$, and compound
propositions such as $\mathsf{And}\,P\,Q$ and $\mathsf{And}\,P'\,Q'$ are
definitionally equal only when $P \defeq P'$ and $Q \defeq Q'$.

In the bidirectional systems $\mathcal{K}_0$ and $\mathcal{K}_1$
(Definition~\ref{def:mltt-k0}), the application $\irrel\,P\,p\,q$ synthesizes
$\Id\,P\,p\,q$ via constant lookup and \tttrule{Inf-App}, while
Rule~\tttrule{Chk-Refl} checks $\refl\,a$ against an identity type only when
both endpoints are joinable with $a$. (The current kernel implementation instead
checks $\refl\,a$ against $\Id\,P\,p\,q$ for $P : \Prop$ without comparing the
endpoints $p$ and $q$, corresponding to Rule~\tttrule{PI-Refl} discussed at the
end of this section; aligning the implementation with the axiom $\irrel$ is
recorded in Volume~II.) By contrast, Lean's kernel theory builds proof
irrelevance directly into conversion ($p \defeq q$ for all $p, q : P$ with
$P : \Prop$), identifying all proofs of a proposition during type checking and
erasing which specific proof was supplied. Similarly, in the observational type
theories of \citet{pujet-tabareau2022,pujet-tabareau2023}, proofs of strict
propositions are never reduced and are convertible whenever they share a type.
In the set-theoretic model of Section~\ref{sec:mltt-metatheory}, every proof
denotes $\emptyset$, so the model validates both the axiom $\irrel$ and
definitional proof irrelevance; Section~\ref{sec:mltt-metatheory} also reviews
the normalization and transitivity failures caused by putting proof irrelevance
into conversion, which provide additional metatheoretic confirmation for keeping
it as an axiom. Outside the type theory itself, Eigenius records which concrete
proof witness establishes each proposition, and Chapter~\ref{ch:institutions}
defines the proof set of a sentence as the set of these stored witnesses.

\paragraph{Interaction with large elimination.} Combining proof irrelevance with
unrestricted large elimination from $\Prop$ is inconsistent; this is why
Definition~\ref{def:mltt-recursor} restricts large elimination from $\Prop$ to
subsingletons.

\begin{proposition}\label{prop:mltt-pi-large-elim}
Extend an environment declaring $\irrel$ with the two-constructor proposition
$\mathsf{inductive}\ \mathsf{Two} : \Prop$ having constructors
$\mathsf{l}, \mathsf{r} : \mathsf{Two}$ and an unrestricted recursor eliminating
into $\Srt\,v$ for every $v$. Then $\tttFalse$ is inhabited.
\end{proposition}

\begin{proof}
Let $\tttBool : \Setu$ be the standard Boolean type with constructors
$\mathsf{true}$ and $\mathsf{false}$, and define the large-elimination predicate
$\mathsf{U} := \rec_{\tttBool}\,(\tttlam{\_}{\tttBool}\Prop)\,\top\,\tttFalse :
\tttBool \to \Prop$, where $\top$ is the unit proposition with proof $\ast$.
Using the unrestricted recursor for $\mathsf{Two}$, define
$f := \rec_{\mathsf{Two}}\,(\tttlam{\_}{\mathsf{Two}}\tttBool)\,\mathsf{true}\,\mathsf{false}
: \mathsf{Two} \to \tttBool$. By Rule~\tttrule{PI}, we have
$\irrel\,\mathsf{Two}\,\mathsf{l}\,\mathsf{r} : \Id\,\mathsf{Two}\,\mathsf{l}\,\mathsf{r}$.
Eliminating this equality via $\Jelim$ at base point $\mathsf{l}$ with motive
\[
\tttlam{y}{\mathsf{Two}}\tttlam{\_}{\Id\,\mathsf{Two}\,\mathsf{l}\,y}\,
  \Id\,\tttBool\,(f\,\mathsf{l})\,(f\,y)
\]
transports $\refl\,(f\,\mathsf{l})$ to a proof of
$\Id\,\tttBool\,(f\,\mathsf{l})\,(f\,\mathsf{r})$; the type of this proof
$\iota$-reduces to $\Id\,\tttBool\,\mathsf{true}\,\mathsf{false}$. A second elimination via $\Jelim$
at base point $f\,\mathsf{l}$ with motive
\[
\tttlam{y}{\tttBool}\tttlam{\_}{\Id\,\tttBool\,(f\,\mathsf{l})\,y}\,\mathsf{U}\,y
\]
then transports $\ast : \mathsf{U}\,\mathsf{true}$ to an inhabitant of
$\mathsf{U}\,\mathsf{false}$, which $\iota$-reduces to $\tttFalse$.
\end{proof}

For a subsingleton family, by contrast, every constructor argument is either a
proof or fixed by the result indices; consequently, any two constructor terms at
the same parameters and indices denote the same element in the model of
Section~\ref{sec:mltt-metatheory}, and large elimination produces identical
results on them. Subsingleton elimination is therefore consistent with proof
irrelevance (Theorem~\ref{thm:mltt-consistency-lean} and
Theorem~\ref{thm:mltt-declarative-consistency}). Notice, however, that when
proof irrelevance is a rule of conversion, an open or non-canonical proof term
becomes definitionally equal to a constructor term on which Rule~\tttrule{$\iota$}
fires, triggering the normalization and transitivity failures of
Theorem~\ref{thm:mltt-failure}; when proof irrelevance is instead the axiom
$\irrel$, Rule~\tttrule{$\iota$} fires solely on syntactic constructor terms.

\paragraph{Consequences.} Because eliminators from non-subsingleton propositions
cannot target data universes and any two proofs of a proposition are provably
equal, proofs carry no computational content: no term of a data type can branch
on which proof of a proposition was supplied. Evidence can therefore be indexed
and compared by the proposition it proves, while individual proof terms remain
distinct objects stored by the system (Chapter~\ref{ch:institutions}).
Chapter~\ref{ch:justification-logic} builds on this separation to distinguish
the justification term recorded for a claim from the proposition it justifies,
and Proposition~\ref{prop:mltt-uip} is its specialization to propositional
equality.

\paragraph{Alternative: proof irrelevance via $\refl$.} Because $\irrel$ has no
computation rule, $\Jelim$ does not reduce when applied to a proof built from
$\irrel$, and the inhabitant of axiom~K in Proposition~\ref{prop:mltt-uip} does
not compute to its minor premise on $\refl\,a$. An alternative rule, currently
used by the kernel's checker, allows $\refl\,r$ itself to witness the equality
of any two proofs $p, q : P$ of a proposition:
\begin{mathpar}
\inferrule*[right=\tttrule{PI-Refl}]
  {\ctx \vdash P : \Prop \\ \ctx \vdash p : P \\ \ctx \vdash q : P \\
   \ctx \vdash r : P}
  {\ctx \vdash \refl\,r : \Id\,P\,p\,q}
\end{mathpar}
Under \tttrule{PI-Refl}, Rule~\tttrule{Id-$\iota$} fires on every reflexivity
proof of a propositional equation, and the inhabitant of axiom~K (using
$\refl\,(\refl\,a)$ as the equality witness) reduces directly to its minor
premise.

However, allowing $\refl\,r$ to inhabit $\Id\,P\,p\,q$ when $p$ and $q$ are not
joinable breaks subject reduction for untyped $\iota$-reduction. Specifically, in
the context $P{:}\Prop,\ d{:}P,\ y{:}\top$ with motive
\[
C := \tttlam{z}{\top}\tttlam{\_}{\Id\,\top\,\ast\,z}\,
  \rec_{\top}\,(\tttlam{\_}{\top}\Prop)\,P\,z ,
\]
the term $\Jelim\,\ast\,C\,d\,y\,(\refl\,\ast)$ has type $C\,y\,(\refl\,\ast)$,
which $\beta$-reduces to the neutral type
$\rec_{\top}\,(\tttlam{\_}{\top}\Prop)\,P\,y$, whereas the term itself reduces
by \tttrule{Id-$\iota$} to $d : P$, whose type $P$ is not joinable with that
neutral type.

Although the set-theoretic model of Section~\ref{sec:mltt-metatheory} validates
\tttrule{PI-Refl} (since every proof denotes $\emptyset$), the consistency proof
of Section~\ref{sec:mltt-consistency} does not extend to it for two reasons:
\begin{enumerate}
\item The stratified model proves semantic invariance under annotated reduction by
  relying on subject reduction (Lemma~\ref{lem:mltt-k0-annotated}(3)), which fails
  under \tttrule{PI-Refl} as just shown.
\item Passing to a model of a typed definitional equality (see the remark after
  Theorem~\ref{thm:mltt-k0-consistency}) would require adding typed equations
  between distinct proofs: for proofs $p, q, r : P$, a proof $h : \Id\,P\,p\,p$,
  and motive $C' := \tttlam{y}{P}\tttlam{\_}{\Id\,P\,p\,y}\,\Id\,P\,p\,y$, both
  $\Jelim\,p\,C'\,h\,q\,(\refl\,p)$ and $\Jelim\,p\,C'\,h\,q\,(\refl\,r)$ check
  against $\Id\,P\,p\,q$ and reduce to $h$, so a typed equality can equate them
  with $h$ only if it equates distinct proofs, reintroducing the non-termination
  phenomena of Theorem~\ref{thm:mltt-failure}(1).
\end{enumerate}
Whether the calculus with \tttrule{PI-Refl} and untyped conversion can be shown
consistent via a model that assigns every term and its reducts the same sort
(proof or data) without subject reduction remains an open question.

\section{Metatheory}
\label{sec:mltt-metatheory}

The first two subsections review metatheoretic results from the literature on
closely related type theories and on the failure of normalization and
transitivity under definitional proof irrelevance. The third subsection establishes
the consistency of the combined theory of this chapter
(Conjecture~\ref{conj:mltt-consistency} and
Theorem~\ref{thm:mltt-declarative-consistency}) relative to one hypothesis on
the global environment: that its declared axioms have elements in the model
(Definition~\ref{def:mltt-obligation}).

\paragraph{Design principle.} The choices that set the theory apart from
Lean's follow a single thread (Chapter~\ref{ch:introduction}): the store keeps
proofs as terms, so conversion may not identify them. Proof irrelevance is
therefore the propositional axiom $\irrel$ of Section~\ref{sec:mltt-prop}
rather than a rule of conversion. With proof irrelevance out of conversion, the
remaining reason for conversion to consult the type at which two terms
are compared is $\eta$, so the theory drops $\eta$ as well
(Section~\ref{sec:mltt-pi-sigma}); definitional equality is untyped
joinability (Definition~\ref{def:mltt-defeq}), which the kernel decides
without types. With conversion untyped, the domain information that a
type-directed conversion would supply must reside in the term itself, so
$\lambda$-abstractions carry their domains, and the model of
Section~\ref{sec:mltt-consistency} can be defined by structural recursion on
terms. The failures of Theorem~\ref{thm:mltt-failure} and the gap in the
stratified argument for Lean's theory (discussed in the remark after
Theorem~\ref{thm:mltt-declarative-consistency}) show that this design is also
better-behaved metatheoretically; they confirm the choice rather than
motivating it.

\subsection{Results from related theories}

\begin{theorem}[Consistency of CIC with impredicative $\Prop$]
\label{thm:mltt-consistency-cic}
The Calculus of Inductive Constructions with an impredicative, proof-irrelevant
universe of propositions and a predicative hierarchy of type universes has a
set-theoretic model in ZFC with countably many inaccessible cardinals, in which
$\Prop$ is interpreted as the two truth values and each type universe as a
Grothendieck universe. In particular, no closed term inhabits
$\tttFalse$~\citep{werner1997}.
\end{theorem}

\begin{theorem}[Lean's type theory]\label{thm:mltt-consistency-lean}
The type theory of the Lean kernel---impredicative $\Prop$, non-cumulative
universe polymorphism with $\limax$, inductive families with subsingleton
elimination, definitional proof irrelevance, $\eta$ for functions, and
quotient types---is consistent relative to ZFC with $n$ inaccessible
cardinals for each $n$, via a set-theoretic model~\citep[Corollary~6.8]{carneiro2019}. A
mechanization of the typing judgment and its checker
is in progress~\citep{carneiro2024}.
\end{theorem}

\begin{theorem}[ECC]\label{thm:mltt-ecc}
The Extended Calculus of Constructions---impredicative $\Prop$, a cumulative
predicative hierarchy $\Typeu_0 \preceq \Typeu_1 \preceq \cdots$ with
$\Prop \preceq \Typeu_0$, strong $\Sigma$, and the cumulativity relation
$\preceq$ with subsumption---is Church--Rosser, satisfies subject reduction,
is strongly normalizing, and is consistent; type inference is decidable, and
every typable term has a principal type with respect to
$\preceq$~\citep{luo1994}.
\end{theorem}

Theorem~\ref{thm:mltt-ecc} covers subsumption with identity coercions for
universe inclusions and codomain-covariant $\Pi$, in a theory with
$\beta$-conversion only. For the predicative fragment without $\Prop$,
normalization, subject reduction, and canonicity of the natural numbers hold for
intensional MLTT with inductive families, and conversion with $\eta$ for $\Pi$ is
decidable for a universe hierarchy over $\Pi$ and $\Nat$~\citep{abel-ohman-vezzosi2018}.

\subsection{Known limits of typed definitional equality}

The theory states proof irrelevance as an axiom because the store
keeps proofs as terms (the design principle above). Two properties that type
checkers ordinarily rely on fail for theories whose definitional equality
includes proof irrelevance as a conversion rule---among them Lean's kernel
theory---and those failures confirm the choice:
\begin{itemize}
\item \emph{Normalization} states that every well-typed term reduces to a
  normal form; a checker that decides definitional equality by comparing normal
  forms relies on normalization to terminate.
\item \emph{Transitivity} of definitional equality states that $a \defeq b$ and
  $b \defeq c$ imply $a \defeq c$; a checker that decides definitional equality
  algorithmically relies on the algorithm to compute a transitive relation,
  since otherwise the relation it decides differs from the one the typing rules
  specify.
\end{itemize}
Both failures arise from combining definitional proof irrelevance in a universe
of propositions with elimination from that universe into data; the failure of
normalization also uses impredicativity. The discussion following the theorem
explains what changes when proof irrelevance is an axiom rather than a
conversion rule.

\begin{theorem}[Failure of normalization and transitivity]
\label{thm:mltt-failure}
\leavevmode
\begin{enumerate}
\item In a type theory with an impredicative universe of definitionally
  proof-irrelevant propositions and an identity type in that universe whose
  eliminator targets every universe, there is a well-typed closed term without
  normal form~\citep{abel-coquand2020}.
\item In a type theory with a universe of definitionally proof-irrelevant
  propositions containing $\tttAcc$ with elimination into every universe and
  its computation rule as a definitional equation, conversion is undecidable,
  with or without impredicativity~\citep{gilbert2019sprop}. In Lean's type
  theory, the definitional equality decided by reduction to
  weak-head normal form together with proof irrelevance and $\eta$ is not
  transitive: the accessibility predicate $\tttAcc$, a subsingleton eliminating
  into every universe, yields terms $a \defeq b$ and $b \defeq c$ with
  $a \not\defeq c$ in the algorithmic relation. The declarative relation, in
  which transitivity is a rule, is undecidable, whereas subject reduction fails for
  the algorithmic relation~\citep[Section~3.1]{carneiro2019}. Lifting a function
  from a quotient of a proposition is a second source of non-transitivity.
\end{enumerate}
\end{theorem}

Both counterexamples rely on proof irrelevance as a conversion rule, combined
with large elimination of $\Id$ and impredicative $\Prop$ (part~1) or large
elimination of a subsingleton with a recursive argument (part~2), together with
$\iota$-reduction. The theory of this chapter retains the large eliminations:
$\Jelim$ eliminates into every universe (Figure~\ref{fig:mltt-identity}) and
$\tttAcc$ is an admissible subsingleton. However, it omits the conversion rule:
proof irrelevance is the axiom $\irrel$ of Section~\ref{sec:mltt-prop}, on
which no reduction rule fires, and $\Jelim$ and the recursors reduce only on
constructor terms (\tttrule{Id-$\iota$} and \tttrule{$\iota$}). Consequently,
neither counterexample transfers: the term of part~1 reduces only because
$\Jelim$ is applied to a proof that conversion identifies with $\refl$, and the
unfolding of part~2 proceeds only because the recursor is applied to a proof
that conversion identifies with a constructor term. Definitional equality in
this chapter is joinability under a confluent reduction
(Lemma~\ref{lem:mltt-confluence}), and is therefore transitive; its
decidability on typable terms reduces to normalization of $\tttstep$, which is
expected though not proved here (Section~\ref{sec:mltt-consistency}). By
contrast, normalization of the theory of
\citet[Theorem~4.1]{pujet-tabareau2023} is provable in Martin-L\"of type
theory with finitely many universes because its strict universe adds no
computation; our theory includes large elimination of $\tttAcc$, through which
impredicativity adds computation as in CIC, closing off that proof route. An
alternative way to avoid both failures---which does admit a proof of
decidability of conversion~\citep{gilbert2019sprop}---is to restrict
definitional irrelevance to a separate universe of strict propositions whose
eliminators target only strict propositions (except for the empty type).

To see the mechanisms in more detail, consider the two parts in turn. Part~1
concerns the identity type. When $\Id$ lies in a definitionally
proof-irrelevant $\Prop$, all proofs of an equation are definitionally equal,
including open proofs not of the form $\refl\,a$; because $\Jelim$ eliminates
into every universe, such proofs act on data. The counterexample of Abel and
Coquand combines this with impredicativity to build a closed term whose
reduction does not terminate. \citet[Section~6]{pujet-tabareau2023} isolate the
underlying mechanism: a transport that reduces without inspecting its proof,
together with a proof-relevant impredicative universe, yields an approximate
fixed-point combinator in an inconsistent context, making conversion of open
terms undecidable both in their theory extended by such a universe and in the
theory of Abel and Coquand. In the theory of this chapter, $\Jelim$ applied to
a variable is simply a neutral term.

Part~2 concerns the accessibility predicate $\tttAcc$. Although $\tttAcc$ has
at most one proof, it eliminates into data, and its recursor unfolds only on a
proof in constructor form. When proof irrelevance is part of conversion, every
proof $p$ of $\tttAcc\,A\,R\,a$ is definitionally equal to a constructor term
$\mathsf{intro}\,a\,h$ with $h$ obtained from $p$ by inversion through the
recursor. The declarative relation therefore unfolds the recursor on every
proof, including one that depends on an assumption under which $R$ is not well
founded, where the unfolding fails to
terminate~\citep{gilbert2019sprop,felicissimo2026acc}. The algorithmic checker,
which reduces to weak-head normal form before applying $\eta$ and proof
irrelevance, can thus identify $a$ with $b$ and $b$ with $c$ by different
routes without identifying $a$ with $c$. With transitivity included as a
declarative rule, conversion is transitive but undecidable; with the
algorithmic relation alone, conversion is decidable but subject reduction
fails, so reducing a well-typed term can produce a term that no longer
type-checks.

Theories with strict propositions recover the two large eliminations in other
forms. Observational equality eliminates an equation into data via a cast
operator that computes by cases on the two types rather than on the proof, keeping
conversion decidable in the presence of impredicativity and inductive
families~\citep{pujet-tabareau2022,pujet-tabareau2023,pujet-leray-tabareau2025}.
For $\tttAcc$ in the strict universe, stating the computation rule of the
eliminator as a propositional equation keeps conversion decidable; stating it
as a definitional equation makes conversion undecidable while preserving
canonicity, and every strict proposition provable in the definitional variant
remains provable in the decidable theory~\citep{felicissimo2026acc}.
\citet[Section~7]{pujet-tabareau2023} describe definitional irrelevance and
observational equality as a trade-off: the former provides a $\Jelim$ that
reduces on every proof of a reflexive equation (open proofs included), whereas
the latter provides extensionality principles instead. The theory of this
chapter takes neither side: it retains the eliminator $\Jelim$ of $\Id$ and the
computation rule of $\tttAcc$ as reductions on constructor terms, adopts no
extensionality principle as a conversion rule (Section~\ref{sec:mltt-pi-sigma}),
and moves proof irrelevance from conversion to an axiom. The decision procedure
of Chapter~\ref{ch:checking} then decides definitional equality directly on
every term whose reduction terminates.

\subsection{Consistency of the combined theory}
\label{sec:mltt-consistency}

The theory of this chapter combines ingredients that the existing literature
treats separately: Theorem~\ref{thm:mltt-consistency-lean} covers inductive
families, subsingleton elimination, proof irrelevance, and $\eta$ in a
non-cumulative setting, while Theorem~\ref{thm:mltt-ecc} covers cumulativity
with $\Prop \preceq \Typeu_0$ and codomain-covariant $\Pi$ subsumption. No
single cited result covers the subsumption judgment together with inductive
families and large elimination from $\Prop$:

\begin{conjecture}[Consistency]\label{conj:mltt-consistency}
No closed term inhabits $\tttFalse$ in the theory of this chapter.
\end{conjecture}

Specifically, ECC has neither inductive families nor proof irrelevance, and UTT
has inductive types in the predicative universes but no inductive propositions
with elimination into data. Lean's theory
(Theorem~\ref{thm:mltt-consistency-lean}) includes the large eliminations of
$\Id$ and $\tttAcc$ together with proof irrelevance, and its definitional
equality contains the one of this chapter, so its model validates this
chapter's rules without subsumption. Subsumption is what the theorems below add,
proving Conjecture~\ref{conj:mltt-consistency} for every environment that meets
its axiom obligation (Definition~\ref{def:mltt-obligation}).

We establish consistency first for the syntax-directed bidirectional systems
$\mathcal{K}_0$ and $\mathcal{K}_1$ (Figure~\ref{fig:mltt-bidir}), which match
the structure of a type checker:
\begin{enumerate}
\item A stratified set-theoretic model interprets $\mathcal{K}_0$, the fragment
  without Rule~\tttrule{S-$\Pi$}, using untyped joinability as conversion, and
  Theorem~\ref{thm:mltt-k0-consistency} proves $\mathcal{K}_0$ consistent.
\item Theorem~\ref{thm:mltt-k1-consistency} extends the model to
  $\mathcal{K}_1$ by adding \tttrule{S-$\Pi$}, which is interpreted as set
  inclusion.
\item Proposition~\ref{prop:mltt-k1-complete} and
  Lemma~\ref{lem:mltt-env-annotation} translate declarative derivations and
  environments into $\mathcal{K}_1$, and
  Theorem~\ref{thm:mltt-declarative-consistency} concludes for the declarative
  system.
\end{enumerate}
Every theorem assumes that each axiom of the global environment has an element
in the model (Definition~\ref{def:mltt-obligation}). Because abstractions carry
their domains, the interpretation of a term can be defined by recursion on the
term itself; the remark after Theorem~\ref{thm:mltt-k0-consistency} contrasts
this with models of typed definitional equality. The model construction does
not depend on normalization, since a model validates a typing judgment whether
or not that judgment is decidable.

\paragraph{The algorithmic fragment.} One-step reduction $\tttstep$ on terms is
the compatible closure of the following rules: $\beta$-reduction
$(\tttlam{x}{A} b)\,a \tttstep b[a/x]$, which does not inspect the domain
$A$; erasure of annotations
$\tttann{a}{A} \tttstep a$; $\delta$-unfolding of a non-opaque definition in
the global environment; $\iota$-reduction for recursors applied to constructor
terms; $\Jelim\,a\,C\,d\,b\,(\refl\,a') \tttstep d$; and projections of pairs.
In the $\iota$-rules, the parameter and index arguments are matched as pattern
variables independently of the constructor term (following the reading fixed
after Definition~\ref{def:mltt-recursor}), and in the $\Jelim$-rule the base
point $a$ and endpoint $b$ are matched independently of the argument $a'$ of
$\refl$. Every rule is therefore left-linear, and no left-hand side overlaps
another; a $\Jelim$-rule that required the base point or endpoint to match the
argument of $\refl$ would fail left-linearity.
Consequently, the rules form a regular combinatory reduction system
\citep[Definition~II.1.16]{klop1980}, and $\tttstep$ is
confluent~\citep[Theorem~II.3.11]{klop1980}.
Definitional equality $\defeq$ (Definition~\ref{def:mltt-defeq}) is
joinability under $\tttstep$ up to level equivalence; in this section we write
$\defeq_K$ where the contrast with a typed equality judgment matters.

\begin{lemma}[Confluence]\label{lem:mltt-confluence}
$\tttstep$ is confluent. Hence $\defeq$ is an equivalence relation closed under
every term former, under substitution, and under reduction; a reduct of
$\Pi x{:}A.\,B$ is $\Pi x{:}A'.\,B'$ with $A \tttstep^* A'$ and
$B \tttstep^* B'$, so $\Pi x{:}A.\,B \defeq \Pi x{:}A'.\,B'$ implies
$A \defeq A'$ and $B \defeq B'$; likewise for $\Sigma$, $\Id$, applied
families, and abstractions, so that
$\tttlam{x}{A} b \defeq \tttlam{x}{A'} b'$ implies $A \defeq A'$ and
$b \defeq b'$; and two types with different head formers, or two universes at
inequivalent levels, are not definitionally equal.
\end{lemma}

\begin{proof}
Confluence is Klop's theorem for regular combinatory reduction systems, since
the rules are left-linear and non-overlapping, with level equivalence a
congruence that commutes with every rule. The closure properties follow from
confluence; the injectivity and separation properties hold because no rule
has a type former or an abstraction at the head of its left-hand side.
\end{proof}

These injectivity and separation properties are precisely what a typed equality
with proof irrelevance in conversion lacks (see the remark after
Theorem~\ref{thm:mltt-declarative-consistency}).

\begin{definition}[The fragment $\mathcal{K}_0$]\label{def:mltt-k0}
$\mathcal{K}_0$ is the bidirectional system of Figure~\ref{fig:mltt-bidir},
together with the inferring rules for the remaining formers, over a global
environment of Definition~\ref{def:mltt-environment}, with three specifications:
\begin{enumerate}
\item every premise $C \defeq D$ is read as untyped joinability $C \defeq_K D$,
  which is its definition;
\item subsumption $A \subty_K B$ holds if $A \defeq_K B$, or if
  $A \defeq_K \Srt\,m$ and $B \defeq_K \Srt\,n$ with $m \cumle n$;
\item \tttrule{Id-I} is replaced by the checking rule
  \[
  \inferrule*[right=\tttrule{Chk-Refl}]
    {C \defeq_K \Id\,A\,x\,y \\ \ctx \vdash a \chk A \\
     x \defeq_K a \defeq_K y}
    {\ctx \vdash \refl\,a \chk C}
  \]
  and the inferring rule for $\Jelim\,a\,C\,d\,b\,p$ checks the base point
  $a \chk A$ against the domain $A$ of the reduct
  $\Pi y{:}A.\,\Pi h{:}\Id\,A'\,a'\,y.\,D$ named for the type of the motive $C$.
\end{enumerate}
$\mathcal{K}_0$ differs from the declarative system (without \tttrule{S-$\Pi$})
in presentation only: subsumption acts solely at checking positions, and the
equality it consults is that of Definition~\ref{def:mltt-defeq}. The fragment
$\mathcal{K}_1$ is $\mathcal{K}_0$ with subsumption extended by the
syntax-directed $\Pi$-case: $A \subty_K B$ also holds if
$A \defeq_K \Pi x{:}A_1.\,B_1$, $B \defeq_K \Pi x{:}A_2.\,B_2$,
$A_1 \defeq_K A_2$, and $\ctx, x{:}A_2 \vdash B_1 \subty_K B_2$. Transitivity
holds for $\subty_K$ in both fragments (Lemma~\ref{lem:mltt-k1-trans}), which
Proposition~\ref{prop:mltt-k1-complete} uses to translate \tttrule{S-Trans}.

Three conventions fix the derivations that the model interprets:
\begin{itemize}
\item \emph{Well-formed reducts:} The reduct named in a premise
  $C \defeq_K \Pi x{:}A.\,B$ (and likewise for $\Sigma$, $\Id$, and declared
  families) is a type joinable with $C$, accompanied by its well-formedness
  derivation, so that $A$ and $B$ are themselves typable; a type already headed
  by $\Pi$ is its own reduct, and an abstraction is checked by
  \tttrule{Chk-Lam}.
\item \emph{Annotated substitution:} A substitution written $B[a/x]$ in a rule
  whose premise checks $a$ against $A$ is read as $B[\tttann{a}{A}/x]$, ensuring
  that the result type of an application remains typable in $\mathcal{K}_0$.
\item \emph{Well-formedness of types:} Every type that enters a derivation---the
  expected type of a checking judgment, the inferred type $C$ of the head in
  \tttrule{Inf-App}, and any reduct named in a premise---is accompanied by a
  derivation of its well-formedness, so that both sides of every joinability
  premise between types are well-formed types, as required by
  Lemma~\ref{lem:mltt-k0-strat}(3).
\end{itemize}
By Lemma~\ref{lem:mltt-k0-annotated}, the conventions on reducts and on
annotated substitution do not change the set of derivable judgments. The third
convention requires every type entering a derivation to be typable in
$\mathcal{K}_0$. A declarative type need not be directly typable without
annotations: for instance, in
$\Id\,\Nat\,(\fst\,((\tttlam{y}{\Nat}\tttpair{y}{y})\,\tttzero))\,\tttzero$,
the abstraction stands in the head position of an application while its body
only checks, so the type infers only after the abstraction is annotated.
Proposition~\ref{prop:mltt-k1-complete} supplies such an annotated counterpart
with the same erasure.
\end{definition}

Every derivation in $\mathcal{K}_0$ is determined by the term and the mode, up to
the choice of reducts in joinability premises and the well-formedness
derivations of types. In particular, two derivations of $\ctx \vdash a \infr A$ and
$\ctx \vdash a \infr A'$ infer joinable types (in both $\mathcal{K}_0$ and
$\mathcal{K}_1$), because two reducts named for a single type are joinable and have
injective head formers; the universe at which a type is formed is unique up to level
equivalence; and because $\Prop$ lies below no type universe, a type checked
against $\Srt\,\ell$ infers a universe whose level is $\lzero$ at exactly the
valuations where $\ell$ is $\lzero$. Lemma~\ref{lem:mltt-k0-sorts} extends this
last property to types with the same erasure and to joinable types, showing that two
typed terms with the same erasure are either both proofs or both non-proofs. Moreover,
a data abstraction denotes a function on its own syntactic domain regardless of the
type against which it is checked, so the semantic value of a term does not depend
on its derivation.
Chapter~\ref{ch:checking} presents the procedures that decide $\mathcal{K}_0$ and
$\mathcal{K}_1$, whose subsumption relations are syntax-directed; Volume~II relates
the implemented kernel to them, and to $\mathcal{K}_1$ once structural $\Pi$
subsumption is added at its check boundary.

\paragraph{The model.} The model is constructed in ZFC with an increasing
sequence of inaccessible cardinals $\kappa_0 < \kappa_1 < \cdots$, the hypothesis
of Werner's model of CIC~\citep{werner1997}; write $V_n$ for $V_{\kappa_n}$. The
sets $V_n$ are Grothendieck universes satisfying $V_n \in V_{n+1}$, as
\citet{felicissimo2026acc} postulate for their model, and the theorems below hold
relative to this hypothesis.
Each $V_n$ is transitive, so $V_0 \subseteq V_1 \subseteq \cdots$. For derivations
whose universe levels lie below a finite bound $N$ in the empty environment or the
environment declaring $\tttFalse$, $N$ inaccessible cardinals suffice, and for
$N \le 3$ (universes $\Srt\,0$ through $\Srt\,3$) Lean's own universe hierarchy
discharges the hypothesis unconditionally (Appendix~\ref{app:formalization}). The
model, defined after the auxiliary lemmas on annotated reduction and sort agreement,
interprets $\mathcal{K}_0$ and $\mathcal{K}_1$ in these universes.

\paragraph{Annotated reduction.} A plain $\tttstep$-reduct of a typable term need not
 remain typable in $\mathcal{K}_0$: erasing the annotation of
$\tttann{\tttlam{x}{A} \tttpair{c}{d}}{\Pi x{:}A.\,B}\,a$ leaves a $\beta$-redex
whose head cannot infer a type, and substituting such an abstraction for a
variable in head position creates the same obstacle. \emph{Annotated reduction}
$\tttstep_a$ avoids this by keeping every term that lands in an inference position
annotated. It is the compatible closure of
\begin{gather*}
(\tttlam{x}{A} b)\,a \tttstep_a b[\tttann{a}{A}/x], \qquad
\tttinst{c}{\bar\ell} \tttstep_a \tttann{t[\bar\ell/\bar u]}{A[\bar\ell/\bar u]}, \\
\tttann{\tttlam{x}{A} b}{V}\,a \tttstep_a
  \tttann{b[\tttann{a}{A}/x]}{B[\tttann{a}{A}/x]}
  \quad (V \defeq_K \Pi x{:}A'.\,B), \\
\fst\,\tttann{\tttpair{a}{b}}{V} \tttstep_a \tttann{a}{A}
  \quad (V \defeq_K \Sigma x{:}A.\,B), \\
\snd\,\tttann{\tttpair{a}{b}}{V} \tttstep_a
  \tttann{b}{B[\tttann{a}{A}/x]} \quad (V \defeq_K \Sigma x{:}A.\,B), \\
\Jelim\,a\,C\,d\,b\,(\refl\,a') \tttstep_a \tttann{d}{C\,a\,(\refl\,a)}, \qquad
\tttann{u}{U} \tttstep_a u,
\end{gather*}
where:
\begin{itemize}
\item the $\delta$-rule applies to a transparent definition
  $\mathsf{def}\ \tttinst{c}{\bar u} : A := t$;
\item the rules with an annotated head read the annotation $V$ through a reduct
  that is a well-formed type, just as \tttrule{Chk-Lam} and \tttrule{Chk-Pair}
  read the expected type;
\item the erasure rule $\tttann{u}{U} \tttstep_a u$ applies either when the
  position is a checking position or when $u$ infers a type $U' \defeq_K U$; and
\item Rule~\tttrule{$\iota$} is included with its arguments annotated: the
  pattern context of its template, instantiated at the levels of the redex, has
  its variables substituted by the corresponding arguments annotated with their
  pattern-context entries (just as \tttrule{Inf-App} substitutes the annotated
  argument), so that the minor premise and constructor fields enter the reduct
  annotated by their types.
\end{itemize}
Erasure in an inference position requires $U' \defeq_K U$ rather than merely
$U' \subty_K U$: under $z{:}\Srt\,1$, the body
$\tttann{\Pi x{:}\Srt\,0.\,\Srt\,0}{\Srt\,5}$ infers $\Srt\,5$ whereas its
erasure infers $\Srt\,1$, so erasing the annotation inside an unannotated
abstraction would change the inferred type from $\Pi z{:}\Srt\,1.\,\Srt\,5$ to
$\Pi z{:}\Srt\,1.\,\Srt\,1$, which $\subty_K$ does not relate in
$\mathcal{K}_0$. Retaining such annotations never blocks a redex: any
annotations standing between the head of a redex and its pattern lie on terms
whose types are $\Pi$- or $\Sigma$-types---where $\subty_K$ coincides with
$\defeq_K$ in $\mathcal{K}_0$---and the innermost annotation is read directly by
the reduction rule. Write $|t|$ for the erasure of $t$, which deletes every
annotation while leaving the domains of abstractions in place.

\begin{lemma}[Annotated reduction]\label{lem:mltt-k0-annotated}
In $\mathcal{K}_0$:
\begin{enumerate}
\item if $\ctx \vdash a \chk A$ and $A \defeq_K A'$ with $A'$ well formed,
  then $\ctx \vdash a \chk A'$;
\item if $\ctx, x{:}A, \Delta \vdash t \chk T$ and $\ctx \vdash a \chk A$, then
  $\ctx, \Delta\sigma \vdash t\sigma \chk T\sigma$ for
  $\sigma = [\tttann{a}{A}/x]$, and likewise in inference mode;
\item if $\ctx \vdash t \infr T$ and $t \tttstep_a t'$, then
  $\ctx \vdash t' \infr T'$ with $T' \defeq_K T$; if
  $\ctx \vdash t \chk T$ and $t \tttstep_a t'$, then $\ctx \vdash t' \chk T$;
\item if $t$ is typable and $|t| \tttstep s$ by a rule other than erasure,
  then $t \tttstep_a^* t'$ for some $t'$ with $|t'| = s$; conversely,
  $t \tttstep_a t'$ implies $|t| \tttstep^{=} |t'|$;
\item two typable terms are joinable iff their erasures are joinable by the
  rules other than erasure;
\item annotated reduction is confluent up to erasure: if $t \tttstep_a^* t_1$
  and $t \tttstep_a^* t_2$, then $|t_1|$ and $|t_2|$ have a common reduct.
\end{enumerate}
\end{lemma}

\begin{proof}
(1) By induction on the derivation. Every rule reads the expected type through a
joinability premise or through $\subty_K$, and $\defeq_K$ is an equivalence;
for \tttrule{Chk-Sub} the inferred type satisfies $\subty_K A \defeq_K A'$, and
$\subty_K$ composes with $\defeq_K$.

(2) By induction on the derivation. The variable $x$ infers $A$ by
\tttrule{Inf-Var} and $\tttann{a}{A}$ infers $A$ by \tttrule{Inf-Ann}, so every
occurrence of $x$ is matched, and $\defeq_K$ is closed under substitution.

(3) By induction on the derivation:
\begin{itemize}
\item For the $\beta$-rule with an unannotated head, the head infers
  $\Pi x{:}A.\,B_0$ by \tttrule{Inf-Lam}, so $b \infr B_0$ under $x{:}A$. The
  premise of \tttrule{Inf-App} names a reduct $\Pi x{:}A_2.\,B_2$ with
  $a \chk A_2$, and Lemma~\ref{lem:mltt-confluence} gives $A \defeq_K A_2$ and
  $B_0 \defeq_K B_2$. By~(1) we have $a \chk A$, and by~(2) we obtain
  $b\sigma \infr B_0\sigma$ with
  $B_0\sigma \defeq_K B_2[\tttann{a}{A_2}/x]$, which is the inferred type of the
  redex.
\item For an annotated head, \tttrule{Chk-Lam} gives $A \defeq_K A'$ and
  $b \chk B$ under $x{:}A$ for a reduct $\Pi x{:}A'.\,B$ of the annotation,
  while the argument checks against the domain of a reduct of the annotation.
  By~(1) we have $a \chk A$, by~(2) we have $b\sigma \chk B\sigma$, and the
  reduct infers $B\sigma$ by \tttrule{Inf-Ann}. The cases for $\delta$ and for
  projections follow by the same argument using \tttrule{E-Def} and
  \tttrule{Chk-Pair}.
\item For $\iota$, the template is typed once in the pattern context: the minor
  premise (a variable) applied to the field variables infers its type
  instantiated by the fields, and each induction-hypothesis term---the recursor
  applied to the parameters, the motive, the minor premises, the indices of the
  recursive field, and the field applied under its telescope---checks against
  the corresponding hypothesis. In a redex, the arguments are typed against the
  recursor's telescope and the fields of the major premise against the
  constructor's telescope; the two agree at the parameters, indices, and levels
  by the injectivity of applied families (Lemma~\ref{lem:mltt-confluence}). Any
  field read off an index argument agrees with that argument, which checks
  against the type of the index at the earlier index arguments---a type
  definitionally equal to the field's type at the earlier fields by the
  admissibility condition on its type. Thus the arguments, annotated with the
  entries of the pattern context, infer those entries; instantiating the
  template judgment by~(2) yields a reduct whose type is the motive applied to
  the constructor's result indices and the major premise, joinable with the type
  of the redex.
\item For $\Jelim\,a\,C\,d\,b\,(\refl\,a')$, Rule~\tttrule{Chk-Refl} gives
  $a \defeq_K a' \defeq_K b$, the minor premise $d$ checks against
  $C\,a\,(\refl\,a)$, and the reduct $\tttann{d}{C\,a\,(\refl\,a)}$ infers
  $C\,a\,(\refl\,a) \defeq_K C\,b\,(\refl\,a')$.
\item Erasure in an inference position replaces $\tttann{u}{U} \infr U$ by
  $u \infr U'$ with $U' \defeq_K U$, and the enclosing rule reads $U$ through
  $\defeq_K$ or $\subty_K$, both of which compose with $\defeq_K$; erasure in a
  checking position follows from~(1).
\end{itemize}

(4) Any rule other than erasure fires on a constant, an application, a
projection, or $\Jelim$. The head of an application, the argument of a
projection, and the partial applications of a constant infer $\Pi$- or
$\Sigma$-types; their annotations therefore erase in inference position down to
the last annotation, which the reduction rule reads through its reduct whenever
the annotated term is an abstraction or a pair. The major premise of $\Jelim$
and the arguments of a redex stand in checking positions, so their annotations
erase outright. The reduction rule then applies, and its reduct erases to $s$.
The converse holds by inspection of the rules.

(5) Follows from~(4), since annotated steps preserve typability by~(3) and
erasure steps leave erasures unchanged.

(6) By the converse of~(4), $|t| \tttstep^* |t_1|$ and $|t| \tttstep^* |t_2|$,
which Lemma~\ref{lem:mltt-confluence} joins. (Confluence of $\tttstep_a$ itself
is not claimed: the two reducts $\tttann{u}{U}$ and $\tttann{u}{U'}$ of a nested
annotation $\tttann{\tttann{u}{U}}{U'}$ carry joinable types, and~(4) lifts the
reductions joining $|U|$ and $|U'|$ to $\tttstep_a$ only up to the annotations
inside those reducts, which need not coincide.)
\end{proof}

\begin{lemma}[Sort agreement]\label{lem:mltt-k0-sorts}
In $\mathcal{K}_0$:
\begin{enumerate}
\item If $\ctx \vdash A \infr D$ and $\ctx \vdash A' \infr D'$ with
  $D \defeq_K \Srt\,m$ and $D' \defeq_K \Srt\,m'$, and the erasures of $A$ and
  $A'$ are equal up to level equivalence or $A \defeq_K A'$, then $m$ and $m'$
  are $\lzero$ at the same valuations of the level parameters: $A$ is a
  proposition exactly when $A'$ is.
\item If $\ctx \vdash t \chk T$ and $\ctx \vdash t' \chk T'$, the erasures of
  $t$ and $t'$ are equal up to level equivalence, and $\ctx \vdash T \infr D$
  and $\ctx \vdash T' \infr D'$ with $D \defeq_K \Srt\,m$ and
  $D' \defeq_K \Srt\,m'$, then $m$ and $m'$ are $\lzero$ at the same valuations:
  $t$ is a proof exactly when $t'$ is.
\end{enumerate}
\end{lemma}

\begin{proof}
(1) An \emph{observation} of a term $u$ at a type $X$ follows a path through $X$ and
ends at a type joinable with a universe. Each step is either:
\begin{itemize}
\item an application to an argument $c$ checked against the domain of a
  $\Pi$-type to which $X$ reduces, continuing at $u\,c$ and the codomain at $c$;
\item a first projection, continuing at $\fst\,u$ and the first component of a
  $\Sigma$-type to which $X$ reduces; or
\item a second projection, continuing at $\snd\,u$ and the second component at
  $\fst\,u$.
\end{itemize}
By induction on the typing derivation of a term $t$, an observation of $t$ at
its type and an observation of a term $t'$ (checked against any type) whose
erasure equals that of $t$ up to level equivalence, along paths with joinable
arguments, reach universes that are $\lzero$ at the same valuations; the
induction hypothesis is stated under simultaneous substitutions of joinable
terms for the context variables. Outermost annotations on $t'$ are stripped
first: an annotation changes the type of its term only by subsumption, and in
the cumulativity order the only universe related to $\Prop$ is $\Prop$ itself,
so the observation passes directly to the type of the underlying term.
Variables, universes, constants, identity types, and $\Jelim$ have types of
equal erasure on both sides. An application is observed through its head at the
argument, a projection through the projected term, a $\Pi$-type or $\Sigma$-type
through the universes of its components, and a pair through its components. An
abstraction observed at an argument is observed through its body with the
argument (annotated by the domain) substituted for the bound variable; this is
the case that requires the simultaneous substitutions in the induction
hypothesis. Finally, for joinable types $A \defeq_K A'$, the erasures of $A$ and
$A'$ reduce to terms equal up to level equivalence. Because the erasure of a
$\delta$- or $\iota$-redex is a redex whose reduct has the same erasure as the
original reduct, a reduction step on a term corresponds to a sequence of steps
on its erasure, each followed by an erasure. By Lemma~\ref{lem:mltt-k0-annotated}(4),
these reductions lift to annotated reductions of $A$ and $A'$, which preserve
their universes by Lemma~\ref{lem:mltt-k0-annotated}(3) and reach types whose
erasures are equal up to level equivalence.

(2) The same induction applies, using observations that end at a term whose type
is joinable with a type formed at a universe; the endpoints of two observations
at joinable types are compared using part~(1) for joinable types. An annotation
changes the type of its term either by joinability or (when the term is a type)
from one universe to another, and the type of a universe is never $\Prop$; thus
outermost annotations on $t'$ strip as before, and a type is never a proof. An
abstraction, a pair, and a reflexivity proof can now be observed at an empty
path. Because a $\Pi$-type is a proposition iff its codomain is, and a
$\Sigma$-type is a proposition iff both of its components are---and because
substitution does not alter the universe at which a type is formed---these cases
follow from the body of the abstraction and the components of the pair, compared
without substitution. An identity type is always a proposition.
\end{proof}

\paragraph{The stratified model.} The model interprets
$\mathcal{K}_0$ with untyped joinability as its conversion, following the
proof-irrelevant construction of \citet{miquel-werner2003}. For the Calculus of
Constructions, Miquel and Werner show that the soundness of $\beta$-reduction
required by conversion fails whenever the code of a proof-valued function loses
its domain (both under their encoding and under Aczel's trace coding), while
restricting $\beta$-soundness to typed terms makes it circular with typing
soundness; they resolve this by separating proofs from data according to the
sort of a term. In the model:
\begin{itemize}
\item Every proof denotes $\emptyset$.
\item A data abstraction denotes the code
  $\mathrm{lam}(A, f) = \{(x, f(x)) \mid x \in \mathrm{El}(A)\}$, the set-theoretic
  graph of $f$ on $\mathrm{El}(A)$, which records the domain $A$ as the set of
  first components; application $\mathrm{app}(\mathrm{lam}(A, f), x)$ equals
  $f(x)$ for $x \in \mathrm{El}(A)$ and is undefined otherwise.
\item A semantic type is simply the set of its elements, $\mathrm{El}(A) = A$,
  with no tag, and propositions are the subsets of $\{\emptyset\}$.
\item The propositional product $\widehat\Pi_{\mathrm{p}}(A, F)$ is the
  proposition $\{y \in \{\emptyset\} \mid \forall x \in \mathrm{El}(A).\
  \emptyset \in F(x)\}$, whereas the data product
  $\widehat\Pi_{\mathrm{d}}(A, F)$ is the set of codes $\mathrm{lam}(A, f)$ with
  $f(x) \in \mathrm{El}(F(x))$ for all $x \in \mathrm{El}(A)$.
\item Likewise, $\widehat\Sigma_{\mathrm{p}}(A, F)$ is the proposition
  $\{y \in \{\emptyset\} \mid \exists x \in \mathrm{El}(A).\ \emptyset \in F(x)\}$,
  and $\widehat\Sigma_{\mathrm{d}}(A, F)$ is the set of pairs $(a, b)$ with
  $a \in \mathrm{El}(A)$ and $b \in \mathrm{El}(F(a))$.
\item $\Prop$ denotes the set $\mathcal{P}(\{\emptyset\})$ of propositions,
  $\Srt\,(\lsucc\,\ell)$ denotes the Grothendieck universe $V_\ell$, and
  $\Id\,A\,a\,b$ denotes $\{\emptyset\}$ if $a$ and $b$ have the same value and
  $\emptyset$ otherwise (inductive families are described below).
\end{itemize}
The choice between the propositional and data formers is governed by the sort of
the type. Because a data $\Pi$-type has its codomain in a predicative type
universe and lies in the universe at the maximum of the domain and codomain
levels, the function codes $\mathrm{lam}(A, f)$ that record their domains never
encounter impredicativity.

Whereas Miquel and Werner read the sort of a term off sort-annotated variables,
\citet{felicissimo2026acc} determine it by a syntactic annotation (which suffices
without cumulativity), and \citet[Section~6.1]{carneiro2019} separates propositional
from relevant products syntactically via unique typing. In $\mathcal{K}_0$, the sort
of a term is not purely syntactic: the sort of $\fst\,t$ depends on the type of $t$,
and large elimination makes whether a type is a proposition depend on a term value.
Sorts are therefore read off typing derivations. Specifically, a term $t$ is a
\emph{proof} in $\ctx$ at a valuation of the level parameters if some term with the
same erasure (up to level equivalence) checks, in a well-formed context joinable
with $\ctx$ entry by entry, against a type formed at a universe whose level evaluates
to $\lzero$; similarly, $t$ is a \emph{proposition} if some term with the same
erasure infers such a universe. By Lemma~\ref{lem:mltt-k0-sorts}, all such
derivations for a typable term agree, so both notions depend on the term only through
its erasure up to level equivalence and on the context only up to joinability;
moreover, a typable term is a proof iff its type is a proposition, and no type is a
proof.

The interpretation $\llbracket t \rrbracket_\ctx\rho$ is a partial function of
the term, the context, and the variable environment $\rho$, defined by recursion
on $t$:
\begin{itemize}
\item Any variable, constant, abstraction, application, pair, projection, or
  $\Jelim$-term that is a proof denotes $\emptyset$, as does $\refl\,a$.
\item Otherwise, a variable denotes its environment entry $\rho(x)$, and a
  constant denotes the value assigned to it by the model of the global
  environment;
  $\llbracket \tttlam{x}{A} b \rrbracket_\ctx\rho =
  \mathrm{lam}(\llbracket A \rrbracket_\ctx\rho,\ v \mapsto
  \llbracket b \rrbracket_{\ctx, x{:}A}\rho[x \mapsto v])$, defined whenever the
  domain is defined and the body is defined at every element of the domain; an
  application denotes $\mathrm{app}$ of the head and argument values; a pair
  denotes the ordered pair of its component values; a projection denotes the
  corresponding component; and $\Jelim\,a\,C\,d\,b\,p$ denotes the value of $d$.
\item A $\Pi$-type denotes $\widehat\Pi_{\mathrm{p}}$ if it is a proposition and
  $\widehat\Pi_{\mathrm{d}}$ otherwise, and similarly for a $\Sigma$-type;
  universes and identity types are interpreted as above; and an annotation
  $\tttann{u}{U}$ denotes the value of $u$.
\end{itemize}
Each clause is defined whenever the subterm values it consults are defined. The
resulting value depends on the term only through its erasure up to level
equivalence and on the context only up to joinability; the value of a derivation
is the value of its subject term, with \tttrule{Chk-Sub} acting as the identity.
An \emph{environment $\rho$ of a context $\ctx$} assigns to each declaration
$x{:}A$ an element of $\mathrm{El}\,\llbracket A \rrbracket_\ctx\rho$ (which is a
proposition whenever $A$ is one). Likewise, the model of the global environment
assigns to each constant, at each valuation of its level parameters, an element
of the interpretation of its closed declared type: for an axiom, an element
guaranteed by Definition~\ref{def:mltt-obligation}; for a definition, the value
of its body.

At fixed values of its parameters, an inductive family denotes the least fixed point
of the monotone operator determined by its constructors on families of sets indexed
by the values of its indices. In a data family, a constructor application denotes
the tuple encoding the constructor's tag together with the values of its arguments
(where a recursive argument ranges over codes of functions into the fibers); in a
family of propositions, every constructor application denotes $\emptyset$, so an
inductive proposition denotes either $\{\emptyset\}$ or $\emptyset$ at each index
tuple. For a data family, the fibers lie in the Grothendieck universe of its level:
the fixed-point approximants stabilize at a regular cardinal above the constructor
arities, which lies below the inaccessible cardinal of that universe, and the
argument types of the constructors lie in the universe by the universe constraint.
A recursor denotes the function whose graph is the least relation closed under its
$\iota$-equations. This relation is functional on the fixed point: for a data family
because constructor encodings are injective, and for a propositional family that
eliminates into arbitrary universes because the declaration is then a subsingleton
whose arguments are either proofs or result-index terms determined by the
indices~\citep[Section~5]{felicissimo2026acc}.

The $\iota$-rules preserve the defined value of a typed redex in every variable
environment. For a data family, the major premise denotes a constructor code that
fixes the values of all fields, and any field read off an index argument has that
same value by membership of the code in the fiber. For a propositional family, the
redex is itself a proof unless the family is a subsingleton, whose fields are either
proofs (with value $\emptyset$) or data arguments that Rule~\tttrule{$\iota$} reads
off the index arguments fixed by the redex. Because Lemma~\ref{lem:mltt-k0-strat}(2)
is proved prior to typing soundness, it is stated for arbitrary variable environments
$\rho$, and this index-reading formulation of Rule~\tttrule{$\iota$} is essential: a
reduct that instead applied the minor premise to a data argument taken from the
constructor term---such as $(\tttlam{z}{B} a)\,y$ in an environment where
$\rho(y) \notin \mathrm{El}\,\llbracket B \rrbracket\rho$---could be undefined even
when the redex (whose major premise denotes $\emptyset$) is defined.

\begin{definition}[Obligation of an environment]\label{def:mltt-obligation}
A well-formed environment $\tttEnv$ \emph{meets its obligation} if for every
axiom $\mathsf{axiom}\ \tttinst{c}{\bar u} : A$ of $\tttEnv$ and every
valuation of $\bar u$, $\llbracket A \rrbracket$ is defined and
$\mathrm{El}\,\llbracket A \rrbracket \neq \emptyset$, the interpretation being
taken in the environment preceding the axiom; the
model then assigns $c$ an element of this set. Definitions and inductive
declarations carry no obligation: the body of a definition inhabits its type by
the theorem below, and an admissible declaration denotes its fixed point.
\end{definition}

The obligation is a condition on each axiom separately. For a propositional
axiom $a : \varphi$, which Chapter~\ref{ch:institutions} places among the
hypotheses $\Gamma_E$ of the theory of the environment, it states that
$\varphi$ is true in the model; a set of propositional axioms with no model,
such as $\varphi$ and $\neg\varphi$, fails it, and the consistency statement is
relative to the axioms, as every judgment made in the environment is. The
axiom $\irrel$ of Section~\ref{sec:mltt-prop} meets it in every model: a
proposition denotes a set with at most one element, so the function with value
$\emptyset$ is an element of the interpretation of its type. For an
axiom whose type is a type, it states that the model assigns the constant an
element. The constants of a knowledge graph, such as $\Entity : \Setu$, the
classes, and the resources, are bound by a declaration of their own, the class
system of the graph (Definition~\ref{def:kg-theory}), and
Theorem~\ref{thm:kg-model} assigns them elements of their types for a
well-formed graph.

\begin{lemma}[The stratified interpretation]\label{lem:mltt-k0-strat}
Let $\tttEnv$ be as in Theorem~\ref{thm:mltt-k0-consistency}, let $\ctx$ be a
context whose entries are types, and let $\rho$ assign values to the
variables.
\begin{enumerate}
\item If $\ctx, x{:}A, \Delta \vdash t \chk T$ or
  $\ctx, x{:}A, \Delta \vdash t \infr T$, and $\ctx \vdash a \infr A$ with
  $\llbracket a \rrbracket_\ctx\rho$ defined, then $t[a/x]$ has the value of
  $t$ at the environment that gives $x$ the value of $a$, the two being
  defined together; likewise a weakened term has the value of the term, and
  $t[\bar\ell/\bar u]$ at a valuation of the level parameters has the value of
  $t$ at the valuation that gives $\bar u$ the values of $\bar\ell$.
\item If $\ctx \vdash t \infr T$ or $\ctx \vdash t \chk T$, $t \tttstep_a t'$,
  and $\llbracket t \rrbracket_\ctx\rho$ is defined, then
  $\llbracket t' \rrbracket_\ctx\rho = \llbracket t \rrbracket_\ctx\rho$.
\item If $\ctx \vdash t \chk T$, $\ctx \vdash t' \chk T'$, $t \defeq_K t'$, and
  both values are defined, then they are equal.
\end{enumerate}
\end{lemma}

\begin{proof}
Parts~(1) and~(2) proceed by induction on the derivation of $t$, so that every
subterm carries a typing derivation.

(1) The substituted, weakened, and level-instantiated derivations type the term
at a type formed at the same universe (up to the level valuation), so being a
proof and being a proposition commute with substitution, weakening, and level
instantiation. A variable that is a proof receives the value of $a$, which is
then also a proof, and any defined value of a proof is $\emptyset$.

(2) By Lemma~\ref{lem:mltt-k0-annotated}(3) and Lemma~\ref{lem:mltt-k0-sorts},
an annotated reduction step preserves whether a term is a proof, and a proof
denotes $\emptyset$ on both sides. When the term is not a proof, the cases
follow the structure of Lemma~\ref{lem:mltt-k0-annotated}(3). At a
$\beta$-redex, the head abstraction is not a proof because an application is a
proof iff its head is; since the value of the application is defined, the value
of the argument lies in the domain recorded by the code of the abstraction, and
the reduct has the value of the body at that argument by~(1). Projections of a
pair, $\Jelim$ at $\refl$, and erasure of an annotation preserve the value by
definition of the interpretation clauses, and $\iota$ preserves it by the model
of the global environment; $\delta$ preserves it because a definition denotes
the value of its body, which by~(1) agrees with the value of the level-instantiated
body in any context (a closed term being independent of the context). A step
inside a subterm preserves the value by the induction hypothesis---under a
binder, at every element of the domain, which are the only values on which a
function code depends.

(3) The reductions of $t$ and $t'$ to a common reduct are reductions of their
erasures, which lift by Lemma~\ref{lem:mltt-k0-annotated}(4) to annotated
reductions of $t$ and $t'$ reaching terms whose erasures are equal up to level
equivalence and which therefore have the same value; by~(2), those annotated
reductions preserve the values.
\end{proof}

Notice that statement~(2) is directional: an annotated reduction step preserves
any value that is already defined. Because typing validity does not enter~(2),
the circularity between typing soundness and the soundness of $\beta$-reduction
does not arise; the definedness of the value of a $\beta$-redex substitutes for
typing its argument.

\begin{theorem}[Consistency of $\mathcal{K}_0$]\label{thm:mltt-k0-consistency}
Let $\tttEnv$ be a well-formed global environment that meets its obligation in
the stratified model, whose declared types are types of $\mathcal{K}_0$ and whose
definitions and inductive declarations are typed in $\mathcal{K}_0$, each over the
declarations before it. If $\ctx \vdash a \chk A$ or $\ctx \vdash a \infr A$ in
$\mathcal{K}_0$, in a context whose entries are types, and $\rho$ is an
environment of $\ctx$, then in the stratified model
$\llbracket a \rrbracket_\ctx\rho$ and $\llbracket A \rrbracket_\ctx\rho$ are
defined and
$\llbracket a \rrbracket_\ctx\rho \in \mathrm{El}\,\llbracket A \rrbracket_\ctx\rho$.
In particular, no closed term checks against $\tttFalse$.
\end{theorem}

\begin{proof}
By induction on the typing derivation, extending the statement for an inferring
$\Sigma$-type with the condition that its components denote propositions
whenever they are propositions (which the projection rules use). Any joinability
premise between two well-formed types yields equal semantic values by
Lemma~\ref{lem:mltt-k0-strat}(3), as do the endpoints in \tttrule{Chk-Refl};
\tttrule{Chk-Sub} for universe cumulativity holds because
$V_0 \subseteq V_1 \subseteq \cdots$.

When the subject term is a proof, its type is a proposition (whose only possible
element is $\emptyset$), so it suffices to show that $\emptyset$ belongs to the
interpretation of the type:
\begin{itemize}
\item for a variable, by the context environment $\rho$, and for a constant, by
  the model of the global environment;
\item for an abstraction, because its body denotes $\emptyset$ at every element
  of the domain, placing $\emptyset$ in $\widehat\Pi_{\mathrm{p}}$;
\item for an application, because its head denotes $\emptyset$ in a
  propositional $\Pi$-type, placing $\emptyset$ in every fiber of the codomain;
\item for a pair and for projections, through the component conditions of
  $\widehat\Sigma_{\mathrm{p}}$ or $\widehat\Sigma_{\mathrm{d}}$; and
\item for $\Jelim\,a\,C\,d\,b\,p$, because its minor premise $d$ is a proof of
  $C\,a\,(\refl\,a)$, the two instances of the motive are types in the codomain
  universe, and $C\,a\,(\refl\,a)$ has the same value as $C\,b\,p$ since the
  premise on $p$ gives $\llbracket a \rrbracket = \llbracket b \rrbracket$ and
  $\llbracket p \rrbracket = \emptyset$.
\end{itemize}

When the subject term is not a proof:
\begin{itemize}
\item \tttrule{Inf-Lam} and \tttrule{Chk-Lam} hold because the abstraction
  denotes the code $\mathrm{lam}(\llbracket A \rrbracket\rho, g)$ with
  $g(v) = \llbracket b \rrbracket\rho[x \mapsto v] \in
  \mathrm{El}\,\llbracket B \rrbracket\rho[x \mapsto v]$ for all
  $v \in \mathrm{El}\,\llbracket A \rrbracket\rho$ by the induction hypothesis,
  which is an element of $\widehat\Pi_{\mathrm{d}}$;
\item \tttrule{Inf-App} holds because the head inhabits a data $\Pi$-type and is
  therefore a code $\mathrm{lam}(\llbracket A \rrbracket\rho, g)$, so
  $\mathrm{app}$ is defined at the argument value
  $\llbracket a \rrbracket\rho \in \mathrm{El}\,\llbracket A \rrbracket\rho$ and
  yields $g(\llbracket a \rrbracket\rho) \in
  \mathrm{El}\,\llbracket B[\tttann{a}{A}/x] \rrbracket\rho$ by
  Lemma~\ref{lem:mltt-k0-strat}(1);
\item \tttrule{$\Pi$-F} and \tttrule{$\Sigma$-F} hold because a data type former
  lies in the universe at the maximum of the two levels, whereas a propositional
  one is a proposition regardless of the size of its domain (impredicativity);
\item pairs and projections hold via $\widehat\Sigma_{\mathrm{d}}$;
\item \tttrule{Id-F} holds because every identity type is a proposition; and
\item constants hold by the model of the global environment, whose recursors
  satisfy their typing rules by the fixed-point construction.
\end{itemize}
\end{proof}

\paragraph{Remark: the typed-equality route.} A model that encodes functions by their
traces~\citep{aczel1999relating},
$\mathrm{lam}(f) = \bigcup_{(x,y) \in f} \{x\} \times y$, does not validate the
untyped $\beta$-rule outside the domain of a code. Under trace coding,
$\mathrm{lam}(f)$ does not determine the domain of
$f$~\citep[Remark~3.4]{lee-werner2011}: $\tttlam{x}{\tttFalse} x$ denotes the code of
a function on $\emptyset$, which is $\emptyset$---just like the code of every function
whose values are $\emptyset$---so no condition on the semantic values of a term
places the argument of a redex in the domain of its abstraction, and
$(\tttlam{x}{\tttFalse} x)\,a$ with $\llbracket a \rrbracket\rho = \{\emptyset\}$
denotes $\emptyset$ whereas its reduct $a$ denotes $\{\emptyset\}$.
\citet{lee-werner2011} avoid this obstacle by taking a typed definitional equality
$\ctx \vdash a \defeq b : A$ as primitive, whose $\beta$-rule checks the argument
against the domain of the abstraction; Carneiro's model of Lean tags its semantic
types, recovering the injectivity of the type
formers~\citep[Section~6.4]{carneiro2019}. For $\mathcal{K}_0$, whose conversion
is untyped joinability, using such a model requires the equivalence of untyped and
typed conversion on typable terms: two joinable terms checked against a common type
must be equal at that type, and two joinable types must be equal at a universe. This
equivalence is established for pure type
systems~\citep{adams2006,siles-herbelin2010} but remains open for theories with
cumulative universes~\citep[Section~5]{lee-werner2011}; \citet{barras2010} builds his
set-theoretic model of the Calculus of Constructions for the typed presentation for
precisely this reason, recording the extension by $\Prop \subseteq \mathsf{Kind}$ with
untyped conversion as open. Proving the equivalence through subject reduction of the
typed system (where each step of a typable term yields a typed equation) requires the
typed equality to be injective on $\Pi$ and to separate distinct type formers (see the
remark after Theorem~\ref{thm:mltt-declarative-consistency}). Subject reduction for
$\mathcal{K}_0$ (Lemma~\ref{lem:mltt-k0-annotated}(3)) establishes the typability of
each reduct rather than a typed equation between a term and its reduct, and the
normalization-by-evaluation procedure of Chapter~\ref{ch:checking} decides untyped
joinability rather than typed equality.

Conversely, a function code that records its domain is nonempty, and under
impredicativity such codes cannot be sets: a proof of $\Pi P{:}\Prop.\,P \to P$
would have to record $\llbracket \Prop \rrbracket$ while simultaneously belonging to
an element of $\mathrm{El}\,\llbracket \Prop \rrbracket$, violating the foundation
axiom of ZFC. The stratified model avoids both obstacles in the manner of
\citet{miquel-werner2003}: only data functions record their domains, every proof
denotes $\emptyset$ by its sort, and every data $\Pi$-type is predicative. Thus the
codes that record a domain are ordinary sets, and partial application (defined only on
the domain of a code) keeps $\beta$-reduction sound for all defined values
(Lemma~\ref{lem:mltt-k0-strat}(2)). This separation requires the sort of a term to be
stable under substitution, which would fail if $\Prop$ lay below a type universe
(since $\tttlam{x}{X} x$ would be a proof when $X$ is a proposition and a data
function when $X$ is a type); Miquel and Werner note that $\Prop \preceq \Typeu_0$
cannot be accommodated easily in their setting. Finally, making the domain of an
abstraction part of the term syntax ensures that the interpretation is a function of
the term alone: with unannotated abstractions, the semantic value of an abstraction
would depend on the expected type against which it is checked. In Lean's kernel,
abstractions carry no domain when compared by conversion, so our syntax represents a
small adjustment to the kernel (recorded in Volume~II): the abstraction node carries a
domain that the checker compares against the expected domain at Rule~\tttrule{Chk-Lam}
and that conversion compares structurally, allowing an abstraction to infer its type
whenever its body does.

\paragraph{Adding \tttrule{S-$\Pi$}.} The $\Pi$-case of $\mathcal{K}_1$
compares domains by joinability, so every subsumption denotes an inclusion of
semantic types: joinable typed domains denote the same semantic type, and any
function on that domain taking values in the codomain of the first $\Pi$-type
also takes values in the codomain of the second. Rule~\tttrule{Chk-Sub} still
acts as the identity and an annotation still denotes the value of its underlying
term, so the results for $\mathcal{K}_0$ lift to $\mathcal{K}_1$ by adding the
$\Pi$-case to their inductions. Rule~\tttrule{S-Trans} of the declarative system
becomes the following admissibility lemma.

\begin{lemma}[Transitivity of subsumption]\label{lem:mltt-k1-trans}
In $\mathcal{K}_0$ and in $\mathcal{K}_1$, if $\ctx \vdash A \subty_K B$ and
$\ctx \vdash B \subty_K C$, then $\ctx \vdash A \subty_K C$; and if
$\ctx \vdash a \chk A$ and $\ctx \vdash A \subty_K B$ for a type $B$, then
$\ctx \vdash a \chk B$.
\end{lemma}

\begin{proof}
The first statement is proved by induction on the derivation of
$\ctx \vdash A \subty_K B$, with a case analysis on $\ctx \vdash B \subty_K C$.
Joinability composes with the second subsumption because $\subty_K$ respects
$\defeq_K$ on both sides. A universe cumulativity step can meet only a
joinability or another cumulativity step (since any type above a universe is
itself a universe), where the level order $\cumle$ is transitive; it cannot meet
a $\Pi$-case because universe and $\Pi$ head formers are distinct. When two
$\Pi$-cases meet at the middle type $B$, injectivity of $\defeq_K$ on $\Pi$
makes all three domains joinable and the two middle codomains joinable. The
codomain premise of the second step, derived under the third domain, holds under
the second domain by context conversion (replacing a context entry with a
joinable type); the induction hypothesis then composes it with the codomain
premise of the first step, and context conversion returns the result to the
third domain.

The second statement follows by induction on the checking derivation:
\tttrule{Chk-Sub} composes the two subsumptions; at \tttrule{Chk-Lam} the
expected type $A$ is a $\Pi$-type, so $B$ is joinable with a $\Pi$-type whose
domain is joinable with the abstraction's domain and whose codomain lies above
the expected codomain, against which the body checks by the induction
hypothesis; and at a $\Sigma$-type or identity type, subsumption reduces to
joinability.
\end{proof}

With contravariant domains, composing two $\Pi$-subsumptions would require
narrowing the codomain premise of the first step from the second domain to the
third. In a dependent type theory, narrowing a derivation also narrows the
well-formedness derivations it carries, inserting subsumptions at every
occurrence of the narrowed variable; neither the height of a derivation nor the
nesting depth of its $\Pi$-cases bounds the resulting derivation, which is why
transitivity elimination for dependent calculi with contravariant domains is
proved via normal forms~\citep{aspinall-compagnoni2001,chen1997}. Following ECC
and Coq~\citep{luo1994}, $\mathcal{K}_1$ requires convertible domains.

\paragraph{Annotated reduction in $\mathcal{K}_1$.} In $\mathcal{K}_1$, the
type inferred by a head $u$ may lie strictly below its annotation $V$ (via a
larger codomain in $V$), in which case the annotation cannot simply erase in
inference position. Annotated reduction in $\mathcal{K}_1$ therefore adds one
rule that pushes the annotation on a head $u$ outward across an application:
\[
\tttann{u}{V}\,a \tttstep_a \tttann{u\,\tttann{a}{A'}}{B[\tttann{a}{A'}/x]}
  \quad (V \defeq_K \Pi x{:}A'.\,B),
\]
reading the annotation $V$ through a well-formed $\Pi$-reduct just as the other
annotated-head rules do. Repeated applications of this rule move stacked
annotations past the head of a $\beta$-redex to expose the underlying
abstraction, and move annotations on partial applications of a recursor outward
to expose an $\iota$-redex.

\begin{lemma}[Annotated reduction in $\mathcal{K}_1$]\label{lem:mltt-k1-annotated}
Lemma~\ref{lem:mltt-k0-annotated} holds in $\mathcal{K}_1$, with annotated
reduction extended by the rule above.
\end{lemma}

\begin{proof}
Parts~(1), (2), (5), and~(6) are proved just as in $\mathcal{K}_0$, adding the
$\Pi$-case to the inductions on subsumption (which still composes with
$\defeq_K$ on both sides and is closed under substitution). Two inversion
properties for $\subty_K$ in $\mathcal{K}_1$ replace the simpler inversion in
$\mathcal{K}_0$: first, at any type headed by $\Sigma$, by $\Id$, or by an
irreducible constant, subsumption reduces to joinability, because the base cases
of $\subty_K$ relate joinable types or universes while the $\Pi$-case relates
$\Pi$-types; second, if one side of a subsumption is joinable with a well-formed
$\Pi$-type, then so is the other, with a joinable domain and with codomains
related by $\subty_K$ in the same direction under either domain.

(3) An annotation in a checking position erases by Lemma~\ref{lem:mltt-k1-trans},
just as it does in $\mathcal{K}_0$ by~(1). For the $\beta$-rule with an
annotated head, the abstraction checks against $V$ either directly by
\tttrule{Chk-Lam}, or by \tttrule{Chk-Sub} after \tttrule{Inf-Lam} at a type
below $V$ whose domain is joinable with that of $V$; in both cases the body
checks against the codomain of $V$, and the reduct infers the type of the redex
as in $\mathcal{K}_0$. For the new annotation-pushing rule, $u$ infers a type
$S \subty_K V$, so $S \defeq_K \Pi x{:}A_1.\,B_1$ with $A_1 \defeq_K A'$ and
$B_1 \subty_K B$ under $x{:}A'$. Then $\tttann{a}{A'}$ checks against $A_1$, and
the inner application $u\,\tttann{a}{A'}$ infers
$B_1[\tttann{\tttann{a}{A'}}{A_1}/x]$, which lies below $B[\tttann{a}{A'}/x]$ by
substitution into the codomain premise; hence the outer annotated reduct infers
$B[\tttann{a}{A'}/x]$, matching the type of the redex. The remaining cases are
identical to those of $\mathcal{K}_0$ (at $\iota$, the major premise infers an
applied inductive family, where subsumption reduces to joinability).

(4) Annotations on the head of a $\beta$-redex push outward across the
application down to the innermost annotation, which the annotated-head
$\beta$-rule reads. The argument of a projection infers a type joinable with a
$\Sigma$-type, where subsumption is joinability, so its annotations erase in
inference position down to the innermost one as in $\mathcal{K}_0$. At an
$\iota$-redex, annotations on the partial applications of the recursor push
outward along its spine and across the application to the major premise; beneath
them, annotations at the top of the major premise erase in checking position, as
do annotations on partial applications of the constructor once pushed out of its
spine. The major premise of $\Jelim$ stands in a checking position, so its
annotations erase outright. The reduction rule then fires beneath the outer
annotations, and its reduct erases to $s$. The converse holds by inspection of
the rules, since pushing an annotation leaves the erasure unchanged.
\end{proof}

\begin{lemma}[Sort agreement in $\mathcal{K}_1$]\label{lem:mltt-k1-sorts}
Lemma~\ref{lem:mltt-k0-sorts} holds in $\mathcal{K}_1$.
\end{lemma}

\begin{proof}
The proof of Lemma~\ref{lem:mltt-k0-sorts} carries over once we allow an
application step in an observation to check its argument against a type below
the domain and add the $\Pi$-case to the passage of an observation across
subsumption. Specifically, an observation at $\Pi x{:}A_2.\,B_2$ whose argument
$c$ checks against a type $A'$ below $A_2$ transfers to $\Pi x{:}A_1.\,B_1$
(with $A_1 \defeq_K A_2$) at the same argument $c$, since $A'$ also lies below
$A_1$; the remainder of its path transfers from $B_2$ to $B_1$ at $c$ by the
codomain premise instantiated with $\tttann{\tttann{c}{A'}}{A_2}$, by induction
on the path length. For part~(2), an observation may end at a $\Pi$-subsumption
step: the two $\Pi$-types are formed at universes that evaluate to $\lzero$
exactly where the universes of their codomains do, and those codomains are
compared under $x{:}A_2$, where $B_1$ is formed at its own universe by context
conversion. For joinable types, the reductions lift by
Lemma~\ref{lem:mltt-k1-annotated}.
\end{proof}

Consequently, the sorts of the stratified model---read off derivations in
$\mathcal{K}_1$ in the same way---still depend on a term only through its
erasure up to level equivalence and on the context only up to joinability, and
are preserved by annotated reduction in $\mathcal{K}_1$.

\paragraph{The $\Pi$-case in the stratified model.} Because the two domains in a
$\Pi$-subsumption step are joinable well-formed types,
Lemma~\ref{lem:mltt-k0-strat}(3) ensures that they denote the same semantic type
$A$. Any function code in $\widehat\Pi_{\mathrm{d}}(A, F)$, which records $A$,
is therefore also a code in $\widehat\Pi_{\mathrm{d}}(A, F')$ whenever
$\mathrm{El}\,F(x) \subseteq \mathrm{El}\,F'(x)$ for all $x \in \mathrm{El}(A)$.
Moreover, the two $\Pi$-types are either both propositions or both data types,
because subsumption relates a proposition only to another proposition: the base
cases of $\subty_K$ relate joinable types (which are propositions together by
Lemma~\ref{lem:mltt-k1-sorts}) or two universes (neither of which is a
proposition), and a $\Pi$-type is a proposition iff its codomain is. When both
$\Pi$-types are propositions, $\emptyset$ belongs to the first only if it
belongs to every $F(x)$, and hence to every $F'(x)$. Every subsumption between
types therefore denotes a set inclusion of their interpretations, and no
coercion enters the semantics.

\begin{theorem}[Consistency of $\mathcal{K}_1$]\label{thm:mltt-k1-consistency}
Theorem~\ref{thm:mltt-k0-consistency} holds for $\mathcal{K}_1$: let
$\tttEnv$ be as there, with $\mathcal{K}_1$ in place of $\mathcal{K}_0$. If
$\ctx \vdash a \chk A$ or $\ctx \vdash a \infr A$ in $\mathcal{K}_1$, in a
context whose entries are types, and $\rho$ is an environment of $\ctx$, then
in the stratified model $\llbracket a \rrbracket_\ctx\rho$ and
$\llbracket A \rrbracket_\ctx\rho$ are defined and
$\llbracket a \rrbracket_\ctx\rho \in \mathrm{El}\,\llbracket A \rrbracket_\ctx\rho$,
and a subsumption between types is an inclusion of their interpretations. In
particular, no closed term checks against $\tttFalse$.
\end{theorem}

\begin{proof}
The induction of Theorem~\ref{thm:mltt-k0-consistency} carries over with sorts
read off derivations in $\mathcal{K}_1$, using Lemma~\ref{lem:mltt-k0-strat} for
$\mathcal{K}_1$ (via Lemmas~\ref{lem:mltt-k1-annotated}
and~\ref{lem:mltt-k1-sorts}), noting that pushing an annotation preserves
semantic values because an annotation denotes the value of its underlying term,
and handling the $\Pi$-case of subsumption by the preceding paragraph. The
induction hypothesis for an inferring $\Pi$-type is strengthened, just like that
for a $\Sigma$-type, to state that its domain and codomain denote propositions
whenever they are propositions, ensuring that in the $\Pi$-case of subsumption
the codomain premise is interpreted in a valid environment of the context
extended by the domain.
\end{proof}

\paragraph{From $\mathcal{K}_1$ to the declarative theory.} The declarative
system of this chapter and $\mathcal{K}_1$ share the same definitional equality
(Definition~\ref{def:mltt-defeq}) and differ only in presentation: the
declarative rule \tttrule{Sub} can apply at any position and the subsumption
judgment includes \tttrule{S-Trans}, whereas $\mathcal{K}_1$ restricts
subsumption to checking positions via a syntax-directed relation and checks any
abstraction whose body does not infer against an expected type supplied by an
annotation. Because the validity of the stratified model is proved by induction
on derivations in $\mathcal{K}_1$---whose annotations supply the types of
application heads and checked abstractions---we translate declarative
derivations into $\mathcal{K}_1$ rather than interpreting them directly; since a
term and its annotated counterpart share the same erasure, they have the same
semantic value.

\begin{proposition}[$\mathcal{K}_1$ is complete for the declarative system]
\label{prop:mltt-k1-complete}
Let the declared types of $\tttEnv$ be types of $\mathcal{K}_1$ in the empty
context. If $\ctx \vdash a : A$ in the declarative system, then there are a
context $\ctx'$ and terms $a'$ and $A'$, obtained from $\ctx$, $a$, and $A$ by
inserting annotations, with $|a'| = |a|$, $|A'| = |A|$, $a \defeq a'$,
$A \defeq A'$, and $\ctx' \vdash a' \chk A'$ in $\mathcal{K}_1$; $\ctx'$ may
be any annotation of $\ctx$ whose entries are types of $\mathcal{K}_1$, and
one exists. If $\ctx \vdash A \subty B$, then $\ctx' \vdash A' \subty_K B'$
for terms $A'$ and $B'$ obtained from $A$ and $B$ in the same way, which are
types of $\mathcal{K}_1$.
\end{proposition}

\begin{proof}
First, $\subty_K$ contains declarative subsumption up to annotation:
\tttrule{S-Conv} and \tttrule{S-Cum} are its base cases, \tttrule{S-$\Pi$} is
its $\Pi$-case (with the typing premises translated into the well-formedness
derivations of the reducts), and \tttrule{S-Trans} follows from
Lemma~\ref{lem:mltt-k1-trans} because the two premises translate the middle type
into definitionally equal annotations and $\subty_K$ composes with $\defeq_K$ on
either side (Lemma~\ref{lem:mltt-k0-annotated}). Moreover, checking in
$\mathcal{K}_1$ is closed under $\subty_K$ up to annotation: if
$\ctx \vdash a' \chk A$ and $A \subty_K B$ for a type $B$, then
$\ctx \vdash \tttann{a'}{A} \chk B$ by \tttrule{Inf-Ann} and \tttrule{Chk-Sub}.

The translation proceeds by induction on the declarative derivation,
quantifying simultaneously over all annotations $\ctx'$ of $\ctx$ whose entries
are types of $\mathcal{K}_1$ so that the premises of a rule translate into a
common context; such a context annotation exists by the same induction. Types
are translated alongside terms: the conclusion type of each rule is formed from
the translated premises, while \tttrule{Conv} and \tttrule{Sub} obtain the
annotation $B'$ of the target type from the translation of their premise
$\ctx \vdash B : \Srt\,\ell$, which is a type of $\mathcal{K}_1$ because any
term checking against a universe infers a universe. Specifically:
\begin{itemize}
\item \tttrule{Var}, \tttrule{Sort}, \tttrule{Const}, and the type-formation
  rules translate directly to their inferring counterparts with checked
  premises;
\item \tttrule{Conv} is handled by Lemma~\ref{lem:mltt-k0-annotated}(1) at the
  annotation $B'$ of $B$, and \tttrule{Sub} by closure under $\subty_K$;
\item \tttrule{$\Pi$-I} translates to \tttrule{Inf-Lam} when the translated body
  infers, and to \tttrule{Chk-Lam} at the translated conclusion type otherwise;
\item for \tttrule{$\Pi$-E}, the translated head $f'$ checks against the
  translation $\Pi x{:}A'.\,B'$ of its $\Pi$-type, so
  $\tttann{f'}{\Pi x{:}A'.\,B'}$ infers that type by \tttrule{Inf-Ann} and the
  application $\tttann{f'}{\Pi x{:}A'.\,B'}\,a'$ infers $B'[a'/x]$ by
  \tttrule{Inf-App};
\item the remaining elimination rules translate to inferring rules with checked
  premises and similarly annotated heads, while \tttrule{Id-I} translates to
  \tttrule{Chk-Refl}.
\end{itemize}
Because inserting annotations is the only modification to the term and type, we
have $|a'| = |a|$ and $|A'| = |A|$, and therefore $a \defeq a'$ and
$A \defeq A'$ by \tttrule{Ann-$\tttstep$}.
\end{proof}

The hypothesis of Proposition~\ref{prop:mltt-k1-complete} on the global
environment holds automatically for every environment accepted by the kernel,
which infers a universe for each declared type and checks each definition body
against its declared type. An arbitrary well-formed declarative environment
(Definition~\ref{def:mltt-environment}) is brought into this form by the
following lemma.

\begin{lemma}[Annotation of an environment]\label{lem:mltt-env-annotation}
Let $\tttEnv$ be a well-formed environment. There is an environment
$\tttEnv'$ of the same constants, whose declared types and bodies are obtained
from those of $\tttEnv$ by inserting annotations, such that
\begin{enumerate}
\item every declared type of $\tttEnv'$ is a type of $\mathcal{K}_1$ in the
  empty context, and every body checks against its type in $\mathcal{K}_1$,
  each over the declarations before it;
\item definitional equality over $\tttEnv'$ is definitional equality over
  $\tttEnv$;
\item every judgment derivable over $\tttEnv$ is derivable over $\tttEnv'$,
  and $\tttEnv'$ is well formed;
\item $\tttEnv'$ meets its obligation (Definition~\ref{def:mltt-obligation})
  if and only if $\tttEnv$ does, and the two environments have the same
  model.
\end{enumerate}
\end{lemma}

\begin{proof}
By induction on the derivation of $\tttEnv \tttwf$, translating the declarations
in order. Let $\tttEnv = \tttEnv_0, d$, with $\tttEnv_0'$ given by the induction
hypothesis.

(1) We consider the three declaration forms:
\begin{itemize}
\item For \tttrule{E-Ax}, the premise
  $\tttEnv_0; \cdot \vdash_{\bar u} A : \Srt\,\ell$ holds over $\tttEnv_0'$
  by~(3). Applying Proposition~\ref{prop:mltt-k1-complete} over $\tttEnv_0'$
  (whose declared types are types of $\mathcal{K}_1$ by~(1) of the induction
  hypothesis) yields an annotation $A'$ of $A$ that checks against a universe,
  and therefore infers one: \tttrule{Chk-Sub} is the only rule that checks
  against a universe, and any type $\subty_K$-below a universe is itself a
  universe.
\item For \tttrule{E-Def}, the body $t$ is translated in the same way against an
  annotation $A''$ of $A$, and then checks against $A'$ by
  Lemma~\ref{lem:mltt-k0-annotated}(1) since $A'' \defeq A'$.
\item For \tttrule{E-Ind}, the entries of the parameter, index, and constructor
  telescopes and the index terms are translated by
  Proposition~\ref{prop:mltt-k1-complete} over the preceding translated
  declarations, each against the translation of its type; the head $D$ of each
  recursive argument type and of each constructor's result is left unannotated,
  since \tttrule{Const} infers its declared type and \tttrule{Inf-App} applies
  directly to it. Because every component of the translated declaration has the
  same erasure as the original, the declaration remains strictly positive by
  Definition~\ref{def:mltt-positivity} (which inspects annotations through their
  underlying terms). The remaining conditions of
  Definition~\ref{def:mltt-admissible} and the elimination restriction of
  Definition~\ref{def:mltt-recursor} consist of typing judgments---which hold
  over the translated declarations by~(3) for $\tttEnv_0$ together with
  \tttrule{Conv} at $D$---and variable-occurrence conditions in the result
  indices, which inserting annotations cannot disturb. The declared types of the
  inductive family and its constructors are types of $\mathcal{K}_1$: the
  family's type just as for \tttrule{E-Ax}, and each constructor's type by
  \tttrule{Const} and \tttrule{Inf-App} over its translated telescope. Finally,
  the type of the recursor is a type of $\mathcal{K}_1$ by following the
  derivation of Definition~\ref{def:mltt-recursor} while tracking inferred types
  up to erasure: because \tttrule{Inf-App} substitutes annotated arguments, at
  each application the inferred type of the head is well formed, its reduct
  beneath outermost annotations is a $\Pi$-type, and the argument checks against
  the domain by \tttrule{Chk-Sub} since the two erasures coincide.
\end{itemize}

(2) A $\delta$-step unfolds $\tttinst{c}{\bar\ell}$ to $t[\bar\ell/\bar u]$ over
$\tttEnv$ and to $t'[\bar\ell/\bar u]$ over $\tttEnv'$, and both reduce via
\tttrule{Ann-$\tttstep$} to $|t|[\bar\ell/\bar u]$ because erasure commutes with
level substitution; all other reduction rules are identical in both
environments. Thus a one-step reduction over either environment is a joinability
over the other, and since joinability over each environment is a congruence
equivalence (Lemma~\ref{lem:mltt-confluence}, whose hypotheses $\tttEnv'$
satisfies because its definition bodies are closed), the two definitional
equalities coincide.

(3) By induction on the derivation over $\tttEnv$. Rule~\tttrule{Const} reads
$A'$ in place of $A$; we have $A'[\bar\ell/\bar u] \defeq A[\bar\ell/\bar u]$,
and the latter is a well-formed type over $\tttEnv'$ by~(3) for $\tttEnv_0$
together with Lemma~\ref{lem:mltt-env-weakening}, level instantiation, and
weakening, so \tttrule{Conv} restores the original type. Rules~\tttrule{Conv}
and \tttrule{S-Conv} consult $\defeq$, which is the same relation by~(2), and no
other rule consults the environment. Lastly, $\tttEnv'$ is well formed because
every $\mathcal{K}_1$ derivation in a well-formed context is also a declarative
derivation (Section~\ref{sec:mltt-pi-sigma}), typing $A'$ and $t'$ over the
preceding translated declarations.

(4) Because a term and its erasure have the same semantic value, the axiom
types, definition bodies, and constructor types denote the exact same sets over
$\tttEnv$ and $\tttEnv'$; both the environment obligation and the model depend
on the declarations solely through these semantic values.
\end{proof}

\begin{theorem}[Consistency of the declarative theory]
\label{thm:mltt-declarative-consistency}
Under the hypotheses of Theorem~\ref{thm:mltt-k1-consistency}, no closed term
inhabits $\tttFalse$ in the declarative system of this chapter.
\end{theorem}

\begin{proof}
Lemma~\ref{lem:mltt-env-annotation} translates $\tttEnv$ into an annotated
environment $\tttEnv'$ that is well formed, meets its obligation, has the same
model, and over which $\cdot \vdash a : \tttFalse$ remains derivable.
Proposition~\ref{prop:mltt-k1-complete} over $\tttEnv'$ then yields
$\cdot \vdash a' \chk \tttFalse'$ in $\mathcal{K}_1$ with
$|\tttFalse'| = |\tttFalse| = \tttFalse$. Since a term and its erasure have the
same semantic value, Theorem~\ref{thm:mltt-k1-consistency} over $\tttEnv'$ gives
$\llbracket a' \rrbracket \in \mathrm{El}\,\llbracket \tttFalse \rrbracket
= \emptyset$, a contradiction.
\end{proof}

Theorem~\ref{thm:mltt-declarative-consistency} establishes
Conjecture~\ref{conj:mltt-consistency} for every environment that meets its
axiom obligation. Any environment whose declarations the kernel has already
checked in $\mathcal{K}_1$ (each over the preceding declarations) is its own
annotation, so the theorem applies to it directly without invoking
Lemma~\ref{lem:mltt-env-annotation}. This model construction relies on
abstractions carrying their domains in the term syntax (as noted in the remark
after Theorem~\ref{thm:mltt-k0-consistency}); Lean's raw kernel terms omit
abstraction domains in conversion, and Volume~II records the corresponding
kernel adjustment. The alternative partial-equivalence-relation model sketched
below interprets terms as untyped realizers and does not require domain
annotations on abstractions.

\paragraph{Remark: typed equality with proof irrelevance in conversion.} When
definitional equality is a typed judgment that includes proof irrelevance as a
conversion rule---as in Lean's kernel theory---establishing its equivalence with
untyped conversion is substantially harder: a proof via subject reduction
requires the typed equality to be injective on $\Pi$-types and to separate
distinct type formers, and both properties remain open in that setting.
Injectivity of $\Pi$ is ordinarily derived from normalization or from a logical
relation for conversion~\citep{abel-ohman-vezzosi2018}, but any theory combining
impredicative $\Prop$, definitional proof irrelevance, and large elimination of
$\Id$ admits the non-normalizing counterexample of
Theorem~\ref{thm:mltt-failure}(1), closing off both routes.
\citet[Section~4]{carneiro2019} proposes a stratification argument for Lean's
theory that avoids normalization: definitional equality is factored into a
reduction $\tttstep_\kappa$ (equipped with a $K$-like rule for subsingleton
eliminators) and an equivalence $\defeq_p$ (collecting proof irrelevance,
universe cumulativity, and $\eta$), so that $\tttstep_\kappa$ is Church--Rosser
modulo $\defeq_p$. The desired properties are then argued by induction on the
alternation depth $n$ between typing and conversion in a derivation, where
$\vdash_n$ denotes typing with conversions at stratum $n$ and $\defeq_{n+1}$
denotes equality whose typing premises lie at stratum $n$. However, the
substitution lemmas in that argument (Lemmas~4.8(4) and~4.6(4) of
\citealp{carneiro2019}) require $\vdash_n$ to be closed under substituting terms
typed at stratum $n$, whereas it is not: substituting a stratum-$n$ term into
the typing premises of a stratum-$n$ conversion raises that conversion to
stratum $n+1$. The following concrete instance, written in the notation of this
chapter extended with the conversion rule $p \defeq q$ for proofs, refutes the
triangle lemma of that argument (Lemma~4.10 of \citealp{carneiro2019}) at the
second stratum (and transfers immediately to Lean's theory with $\mathsf{Eq}$ in
place of $\Id$). Let
\[
\begin{gathered}
\ctx = A{:}\Prop,\ a{:}A,\ z{:}(\tttlam{X}{\Prop} X)\,A,\ H{:}\Pi x{:}A.\,\Id\,A\,a\,x, \\
C = \tttlam{y}{A}\tttlam{h}{\Id\,A\,a\,y} \Nat,
\qquad
t = (\tttlam{x}{A} \Jelim\,a\,C\,\tttzero\,x\,(H\,x))\,z .
\end{gathered}
\]
Here $\ctx \vdash_1 z : A$ holds by \tttrule{Conv} using the $\beta$-equation
$(\tttlam{X}{\Prop} X)\,A \defeq_1 A$ (whose premises lie at stratum~0), whereas
$z : A$ is not derivable at stratum~0. Consequently, $a \defeq_1 z : A$ is not
derivable because applying proof-irrelevance conversion at stratum~1 requires
$z : A$ at stratum~0, even though $a \defeq_2 z : A$ holds. In the extended
context $\ctx, x{:}A$, the $K$-like rule fires on
$\Jelim\,a\,C\,\tttzero\,x\,(H\,x)$ because $x \defeq_1 a : A$ has premises at
stratum~0; thus the complete development of $t$ is $\tttzero$, while $t$ reduces
by $\beta$ alone to $t' = \Jelim\,a\,C\,\tttzero\,z\,(H\,z)$ with $t : \Nat$ at
stratum~1. The triangle lemma requires $t' \tttstep_\kappa t'' \defeq_p \tttzero$
for some $t''$. Yet the $K$-like rule cannot fire on $t'$ because its side
condition $z \defeq_1 a : A$ fails; no other reduction rule applies to $t'$; and
$t' \defeq_p \tttzero$ fails because $\Nat$ is not a proposition and neither
side is an abstraction. Accordingly, the Lean~4 mechanization of the metatheory
states $\Pi$-injectivity and type-former separation as axioms and leaves them
unproved, with its experimental directory exploring sixteen alternative files
without a complete proof~\citep{lean4lean}. As \citet[Section~5]{lee-werner2011}
note following \citet{adams2006}, the equivalence of typed and untyped
conversion---from which these properties would follow--- remains open for
theories with cumulative universes. The theory of this chapter avoids the entire
impasse by taking untyped joinability as its definition of conversion, which the
kernel decides directly; the equivalence question would arise only if one sought
a model of typed equality, which the stratified model does not require.

\paragraph{Alternative: a model of partial equivalence relations.} In a
partial-equivalence-relation (PER) model, each type denotes a PER on a universal
domain of untyped realizers, and each proposition denotes a PER with at most one
equivalence class, validating \tttrule{PI} and \tttrule{Unit-$\eta$} (whereas
bare untyped realizers without PERs do not, since distinct proofs can have
distinct realizers). Subsumption is interpreted as inclusion of PERs, which
validates \tttrule{S-$\Pi$}: whenever $B \subty B'$, two realizers related at
$\Pi x{:}A.\,B$ are also related at $\Pi x{:}A.\,B'$, and because the PER
definition quantifies over the domain in a negative position, it would validate
contravariant domain subtyping as well. Because realizers are untyped and the
realizer of $\tttlam{x}{A} b$ is an untyped abstraction that ignores $A$,
declarative derivations can be interpreted directly without prior annotation.
Models of this kind have been constructed for the Implicit Calculus of
Constructions, which features cumulative universes and domain-contravariant
$\Pi$ subtyping~\citep{miquel2001,miquel2000}. Moreover, the realizer of a term
is defined by recursion on the term syntax, so the interpretation of a judgment
is independent of its derivation; and because a product PER is full or empty
whenever its codomain PERs are, the universe at which a product is formed does
not enter its interpretation. Finally, untyped realizers validate every
reduction rule without typing side conditions, so assigning PERs to type codes
up to joinability interprets untyped conversion directly. Large elimination from
$\Prop$ can be interpreted similarly to the stratified model: an inductive
family in $\Prop$ denotes the full or empty relation according to whether its
index tuple belongs to the least fixed point of its rule set on truth values,
$\Jelim\,a\,C\,d\,b\,p$ denotes the realizer of $d$, and the eliminator of a
subsingleton with recursive arguments is interpreted via a fixed-point
combinator in the realizer domain, typed by induction on the ordinal stage at
which the index tuple enters the fixed point. The remaining steps not carried
out along this route are the construction of the universe hierarchy as PERs on
type codes with cumulativity as inclusion---along the lines of Luo's
$\omega$-set model of ECC~\citep{luo1990} and Miquel's model---and the
domain-specific base cases that Chapter~\ref{ch:knowledge-graphs} adds to the
subsumption judgment.

\paragraph{Alternative: translation into Lean.} A syntactic translation into
Lean's kernel theory can eliminate subsumption by inserting explicit coercions:
$\mathsf{ULift}$ for \tttrule{S-Cum}, and
$\tttlam{f}{\Pi x{:}A.\,B}\tttlam{x}{A'} \kappa_B\,(f\,x)$ for \tttrule{S-$\Pi$},
where $\kappa_B$ is the coercion for the codomain premise. Such a translation
maps $\tttFalse$ to $\tttFalse$ and closed terms to closed terms, reducing
Conjecture~\ref{conj:mltt-consistency} to Theorem~\ref{thm:mltt-consistency-lean}
provided the coercions satisfy \emph{coherence}: the translation of a judgment
must not depend on the chosen derivation, meaning that the coercions produced by
any two derivations of $A \subty B$ are definitionally equal and commute with
$\defeq$, preserving \tttrule{Conv}, \tttrule{S-Conv}, and \tttrule{Sub}. In the
framework of \citet{luo-soloviev-xue2013}, the subsumption judgment corresponds
to the coercive subtyping extension $T[C]$ of Lean's theory $T$ by the coercion
set $C$, and coherence of $C$ (their Definition~2.2) makes $T[C]$ a conservative
extension of $T$ (their Theorem~3.9). Two gaps remain on this route as a
consistency proof: first, their conservativity theorem is proved for theories
formulated in Luo's logical framework, which does not directly encompass Lean's
type theory; second, the theory of this chapter leaves terms unchanged under
subsumption, so erasing the inserted coercions must commute with conversion. In
particular, because $\mathsf{ULift}$ in Lean is a structure,
$\mathsf{ULift}\,(\mathsf{ULift}\,A)$ is isomorphic to $\mathsf{ULift}\,A$
without being definitionally equal to it, so a naive translation that composes
one $\mathsf{ULift}$ per step of \tttrule{S-Trans} is not coherent; instead, the
translation must insert a single lift from the inferred universe of a type to
the expected universe at Rule~\tttrule{Chk-Sub}. Volume~II constructs this
translation for the Lean~4 bridge, and coherence of the cumulativity fragment is
used in Chapter~\ref{ch:institutions}; both uses rely on coherence independently
of the consistency proof given above.

\subsection{Position among related theories}

Table~\ref{tab:mltt-comparison} lists the differences between the
theory of this chapter and the four theories it draws on. MLTT refers to the
presentation of~\citet{martinlof1984}, which has extensional identity types;
CIC refers to the theory of the Coq proof assistant; the Lean~4 column refers to
the kernel theory~\citep{lean4,carneiro2024}.

\begin{table}[t]
\centering\footnotesize
\setlength{\tabcolsep}{4pt}
\begin{tabularx}{\textwidth}{@{}>{\raggedright\arraybackslash}p{0.17\textwidth}
  *{5}{>{\raggedright\arraybackslash}X}@{}}
\toprule
& MLTT (1984) & ECC & CIC (Coq) & Lean~4 kernel & This chapter \\
\midrule
Impredicative $\Prop$ & no & yes & yes & yes & yes \\
Cumulativity & yes ($U_n$) & yes, $\Prop \preceq \Typeu_0$ & yes & no & yes, from $\Setu$ upward \\
Structural subsumption & none & $\Pi$ codomain; $\Sigma$ both & $\Pi$ codomain; cumulative
inductives & none & $\Pi$ codomain; no $\Sigma$ \\
Universe polymorphism & no & no & yes & yes & yes \\
Inductive families & fixed set formers, W-types & no & yes & yes & yes, strictly positive, not
nested \\
Identity type & primitive, extensional & none (Leibniz equality) & inductive, in $\Prop$ &
inductive, in $\Prop$ & primitive, in $\Prop$ \\
Abstraction & unannotated & carries its domain & carries its domain & carries its domain & carries
its domain \\
Definitional equality & typed judgment & untyped conversion & untyped conversion & typed judgment &
untyped conversion \\
Proof irrelevance & no & no & definitional, for $\mathsf{SProp}$ only & definitional &
propositional, the axiom $\irrel : \Id\,P\,p\,q$, so that proofs stay distinct terms \\
$\eta$ & derivable (extensional) & no & $\Pi$; records & $\Pi$; structures & none \\
Quotients & no & no & no & primitive & no \\
\bottomrule
\end{tabularx}
\caption{Differences between the theory of this chapter and the
theories it draws on.}
\label{tab:mltt-comparison}
\end{table}

\section{Scope}
\label{sec:mltt-scope}

The theory of this chapter is the logical core of the language the kernel
decides. Table~\ref{tab:mltt-scope} lists the constructs that the full language
adds and the chapters that treat them. Each addition is either a conservative
extension by new constants and rules whose metatheory is stated where the
construct is introduced, or a construct of the program layer outside the
type theory.

\begin{table}[h]
\centering\small
\begin{tabularx}{\textwidth}{@{}>{\raggedright\arraybackslash}p{0.28\textwidth}
  >{\raggedright\arraybackslash}X >{\raggedright\arraybackslash}p{0.22\textwidth}@{}}
\toprule
Left out & Constructs & Treated in \\
\midrule
Program constructs & component calls, mapping and reduction combinators,
  templates, institution invocation & Volume~IV (ESL, the runtime) \\
Knowledge-graph types & record types, refinement types, class types,
  primitive types and literals, property access, record construction,
  native decision procedures, apartness & Chapter~\ref{ch:knowledge-graphs} \\
Institution-facing references & axioms anchored on the chain, references to
  externally checked propositions & Chapter~\ref{ch:institutions}; Volume~II \\
General recursion & recursive definitions not reducible to recursors &
  excluded from the theory \\
Coercive subtyping with computing coercions & dot-types, changes of
  representation & Volume~III, as an elaboration-level extension \\
\bottomrule
\end{tabularx}
\caption{Constructs outside the theory of this chapter.}
\label{tab:mltt-scope}
\end{table}

\chapter{Justification Logic}
\label{ch:justification-logic}

\providecommand{\jlcs}{\mathit{cs}}
\providecommand{\jlCS}{\mathit{CS}}
\providecommand{\jlLeaf}{\mathsf{Leaf}}
\providecommand{\jlName}{\mathsf{Name}}
\providecommand{\jlCode}{\mathsf{Code}}
\providecommand{\jlq}[1]{\ulcorner #1\urcorner}
\providecommand{\jlres}{\mathrm{res}}
\providecommand{\jlleaves}{\mathrm{leaves}}
\providecommand{\jlkind}{\mathrm{kind}}
\providecommand{\jlsuppz}{\supp_{0}}
\providecommand{\jlWhy}{\mathrm{Why}}
\providecommand{\jlPos}{\mathrm{PosBool}}
\providecommand{\jlLin}{\mathrm{Lin}}
\providecommand{\jlNX}{\mathbb{N}[X]}
\providecommand{\jlprov}{\mathrm{prov}}
\providecommand{\jlMin}{\mathrm{Min}}
\providecommand{\jlDer}{\mathrm{Der}}
\providecommand{\jlWF}{\mathrm{WF}}
\providecommand{\jlchk}{\mathord{!}}
\providecommand{\jlfv}{\mathsf{fully\_verified}}
\providecommand{\jlsw}{\mathsf{survives\_without}}
\providecommand{\jllo}{\mathsf{leaves\_of}}
\providecommand{\jlcited}{\mathsf{cited}}
\providecommand{\jlclaims}{\mathsf{claims}}
\providecommand{\jlG}{\mathsf{G}}
\providecommand{\jlimp}{\rightarrow}
\providecommand{\jlvd}{\vdash_{\jlcs}}
\providecommand{\tttFalse}{\mathsf{False}}
\providecommand{\tttBool}{\mathsf{Bool}}

This chapter treats justification logic as the theory of \emph{grounds}: the record of what a
claim rests on, kept formally separate from \emph{proofs}, which establish the claim. The first
part (Sections~\ref{sec:jl-explicit} and~\ref{sec:jl-basic}) states the standard systems and their
semantics. The second part (Sections~\ref{sec:jl-proofs-grounds}--\ref{sec:jl-wellfounded}) defines
the Eigenius fragment: an untyped algebra of justification terms $\JTerm$ in $\Setu$, a
proposition $\JB(j, P)$ decided by a computable evidence function $\Evid$, and the support of a
term relative to a proposition. The third part (Sections~\ref{sec:jl-logic}--\ref{sec:jl-checker})
places the fragment among the logics of Chapter~\ref{ch:institutions}, analyses the Curry-style
and Church-style presentations of justification terms, and states the extension by the checker
operator~$\jlchk$. The type theory is that of Chapter~\ref{ch:mltt}; the implementation is treated
in Volume~II.

\section{From knowledge modalities to explicit justifications}
\label{sec:jl-explicit}

Justification logic replaces the modality $\Box F$, read ``$F$ is known'' or ``$F$ is provable'',
by formulas $t : F$, read ``$t$ is a justification for $F$''
\citep{artemov2008justification,artemov-fitting2019}. The
justification $t$ is a term of the object language, built from constants and variables by a fixed
set of operations. The expression $t : F$ is a formula: it can occur under negation
($\neg\, t : F$, ``$t$ does not justify $F$''), as a hypothesis ($t : F \jlimp G$), and nested
inside another justification ($s : (t : F)$).

\paragraph{The Logic of Proofs and realization.}
The first justification logic is Artemov's Logic of Proofs $\mathsf{LP}$, introduced to give
the modal logic $\mathsf{S4}$ a provability semantics in the sense of G\"odel's 1933 program: $t
: F$ is read as ``$t$ is a proof of $F$'' in an arithmetical proof predicate~\citep{artemov2001}.
The connection between
$\mathsf{LP}$ and $\mathsf{S4}$ is the \emph{realization theorem}: the forgetful projection that
replaces every term by $\Box$ maps the theorems of $\mathsf{LP}$ onto the theorems of
$\mathsf{S4}$, and conversely every theorem of $\mathsf{S4}$ has a \emph{realization}---an
assignment of justification terms to its occurrences of $\Box$ that yields a theorem of
$\mathsf{LP}$~\citep{artemov2001,artemov2008justification}.
Analogous realization theorems hold for the pairs $\mathsf{J}/\mathsf{K}$,
$\mathsf{J4}/\mathsf{K4}$, and $\mathsf{JT}/\mathsf{T}$ of
Section~\ref{sec:jl-basic}~\citep{artemov-fitting2019}. The Eigenius fragment is not a realization
of a modal logic; instead, it uses justification logic directly as an explicit calculus of grounds.

\paragraph{Terms and grades.}
A \emph{grade} summarizes the evidence for a claim by mapping it to a finite ordered set, such as
$\{\mathit{Declared} < \mathit{Verified}\}$. While a stored scalar grade answers one fixed
ordering question, it forgets the individual reasons on which the claim rests. By contrast, a
justification term is an element of a free algebra over its leaves.
Section~\ref{sec:jl-support} shows that every grade is a homomorphic image of the support of a
term (Theorem~\ref{thm:jl-lattice-hom}), whereas a grade alone cannot answer counterfactual
queries of the form ``does the claim still stand without the resource $r$?''
(Proposition~\ref{thm:jl-grade-loses}). The system therefore retains the justification term and
computes grades from it on demand.

\section{The basic systems}
\label{sec:jl-basic}

\begin{definition}[Language]\label{def:jl-language}
The \emph{justification terms} of the basic systems are generated by
\[
  t \;::=\; c \mid x \mid t \cdot t \mid t + t \mid \jlchk t
\]
where $c$ ranges over a countable set of \emph{justification constants} and $x$ over a countable
set of \emph{justification variables}. The \emph{formulas} are generated by
\[
  F \;::=\; p \mid \bot \mid F \jlimp F \mid t : F
\]
where $p$ ranges over propositional letters. The remaining boolean connectives are defined as usual.
The checker operation $\jlchk$ belongs to the language of $\mathsf{J4}$ and $\mathsf{LP}$ only.
\end{definition}

\begin{definition}[Axioms]\label{def:jl-axioms}
The axiom schemata and inference rules of the basic systems are:

\begin{center}
\small
\begin{tabular}{@{}lll@{}}
  \toprule
  \textbf{Name} & \textbf{Schema} & \\
  \midrule
  $\mathrm{A0}$ & all instances of classical propositional tautologies & \\
  Application & $s : (F \jlimp G) \jlimp (t : F \jlimp s \cdot t : G)$ & \\
  Sum & $s : F \jlimp (s + t) : F$, \quad $t : F \jlimp (s + t) : F$ & \\
  Positive introspection & $t : F \jlimp \jlchk t : (t : F)$ & \\
  Factivity & $t : F \jlimp F$ & \\
  \midrule
  Modus ponens & from $F$ and $F \jlimp G$ infer $G$ & \\
  Axiom necessitation & infer $c : A$ for $c : A \in \jlCS$ & \\
  \bottomrule
\end{tabular}
\end{center}
\end{definition}

\begin{definition}[The hierarchy]\label{def:jl-hierarchy}
$\mathsf{J}_0$ has $\mathrm{A0}$, Application, Sum, and modus ponens. For a constant
specification $\jlCS$ (Definition~\ref{def:jl-cs}), $\mathsf{J}_{\jlCS}$ adds axiom necessitation
for $\jlCS$. $\mathsf{J4}_{\jlCS}$ adds positive introspection, $\mathsf{JT}_{\jlCS}$ adds
factivity, and $\mathsf{LP}_{\jlCS} = \mathsf{J4}_{\jlCS} + \text{Factivity}$. The systems
$\mathsf{J}$,
$\mathsf{J4}$, $\mathsf{JT}$, and $\mathsf{LP}$ are the instances with the total constant
specification.
\end{definition}

\begin{definition}[Constant specification]\label{def:jl-cs}
A \emph{constant specification} for a system $\mathsf{L}$ is a set $\jlCS$ of formulas
$c_n : c_{n-1} : \cdots : c_1 : A$, with $n \geq 1$, $A$ an axiom of $\mathsf{L}$, and each $c_i$ a
constant, that is closed under taking suffixes: if $c_n : c_{n-1} : \cdots : c_1 : A \in \jlCS$
then $c_{n-1} : \cdots : c_1 : A \in \jlCS$. It is \emph{axiomatically appropriate} if every
axiom $A$, and every formula in $\jlCS$, is justified by some constant in $\jlCS$. It is
\emph{total} if it contains $c : A$ for every constant $c$ and every axiom~$A$. A formula $c : A
\in \jlCS$ is \emph{self-referential} if $c$ occurs in $A$.
\end{definition}

The constant specification records which constants justify which axioms; constants are
otherwise uninterpreted. Two metatheoretic properties depend on it:

\begin{itemize}
\item \textbf{Internalization.} If $\jlCS$ is axiomatically appropriate and
  $\mathsf{J}_{\jlCS} \vdash F$, then $\mathsf{J}_{\jlCS} \vdash t : F$ for some term $t$ built
  from constants by application and sum; the analogous statement holds for the other systems
  \citep{artemov2008justification,artemov-fitting2019}.
  Internalization is the justification-logic counterpart of the necessitation rule of modal logic.
\item \textbf{Self-reference.} Realizing $\mathsf{S4}$ in $\mathsf{LP}$ requires
  self-referential constant specifications: some theorems of $\mathsf{S4}$ have no realization
  over any constant specification lacking formulas of the form $c : A(c)$~\citep{kuznets2010,artemov-fitting2019}.
\end{itemize}

\subsection{Semantics}
\label{sec:jl-semantics}

The semantics of justification logic is organized around the \emph{evidence function}, which
assigns to each justification term the set of formulas it justifies.

\begin{definition}[Evidence function]\label{def:jl-evidence-function}
An \emph{evidence function} is a map $\Evid$ from terms to sets of formulas such that, for all
terms $s, t$ and formulas $F, G$:
\begin{enumerate}[label=(E\arabic*)]
\item if $(F \jlimp G) \in \Evid(s)$ and $F \in \Evid(t)$ then $G \in \Evid(s \cdot t)$;
\item $\Evid(s) \cup \Evid(t) \subseteq \Evid(s + t)$;
\item for $\mathsf{J4}$ and $\mathsf{LP}$: if $F \in \Evid(t)$ then $(t : F) \in \Evid(\jlchk t)$.
\end{enumerate}
It \emph{respects} $\jlCS$ if $c : A \in \jlCS$ implies $A \in \Evid(c)$.
\end{definition}

\begin{definition}[Mkrtychev model]\label{def:jl-mkrtychev}
A \emph{Mkrtychev model} for $\mathsf{J}_{\jlCS}$ is a pair $\mathcal{M} = (V, \Evid)$ of a
valuation $V$ of the propositional letters and an evidence function $\Evid$ respecting $\jlCS$.
Truth commutes with the boolean connectives, and $\mathcal{M} \models t : F$ iff $F \in \Evid(t)$.
For $\mathsf{JT}_{\jlCS}$ and $\mathsf{LP}_{\jlCS}$ the model must in addition be
\emph{factive}: $F \in \Evid(t)$ implies $\mathcal{M} \models F$~\citep{mkrtychev1997}.
\end{definition}

\begin{definition}[Fitting model]\label{def:jl-fitting}
A \emph{Fitting model} is a tuple $(W, R, \Evid, V)$ of a Kripke frame $(W, R)$, a valuation $V$,
and an evidence function $\Evid(t, w)$ indexed by worlds, satisfying the closure conditions of
Definition~\ref{def:jl-evidence-function} at each world and monotone along $R$ whenever the system
contains positive introspection. Truth at a world is given by $w \models t : F$ iff
$F \in \Evid(t, w)$ and $v \models F$ for every $v$ with $w R v$~\citep{fitting2005}. A Mkrtychev
model is a single-world Fitting model (with the world related to itself for the
factive systems).
\end{definition}

\emph{Basic models}~\citep{artemov-fitting2019} take the evidence
function as the primary semantic datum and dispense with the Kripke structure. For $\mathsf{J}$
they consist of the same data as Mkrtychev models: a truth assignment to the propositional letters
together with an evidence function satisfying (E1) and (E2).

\begin{theorem}[Soundness and completeness, cited]\label{thm:jl-completeness-cited}
For each of $\mathsf{J}$, $\mathsf{J4}$, $\mathsf{JT}$, $\mathsf{LP}$ and each constant
specification $\jlCS$ in the class for which the cited results are stated, $\mathsf{L}_{\jlCS}
\vdash F$ iff $F$ holds in every Mkrtychev model for $\mathsf{L}_{\jlCS}$, iff $F$ holds in every
Fitting model for $\mathsf{L}_{\jlCS}$~\citep{mkrtychev1997,fitting2005,kuznets-studer2019}.
\end{theorem}

The Eigenius decision procedure (Section~\ref{sec:jl-evidence}) computes the least evidence
function respecting the leaf specification.

\begin{lemma}[Minimal evidence function]\label{thm:jl-minimal-evidence}
For each constant specification $\jlCS$ there is a least evidence function $\Evid_{\jlCS}$
respecting $\jlCS$, ordered pointwise by inclusion.
\end{lemma}

\begin{proof}
The pointwise intersection of any non-empty family of evidence functions respecting $\jlCS$
satisfies (E1)--(E3) and respects $\jlCS$, because each condition is a Horn closure condition.
The family is non-empty since the function assigning every formula to every term qualifies;
$\Evid_{\jlCS}$ is the intersection of the entire family.
\end{proof}

\begin{proposition}\label{thm:jl-minimal-characterizes}
Let $\mathsf{L}$ be $\mathsf{J}$ or $\mathsf{J4}$, and let $\jlCS$ be a constant specification
for which $\mathsf{L}_{\jlCS}$ is complete with respect to Mkrtychev models. Then
$\mathsf{L}_{\jlCS} \vdash t : F$ iff $F \in \Evid_{\jlCS}(t)$.
\end{proposition}

\begin{proof}
For any valuation $V$, the pair $(V, \Evid_{\jlCS})$ is a Mkrtychev model. If $\mathsf{L}_{\jlCS}
\vdash t : F$, soundness gives $(V, \Evid_{\jlCS}) \models t : F$, that is, $F \in \Evid_{\jlCS}(t)$.
Conversely, if $F \in \Evid_{\jlCS}(t)$, then $F \in \Evid(t)$ for every evidence function
$\Evid$ respecting $\jlCS$ by Lemma~\ref{thm:jl-minimal-evidence}, so $t : F$ holds in every
Mkrtychev model, and completeness gives $\mathsf{L}_{\jlCS} \vdash t : F$.
\end{proof}

\paragraph{Curry-style terms and the role of Sum.}
Terms of the basic systems carry no intrinsic formula type: a single term justifies a set of
formulas, and by (E2) we have $\Evid(s + t) \supseteq \Evid(s) \cup \Evid(t)$. Sum is thus the
operation of \emph{aggregation}, combining evidence for possibly different formulas into a single
term that justifies each of them. In particular, the realization procedure for $\mathsf{S4}$ uses
Sum to merge terms obtained for distinct occurrences of the same modal subformula into a single
term covering all of them~\citep{artemov2001}. Sum is \emph{monotone}: adding a summand never
removes a justified formula, and the second summand of $s + t$ need not justify anything at all.

\paragraph{Quantified justification logic.}
First-order extensions of $\mathsf{LP}$ and of justification
logic~\citep{artemov-yavorskaya2011,fitting2014} add first-order quantifiers to the
formula language and treat object variables in terms explicitly. Universal instantiation $\forall
x.\, A \jlimp A[a/x]$ is a logical axiom whose instances are justified by constants. The
Eigenius fragment instead internalizes instantiation as a term former $\jinst$ that records the
instantiating term (Definition~\ref{def:jl-jterm}), while omitting any term former for
generalization.

\section{Proofs and grounds are different objects}
\label{sec:jl-proofs-grounds}

A proof of $P$ and a record of grounds for $P$ differ in what they establish, in factivity, and
in transportability:

\begin{center}
\small
\begin{tabular}{@{}lll@{}}
  \toprule
  & \textbf{Proof} & \textbf{Record of grounds} \\
  \midrule
  Establishes & $t : P$, checked & that a claim to $P$ rests on given materials \\
  Factive & yes & no \\
  Transportable & yes & no \\
  \bottomrule
\end{tabular}
\end{center}

A proof is factive: a well-typed term $t : P$ entails that $P$ holds. A proof is also
transportable: any party holding $t$ and a checker for the logic reaches the same verdict. By
contrast, a record of grounds reports which materials a claim rests on; a party that does not
trust an author or an instrument re-evaluates those materials rather than inheriting a verdict.
Were both objects given the same representation, a derivation from ``these are the grounds for
$P$'' to ``$P$ is proved'' would become expressible. The Eigenius fragment therefore places proofs
and grounds at two distinct levels of the type theory and provides no coercion or elimination rule
from grounds to proofs.

\subsection{The two levels}

\begin{itemize}
\item \textbf{Propositions and proofs.} $P : \Prop$ is a proposition of the kernel type theory
  (Chapter~\ref{ch:mltt}), and a proof is a term $t$ with $\vdash t : P$. The universe $\Prop$ is
  proof-irrelevant: any two proofs of $P$ are provably identical (Section~\ref{sec:mltt-prop}),
  and no data computation depends on which proof is supplied. Every proof is factive.
\item \textbf{Justification terms.} $j : \JTerm$ is an element of an inductive data type in
  $\Setu$ (Definition~\ref{def:jl-jterm}). Because $\JTerm$ lives in $\Setu$, justification terms
  are data: two justification terms are equal only when they are the same term. Each leaf $\lambda$
  of $j$ names a resource of the chain and carries one of three leaf kinds---$\jdecl$ (an agent's
  assertion), $\jobs$ (a recorded observation), or $\jver$ (a checked proof)---and a chain-supplied
  constant specification $\jlcs$ assigns to each admitted leaf $\lambda$ the syntactic code
  $\jlq{P_\lambda}$ of the closed proposition $P_\lambda$ attached to its resource
  (Definition~\ref{def:jl-leaf-cs}).
\item \textbf{Justification assertions.} $\JB(j, P) : \Prop$ represents the formula $j : P$ of
  justification logic. It is a proposition \emph{about} $j$ and $P$, decided by testing whether
  $P$ belongs to the finite evidence set $\Evid(j)$ computed from $j$ and $\jlcs$
  (Section~\ref{sec:jl-evidence}); on an admitted leaf $\lambda$, $\Evid(\lambda) = \{P_\lambda\}$.
\end{itemize}

Proof irrelevance applies to the third bullet ($\JB(j, P) : \Prop$) but not to the second
($j : \JTerm$): any two proofs of $\JB(j, P)$ are provably identical, whereas the justification
term $j$ is a data index of the proposition, so distinct justification terms yield distinct
propositions.

\subsection{Codes and hyperintensionality}
\label{sec:jl-codes}

The second argument of $\JB$ is a syntactic code of a proposition rather than an inhabitant of
$\Prop$.

\begin{definition}[Codes]\label{def:jl-code}
$\jlCode : \Setu$ is the inductive type of labeled binary trees, with constructors
$\mathsf{leaf} : \Nat \to \jlCode$ and $\mathsf{node} : \Nat \to \jlCode \to \jlCode \to
\jlCode$, into which the syntax of closed terms of the kernel type theory is encoded (with each term
former assigned a numeric label). For a closed term $a$, we write $\jlq{a} : \jlCode$ for its code.
Codes are compared as data; the decision procedure of Section~\ref{sec:jl-evidence} decodes them and
compares the decoded propositions up to definitional equality $\defeq$. The predicate $\JB$ has type
$\JTerm \to \jlCode \to \Prop$, and we write $\JB(j, P)$ as an abbreviation for $\JB(j, \jlq{P})$.
\end{definition}

\begin{proposition}[Hyperintensionality]\label{thm:jl-hyperintensional}
There exist a constant specification, a term $j$, and closed propositions $P, Q$ over a kernel
environment $\Sigma$ such that $P \leftrightarrow Q$ is provable, $\JB(j, P)$ holds, and
$\JB(j, Q)$ fails. Consequently, in any type theory with propositional extensionality, no family
$\JB' : \JTerm \to \Prop \to \Prop$ indexed directly by propositions can be decided by membership in
$\Evid(j)$.
\end{proposition}

\begin{proof}
Let $P$ be a propositional constant of $\Sigma$, take a leaf $\lambda$ admitted by $\jlcs$ with
$\jlcs(\lambda) = \jlq{P}$, and set $Q = P \wedge P$. By Definition~\ref{def:jl-evidence},
$\Evid(\lambda) = \{P\}$, so $\JB(\lambda, P)$ holds whereas $\JB(\lambda, Q)$ fails because
$P \wedge P \not\defeq P$. Yet the logical equivalence $P \leftrightarrow P \wedge P$ is provable.
Under propositional extensionality, $P$ and $Q$ would be equal in $\Prop$, forcing
$\JB'(\lambda, P) = \JB'(\lambda, Q)$, so $\JB'$ could not coincide with membership in $\Evid$.
\end{proof}

Justification assertions are therefore hyperintensional in their proposition argument: they
respect definitional equality $\defeq$ but not logical equivalence. Indexing $\JB$ by syntactic
codes in $\jlCode$ keeps this hyperintensionality consistent with any extensionality principle
that might be assumed for $\Prop$. In fact, the types of both arguments of $\JB$ follow from the
central thesis of Chapter~\ref{ch:introduction}:
\begin{enumerate}
\item Because $\JB(j, P)$ is decided by inspecting the structure of $j$, two distinct
  justifications of the same proposition must remain distinguishable to the kernel. Were a
  justification represented as a proof of $P$, a theory with definitional proof irrelevance would
  identify $j_1 \defeq j_2$ for any $j_1, j_2 : P$ and hence collapse $\JB(j_1, P) \defeq
  \JB(j_2, P)$. Making the first argument an element of the inductive data type $\JTerm : \Setu$
  keeps distinct justification terms distinct. (Moreover, because the kernel's own proof
  irrelevance for $\Prop$ is propositional rather than definitional, the proofs stored by the
  system remain distinct terms as well; see Section~\ref{sec:mltt-prop} and
  Chapter~\ref{ch:institutions}.)
\item Because $\JB(j, P)$ distinguishes logically equivalent propositions that are not
  definitionally equal, a theory with propositional extensionality would collapse
  $\JB'(\lambda, P) = \JB'(\lambda, P \wedge P)$ if the second argument had type $\Prop$. Making
  the second argument a syntactic code $\jlq{P} : \jlCode$ insulates $\JB$ from extensionality in
  $\Prop$.
\end{enumerate}

\subsection{Non-factivity}
\label{sec:jl-nonfactivity}

\begin{theorem}[Non-factivity]\label{thm:jl-nonfactive}
Let $\jlcs$ be a constant specification (Definition~\ref{def:jl-leaf-cs}) with an admitted leaf
$\lambda$ whose proposition $P_0 = P_\lambda$ does not mention $\JB$ and is not provable without
$\JB$. Then the implication $\JB(\lambda, P_0) \jlimp P_0$ is not provable, and no closed term of
type $\Pi (j : \JTerm)(c : \jlCode).\, \JB(j, c) \jlimp \mathrm{Dec}(c)$ exists for any family
$\mathrm{Dec}$ satisfying $\mathrm{Dec}(\jlq{P_0}) \defeq P_0$.
\end{theorem}

\begin{proof}
Since $P_0 \in \Evid(\lambda)$, the assertion $\JB(\lambda, P_0)$ is provable from the decision
rules (Theorem~\ref{thm:jl-decision}). Applying a hypothetical proof $f$ of $\JB(\lambda, P_0)
\jlimp P_0$ to that decision yields a proof of $P_0$ in the theory extended by $\JB$. Because
$P_0$ does not mention $\JB$, conservativity of the extension by $\JB$
(Proposition~\ref{thm:jl-conservative}) yields a proof of $P_0$ without $\JB$, contradicting the
hypothesis. Finally, instantiating any term of the universally quantified type at $\lambda$ and
$\jlq{P_0}$ would produce such an $f$.
\end{proof}

The hypothesis of Theorem~\ref{thm:jl-nonfactive} holds whenever the constant specification
admits a $\jdecl$ or $\jobs$ leaf for an unprovable proposition---for instance, an agent's
assertion of $\tttFalse$. The theorem formalizes the boundary between the two levels: deciding
$\JB(j, P)$ never inspects whether $P$ is inhabited, so a positive decision of $\JB(j, P)$
entails nothing about the truth of $P$ itself. A proposition acquires the status \emph{Verified}
only from a checked proof of $P$, never from a proof of $\JB(j, P)$. Conversely,
Theorem~\ref{thm:jl-verified-factive} below identifies the exact condition under which a proof of
$P$ can be reconstructed from its grounds: when $\supp(j, P)$ contains an alternative whose leaves
are all $\jver$.

Because $\JB(j, P)$ belongs to $\Prop$, its negation $\neg\JB(j, P)$ is likewise a first-class
proposition, decided by the same procedure and read as ``the evidence $j$ does not support $P$''.
Justification assertions can likewise serve as hypotheses ($\JB(j, P) \jlimp Q$) and, under the
checker extension of Section~\ref{sec:jl-checker}, can be nested ($\JB(\jlchk j, \jlq{\JB(j,
P)})$).

\section{The Eigenius term algebra}
\label{sec:jl-term-algebra}

\begin{definition}[Justification terms]\label{def:jl-jterm}
Let $\jlName : \Setu$ be the type of resource names. $\JTerm : \Setu$ is the inductive type
generated by the six constructors
\[
  j \;::=\; \jdecl(r) \mid \jobs(r) \mid \jver(r) \mid \japp(j, j) \mid \jsum(j, j) \mid
  \jinst(j, \jlq{a})
\]
with $r : \jlName$ and $\jlq{a} : \jlCode$. The terms $\jdecl(r)$, $\jobs(r)$, and $\jver(r)$ are
called \emph{leaves}; we write $\jlLeaf$ for the set of leaves, and for a leaf $\lambda$ with
resource $r$ we write $\jlres(\lambda) = r$ and $\jlkind(\lambda) \in \{\jdecl, \jobs, \jver\}$,
extending $\jlres$ to sets of leaves elementwise. We write $\jlleaves(j)$ for the set of leaves
occurring in $j$.
\end{definition}

The six core formers (together with the optional conjunction-introduction former $\jpair(s, t)$
analyzed in Section~\ref{sec:jl-curry-church}) have the following readings:

\begin{center}
\small
\begin{tabular}{@{}lp{0.62\textwidth}@{}}
  \toprule
  \textbf{Former} & \textbf{Reading} \\
  \midrule
  $\jdecl(r)$ & the agent responsible for $r$ asserted the proposition of $r$ \\
  $\jobs(r)$ & a recording $r$ occurred, with the proposition that describes it \\
  $\jver(r)$ & a checked proof of the proposition of $r$ exists \\
  $\japp(s, t)$ & Application: $s$ justifies an implication, $t$ its antecedent \\
  $\jsum(s, t)$ & Sum, with arbitrary summands \\
  $\jinst(j, \jlq{a})$ & instantiation of a universal proposition at the instance $a$ \\
  $\jpair(s, t)$ & conjunction introduction; optional (Section~\ref{sec:jl-curry-church}) \\
  \bottomrule
\end{tabular}
\end{center}

$\JTerm$ is Curry-style: a term carries no proposition in its type, and the propositions it
justifies are computed by the evidence function (Section~\ref{sec:jl-evidence}). Because its
constructors take only resource names, syntactic codes, and subterms, $\JTerm$ is an ordinary
first-order inductive type in $\Setu$ in the sense of Chapter~\ref{ch:mltt}, and no constructor
quantifies over a universe.

\subsection{Leaves and the constant specification}

\begin{definition}[Leaf specification]\label{def:jl-leaf-cs}
A \emph{constant specification} for $\JTerm$ over a kernel environment $\Sigma$ is a partial map
$\jlcs : \jlLeaf \rightharpoonup \jlCode$ such that:
\begin{enumerate}[label=(CS\arabic*)]
\item each $\jlcs(\lambda)$ decodes to a closed proposition $P_\lambda$ with $\Sigma \vdash
  P_\lambda : \Prop$;
\item each resource carries at most one proposition: if $\jlcs(\lambda)$ and $\jlcs(\mu)$ are
  defined and $\jlres(\lambda) = \jlres(\mu)$, then $\jlcs(\lambda) = \jlcs(\mu)$;
\item if $\jlcs(\jver(r))$ is defined, a proof of $P_{\jver(r)}$ has been checked in $\Sigma$:
  there is a closed term $t$ with $\Sigma \vdash t : P_{\jver(r)}$.
\end{enumerate}
A leaf $\lambda$ is \emph{admitted} if $\jlcs(\lambda)$ is defined.
\end{definition}

The constant specification $\jlcs$ is fixed by the chain rather than by the author of a term. A
leaf is admitted only when the chain records a witness of the corresponding kind for its resource:
an attributed assertion for $\jdecl$, an observation record for $\jobs$, or a checked proof
judgment for $\jver$ (implemented by the reasoning institution of Volume~II). An author who writes
$\jver(r)$ for a resource that lacks a checked proof obtains an unadmitted leaf whose evidence set
is empty. Whereas $\jdecl$ and $\jobs$ leaves are postulated---attribution and occurrence are
empirical facts recorded on the chain rather than proved in the kernel---each admitted $\jver$ leaf
is discharged by a kernel proof via~(CS3). Because every decision of $\JB$ consults $\jlcs$, the
map $\jlcs$ is part of the trusted computing base.

Compared with the standard constant specification of Definition~\ref{def:jl-cs}, $\jlcs$ has two
structural differences: first, each admitted leaf justifies a single closed proposition, which need
not be a logical axiom; second, leaves are not iterated, so a leaf never justifies a nested formula
$\lambda : A$ unless $\JB(\lambda, A)$ is itself the proposition of the cited resource.

\subsection{Restriction: no internalized logical axioms}
\label{sec:jl-no-internalization}

Because the Eigenius fragment provides no built-in constants for the axioms of propositional or
first-order logic, the general internalization property of Section~\ref{sec:jl-basic} does not
hold automatically. Instead, any logical or domain-specific implication used in an Application step
must enter as an explicit leaf, called a \emph{bridge}:

\begin{itemize}
\item a bridge whose implication has a checked proof enters as a $\jver$ leaf;
\item every unproved bridge enters as a $\jdecl$ leaf attributed to an agent.
\end{itemize}

For example, a vocabulary mapping, an assertion that an analysis plan denotes a function
$I \jlimp O$, or an implication linking two differently phrased domain claims enters as a
$\jdecl$ bridge unless proved. Every unproved bridge thus has an accountable owner. Conversely, a
tautology or logical law that has a kernel proof---such as conjunction introduction
$\Pi (A\, B : \Prop).\, A \jlimp B \jlimp A \wedge B$---needs no owner: it can be committed once
together with its proof and cited thereafter as a $\jver$ leaf. In this way, the fragment satisfies
a relative form of internalization:

\begin{proposition}[Relative internalization]\label{thm:jl-relative-internalization}
If $\Sigma \vdash t : P$ for a closed proposition $P$, then extending $\jlcs$ by a fresh leaf
$\jver(r)$ with $\jlcs(\jver(r)) = \jlq{P}$ satisfies (CS1)--(CS3), and $\JB(\jver(r), P)$ holds
in the extension.
\end{proposition}

\begin{proof}
Conditions~(CS1) and~(CS3) hold by hypothesis, and~(CS2) holds because $r$ is fresh;
$\Evid(\jver(r)) = \{P\}$ then follows by Definition~\ref{def:jl-evidence}.
\end{proof}

Whereas internalization in $\mathsf{J}_{\jlCS}$ constructs a compound justification term from a
Hilbert-style derivation of $F$, relative internalization takes a checked kernel proof of $P$ and
introduces a single $\jver$ leaf: the justification term records that a checked proof exists
without duplicating the proof term itself.

\section{The evidence function of the fragment}
\label{sec:jl-evidence}

Throughout this section, propositions are taken up to definitional equality: for a finite set $S$
of propositions, $P \in S$ means $P \defeq Q$ for some $Q \in S$. We assume that definitional
equality is decidable on the propositions under consideration and that every such proposition
reduces to a weak-head normal form (Chapter~\ref{ch:mltt} reviews these properties for the kernel
type theory). As usual, an implication $A \jlimp B$ is a non-dependent product $\Pi x{:}A.\, B$
with $x$ not free in $B$.

\begin{definition}[Evidence function of the fragment]\label{def:jl-evidence}
For a constant specification $\jlcs$, the evidence function $\Evid = \Evid_{\jlcs}$ assigns to
each $j : \JTerm$ a finite set of closed propositions by structural recursion:
\begin{align*}
  \Evid(\lambda) &= \{P_\lambda\} \text{ if $\lambda$ is admitted, else } \emptyset
    && \lambda \in \jlLeaf \\
  \Evid(\japp(s, t)) &= \{B \mid (A \jlimp B) \in \Evid(s),\ A \in \Evid(t)\} \\
  \Evid(\jsum(s, t)) &= \Evid(s) \cup \Evid(t) \\
  \Evid(\jinst(j, \jlq{a})) &= \{B[a/x] \mid (\Pi x{:}T.\, B) \in \Evid(j),\ \Sigma \vdash a : T\}
\end{align*}
In the clauses for $\japp$ and $\jinst$, the members of $\Evid(s)$ and $\Evid(j)$ are inspected
in weak-head normal form.
\end{definition}

\begin{lemma}[Finiteness]\label{thm:jl-finite}
For every $j$, $|\Evid(j)| \leq |\jlleaves(j)|_{\#}$, where $|\cdot|_{\#}$ counts leaf
occurrences with multiplicity.
\end{lemma}

\begin{proof}
By structural induction on $j$. A leaf contributes at most one proposition. Each member of
$\Evid(\japp(s, t))$ is the codomain of a member of $\Evid(s)$, so $|\Evid(\japp(s,t))| \leq
|\Evid(s)|$; similarly, $|\Evid(\jinst(j, \jlq{a}))| \leq |\Evid(j)|$ since $a$ is fixed. Finally,
$|\Evid(\jsum(s, t))| \leq |\Evid(s)| + |\Evid(t)|$.
\end{proof}

\begin{definition}[Justification assertion]\label{def:jl-jb}
$\JB(j, \jlq{P})$ holds iff $P$ is a closed proposition and $P \in \Evid(j)$. The kernel decides
it via \emph{decision rules}: for closed terms $j : \JTerm$ and $c : \jlCode$ whose reduction by
$\beta$, $\delta$, and $\iota$ reaches closed constructor trees $j_0$ and $c_0$ on which the
computation of $\Evid(j_0)$ terminates, the kernel admits a proof of $\JB(j, c)$ if $c_0$ decodes
to a closed proposition $P$ with $P \in \Evid(j_0)$, and admits a proof of $\neg\JB(j, c)$
otherwise. The constant $\JB$ has no $\delta$-rule and no eliminator, and no kernel rule relates
$\JB(j, c)$ and $\JB(j, c')$ for distinct codes $c$ and $c'$.
\end{definition}

The decision rules act as axioms at closed data. Because the computation of $\Evid$ is
deterministic, no pair $(j_0, c_0)$ of constructor trees ever receives both a positive and a
negative decision axiom, regardless of whether the conversion test used during the computation is
complete.

\subsection{A rule system and its decision procedure}

\begin{definition}[Derivability]\label{def:jl-rules}
The judgment $\jlvd \JB(j, P)$ is generated by the rules
\begin{mathpar}
\inferrule*[right=Leaf]{\jlcs(\lambda) = \jlq{P}}{\jlvd \JB(\lambda, P)}
\and
\inferrule*[right=App]{\jlvd \JB(s, A \jlimp B) \\ \jlvd \JB(t, A)}{\jlvd \JB(\japp(s, t), B)}
\\
\inferrule*[right=SumL]{\jlvd \JB(s, P)}{\jlvd \JB(\jsum(s, t), P)}
\and
\inferrule*[right=SumR]{\jlvd \JB(t, P)}{\jlvd \JB(\jsum(s, t), P)}
\\
\inferrule*[right=Inst]{\jlvd \JB(j, \Pi x{:}T.\, B) \\ \Sigma \vdash a : T}
  {\jlvd \JB(\jinst(j, \jlq{a}), B[a/x])}
\and
\inferrule*[right=Conv]{\jlvd \JB(j, P) \\ P \defeq P'}{\jlvd \JB(j, P')}
\end{mathpar}
\end{definition}

Rules Leaf, App, SumL, and SumR are the inference-rule counterparts of axiom necessitation,
Application, and Sum from Definition~\ref{def:jl-axioms}, restricted to atomic justification
assertions; Inst is the rule counterpart of justified universal instantiation with the instance
recorded in the term; and Conv makes derivability invariant under definitional equality.

\begin{theorem}[Soundness and completeness of the decision procedure]\label{thm:jl-decision}
Assume that definitional equality is decided by a transitive relation and that $\Pi$-types are
injective under it. Then for every constant specification $\jlcs$, term $j$, and closed
proposition $P$:
\[
  \jlvd \JB(j, P) \iff P \in \Evid_{\jlcs}(j).
\]
\end{theorem}

\begin{proof}
($\Rightarrow$) By induction on the derivation:
\begin{itemize}
\item Leaf: $\Evid(\lambda) = \{P\}$.
\item App: by the induction hypothesis, $(A \jlimp B) \in \Evid(s)$ and $A \in \Evid(t)$. Any
  member of $\Evid(s)$ definitionally equal to $A \jlimp B$ has as its weak-head normal form an
  implication $A' \jlimp B'$ with $A' \defeq A$ and $B' \defeq B$ by $\Pi$-injectivity under
  $\defeq$; hence $B \in \Evid(\japp(s,t))$.
\item SumL and SumR: $\Evid(s), \Evid(t) \subseteq \Evid(\jsum(s,t))$.
\item Inst: analogous to App, since substitution respects $\defeq$.
\item Conv: membership in $\Evid(j)$ is defined up to $\defeq$.
\end{itemize}

($\Leftarrow$) By structural induction on $j$. If $P \in \Evid(\lambda)$, then $\lambda$ is
admitted with $P_\lambda \defeq P$, so Leaf followed by Conv derives $\JB(\lambda, P)$. If $P \in
\Evid(\japp(s,t))$, there exist $(A \jlimp B) \in \Evid(s)$ and $A' \in \Evid(t)$ with $A' \defeq
A$ and $B \defeq P$; the induction hypothesis yields $\jlvd \JB(s, A \jlimp B)$ and
$\jlvd \JB(t, A')$, Conv yields $\jlvd \JB(t, A)$, and App followed by Conv derives
$\jlvd \JB(\japp(s,t), P)$. The cases for $\jsum$ and $\jinst$ follow similarly using SumL, SumR,
and Inst.
\end{proof}

The hypotheses of Theorem~\ref{thm:jl-decision} hold for the kernel type theory up to termination:
definitional equality is untyped joinability under a confluent reduction and is therefore
transitive with injective $\Pi$-types (Lemma~\ref{lem:mltt-confluence}), and the kernel's
conversion procedure decides it on all normalizing terms (Theorem~\ref{thm:checking-conv}), where
normalization is expected on typable terms though not proved here
(Hypothesis~\ref{hyp:checking-normalization}); Corollary~\ref{cor:checking-jl} records the
resulting guarantee. Notice that soundness ($P \in \Evid_{\jlcs}(j) \Rightarrow \jlvd \JB(j, P)$)
requires only the soundness of the conversion test, whereas completeness uses transitivity and
$\Pi$-injectivity and holds wherever the conversion test terminates.

\begin{corollary}[Decidability]\label{thm:jl-decidable}
Given decidable definitional equality, decidable type checking of instances, and weak-head
normalization of the propositions in the range of $\jlcs$, $\jlvd \JB(j, P)$ is decidable using
$O(n^2)$ conversion tests and $O(n^2)$ type-checking checks for a term $j$ with $n$ nodes.
\end{corollary}

\begin{proof}
Compute $\Evid(j)$ recursively by Definition~\ref{def:jl-evidence} and test $P \in \Evid(j)$. By
Lemma~\ref{thm:jl-finite}, evaluating $\japp(s,t)$ compares at most $|\jlleaves(s)|_{\#} \cdot
|\jlleaves(t)|_{\#}$ pairs; summed over all $\japp$ nodes in $j$, this counts pairs of leaf
occurrences by their lowest common ancestor and is therefore bounded by $n^2$. Evaluating $\jinst$
checks the instance $a$ against at most $n$ domains at each of at most $n$ nodes, and the final
membership test performs at most $n$ comparisons.
\end{proof}

\begin{proposition}[Minimality]\label{thm:jl-minimal}
Call a map $\Evid'$ from $\JTerm$ to sets of propositions \emph{admissible} for $\jlcs$ if
$P_\lambda \in \Evid'(\lambda)$ for every admitted $\lambda$ and $\Evid'$ is closed under the
clauses of Definition~\ref{def:jl-evidence} read as inclusions $\subseteq$. Then $\Evid_{\jlcs}$ is
admissible and $\Evid_{\jlcs}(j) \subseteq \Evid'(j)$ for every admissible $\Evid'$ and every
$j$.
\end{proposition}

\begin{proof}
$\Evid_{\jlcs}$ satisfies each clause with equality. For any admissible $\Evid'$, we proceed by
induction on $j$: the leaf case holds by the base condition; for $\japp(s,t)$, any $B \in
\Evid_{\jlcs}(\japp(s,t))$ arises from $(A \jlimp B) \in \Evid_{\jlcs}(s) \subseteq \Evid'(s)$ and
$A \in \Evid_{\jlcs}(t) \subseteq \Evid'(t)$, so closure gives $B \in \Evid'(\japp(s,t))$; the
cases for $\jsum$ and $\jinst$ are identical.
\end{proof}

Thus $\Evid_{\jlcs}$ is the minimal evidence function of Lemma~\ref{thm:jl-minimal-evidence},
carried over to kernel propositions modulo $\defeq$ and extended by the $\jinst$ clause.
Theorem~\ref{thm:jl-decision} is the fragment's counterpart to
Proposition~\ref{thm:jl-minimal-characterizes}, turning the Mkrtychev semantics with the minimal
evidence function into an executable decision procedure.

\begin{corollary}\label{thm:jl-j-embedding}
Let $\jlcs$ range over propositions built from atomic propositions by $\jlimp$ alone, with $\defeq$
coinciding with syntactic identity on them, and let $\jlCS = \{\lambda : P_\lambda \mid \lambda
\text{ admitted}\}$ with leaves read as justification constants. For any $\jinst$-free term $j$,
$\jlvd \JB(j, P)$ implies $\mathsf{J}_{0} + \jlCS \vdash j : P$ (reading $\japp$ as $\cdot$ and
$\jsum$ as $+$), and the converse holds whenever $\mathsf{J}_0 + \jlCS$ is complete for Mkrtychev
models respecting $\jlCS$.
\end{corollary}

\begin{proof}
Every rule of Definition~\ref{def:jl-rules} other than Inst and Conv is derivable in
$\mathsf{J}_0 + \jlCS$ via an axiom instance and modus ponens. The converse follows from
Proposition~\ref{thm:jl-minimal} together with the proof of
Proposition~\ref{thm:jl-minimal-characterizes}.
\end{proof}

\paragraph{Which system.}
In summary, the Eigenius fragment corresponds to $\mathsf{J}$ restricted to atomic justification
assertions, with unrestricted Sum, a chain-controlled constant specification assigning at most one
proposition per leaf and containing no constants for logical axioms, and universal instantiation as
an explicit term former. Positive introspection is provided by the checker extension of
Section~\ref{sec:jl-checker}. General factivity fails for $\JB(j, P)$
(Theorem~\ref{thm:jl-nonfactive}), holding instead for propositions equipped with a kernel proof
and---as Theorem~\ref{thm:jl-verified-factive} shows---for propositions supported by an alternative
consisting entirely of $\jver$ leaves.

\section{Support relative to a proposition}
\label{sec:jl-support}

Whereas the evidence function $\Evid(j)$ determines \emph{which} propositions a justification term
$j$ justifies, the support function $\supp(j, P)$ determines \emph{which leaves} a justification of
$P$ by $j$ rests on.

\begin{definition}[Support]\label{def:jl-support}
For $j : \JTerm$ and a closed proposition $P$, $\supp(j, P)$ is a finite set of finite sets of
leaves, called the \emph{alternatives} for $P$ in $j$:
\begin{align*}
  \supp(\lambda, P) &= \{\{\lambda\}\} \text{ if $\lambda$ is admitted and $P_\lambda \defeq P$,
    else } \emptyset \\
  \supp(\japp(s, t), P) &= \bigcup_{A}\, \{a \cup b \mid a \in \supp(s, A \jlimp P),\
    b \in \supp(t, A)\} \\
  \supp(\jsum(s, t), P) &= \supp(s, P) \cup \supp(t, P) \\
  \supp(\jinst(j, \jlq{a}), P) &= \bigcup\, \bigl\{\supp(j, \Pi x{:}T.\, B) \;\big|\;
    (\Pi x{:}T.\, B) \in \Evid(j), \\[-2pt]
  &\qquad\qquad \Sigma \vdash a : T,\ B[a/x] \defeq P\bigr\}
\end{align*}
where $A$ ranges over the $\defeq$-classes satisfying $(A \jlimp P) \in \Evid(s)$ and $A \in
\Evid(t)$.
\end{definition}

The index sets in the unions are finite by Lemma~\ref{thm:jl-finite}, and structural induction on
$j$ shows that $\supp(j, P)$ depends on $P$ only up to $\defeq$.

\begin{lemma}[Support detects justification]\label{thm:jl-supp-nonempty}
$\supp(j, P) \neq \emptyset$ iff $P \in \Evid(j)$. Moreover, every alternative $a \in \supp(j, P)$
consists of admitted leaves occurring in $j$.
\end{lemma}

\begin{proof}
By structural induction on $j$. For a leaf, the equivalence holds by definition. For $\japp(s, t)$,
the union is non-empty iff some class $A$ in the index set has both $\supp(s, A \jlimp P)$ and
$\supp(t, A)$ non-empty; by the induction hypothesis this holds iff some $A$ satisfies $(A \jlimp P)
\in \Evid(s)$ and $A \in \Evid(t)$, which is the definition of $P \in \Evid(\japp(s,t))$. The
$\jsum$ case is the union of the two branches, and $\jinst$ is analogous to $\japp$. The second
claim follows immediately from the recursive clauses.
\end{proof}

\begin{definition}[Term support]\label{def:jl-term-support}
The \emph{term support} $\jlsuppz(j)$ collects leaf alternatives without reference to a target
proposition:
$\jlsuppz(\lambda) = \{\{\lambda\}\}$, $\jlsuppz(\japp(s,t)) = \{a \cup b \mid a \in
\jlsuppz(s), b \in \jlsuppz(t)\}$, $\jlsuppz(\jsum(s,t)) = \jlsuppz(s) \cup \jlsuppz(t)$, and
$\jlsuppz(\jinst(j, \jlq{a})) = \jlsuppz(j)$.
\end{definition}

\begin{proposition}[Refinement]\label{thm:jl-refines}
$\supp(j, P) \subseteq \jlsuppz(j)$ for every $P$.
\end{proposition}

\begin{proof}
By structural induction on $j$: in each clause, every alternative of $\supp(j, P)$ is formed by the
same set operation as the corresponding element of $\jlsuppz(j)$ from alternatives that belong to
$\jlsuppz$ of the immediate subterms by the induction hypothesis.
\end{proof}

\subsection{The phantom-summand property}

Because $\jlsuppz(j)$ ignores the target proposition $P$, it counts every summand of a $\jsum$
node---including summands that justify nothing at all, or that justify a proposition other than
$P$. Evaluating support relative to $P$ filters out such irrelevant summands automatically:

\begin{theorem}[Phantom summands]\label{thm:jl-phantom}
If $P \notin \Evid(t)$, then $\supp(\jsum(s, t), P) = \supp(s, P)$. In particular, no leaf that
occurs in $t$ but not in $s$ belongs to any alternative of $\supp(\jsum(s, t), P)$.
\end{theorem}

\begin{proof}
We have $\supp(t, P) = \emptyset$ by Lemma~\ref{thm:jl-supp-nonempty}, and the second claim follows
from the second part of Lemma~\ref{thm:jl-supp-nonempty} applied to $s$.
\end{proof}

\begin{example}[A phantom summand]\label{def:jl-phantom-example}
Let $\lambda$ be an admitted leaf with $P_\lambda = P$, and let $\mu = \jdecl(r')$ be a leaf naming
a non-existent resource $r'$ (so that $\mu$ is not admitted). For $j = \jsum(\lambda, \mu)$, we
have $\Evid(j) = \{P\}$, $\jlsuppz(j) = \{\{\lambda\}, \{\mu\}\}$, and $\supp(j, P) =
\{\{\lambda\}\}$. If counterfactual queries were evaluated on $\jlsuppz(j)$, the query ``does $P$
still stand without $\jlres(\lambda)$?'' would be answered \emph{yes} via the phantom alternative
$\{\mu\}$, which justifies nothing. Evaluated on $\supp(j, P)$, the query is answered \emph{no}.
Thus term support yields unsound answers to counterfactual queries, whereas proposition-relative
support $\supp(\cdot, P)$ gives the sound answer in the presence of $\mathsf{LP}$'s unrestricted
Sum without needing to restrict Sum itself.
\end{example}

\subsection{Support as a provenance semiring}
\label{sec:jl-semiring}

Relative to a fixed proposition, $\japp$ acts as a product and $\jsum$ as a sum over alternative
derivations. This is the algebraic structure of the provenance semirings of
\citet{green-karvounarakis-tannen2007}.

\begin{definition}[Semirings of provenance]\label{def:jl-semirings}
Let $X = \jlLeaf$.
\begin{itemize}
\item $\jlNX$ is the commutative semiring of polynomials with natural-number coefficients in the
  indeterminates $X$.
\item $\jlWhy(X)$ is the set of finite sets of finite subsets of $X$, with $S + T = S \cup T$,
  $S \cdot T = \{a \cup b \mid a \in S, b \in T\}$, $0 = \emptyset$, and $1 = \{\emptyset\}$.
\item $\jlPos(X)$ is the set of finite antichains of finite subsets of $X$ under inclusion, with
  $S + T = \jlMin(S \cup T)$ and $S \cdot T = \jlMin\{a \cup b \mid a \in S, b \in T\}$, where
  $\jlMin(S)$ is the set of inclusion-minimal elements of $S$.
\item $\jlLin(X) = \mathcal{P}_{\mathrm{fin}}(X) \cup \{\bot\}$, with $+$ the union and $\cdot$
  the union when both arguments differ from $\bot$, $\bot$ the zero, and $\emptyset$ the unit.
\end{itemize}
\end{definition}

The polynomial semiring $\jlNX$ is the free commutative semiring on $X$: every map from $X$ into a
commutative semiring $K$ extends uniquely to a semiring homomorphism from $\jlNX$ to
$K$~\citep{green-karvounarakis-tannen2007}. The Why-provenance semiring $\jlWhy(X)$ satisfies the
idempotence laws $S + S = S$ and $S \cdot S \supseteq S$, whereas absorption $S + S \cdot T = S$
fails in $\jlWhy(X)$ and holds in $\jlPos(X)$.

\begin{definition}[Provenance polynomial]\label{def:jl-prov}
$\jlprov(j, P) \in \jlNX$ is defined by the recursive clauses of Definition~\ref{def:jl-support},
replacing $\{\{\lambda\}\}$ by the indeterminate $\lambda$, $\emptyset$ by $0$, union $\cup$ of
alternative sets by $+$, and pairwise union $\{a \cup b\}$ by $\cdot$.
\end{definition}

Let $\jlDer(j, P)$ be the set of derivations of $\jlvd \JB(j, P)$ in the rules of
Definition~\ref{def:jl-rules}, with Conv absorbed into the remaining rules by reading their
propositions as $\defeq$-classes. For each derivation $D$, let $m(D) \in \jlNX$ be the monomial
product of the leaves occurring at its Leaf rules, counted with multiplicity.

\begin{proposition}\label{thm:jl-prov-derivations}
$\jlDer(j, P)$ is finite and $\jlprov(j, P) = \sum_{D \in \jlDer(j, P)} m(D)$.
\end{proposition}

\begin{proof}
By structural induction on $j$. A leaf has one derivation of $P_\lambda$ and none of any other
class. A derivation of $\JB(\japp(s,t), P)$ consists of an App step over a pair of derivations of
$\JB(s, A \jlimp P)$ and $\JB(t, A)$ for some class $A$; by Theorem~\ref{thm:jl-decision}, these
pair sets are empty unless $A$ lies in the finite index set of Definition~\ref{def:jl-support}.
Thus $\jlDer(\japp(s,t), P) \cong \coprod_A \jlDer(s, A \jlimp P) \times \jlDer(t, A)$, and $m$ is
multiplicative across the pairs. For $\jsum$, $\jlDer(\jsum(s,t), P) \cong \jlDer(s, P) \sqcup
\jlDer(t, P)$ via SumL and SumR. For $\jinst$, $\jlDer(\jinst(j,\jlq{a}), P) \cong
\coprod_{B} \jlDer(j, \Pi x{:}T.\, B)$ over the index set of Definition~\ref{def:jl-support}. The
clauses of $\jlprov$ are the images of these decompositions under $m$.
\end{proof}

\begin{theorem}[Support is a homomorphic image]\label{thm:jl-supp-hom}
Let $h : \jlNX \to \jlWhy(X)$ be the unique semiring homomorphism satisfying $h(\lambda) =
\{\{\lambda\}\}$. Then $\supp(j, P) = h(\jlprov(j, P))$ for all $j$ and $P$. In $\jlWhy(X)$,
\begin{align*}
  \supp(\jsum(s,t), P) &= \supp(s, P) + \supp(t, P), \\
  \supp(\japp(s,t), P) &= \textstyle\sum_{A} \supp(s, A \jlimp P) \cdot \supp(t, A).
\end{align*}
\end{theorem}

\begin{proof}
The displayed equations restate Definition~\ref{def:jl-support} in the operations of
$\jlWhy(X)$. The identity $\supp = h \circ \jlprov$ follows by induction on $j$, since
$\jlprov$ satisfies the same recursive clauses in $\jlNX$ and $h$ preserves $0$, $+$, and $\cdot$.
\end{proof}

By Proposition~\ref{thm:jl-prov-derivations} and Theorem~\ref{thm:jl-supp-hom}, each alternative
in $\supp(j, P)$ is the set of leaves used by some derivation of $\JB(j, P)$, and every derivation
contributes an alternative; $\jlprov$ retains the multiplicity of derivations and of leaf uses,
whereas $\supp$ forgets both.

\subsection{Queries}
\label{sec:jl-queries}

\begin{definition}[Warrant queries]\label{def:jl-queries}
For $j : \JTerm$, a closed proposition $P$, and a resource name $r$:
\begin{center}
\small
\begin{tabularx}{\textwidth}{@{}l@{\hspace{0.8em}}l@{\hspace{0.8em}}>{\raggedright\arraybackslash}X@{}}
  \toprule
  \textbf{Query} & \textbf{Definition} & \textbf{Reading} \\
  \midrule
  $\jlfv(j, P)$ & $\exists a \in \supp(j, P).\, \forall \lambda \in a.\, \jlkind(\lambda) = \jver$
    & some alternative is all-verified \\
  $\jlsw(j, P, r)$ & $\exists a \in \supp(j, P).\ r \notin \jlres(a)$ & the counterfactual \\
  $\jllo(j, P)$ & $\bigcup \supp(j, P)$ & exposure: agents, recordings, proofs \\
  $\jlcited(j)$ & $\jlleaves(j)$ & the audit surface \\
  $\jlclaims(j)$ & $\Evid(j)$ & which claims the evidence supports \\
  \bottomrule
\end{tabularx}
\end{center}
\end{definition}

Note that $\jlcited(j)$ is the lineage of the term support, $\jlcited(j) = \bigcup \jlsuppz(j)$:
it includes any phantom summands and records every leaf cited anywhere in $j$. By
Proposition~\ref{thm:jl-refines}, $\jllo(j, P) \subseteq \jlcited(j)$.

\begin{definition}[Lattice valuation]\label{def:jl-valuation}
Let $(L, \vee, \wedge, \bot, \top)$ be a bounded distributive lattice and $v : X \to L$. Define
$h_v : \jlWhy(X) \to L$ by $h_v(S) = \bigvee_{a \in S} \bigwedge_{\lambda \in a} v(\lambda)$,
with the empty join $\bot$ and the empty meet $\top$.
\end{definition}

\begin{theorem}[Lattice-valued queries]\label{thm:jl-lattice-hom}
$h_v$ is a semiring homomorphism from $\jlWhy(X)$ to $(L, \vee, \wedge, \bot, \top)$, and
$h_v(S) = h_v(\jlMin(S))$ for every $S$. In particular:
\begin{enumerate}
\item $\jlfv(j, P)$ is $h_v(\supp(j, P)) = \top$ for $L = \{\bot < \top\}$ and $v(\lambda) = \top$
  iff $\jlkind(\lambda) = \jver$;
\item $\jlsw(j, P, r)$ is $h_v(\supp(j, P)) = \top$ for $L = \{\bot < \top\}$ and $v(\lambda) =
  \top$ iff $\jlres(\lambda) \neq r$;
\item for a total order $\jdecl \leq \jobs \leq \jver$, or any other lattice order on leaf kinds,
  the \emph{grade} $h_{\jlkind}(\supp(j, P))$ is the best alternative's weakest leaf.
\end{enumerate}
All three queries are invariant under absorption.
\end{theorem}

\begin{proof}
We have $h_v(\emptyset) = \bot$, $h_v(\{\emptyset\}) = \top$, and $h_v(S \cup T) = h_v(S) \vee
h_v(T)$ by associativity and idempotence of $\vee$. For products,
\[
  h_v(S \cdot T) = \bigvee_{a \in S, b \in T} \bigwedge_{\lambda \in a \cup b} v(\lambda)
  = \bigvee_{a \in S, b \in T} \Bigl(\bigwedge_{a} v \wedge \bigwedge_{b} v\Bigr)
  = h_v(S) \wedge h_v(T),
\]
using idempotence of $\wedge$ on leaves in $a \cap b$ together with distributivity. For absorption,
whenever $a \subseteq a'$ we have $\bigwedge_{a'} v \leq \bigwedge_a v$, so every non-minimal
element of a finite set $S$ is dominated by a minimal element below it and leaves the join
unchanged. Items~(1)--(3) follow by substituting the stated $L$ and $v$ into $h_v$.
\end{proof}

\begin{lemma}[$\jlMin$ is a homomorphism]\label{thm:jl-min-hom}
$\jlMin : \jlWhy(X) \to \jlPos(X)$ is a surjective semiring homomorphism. Moreover, a map $Q$ on
$\jlWhy(X)$ satisfies $Q(S) = Q(\jlMin(S))$ for all $S$ iff it factors through $\jlMin$.
\end{lemma}

\begin{proof}
For finite sets $S, T$, $\jlMin(S \cup T) = \jlMin(\jlMin(S) \cup \jlMin(T))$ because every
element of a finite set lies above a minimal element. For products, every $a \cup b$ with $a \in
S$ and $b \in T$ lies above some $a' \cup b'$ with $a' \in \jlMin(S)$ and $b' \in \jlMin(T)$, and
conversely $\jlMin(S) \cdot \jlMin(T) \subseteq S \cdot T$; hence both products have the same
minimal elements. Surjectivity holds because $\jlMin$ fixes antichains, and the factorization
property follows immediately.
\end{proof}

\begin{proposition}[Lineage is not absorption-invariant]\label{thm:jl-lineage}
The map $S \mapsto \bigcup S$ is a semiring homomorphism $\jlWhy(X) \to \jlLin(X)$ (sending
$\emptyset$ to $\bot$) that does not factor through $\jlMin$. Consequently, $\jllo$ and $\jlcited$
are lineage queries and are not invariant under absorption.
\end{proposition}

\begin{proof}
Preservation of $+$ is immediate; for $\cdot$, $\bigcup (S \cdot T) = \bigcup S \cup \bigcup T$
whenever $S, T \neq \emptyset$, and $S \cdot T = \emptyset$ otherwise. To see that the map does not
factor through $\jlMin$, let $\lambda, \mu$ be admitted leaves with $P_\lambda = P$ and $P_\mu =
P \jlimp P$, and set $j = \jsum(\lambda, \japp(\mu, \lambda))$. Then $\supp(j, P) = \{\{\lambda\},
\{\lambda, \mu\}\}$ has $\jlMin(\supp(j, P)) = \{\{\lambda\}\}$, whereas their unions are
$\{\lambda, \mu\}$ and $\{\lambda\}$.
\end{proof}

Intuitively, $\jlfv$, $\jlsw$, and grades ask \emph{whether} a claim stands under a given
condition on leaves; they factor through minimal witnesses ($\jlPos(X)$), so adding a redundant
leaf to an alternative never changes their value. By contrast, $\jllo$ and $\jlcited$ ask
\emph{what was used}; every alternative, minimal or not, contributes its leaves to the lineage
($\jlLin(X)$).

\begin{proposition}[Polynomial evaluation]\label{thm:jl-poly}
For any homomorphism $g : \jlWhy(X) \to K$ into a semiring with computable operations, $g(\supp(j,
P))$ can be computed by evaluating the recursion of Definition~\ref{def:jl-support} directly in $K$
using $O(n^3)$ semiring operations for a term $j$ with $n$ nodes (in addition to the conversion
tests of Corollary~\ref{thm:jl-decidable}), whereas the cardinality of $\jlMin(\supp(j, P))$ can be
$2^{\Omega(n)}$.
\end{proposition}

\begin{proof}
Tabulate $g(\supp(u, Q))$ for each subterm $u$ of $j$ and each $Q \in \Evid(u)$; by
Lemma~\ref{thm:jl-finite} there are at most $n$ such propositions $Q$ per node across the $O(n)$
nodes. At an $\japp$ node, each of the at most $n$ target propositions sums over at most $n$
antecedent classes $A$, performing $O(n^2)$ semiring operations per node and $O(n^3)$ overall. For
the exponential lower bound on $|\jlMin(\supp(j, P))|$, let $\lambda_i, \mu_i$ be admitted leaves
with $P_{\lambda_i} = P_{\mu_i} = Q_i$ for $1 \leq i \leq m$, and let $\delta$ be admitted with
$P_\delta = Q_1 \jlimp \cdots \jlimp Q_m \jlimp P$. Then $j = \japp(\cdots\japp(\delta,
\jsum(\lambda_1, \mu_1))\cdots, \jsum(\lambda_m, \mu_m))$ has $4m + 1$ nodes, while $\supp(j, P)$
consists of the $2^m$ pairwise incomparable sets $\{\delta\} \cup \{\nu_1, \ldots, \nu_m\}$ with
$\nu_i \in \{\lambda_i, \mu_i\}$.
\end{proof}

Both the absorption-invariant queries and the lineage queries can therefore be evaluated in
polynomial time without materializing the set of alternatives: the former via $h_v$ in a
two-element or finite distributive lattice, and the latter in $\jlLin(X)$.

\subsection{Computed grades}

\begin{proposition}[A grade does not determine the counterfactual]\label{thm:jl-grade-loses}
There exist terms $j_1, j_2$ and a proposition $P$ that have equal grades for $P$ under every
valuation of leaf kinds, yet satisfy $\jlsw(j_1, P, r) \neq \jlsw(j_2, P, r)$ for some resource $r$.
\end{proposition}

\begin{proof}
Let $\jobs(r_1)$ and $\jobs(r_2)$ be admitted for distinct resources $r_1 \neq r_2$ with the same
proposition $P$, and set $j_i = \jobs(r_i)$. Both terms receive grade $v(\jobs)$ for $P$; however,
$\jlsw(j_1, P, r_1)$ is false whereas $\jlsw(j_2, P, r_1)$ is true.
\end{proof}

A stored scalar grade is merely the value of $h_{\jlkind}$ cached at a single chain state. Because
$\Evid$ and $\supp$ are functions of the current constant specification $\jlcs$, any change to
$\jlcs$---such as the withdrawal of a declaration---updates every dependent support set, and with
it every grade and counterfactual query, without modifying any stored justification term.

\begin{example}[Computed and Sampled]\label{def:jl-computed-sampled}
Let $\delta = \jdecl(f)$ be admitted with $P_\delta = I \jlimp O$, declaring that an analysis plan
$f$ maps inputs satisfying $I$ to outputs satisfying $O$, and let $\omega = \jobs(x)$ be admitted
with $P_\omega = I$. The application term $\japp(\delta, \omega)$ justifies $O$ with support
$\supp(\japp(\delta, \omega), O) = \{\{\delta, \omega\}\}$: this is a \emph{Computed} conclusion,
resting jointly on the declared plan and the observed input. By contrast, a bare leaf $\jobs(y)$
with $P_{\jobs(y)} = O$ is a \emph{Sampled} conclusion, where the observation record itself is the
entire evidence. Thus \emph{Computed} and \emph{Sampled} are shapes of justification terms rather
than primitive leaf kinds. If the declaration $\delta$ is later withdrawn from $\jlcs$, then
$O \notin \Evid(\japp(\delta, \omega))$, and every conclusion of the first shape loses its
justification automatically.
\end{example}

\begin{example}[One term, several claims]\label{def:jl-shared}
Let $\omega_1, \omega_2$ be admitted leaves with propositions $P$ and $Q$. The single term $j =
\jsum(\omega_1, \omega_2)$ has $\Evid(j) = \{P, Q\}$; two separate conclusions may cite the same
term $j$, with $\supp(j, P) = \{\{\omega_1\}\}$ and $\supp(j, Q) = \{\{\omega_2\}\}$, so that
$\jlsw(j, P, \jlres(\omega_2))$ is true.
\end{example}

\subsection{Factivity for verified alternatives}

\begin{theorem}[Verified alternatives are factive]\label{thm:jl-verified-factive}
If $a \in \supp(j, P)$ and every leaf of $a$ is a $\jver$ leaf, then $\Sigma \vdash t : P$ for a
proof term $t$ computable from $j$, $a$, and the proofs witnessing~(CS3) for the leaves of $a$. In
particular, $\jlfv(j, P)$ implies that $P$ is provable in $\Sigma$.
\end{theorem}

\begin{proof}
By structural induction on $j$, constructing $t$:
\begin{itemize}
\item Leaf $\jver(r)$: condition~(CS3) supplies $t_r$ with $\Sigma \vdash t_r : P_{\jver(r)}$, and
  $P \defeq P_{\jver(r)}$.
\item $\japp(s, u)$: here $a = a_1 \cup a_2$ with $a_1 \in \supp(s, A \jlimp P)$ and $a_2 \in
  \supp(u, A)$, both consisting solely of $\jver$ leaves; the induction hypothesis yields
  $t_1 : A \jlimp P$ and $t_2 : A$, and we take $t = t_1\, t_2$.
\item $\jsum(s, u)$: $a$ belongs to $\supp(s, P)$ or $\supp(u, P)$, so we apply the induction
  hypothesis to that summand.
\item $\jinst(k, \jlq{b})$: here $a \in \supp(k, \Pi x{:}T.\, B)$ with $\Sigma \vdash b : T$ and
  $B[b/x] \defeq P$; the induction hypothesis yields $t_1 : \Pi x{:}T.\, B$, and we take
  $t = t_1\, b$.
\end{itemize}
\end{proof}

Theorem~\ref{thm:jl-verified-factive} is a metatheorem about the constant specification rather
than an object-language rule from $\JB(j, P)$ to $P$, leaving Theorem~\ref{thm:jl-nonfactive}
intact: the reconstructed term $t$ is a kernel proof of $P$, and $P$ acquires its Verified status
from $t$. Combined with the consistency of the kernel type theory
(Theorem~\ref{thm:mltt-declarative-consistency}), Theorem~\ref{thm:jl-verified-factive} entails
that no all-verified alternative can support $\tttFalse$ in a kernel-checked environment without
axioms (relative to the inaccessible cardinals of Chapter~\ref{ch:mltt}), since the reconstructed
proof $t$ would otherwise inhabit $\tttFalse$.

\section{Well-foundedness}
\label{sec:jl-wellfounded}

Because support is evaluated against the current constant specification $\jlcs$, circular
dependencies could arise across successive commits if unchecked: for instance, a derived resource
$r$ might first be committed using a bare observation $\lambda_0$ or a bridge $\delta$, and a later
commit might introduce a resource that cites $r$ while also supplying an implication back to $P_r$.
Evaluating support at the new chain head would then encounter circular justification alternatives
unless ruled out by a commit-time well-formedness condition.

\begin{definition}[Chain state]\label{def:jl-chain-state}
A \emph{chain state} is a finite set $R$ of resources equipped with a constant specification
$\jlcs$. Each resource $r \in R$ carries a proposition $P_r$ and is classified as either
\emph{primitive} or \emph{derived}. Every derived resource carries a justification term $j_r$ with
$P_r \in \Evid(j_r)$, and any leaf $\lambda$ with $\jlres(\lambda)$ derived depends on the support
of that derived resource.
\end{definition}

\begin{definition}[Well-foundedness]\label{def:jl-wellfounded}
Let $D \subseteq R$ be the subset of derived resources. For $W \subseteq D$, define
\[
  \Phi(W) = \{\, r \in D \mid \exists a \in \supp(j_r, P_r).\ \jlres(a) \cap D \subseteq W \,\}.
\]
Because $\Phi$ is monotone on subsets of $D$, it has a least fixed point $\jlWF \subseteq D$. The
chain state is \emph{well-founded} if $D = \jlWF$; a commit is rejected whenever the resulting
chain state fails to be well-founded.
\end{definition}

Unfolding the least fixed point, a derived resource $r$ belongs to $\jlWF$ iff at least one
alternative in $\supp(j_r, P_r)$ cites only primitive resources together with derived resources
that already entered $\jlWF$ at an earlier stage. Primitive resources have no support term to
inspect: a $\jdecl$ leaf rests on the trust placed in its attributing agent, an $\jobs$ leaf on the
recorded observation, and a $\jver$ leaf on its checked kernel proof.

\begin{lemma}[Failure is circularity]\label{thm:jl-wf-cycle}
If $r \in D \setminus \jlWF$, then every alternative of $\supp(j_r, P_r)$ contains a leaf whose
resource lies in $D \setminus \jlWF$. Following such leaves from $r$ reaches a cycle in $D
\setminus \jlWF$.
\end{lemma}

\begin{proof}
Since $\jlWF = \Phi(\jlWF)$, any $r \in D \setminus \jlWF$ satisfies $r \notin \Phi(\jlWF)$, which
is the first claim. Choosing one such leaf in $D \setminus \jlWF$ for each resource in $D
\setminus \jlWF$ defines an endomap on a non-empty finite set, whose iteration from $r$ is
eventually periodic and therefore enters a cycle.
\end{proof}

\begin{definition}[Edge check]\label{def:jl-edge-check}
The \emph{reference graph} of a chain state has a directed edge $r \to r'$ whenever $r \in D$ and
$r' \in \jlres(\jlcited(j_r)) \cap D$. The \emph{edge check} accepts a chain state iff its
reference graph is acyclic.
\end{definition}

\begin{theorem}[No false rejection from Sum]\label{thm:jl-wf-sum}
\leavevmode
\begin{enumerate}
\item If the edge check accepts a chain state, the state is well-founded.
\item If a chain state is well-founded and $j_r$ is replaced by $\jsum(j_r, t)$ for any term $t$,
  the resulting state remains well-founded.
\item There exist well-founded chain states that the edge check rejects.
\end{enumerate}
\end{theorem}

\begin{proof}
(1) By well-founded induction along the acyclic reference graph. For $r \in D$, we have
$\supp(j_r, P_r) \neq \emptyset$ by Lemma~\ref{thm:jl-supp-nonempty}; any alternative $a \in
\supp(j_r, P_r)$ satisfies $\jlres(a) \cap D \subseteq \jlres(\jlcited(j_r)) \cap D$ by
Proposition~\ref{thm:jl-refines}, and all such target resources already belong to $\jlWF$ by the
induction hypothesis, placing $r \in \Phi(\jlWF) = \jlWF$.

(2) We have $\supp(\jsum(j_r, t), P_r) \supseteq \supp(j_r, P_r)$ and $P_r \in \Evid(\jsum(j_r,
t))$. The monotone operator $\Phi'$ of the updated state therefore satisfies $\Phi(W) \subseteq
\Phi'(W)$ for all $W \subseteq D$, so its least fixed point contains $\jlWF = D$.

(3) Let $r_0$ be primitive, and let $r$ be derived with $j_r = \jsum(\lambda_0, \japp(\delta,
\lambda))$, where $\lambda_0$ is admitted with $\jlres(\lambda_0) = r_0$ and $P_{\lambda_0} = P_r$,
$\lambda$ is admitted with $\jlres(\lambda) = r$ and $P_\lambda = P_r$, and $\delta$ is admitted
with $P_\delta = P_r \jlimp P_r$. The reference graph contains the self-loop $r \to r$ and fails
the edge check, whereas $\{\lambda_0\} \in \supp(j_r, P_r)$ places $r \in \Phi(\emptyset)
\subseteq \jlWF$.
\end{proof}

Part~(2) captures the disjunctive nature of Sum: a circular path through one summand does not
invalidate a conclusion that is independently supported by another summand. Part~(3) shows that a
naive syntactic check on citation edges would falsely reject such terms.

\paragraph{Postulated and derived self-reference.}
A primitive resource may carry a proposition $P_r = A(\jlq{r})$ that refers to its own resource
name, the counterpart of a self-referential constant specification $c : A(c)$ in $\mathsf{LP}$.
As noted in Section~\ref{sec:jl-basic}, self-referential constant specifications are semantically
sound and are required for modal realization in $\mathsf{LP}$. Definition~\ref{def:jl-wellfounded}
accepts self-referential primitive resources because primitive resources have no support term. By
contrast, Lemma~\ref{thm:jl-wf-cycle} shows that derived circularity---where every supporting
alternative for a derived resource passes through ungrounded derived resources---is rejected,
ruling out mutual or circular justification among derived claims.

\paragraph{Stability.}
Well-foundedness is enforced at commit time. If a subsequent withdrawal of a declaration from
$\jlcs$ removes the well-founded alternatives of a derived resource and leaves only circular ones,
the warrant queries of Definition~\ref{def:jl-queries} evaluate over the well-founded alternatives
\[
  \supp^{\jlWF}(j_r, P_r) = \{a \in \supp(j_r, P_r) \mid \jlres(a) \cap D \subseteq \jlWF\}.
\]

\paragraph{Analogy.}
Definition~\ref{def:jl-wellfounded} is a signature-relative well-formedness condition on
justification terms, analogous to strict positivity for inductive declarations in
Chapter~\ref{ch:mltt}: both conditions rule out circular definitions that would undermine soundness,
and both are checked when a declaration is committed.

\section{Justification logic as a logic}
\label{sec:jl-logic}

Chapter~\ref{ch:institutions} formalizes each logic of Eigenius as a proof calculus in the sense of
\citet{meseguer1989}, from which a $\pi$-institution and an institution arise by projection. This
section packages the justification fragment in that form.

\begin{definition}[Signatures, sentences, and witnesses]\label{def:jl-calculus}
\begin{itemize}
\item A \emph{signature} is a pair $(\Sigma, \jlcs)$ consisting of a kernel environment $\Sigma$
  and a constant specification $\jlcs$ over $\Sigma$. A \emph{signature morphism} $\sigma :
  (\Sigma, \jlcs) \to (\Sigma', \jlcs')$ is a kernel signature morphism $\sigma : \Sigma \to
  \Sigma'$ such that every leaf $\lambda$ admitted in $\jlcs$ is also admitted in $\jlcs'$ with
  $P'_\lambda \defeq \sigma(P_\lambda)$.
\item The \emph{sentences} over $(\Sigma, \jlcs)$ are the justification assertions $\JB(j, P)$ with
  $\Sigma \vdash P : \Prop$.
\item The \emph{witnesses} of a sentence are the kernel's decisions of it, recorded as grounds
  judgments (Section~\ref{sec:jl-checker}); the projection $\pi$ maps a witness to the sentence
  $\JB(j, P)$, never to $P$.
\end{itemize}
\end{definition}

The third bullet expresses non-factivity (Theorem~\ref{thm:jl-nonfactive}) at the level of proof
calculi: the witness projection $\pi$ lands in justification assertions $\JB(j, P)$ rather than
underlying propositions $P$. Under the hypotheses of Theorem~\ref{thm:jl-nonfactive}, no witness
transport can map the decision of $\JB(\lambda, P_0)$ to a kernel proof of $P_0$.

\begin{proposition}[Translation along signature morphisms]\label{thm:jl-translation}
If $\sigma : (\Sigma, \jlcs) \to (\Sigma', \jlcs')$ is a signature morphism and $P \in
\Evid_{\jlcs}(j)$, then $\sigma(P) \in \Evid_{\jlcs'}(\sigma(j))$, where $\sigma$ acts on $j$ by
translating the instantiating terms at its $\jinst$ nodes.
\end{proposition}

\begin{proof}
By structural induction on $j$: the leaf case follows from the morphism condition; $\japp$ holds
because $\sigma(A \jlimp B) = \sigma(A) \jlimp \sigma(B)$ and $\sigma$ preserves $\defeq$; $\jsum$
is immediate; and $\jinst$ holds because $\sigma(\Pi x{:}T.\, B) = \Pi x{:}\sigma(T).\, \sigma(B)$,
$\Sigma' \vdash \sigma(a) : \sigma(T)$ by the morphism property on typing, and $\sigma(B[a/x]) =
\sigma(B)[\sigma(a)/x]$.
\end{proof}

Proposition~\ref{thm:jl-translation} establishes the entailment-translation condition for the
atomic justification fragment. Notice that the inclusion $\sigma(\Evid_{\jlcs}(j)) \subseteq
\Evid_{\jlcs'}(\sigma(j))$ can be strict, because $\sigma$ may identify previously distinct
propositions and thereby enable an $\japp$ step that was blocked in the source signature.

\subsection{Comorphisms and warrant transport}

A comorphism translates a claim $P$ of a source logic into a claim $\alpha(P)$ of a target logic.
When both claims are represented as kernel propositions, the translation introduces a bridge leaf
$\beta$ with $P_\beta = P \jlimp \alpha(P)$ (or a universally quantified bridge instantiated via
$\jinst$). Following Section~\ref{sec:jl-no-internalization}, $\beta$ is a $\jver(c)$ leaf if the
comorphism's satisfaction condition for $P$ has a checked proof, and a $\jdecl(c)$ leaf otherwise;
the translated justification term is $\japp(\beta, j)$.

\begin{theorem}[Warrant monotonicity]\label{thm:jl-monotone}
Let $\beta$ be admitted with $P_\beta = P \jlimp \alpha(P)$. Then
\[
  \supp(\japp(\beta, j), \alpha(P)) = \{a \cup \{\beta\} \mid a \in \supp(j, P)\},
\]
and for every lattice valuation $v$, $h_v(\supp(\japp(\beta, j), \alpha(P))) = v(\beta) \wedge
h_v(\supp(j, P)) \leq h_v(\supp(j, P))$. In particular:
\begin{enumerate}
\item the grade of the translated claim is at most the grade of the source claim, with equality
  iff $v(\beta)$ is at least the source grade;
\item $\jlfv$ is preserved iff $\beta$ is a $\jver$ leaf;
\item $\jlsw(\japp(\beta, j), \alpha(P), \jlres(\beta))$ is false whenever $\supp(j, P) \neq
  \emptyset$.
\end{enumerate}
\end{theorem}

\begin{proof}
We have $\Evid(\beta) = \{P \jlimp \alpha(P)\}$. Any antecedent class $A$ in the $\japp$ clause of
Definition~\ref{def:jl-support} satisfies $A \jlimp \alpha(P) \defeq P \jlimp \alpha(P)$, whence
$A \defeq P$ by $\Pi$-injectivity. Since $\supp(\beta, P \jlimp \alpha(P)) = \{\{\beta\}\}$, the
displayed support equation follows. Theorem~\ref{thm:jl-lattice-hom} then gives
$h_v(\{\{\beta\}\} \cdot S) = v(\beta) \wedge h_v(S)$, and items~(1)--(3) are immediate instances.
\end{proof}

Warrant therefore never increases across a comorphism: it either weakens or stays the same. In
particular, any axiom of the source logic that a comorphism adopts without proof in the target
enters as a $\jdecl$ leaf and carries no verified warrant from the source
(Chapter~\ref{ch:institutions}). Chapter~\ref{ch:institutions} classifies the axiom policies that
generate these bridge leaves, using Theorem~\ref{thm:jl-monotone} as its algebra of warrant
transport.

\subsection{Models}

The justification fragment thus carries two complementary semantics:
\begin{itemize}
\item \textbf{Which claims are justified.} $\Evid_{\jlcs}$ is the minimal evidence function of a
  Mkrtychev model over kernel propositions (Proposition~\ref{thm:jl-minimal}): a sentence
  $\JB(j, P)$ is satisfied iff $P \in \Evid_{\jlcs}(j)$, which coincides with derivability by
  Theorem~\ref{thm:jl-decision}. Under the canonical construction of Chapter~\ref{ch:institutions}
  (where models are closed theories and satisfaction is membership), the closed theory of
  $(\Sigma, \jlcs)$ is $\{\JB(j, P) \mid P \in \Evid_{\jlcs}(j)\}$.
\item \textbf{Which grounds support a claim.} $\supp(\cdot, P)$ takes values in the provenance
  semiring $\jlWhy(X)$ (Theorem~\ref{thm:jl-supp-hom}), and every warrant or lineage query is a
  further semiring homomorphic image.
\end{itemize}

\section{Curry-style and Church-style presentations}
\label{sec:jl-curry-church}

Justification terms can be presented either untyped, with the propositions they justify computed by
an external evidence function (\emph{Curry-style}), or as elements of an inductive type family
indexed by the proposition being justified (\emph{Church-style}). Whereas $\JTerm$ in
Section~\ref{sec:jl-term-algebra} is Curry-style, it is instructive to examine what changes under a
Church-style presentation.

\begin{definition}[Church-style justification family]\label{def:jl-church}
A \emph{Church-style justification family} is an inductive family $\jlG : \Prop \to
\Typeu_k$ equipped with a constructor of type $\jlG(P_\lambda)$ for each admitted leaf $\lambda$,
an application constructor
\[
  \japp : \jlG(A \jlimp B) \to \jlG(A) \to \jlG(B),
\]
and constructors for sum and instantiation typed as analyzed below. Every well-typed term
$g : \jlG(P)$ justifies a unique proposition $P$ up to $\defeq$.
\end{definition}

\paragraph{The judgment form.}
In a Church-style family, the assertion ``$g$ justifies $P$'' is the typing judgment $g : \jlG(P)$
rather than a proposition in $\Prop$: consequently, it cannot be negated, assumed as a hypothesis,
or nested inside another justification. The Curry-style proposition $\JB(j, P)$ supports all three.
Moreover, because $\jlG$ is indexed directly by $\Prop$, propositional extensionality would force
$\jlG(P) = \jlG(Q)$ whenever $P \leftrightarrow Q$, sacrificing hyperintensionality
(Proposition~\ref{thm:jl-hyperintensional}).

\begin{proposition}[Sum splits]\label{thm:jl-sum-splits}
Let $\Evid_{\jlG}(g) = \{P \mid g : \jlG(P)\}$, which contains at most one $\defeq$-class. Any
binary constructor $\oplus$ satisfying the Sum inclusion $\Evid_{\jlG}(s) \cup \Evid_{\jlG}(t)
\subseteq \Evid_{\jlG}(s \oplus t)$ is typable only on summands with $\Evid_{\jlG}(s) =
\Evid_{\jlG}(t)$. Consequently, passing from Curry-style to Church-style splits $\mathsf{LP}$'s Sum
into two distinct operations:
\begin{enumerate}
\item \emph{alternatives for a single proposition}, $\oplus : \jlG(P) \to \jlG(P) \to \jlG(P)$; and
\item \emph{aggregation across distinct propositions}, which requires a compound propositional
  index such as conjunction, $\jpair : \jlG(A) \to \jlG(B) \to \jlG(A \wedge B)$.
\end{enumerate}
In addition, every summand of $\oplus$ must be a well-typed member of $\jlG(P)$, so a summand that
justifies nothing has no type and cannot occur.
\end{proposition}

\begin{proof}
Every well-typed term $g$ has $|\Evid_{\jlG}(g)| = 1$ up to $\defeq$. If $s : \jlG(A)$ and $t :
\jlG(B)$, the Sum inclusion places both $A$ and $B$ in the singleton $\Evid_{\jlG}(s \oplus t)$,
forcing $A \defeq B$. Combining $s : \jlG(A)$ and $t : \jlG(B)$ when $A \not\defeq B$ therefore
requires a constructor whose target index differs from both $A$ and $B$, such as $A \wedge B$.
The final claim holds because every argument of a constructor in $\jlG$ must inhabit some fiber
$\jlG(P)$.
\end{proof}

By contrast, the Curry-style presentation retains Sum as a single operation with $\Evid(\jsum(s, t))
= \Evid(s) \cup \Evid(t)$: alternatives for a single proposition and aggregation across distinct
propositions are both instances of $\jsum$, distinguished only by the target proposition $P$ at
which $\supp(\cdot, P)$ is evaluated (Example~\ref{def:jl-shared}). A summand that justifies
nothing remains a well-formed term in $\JTerm$, and Theorem~\ref{thm:jl-phantom} excludes it from
$\supp(\cdot, P)$ automatically.

\paragraph{Universe level and higher-order unification.}
In a Church-style family, a constructor for universal instantiation must quantify over the domain
type $T$:
\[
  \jinst : \Pi (T : \Srt\,\ell)(B : T \to \Prop).\ \jlG(\Pi x{:}T.\, B\, x) \to \Pi (a : T).\,
  \jlG(B\, a).
\]
By the universe constraint on inductive families outside $\Prop$ (Definition~\ref{def:mltt-admissible}),
quantifying over a parameter $T : \Setu$ (which itself has type $\Typeu_1$) forces $\jlG$ to live
in $\Typeu_1$ or higher, and quantifying over $T : \Typeu_m$ forces $\jlG$ into $\Typeu_{m+1}$ or
higher; furthermore, because $\Prop$ is not cumulative into $\Setu$, instantiating universal
quantifiers over proofs ($T : \Prop$) requires a second constructor. In practice, $T$ and the
predicate $B$ are left implicit at use sites, so type-checking $\jlG(B\, a)$ requires recovering
$B$ from $B\, a$ by higher-order unification outside Miller's pattern fragment. The Curry-style
constructor $\jinst(j, \jlq{a})$ avoids all three costs: it stores only the syntactic code
$\jlq{a}$, keeps $\JTerm$ in the lowest data universe $\Setu$, and computes $B[a/x]$ by matching
each $\Pi$-type in $\Evid(j)$ syntactically.

\paragraph{Derivations.}
Whereas a Church-style term of type $\jlG(P)$ built with $\oplus$ and $\jpair$ records a single
typing derivation, Proposition~\ref{thm:jl-prov-derivations} shows that the Curry-style
presentation recovers all derivations of $\JB(j, P)$ as the monomials of $\jlprov(j, P)$ without
baking a single derivation into the term syntax. Related typed justification calculi with explicit
proof terms for justification assertions are studied in the intensional $\lambda$-calculus of
\citet{artemov-bonelli2007} and in the J-Calc system of \citet{pouliasis-primiero2014}.

\subsection{Conjunction introduction as an optional former}

In the Curry-style presentation, aggregating evidence via $\jsum(s, t)$ produces a single term that
justifies both $A$ and $B$ separately, but does not by itself justify the conjunction $A \wedge B$.
Adding an optional pairing former $\jpair(s, t)$ internalizes conjunction introduction directly.

\begin{definition}[Optional former $\jpair$]\label{def:jl-pair}
Extending $\JTerm$ by $\jpair(s, t)$, the evidence and support clauses are
\begin{align*}
  \Evid(\jpair(s, t)) &= \{A \wedge B \mid A \in \Evid(s),\ B \in \Evid(t)\}, \\
  \supp(\jpair(s, t), P) &= \bigcup\, \{\supp(s, A) \cdot \supp(t, B) \mid A \in \Evid(s),\ B \in
    \Evid(t),\ A \wedge B \defeq P\}.
\end{align*}
\end{definition}

With $\jpair$, the results of Sections~\ref{sec:jl-evidence} and~\ref{sec:jl-support} extend by one
clause each: $|\Evid(\jpair(s,t))| \leq |\Evid(s)| \cdot |\Evid(t)|$ (making the size bound of
Lemma~\ref{thm:jl-finite} polynomial rather than linear in the worst case), the inference rule
\[
  \inferrule*[right=Pair]{\jlvd \JB(s, A) \\ \jlvd \JB(t, B)}{\jlvd \JB(\jpair(s,t), A \wedge B)}
\]
is added to Definition~\ref{def:jl-rules}, and $\jpair$ acts as multiplication in $\jlWhy(X)$. Even
without adding $\jpair$ as a primitive constructor, however, conjunction introduction is already
definable using a single verified lemma leaf:

\begin{proposition}[$\jpair$ is definable up to one verified leaf]\label{thm:jl-pair-simulation}
Let $\kappa = \jver(c)$ be admitted with $P_\kappa = \Pi (A\, B : \Prop).\, A \jlimp B \jlimp A
\wedge B$, and let $A \in \Evid(s)$ and $B \in \Evid(t)$. For the term $k =
\japp(\japp(\jinst(\jinst(\kappa, \jlq{A}), \jlq{B}), s), t)$, we have
\[
  A \wedge B \in \Evid(k), \qquad \supp(k, A \wedge B) = \{\{\kappa\}\} \cdot \supp(s, A) \cdot
  \supp(t, B).
\]
Hence $h_v(\supp(k, A \wedge B)) = v(\kappa) \wedge h_v(\supp(s, A) \cdot \supp(t, B))$, and for
$\jlfv$ as well as for $\jlsw$ at any resource $r \neq c$, the query values for $k$ coincide with
those of $\jpair(s, t)$ at $A \wedge B$ whenever $A \wedge B$ has a unique conjunction
decomposition.
\end{proposition}

\begin{proof}
Two instantiations at $A, B : \Prop$ give $\Evid(\jinst(\jinst(\kappa, \jlq{A}), \jlq{B})) =
\{A \jlimp B \jlimp A \wedge B\}$, and two applications yield $A \wedge B \in \Evid(k)$ together
with the stated support product. For the lattice queries, $v(\kappa) = \top$ under both valuations
of Theorem~\ref{thm:jl-lattice-hom}.
\end{proof}

Thus $\jpair(s, t)$ is a syntactic convenience rather than a theoretical necessity: it saves citing
the lemma resource $\kappa$ and writing two $\jinst$ nodes per conjunction, while leaving all
absorption-invariant queries unchanged (and differing on lineage queries only by the presence of
$\kappa$). Because $\kappa$ is a $\jver$ leaf for a proved tautology and therefore needs no
accountable owner, the accountability requirement of Section~\ref{sec:jl-no-internalization} is
satisfied equally well by either choice.

\section{The checker operator and judgments}
\label{sec:jl-checker}

\paragraph{Judgments.}
On the chain, a \emph{judgment} is a record $\mathrm{holds}(L, t, T)$ consisting of a logic $L$, a
witness term $t$, and a sentence or type $T$, recording that the checker for $L$ verified $t$
against $T$. Two kinds of judgments participate in the justification layer:
\begin{itemize}
\item a \emph{proof judgment} $\mathrm{holds}(L, t, P)$ for a proposition $P$, which is factive; and
\item a \emph{grounds judgment} $\mathrm{holds}(\mathrm{kernel}, d, \JB(j, P))$, which records
  the kernel's decision $d$ of a justification assertion and is therefore a proof judgment whose
  target proposition is $\JB(j, P)$.
\end{itemize}
These judgments stratify cleanly: a grounds judgment establishes $\JB(j, P)$, which is a statement
about $j$ and $P$, without establishing $P$ itself:
\[
  \mathrm{holds}(\mathrm{kernel}, d, \JB(j, P)) \quad\longrightarrow\quad \JB(j, P)
  \quad \not\longrightarrow \quad P.
\]
The first arrow reflects the factivity of proof judgments, while the second is barred by
Theorem~\ref{thm:jl-nonfactive}.

\paragraph{The operator $\jlchk$.}
Because $\JB(j, P)$ is a proposition in $\Prop$, $\mathsf{LP}$'s positive-introspection operator
$\jlchk$ becomes expressible: $\jlchk j$ is a justification term for every justification assertion
$\JB(j, P)$ satisfied by $j$, reifying the checker's verdict inside the term algebra.

\begin{definition}[Checker extension]\label{def:jl-checker}
Extend $\JTerm$ by the unary constructor $\jlchk j$, with
\[
  \Evid(\jlchk j) = \{\JB(j, P) \mid P \in \Evid(j)\}, \qquad
  \supp(\jlchk j, Q) = \begin{cases} \{\emptyset\} & \text{if } Q \in \Evid(\jlchk j), \\
  \emptyset & \text{otherwise}, \end{cases}
\]
and add the inference rule $\inferrule{\jlvd \JB(j, P)}{\jlvd \JB(\jlchk j, \JB(j, P))}$.
\end{definition}

Notice that the support of $\jlchk j$ for a valid assertion $\JB(j, P)$ is the multiplicative unit
$1 = \{\emptyset\}$ of $\jlWhy(X)$: the checker's verdict rests on the kernel's decision procedure
(which is part of the trusted computing base) rather than on any external leaf resource.

\begin{proposition}[J4 and factivity of checker terms]\label{thm:jl-j4}
In the checker extension:
\begin{enumerate}
\item for all closed $j$ and $P$, positive introspection is provable:
  \[ \JB(j, P) \jlimp \JB(\jlchk j, \JB(j, P)); \]
\item if $\JB(\jlchk j, Q)$ holds, then $Q \defeq \JB(j, P)$ for some $P$ with
  $\JB(j, P)$ provable (factivity for checker terms);
\item if $\JB(j, P)$ holds, then $\jlfv(\jlchk j, \JB(j, P))$ holds.
\end{enumerate}
\end{proposition}

\begin{proof}
(1) $\JB(j, P)$ is decided by Definition~\ref{def:jl-jb}: if it holds, then $\JB(j, P) \in
\Evid(\jlchk j)$ and the consequent is provable; otherwise $\neg\JB(j, P)$ is provable and the
implication holds vacuously. (2) Follows immediately from the definition of $\Evid(\jlchk j)$ and
Definition~\ref{def:jl-jb}. (3) The empty leaf set $\emptyset \in \supp(\jlchk j, \JB(j, P))$
satisfies the universal condition of $\jlfv$ vacuously.
\end{proof}

With $\jlchk$, the Eigenius fragment becomes a fragment of $\mathsf{J4}$: Application, Sum, and
positive introspection all hold, while factivity holds specifically for checker terms $\jlchk j$
and for all-verified alternatives (Theorem~\ref{thm:jl-verified-factive}), remaining false for
arbitrary justification terms. In particular, a nested assertion $\JB(s, \JB(j, P))$ whose outer
term $s$ is not $\jlchk j$ remains non-factive: for example, a $\jdecl$ leaf with proposition
$\JB(j, P)$ merely records that an agent claimed $j$ justifies $P$, a claim with which the kernel's
own decision of $\JB(j, P)$ may disagree.

Without the $\jlchk$ constructor, a grounds judgment records the checker's verdict as chain metadata
outside $\JTerm$; with $\jlchk$, the verdict can be cited directly inside compound justification
terms, placing the fragment at $\mathsf{J4}$ in the hierarchy of Definition~\ref{def:jl-hierarchy}.

\section{Conservativity}
\label{sec:jl-open}

Because the decision rules of Definition~\ref{def:jl-jb} extend the kernel type theory by axioms, we
conclude by showing that this extension is conservative: it proves no new propositions that do not
mention $\JB$.

\begin{proposition}[Conservativity of the justification extension]\label{thm:jl-conservative}
Let $K$ be the kernel type theory of Chapter~\ref{ch:mltt} with the inductive types $\tttBool$,
$\Nat$, $\JTerm$, and $\jlCode$, and let $K_{\JB}$ extend $K$ by the constant $\JB : \JTerm \to
\jlCode \to \Prop$ and the decision rules of Definition~\ref{def:jl-jb}, for a constant
specification satisfying (CS1)--(CS3), with or without the checker extension of
Definition~\ref{def:jl-checker}. Then for every proposition $\varphi$ of $K$ that does not
mention $\JB$, if $K_{\JB} \vdash t : \varphi$ for some $t$, then $K \vdash t' : \varphi$ for
some $t'$. In particular, $K_{\JB}$ is consistent if $K$ is, and $K$ is consistent by
Theorem~\ref{thm:mltt-declarative-consistency}, relative to the inaccessible cardinals of
Chapter~\ref{ch:mltt}.
\end{proposition}

\begin{proof}
Fix a derivation $\mathcal{D}$ of $K_{\JB} \vdash t : \varphi$. Because $\mathcal{D}$ is finite, it
invokes finitely many decision rules; let $(t_1, c_1), \ldots, (t_n, c_n)$ be the closed
constructor-tree pairs at which $\mathcal{D}$ uses a positive decision rule.

First, we construct a closed Boolean decision term $J : \JTerm \to \jlCode \to \Prop$ inside $K$.
In $K$, structural equality of codes $\mathsf{eq} : \jlCode \to \jlCode \to \tttBool$---defined by
double recursion on $\jlCode$ using label equality on $\Nat$---reduces on any two closed
constructor trees to $\mathsf{true}$ if they are identical and to $\mathsf{false}$ otherwise.
Similarly, a recursive encoding $\mathsf{enc} : \JTerm \to \jlCode$ reduces on any closed
constructor tree of $\JTerm$ to its code injectively. Writing $\langle j, c \rangle$ for
$\mathsf{node}\,0\,(\mathsf{enc}\,j)\,c$, and defining $\mathsf{T} : \tttBool \to \Prop$ by
$\tttBool$-elimination to map $\mathsf{true}$ to $\Pi P{:}\Prop.\, P \to P$ and $\mathsf{false}$ to
$\tttFalse$, we set
\[
  J(j, c) \;=\; \mathsf{T}\bigl(\mathsf{eq}\,\langle j, c \rangle\,\langle t_1, c_1 \rangle
    \mathbin{\|} \cdots \mathbin{\|} \mathsf{eq}\,\langle j, c \rangle\,\langle t_n, c_n \rangle
    \mathbin{\|} \mathsf{false}\bigr) \;:\; \Prop,
\]
where $\|$ denotes Boolean disjunction.

Next, we replace $\JB$ by $J$ throughout $\mathcal{D}$ and replace each decision-axiom application
by the corresponding proof term constructed below. Because $\JB$ has no $\delta$-rule, no
eliminator, and no rule relating distinct codes---and because no rule of $K$ mentions $\JB$---every
rule step of $K$ in $\mathcal{D}$ remains a valid instance of the same rule after substitution. In
the terminology of Chapter~\ref{ch:institutions}, this replacement is a kernel signature morphism
(Definition~\ref{def:inst-kernel-sign}) from the finite extension of $K$ used by $\mathcal{D}$ back
to $K$, so the translated derivation is a valid derivation in $K$
(Lemma~\ref{thm:inst-translation}).

Finally, each decision rule used in $\mathcal{D}$ is discharged in $K$:
\begin{itemize}
\item A positive decision rule at $(j, c)$ with constructor form $(t_i, c_i)$ translates to
  $J(j, c)$, which reduces to $\Pi P{:}\Prop.\, P \to P$ and is inhabited by $\lambda P\, p.\, p$.
\item A negative decision rule at $(j, c)$ with constructor form $(j_0, c_0)$ translates to
  $\neg J(j, c)$. By confluence (Lemma~\ref{lem:mltt-confluence}), $(j_0, c_0)$ is the unique
  constructor form of $(j, c)$, and because the computation of $\Evid$ is deterministic,
  $(j_0, c_0) \neq (t_i, c_i)$ for all $1 \leq i \leq n$. Injectivity of $\mathsf{enc}$ ensures
  that $\langle j_0, c_0 \rangle$ differs from every $\langle t_i, c_i \rangle$, so $J(j, c)$
  reduces to $\tttFalse$ and $\neg J(j, c) = (\tttFalse \to \tttFalse)$ is inhabited by
  $\lambda h.\, h$.
\end{itemize}
Since $\varphi$ does not mention $\JB$, it is left unchanged by the replacement, yielding a
derivation of $K \vdash t' : \varphi$ where $t'$ is $t$ with $J$ substituted for $\JB$. Taking
$\varphi = \tttFalse$ yields relative consistency. Adding the checker operator $\jlchk$ changes
nothing in this argument: $\Evid(\jlchk j)$ is still computed deterministically from closed data,
and codes of propositions mentioning $\JB$ are simply closed constructor trees in $\jlCode$.
\end{proof}

This proof is the type-theoretic counterpart of the finite-definition conservativity argument for a
typed disquotational truth predicate over its base theory~\citep{halbach2011}: any single derivation
uses only finitely many instances, and the predicate restricted to those instances is explicitly
definable in the base theory. Notice that the argument relies crucially on indexing $\JB$ by
syntactic codes in $\jlCode$: if $\JB$ were indexed directly by $\Prop$, it would be subject to
kernel conversion and any propositional extensionality principles at its second argument, so the
decision axioms would no longer be statements about distinct closed constructor trees
(Proposition~\ref{thm:jl-hyperintensional}). Whereas the analogous conservativity result for the
propositional justification logics $\mathsf{J}_{\jlCS}$ follows semantically from completeness for
Mkrtychev models~\citep{artemov-fitting2019}, the kernel type theory has no such Mkrtychev
completeness theorem, making the syntactic finite-definition argument above the natural route.
Moreover, conservativity does not require the kernel's conversion test to be complete: under any
deterministic conversion algorithm, $\JB$ decides a subrelation of $\jlvd$
(Theorem~\ref{thm:jl-decision}), and Proposition~\ref{thm:jl-conservative} applies to that decided
relation.

In sum, whether to include $\jpair$ and $\jlchk$ as primitive constructors of $\JTerm$ are modular
design choices within the theory; Propositions~\ref{thm:jl-pair-simulation} and~\ref{thm:jl-j4}
record their exact consequences.

\chapter{Institution Theory over Constructive Type Theories}
\label{ch:institutions}

\providecommand{\instE}{\mathbb{E}}             
\providecommand{\instP}{\mathbb{P}}             
\providecommand{\instK}{\mathbb{K}}             
\providecommand{\instJ}{\mathcal{J}}            
\providecommand{\instI}{\mathcal{I}}            
\providecommand{\instclo}[1]{#1^{\bullet}}      
\providecommand{\instsign}{\mathit{sign}}       
\providecommand{\instPr}{\mathit{Pr}}
\providecommand{\instPC}{\mathbf{PCalc}}
\providecommand{\instMultCat}{\mathbf{MultCat}}
\providecommand{\instco}{\mathrm{co}}
\providecommand{\instax}[1]{\mathsf{a}_{#1}}    
\providecommand{\instFA}{\mathrm{FA}}           
\providecommand{\insthat}[1]{\widehat{#1}}      
\providecommand{\instThl}{\mathbf{Th}_{\mathrm{lst}}}
\providecommand{\instimp}{\Rightarrow}          
\providecommand{\instdia}{\diamond}
\providecommand{\instG}{\mathsf{G}}
\providecommand{\instF}{\mathsf{F}}
\providecommand{\instU}{\mathsf{U}}
\providecommand{\instJTh}{\mathcal{J}^{\mathrm{th}}}
\providecommand{\tttFalse}{\mathsf{False}}
\providecommand{\tttBool}{\mathsf{Bool}}

Eigenius hosts several logics: the kernel's type theory (Chapter~\ref{ch:mltt}),
justification logic (Chapter~\ref{ch:justification-logic}), Lean's kernel theory
through a hosted checker, and statistical inference. It also translates claims between
them. The logics and translations are formalized at three levels, each obtained from the
one above by forgetting structure:
\begin{enumerate}[label=(\arabic*)]
\item \emph{Witnesses.} Each logic is a proof calculus in the sense of
  \citet{meseguer1989}; its proofs are the witnesses the system stores. A translation
  between logics is a map of proof calculi: a witness transport whose output proves the
  translated sentence.
\item \emph{Provability.} Forgetting witnesses leaves a $\pi$-institution, a closure
  operator per signature \citep{fiadeiro-sernadas1988}. A translation becomes a
  $\pi$-institution comorphism.
\item \emph{Models.} The functor $\instG : \piIns^{\instco} \to \Ins^{\instco}$ of
  \citet{mariano-pinto2017} turns each $\pi$-institution into an institution whose
  models are closed theories and whose satisfaction relation is membership. Each
  comorphism becomes an institution comorphism, and the satisfaction condition holds by
  construction.
\end{enumerate}

The system maintains a proof calculus for each hosted logic and evaluates no models at
runtime. Nevertheless, it needs the vocabulary of institution theory---comorphisms and
the satisfaction condition---in which translations between logics are stated and compared
with the literature. A $\pi$-institution provides that vocabulary without models: its
two defining conditions, structurality of the closure operators and preservation of
consequence by a comorphism (Definitions~\ref{def:inst-pi-institution}
and~\ref{def:inst-pi-comorphism}), are the two provability requirements of
Section~\ref{sec:inst-needs}. Whereas Meseguer's maps of entailment systems act on
theories, $\pi$-institution comorphisms act on signatures, just as institution
comorphisms do; Section~\ref{sec:inst-forgetting} promotes theories to signatures so
that the two agree.

The coreflection of Theorem~\ref{thm:inst-mariano-pinto} then supplies an institution
for every $\pi$-institution, with the satisfaction condition holding by construction, so
no model category needs to be built by hand for every hosted logic. By contrast, a logic
in Meseguer's sense (Definition~\ref{def:inst-logic}) pairs an entailment system with a
semantic institution and a soundness proof; the kernel theory together with the
set-theoretic models of Chapter~\ref{ch:mltt} is such a logic
(Theorem~\ref{thm:inst-kernel-logic}). Because the models produced by $\instG$ are
purely syntactic (Section~\ref{sec:inst-limits}), soundness with respect to the semantic
models of a hosted logic is a separate obligation for each checker
(Section~\ref{sec:inst-real-models}). In summary, the system's operational requirements
are stated at the witness level, while the provability and syntactic-model levels are
derived from it by two forgetful constructions. Section~\ref{sec:inst-status} records
each obligation together with its status.

\paragraph{Notation.}
Meseguer writes $\mathit{sen}$ for the sentence functor; this chapter writes $\Sen$
throughout, and $\Sign$ for a category of signatures. For an entailment system,
$\instclo{\Gamma}$ denotes the set of sentences entailed by $\Gamma$. Composition of
natural transformations is written $\beta \cdot \alpha$, and whiskering is written by
juxtaposition ($\alpha\Phi$).

\section{Institutions and the Requirements of a Witness-Storing System}
\label{sec:inst-requirements}

An institution axiomatizes a logic by its signatures, sentences, models, and a
satisfaction relation invariant under change of signature
\citep{goguen-burstall1984,goguen1992}.

\begin{definition}[Institution]\label{def:inst-institution}
An \emph{institution} $\instI = (\Sign, \Sen, \Mod, \models)$ consists of a category
$\Sign$ of signatures, a functor $\Sen : \Sign \to \SetC$, a functor
$\Mod : \Sign^{\mathrm{op}} \to \Cat$, and for each $\Sigma \in |\Sign|$ a relation
${\models_\Sigma} \subseteq |\Mod(\Sigma)| \times \Sen(\Sigma)$, such that for every
$f : \Sigma \to \Sigma'$, $M' \in |\Mod(\Sigma')|$, and $\varphi \in \Sen(\Sigma)$,
\[
  M' \models_{\Sigma'} \Sen(f)(\varphi)
  \iff \Mod(f)(M') \models_{\Sigma} \varphi
  \qquad\text{(satisfaction condition).}
\]
\end{definition}

This is Definition~3 of \citet{meseguer1989} and Definition~1.1 of
\citet{bittencourt-rios2022}.
For $\Gamma \subseteq \Sen(\Sigma)$, $\Gamma \models_\Sigma \varphi$ holds iff every
$\Sigma$-model satisfying all of $\Gamma$ satisfies $\varphi$; by Proposition~4 of
\citet{meseguer1989}, this relation is an entailment system
(Definition~\ref{def:inst-entailment}).

\begin{definition}[Morphism and comorphism]\label{def:inst-comorphism}
Let $\instI$ and $\instI'$ be institutions.
\begin{enumerate}[label=(\alph*)]
\item An \emph{institution morphism} $(\Phi, \alpha, \beta) : \instI \to \instI'$
  consists of a functor $\Phi : \Sign \to \Sign'$ and natural transformations
  $\alpha : \Sen' \circ \Phi \Rightarrow \Sen$ and
  $\beta : \Mod \Rightarrow \Mod' \circ \Phi^{\mathrm{op}}$ such that
  $m \models_\Sigma \alpha_\Sigma(\varphi') \iff \beta_\Sigma(m) \models'_{\Phi\Sigma}
  \varphi'$.
\item An \emph{institution comorphism} $\langle \Phi, \alpha, \beta \rangle : \instI
  \to \instI'$ consists of a functor $\Phi : \Sign \to \Sign'$ and natural
  transformations $\alpha : \Sen \Rightarrow \Sen' \circ \Phi$ and
  $\beta : \Mod' \circ \Phi^{\mathrm{op}} \Rightarrow \Mod$ such that for all
  $\Sigma$, $m' \in |\Mod'(\Phi\Sigma)|$, and $\varphi \in \Sen(\Sigma)$,
  \[
    m' \models'_{\Phi\Sigma} \alpha_\Sigma(\varphi) \iff
    \beta_\Sigma(m') \models_\Sigma \varphi .
  \]
\end{enumerate}
Institutions with comorphisms form the category $\Ins^{\instco}$; composition is
$\langle \Phi', \alpha', \beta' \rangle \circ \langle \Phi, \alpha, \beta \rangle =
\langle \Phi'\Phi,\ \alpha'\Phi \cdot \alpha,\ \beta \cdot \beta'\Phi^{\mathrm{op}}
\rangle$.
\end{definition}

Both notions are Definition~1.3 of \citet{bittencourt-rios2022}. The direction
of each component determines the use. A morphism translates sentences from target to
source and models from source to target. A comorphism translates sentences from source
to target and models from target to source. Meseguer's map of institutions
(Definition~27 of \citet{meseguer1989}) has the comorphism directions, with $\Phi$
acting on theories; Meseguer's remark following Lemma~28 records the reversal of
$\alpha$ and $\beta$ relative to Goguen and Burstall's institution morphisms.
\citet{goguen-rosu2002} classify these and further variants, among them forward
morphisms and semi-morphisms. Eigenius uses comorphisms: a claim in a source logic is
translated forward into a target logic, checked there, and the result is read back as
evidence about the source claim.

\subsection{Requirements}
\label{sec:inst-needs}

The requirements below are derived from what the system does with claims. Each is
assigned to the level that supplies it.

\begin{center}
\small
\begin{tabularx}{\textwidth}{@{}>{\raggedright\arraybackslash}X
  >{\raggedright\arraybackslash}p{0.38\textwidth} l@{}}
\toprule
Requirement & Occasion & Level \\
\midrule
Provability is preserved under change of vocabulary (structurality,
  Definition~\ref{def:inst-pi-institution}) &
  extension of a theory, renaming, merging of theories & provability \\
Translation between logics preserves provability (the comorphism condition,
  Definition~\ref{def:inst-pi-comorphism}) &
  every comorphism & provability \\
The witness is a first-class object &
  every accepted claim is held together with the witness that established it &
  witnesses \\
Witnesses are transported, and the transported witness is checked &
  every comorphism & witnesses \\
The justification supporting a claim is identified &
  justification terms $j : P$, warrant grades, audit & witnesses \\
Each logic has its own witness kind &
  kernel terms, externally checked proofs, statistical verdicts & witnesses \\
Models &
  not evaluated by the system; required for meta-level soundness statements and
  for the vocabulary of institution theory & models \\
\bottomrule
\end{tabularx}
\end{center}

Four of the seven requirements refer to witnesses. An entailment system or a
$\pi$-institution records only \emph{that} $\Gamma$ entails $\varphi$, not by which
proof; both are therefore insufficient as the primary structure. The witness level is
primary, and the other two levels are derived from it.

\section{Entailment Systems, Proof Calculi, and \texorpdfstring{$\pi$}{pi}-Institutions}
\label{sec:inst-frameworks}

The three levels use three frameworks. Each definition is stated as in its primary
source.

\begin{definition}[Entailment system; Meseguer, Definition~1]
\label{def:inst-entailment}
An \emph{entailment system} $\instE = (\Sign, \Sen, \vdash)$ consists of a category
$\Sign$, a functor $\Sen : \Sign \to \SetC$, and for each $\Sigma$ a relation
${\vdash_\Sigma} \subseteq \mathcal{P}(\Sen(\Sigma)) \times \Sen(\Sigma)$ such that
\begin{enumerate}[label=(\arabic*)]
\item \emph{reflexivity:} $\{\varphi\} \vdash_\Sigma \varphi$;
\item \emph{monotonicity:} $\Gamma \vdash_\Sigma \varphi$ and $\Gamma' \supseteq
  \Gamma$ imply $\Gamma' \vdash_\Sigma \varphi$;
\item \emph{transitivity:} $\Gamma \vdash_\Sigma \varphi_i$ for all $i \in I$ and
  $\Gamma \cup \{\varphi_i \mid i \in I\} \vdash_\Sigma \psi$ imply
  $\Gamma \vdash_\Sigma \psi$;
\item \emph{$\vdash$-translation:} $\Gamma \vdash_\Sigma \varphi$ implies
  $H(\Gamma) \vdash_{\Sigma'} H(\varphi)$ for every $H : \Sigma \to \Sigma'$, where
  $H$ abbreviates $\Sen(H)$.
\end{enumerate}
$\instE$ is \emph{compact} if $\Gamma \vdash_\Sigma \varphi$ implies
$\Gamma_0 \vdash_\Sigma \varphi$ for some finite $\Gamma_0 \subseteq \Gamma$.
\end{definition}

The map $\Gamma \mapsto \instclo{\Gamma} = \{\varphi \mid \Gamma \vdash \varphi\}$ is a
closure function (\citet{meseguer1989}, Remark~2 after Definition~1).

\begin{definition}[Theories; Meseguer, Definition~2]\label{def:inst-theories}
A \emph{theory} is a pair $T = (\Sigma, \Gamma)$ with $\Gamma \subseteq \Sen(\Sigma)$.
A \emph{theory morphism} $H : (\Sigma, \Gamma) \to (\Sigma', \Gamma')$ is a signature
morphism $H : \Sigma \to \Sigma'$ with $\Gamma' \vdash_{\Sigma'} H(\varphi)$ for every
$\varphi \in \Gamma$; theories and theory morphisms form $\Th$. $H$ is
\emph{axiom-preserving} if $H(\Gamma) \subseteq \Gamma'$; theories and
axiom-preserving morphisms form the subcategory $\Thz$. The functor
$\instsign : \Th \to \Sign$ projects to signatures.
\end{definition}

$\Thz$ does not depend on $\vdash$ (\citet{meseguer1989}, Definition~2). Remark~5 after
Definition~2 extends entailment to theories: $\Delta \vdash_{(\Sigma, \Gamma)} \varphi$
iff $\Delta \cup \Gamma \vdash_\Sigma \varphi$.

\begin{definition}[Logic; Meseguer, Definition~6]\label{def:inst-logic}
A \emph{logic} $(\Sign, \Sen, \Mod, \vdash, \models)$ is an entailment system
$(\Sign, \Sen, \vdash)$ and an institution $(\Sign, \Sen, \Mod, \models)$ such that
$\Gamma \vdash_\Sigma \varphi$ implies $\Gamma \models_\Sigma \varphi$
(soundness). It is \emph{complete} if the converse also holds.
\end{definition}

\begin{definition}[Proof calculus; Meseguer, Definition~12]\label{def:inst-proof-calculus}
A \emph{proof calculus} $\instP = (\Sign, \Sen, \vdash, P, \instPr, \pi)$ consists of
an entailment system $(\Sign, \Sen, \vdash)$; a functor $P : \Thz \to
\mathbf{Struct}_{\instP}$, giving each theory $T$ its \emph{proof-theoretic
structure} $P(T)$; a functor $\instPr : \mathbf{Struct}_{\instP} \to \SetC$, with
$\proofs = \instPr \circ P : \Thz \to \SetC$; and a natural transformation
$\pi : \proofs \Rightarrow \Sen$ (with $\Sen$ extended to theories along
$\instsign$) such that for each $T = (\Sigma, \Gamma)$ the image of
$\pi_T : \proofs(T) \to \Sen(T)$ is $\instclo{\Gamma}$.
\end{definition}

A \emph{logical system} (Definition~13 of \citet{meseguer1989}) is a logic together
with a proof calculus on its entailment system. The proof-theoretic structure is
typically a multicategory: Definition~10 of \citet{meseguer1989} defines
multicategories as strict monoidal categories whose monoid of objects is free, and
Example~11 builds $P(T)$ for natural deduction, with composition given by grafting
proof trees at leaves. $P$ is defined on $\Thz$ and not on $\Th$: a theory morphism
sends an axiom to a theorem of the target, and the target proof of that theorem is not
determined (\citet{meseguer1989}, the discussion following Example~11).

\begin{definition}[Map of entailment systems; Meseguer, Definitions~21 and 23]
\label{def:inst-ent-map}
Let $\instE, \instE'$ be entailment systems, $\Phi : \Thz \to \Thz'$ a functor, and
$\alpha : \Sen \Rightarrow \Sen' \circ \Phi$ a natural transformation. Write
$\Phi(\Sigma, \Gamma) = (\Sigma', \Gamma')$ and $\Phi(\Sigma, \emptyset) = (\Sigma',
\emptyset')$. $\Phi$ is \emph{$\alpha$-sensible} if
(1) there is a functor $\Phi^{\instdia} : \Sign \to \Sign'$ with
$\instsign' \circ \Phi = \Phi^{\instdia} \circ \instsign$, and
(2) $\instclo{(\Gamma')} = \instclo{(\emptyset' \cup \alpha_\Sigma(\Gamma))}$.
A \emph{map of entailment systems} $(\Phi, \alpha) : \instE \to \instE'$ is such a
pair with $\Phi$ $\alpha$-sensible and
\[
  \Gamma \vdash_\Sigma \varphi \implies
  \alpha_\Sigma(\Gamma) \vdash'_{\Phi(\Sigma, \emptyset)} \alpha_\Sigma(\varphi).
\]
\end{definition}

By Lemma~22 of \citet{meseguer1989}, condition~(1) forces $\alpha$ to depend only on
signatures: $\alpha$ is obtained by whiskering a natural transformation
$\alpha^{\instdia} : \Sen \Rightarrow \Sen' \circ \Phi^{\instdia}$ with $\instsign$.
Condition~(2) then bridges signature-level and theory-level entailment preservation,
yielding the theory-relative form of Lemma~24 of \citet{meseguer1989}:
$\Gamma \vdash_{(\Sigma, \Delta)} \varphi$ implies
$\alpha_\Sigma(\Gamma) \vdash'_{\Phi(\Sigma, \Delta)} \alpha_\Sigma(\varphi)$.
Maps of entailment systems compose via $\Phi_3 = \Phi_2 \Phi_1$ and
$\alpha_3 = \alpha_2 \Phi_1 \cdot \alpha_1$, forming the category $\Ent$; the composite
remains $\alpha$-sensible because the image of a set of sentences under a map is
determined, up to closure over the axioms added by the map, by the closure of that set.

In practice, theory functors arise from signature translations as follows
(\citet{meseguer1989}, \S4.1). A functor $\Psi : \Sign \to \Thz'$ equipped with a
natural transformation $\alpha : \Sen \Rightarrow \Sen' \circ \instsign' \circ \Psi$
extends to theories by setting
\[
  \Psi(\Sigma, \Gamma) = \big(\instsign'(\Psi\Sigma),\ \Gamma_{\Psi\Sigma} \cup
  \alpha_\Sigma(\Gamma)\big),
\]
called the \emph{$\alpha$-extension} of $\Psi$. A theory functor of this form is
\emph{$\alpha$-simple}, and it is \emph{$\alpha$-plain} when $\Psi$ factors through the
empty-theory inclusion $\Sigma' \mapsto (\Sigma', \emptyset)$ so that
$\Gamma_{\Psi\Sigma} = \emptyset$. Every $\alpha$-simple functor (and hence every
$\alpha$-plain functor) is $\alpha$-sensible.

\begin{definition}[Map of proof calculi; Meseguer, Definition~33]
\label{def:inst-pc-map}
A \emph{map of proof calculi} $(\Phi, \alpha, \gamma) : \instP \to \instP'$ is a map
$(\Phi, \alpha)$ of the underlying entailment systems together with a natural
transformation $\gamma : \proofs \Rightarrow \proofs' \circ \Phi$ such that
\[
  \pi'\Phi \cdot \gamma = \alpha \cdot \pi ,
  \qquad\text{i.e.}\qquad
  \pi'_{\Phi T}(\gamma_T(p)) = \alpha_T(\pi_T(p))
  \text{ for all } T \text{ and } p \in \proofs(T).
\]
Maps of proof calculi compose by composing the entailment maps and pasting the
$\gamma$ components; they form the category $\instPC$, with a forgetful functor
$\mathit{ent} : \instPC \to \Ent$.
\end{definition}

The forgetful functor $\mathit{ent}$ has a right adjoint $(\ )^{\natural}$
(\citet{meseguer1989}, Proposition~34): $\instE^{\natural}$ has $\proofs = \mathit{thm}$,
the functor sending a theory to its set of theorems, and $\pi$ the inclusion
$\mathit{thm} \subseteq \Sen$. $\instE^{\natural}$ is the proof calculus with exactly one
proof per theorem.

\begin{definition}[$\pi$-institution]\label{def:inst-pi-institution}
A \emph{$\pi$-institution} $\instJ = \langle \Sign, \Sen, \{C_\Sigma\}_{\Sigma \in
|\Sign|} \rangle$ consists of a category $\Sign$, a functor $\Sen : \Sign \to \SetC$,
and for each $\Sigma$ a closure operator $C_\Sigma$ on $\mathcal{P}(\Sen(\Sigma))$ such
that for every $f : \Sigma_1 \to \Sigma_2$ and $\Gamma \subseteq \Sen(\Sigma_1)$,
\[
  \Sen(f)(C_{\Sigma_1}(\Gamma)) \subseteq C_{\Sigma_2}(\Sen(f)(\Gamma)).
\]
\end{definition}

This is Definition~1.5 of \citet{bittencourt-rios2022}, after \citet{fiadeiro-sernadas1988}. By
their Remark~1.7, a $\pi$-institution is equivalently a family of Tarskian consequence
relations $\vdash_\Sigma$ for which each $\Sen(f)$ is a logical translation:
$\Gamma \vdash_{\Sigma_1} \varphi$ implies $\Sen(f)[\Gamma] \vdash_{\Sigma_2}
\Sen(f)(\varphi)$. Definition~\ref{def:inst-pi-institution} imposes no compactness;
the original notion of Fiadeiro and Sernadas includes it (\citet{meseguer1989}, Remark~3
after Definition~1).

\begin{definition}[$\pi$-institution comorphism]\label{def:inst-pi-comorphism}
A \emph{comorphism} $\langle \Phi, \alpha \rangle : \instJ \to \instJ'$ of
$\pi$-institutions consists of a functor $\Phi : \Sign \to \Sign'$ and a natural
transformation $\alpha : \Sen \Rightarrow \Sen' \circ \Phi$ such that for all
$\Gamma \cup \{\varphi\} \subseteq \Sen(\Sigma)$,
\[
  \varphi \in C_\Sigma(\Gamma) \implies
  \alpha_\Sigma(\varphi) \in C'_{\Phi\Sigma}(\alpha_\Sigma(\Gamma)).
\]
With composition $\langle \Phi', \alpha' \rangle \circ \langle \Phi, \alpha \rangle =
\langle \Phi'\Phi, \alpha'\Phi \cdot \alpha \rangle$ these form the category
$\piIns^{\instco}$.
\end{definition}

This is Definition~1.6(b) of \citet{bittencourt-rios2022}.

\paragraph{Related frameworks.}
Mossakowski, Goguen, Diaconescu, and Tarlecki define proof-theoretic institutions, in
which an entailment system without models is a legitimate instance~\citep{mossakowski2005logic}.
Goguen and Burstall's charters and
parchments make the sentences of a signature into a category whose morphisms are
proofs~\citep{goguen-burstall1986}; Meseguer
(Remark~1 after Definition~6) observes that an entailment relation is recovered from
such a category only under further assumptions, such as compactness and conjunction.
Categorical logics (\citet{meseguer1989}, Definition~40) give semantic models as
structure-preserving functors out of a classifying object;
Section~\ref{sec:inst-real-models} uses them as a candidate route to model maps
between the kernel theory and Lean's theory and nowhere else.

\section{The Witness Level: Each Logic as a Proof Calculus}
\label{sec:inst-witness}

This section constructs the kernel's type theory as a proof calculus $\instK$, proves
the axioms of Definitions~\ref{def:inst-entailment} and~\ref{def:inst-proof-calculus},
and fixes the shape of the proof calculi of the other hosted logics.

Throughout, $\Sigma; \Delta \vdash \mathcal{J}$ denotes a judgment of the type theory
of Chapter~\ref{ch:mltt} in global environment $\Sigma$ and local context $\Delta$. The
construction uses four standard properties of that theory, stated there:
\begin{enumerate}[label=(T\arabic*)]
\item\label{inst:T-var} \emph{variable rule:} if $\Sigma; \Delta \vdash A : s$ for a
  sort $s$, then $\Sigma; \Delta, x : A \vdash x : A$;
\item\label{inst:T-weak} \emph{weakening:} if $\Sigma; \Delta \vdash \mathcal{J}$ and
  $\Sigma; \Delta, \Delta' \vdash$ is a well-formed context, then
  $\Sigma; \Delta, \Delta' \vdash \mathcal{J}$;
\item\label{inst:T-subst} \emph{substitution:} if $\Sigma; \Delta, x : A, \Delta'
  \vdash \mathcal{J}$ and $\Sigma; \Delta \vdash u : A$, then
  $\Sigma; \Delta, \Delta'[u/x] \vdash \mathcal{J}[u/x]$;
\item\label{inst:T-level} \emph{level instantiation:} judgments are preserved by
  substituting universe levels for universe-level parameters.
\end{enumerate}
All four are properties of the declarative typing relation, in which definitional
equality is an equivalence relation by confluence (Lemma~\ref{lem:mltt-confluence}). None
of them depends on normalization, subject reduction, or decidability of definitional
equality.

\subsection{Signatures, sentences, and entailment}
\label{sec:inst-kernel-ent}

\begin{definition}[Kernel signatures]\label{def:inst-kernel-sign}
A \emph{kernel signature} is a well-formed global environment $\Sigma$
(Chapter~\ref{ch:mltt}): a sequence of constant declarations, each with universe-level
parameters $\vec u$ and a type, and each either opaque (an axiom constant), a
definition with a body, or part of an inductive declaration (type former,
constructors, recursor). A \emph{signature morphism} $\sigma : \Sigma \to \Sigma'$
assigns to each constant $c.\{\vec u\}$ of $\Sigma$ a term $\sigma(c)$ of $\Sigma'$,
with level parameters among $\vec u$ and no free variables, such that, writing
$\insthat{\sigma}$ for the homomorphic extension of $\sigma$ to terms with
$\insthat{\sigma}(c.\{\vec \ell\}) = \sigma(c)[\vec \ell / \vec u]$:
\begin{enumerate}[label=(S\arabic*)]
\item\label{inst:S-type} if $c : A$ in $\Sigma$, then
  $\Sigma'; \cdot \vdash \sigma(c) : \insthat{\sigma}(A)$;
\item\label{inst:S-delta} if $c$ is a definition with body $b$, then
  $\sigma(c) \defeq \insthat{\sigma}(b)$ in $\Sigma'$;
\item\label{inst:S-iota} the two sides of the image under $\insthat{\sigma}$ of each
  computation rule of an inductive declaration of $\Sigma$ are definitionally equal in
  $\Sigma'$.
\end{enumerate}
Identities send $c$ to $c.\{\vec u\}$; composition is
$(\sigma' \circ \sigma)(c) = \insthat{\sigma'}(\sigma(c))$.
\end{definition}

Inclusions of environments, renamings, and instantiations of declared constants by
closed terms are all signature morphisms. When the environment includes the class
systems of Chapter~\ref{ch:knowledge-graphs}---which add base cases to the subsumption
judgment---a signature morphism is additionally required to fix the constants of each
class system and to preserve the class-system ordering in $\Sigma'$, ensuring that those
subsumption base cases remain valid under translation.

\begin{lemma}[Translation]\label{thm:inst-translation}
Let $\sigma : \Sigma \to \Sigma'$ be a signature morphism. If $\Sigma; \Delta \vdash
\mathcal{J}$, then $\Sigma'; \insthat{\sigma}(\Delta) \vdash
\insthat{\sigma}(\mathcal{J})$, for context formation, typing, and subsumption; and
$a \defeq b$ in $\Sigma$ implies $\insthat{\sigma}(a) \defeq \insthat{\sigma}(b)$ in
$\Sigma'$.
\end{lemma}

\begin{proof}
By induction on the derivation. Since each $\sigma(c)$ is closed, $\insthat{\sigma}$
commutes with substitution of terms for variables:
$\insthat{\sigma}(t[u/x]) = \insthat{\sigma}(t)[\insthat{\sigma}(u)/x]$. Every rule
other than the rules for constants is schematic in the environment, so the image of an
instance is an instance of the same rule, and the induction hypothesis applies to the
premises. The constant rule concludes $\Sigma; \Delta \vdash c.\{\vec \ell\} :
A[\vec \ell / \vec u]$; its image follows from~\ref{inst:S-type},
property~\ref{inst:T-level}, and property~\ref{inst:T-weak} applied to the well-formed
context $\insthat{\sigma}(\Delta)$. The $\delta$-rule for a definition follows
from~\ref{inst:S-delta}, the $\iota$-rules from~\ref{inst:S-iota}, each again under
level instantiation and weakening. Definitional equality is joinability: $\insthat{\sigma}$
maps a step of $\beta$, of a projection, of an annotation, or of $\Id$-$\iota$ to a step of
the same rule, since it commutes with substitution, and a $\delta$- or $\iota$-step to a
definitional equality, by~\ref{inst:S-delta} and~\ref{inst:S-iota}; definitional equality
being transitive (Lemma~\ref{lem:mltt-confluence}), $\insthat{\sigma}$ maps joinable terms
to definitionally equal ones. Subsumption has as base cases definitional equality,
universe cumulativity, and, where present, the base cases of
Chapter~\ref{ch:knowledge-graphs}; the first is covered by the induction, the second
does not mention constants, and the third holds in $\Sigma'$ because $\sigma$ fixes the
constants of the class systems. The structural $\Pi$ rule is schematic.
\end{proof}

Lemma~\ref{thm:inst-translation} shows that composition in
Definition~\ref{def:inst-kernel-sign} is well defined:
$\insthat{\sigma'}(\sigma(c))$ satisfies~\ref{inst:S-type}--\ref{inst:S-iota} by the
lemma applied to $\sigma'$. Kernel signatures and signature morphisms form a category
$\Sign_{\instK}$; associativity and identity laws hold because homomorphic extension
commutes with composition of assignments.

\begin{definition}[Kernel sentences]\label{def:inst-kernel-sen}
$\Sen_{\instK}(\Sigma)$ is the set of terms $\varphi$ with no free variables and no
free level parameters such that $\Sigma; \cdot \vdash \varphi : \Prop$. For
$\sigma : \Sigma \to \Sigma'$, $\Sen_{\instK}(\sigma) = \insthat{\sigma}$.
\end{definition}

$\Sen_{\instK}(\sigma)$ lands in $\Sen_{\instK}(\Sigma')$ by
Lemma~\ref{thm:inst-translation}, and functoriality is the composition law of
$\Sign_{\instK}$.

Hypotheses are named canonically. For each $\varphi \in \Sen_{\instK}(\Sigma)$ fix a
variable $\instax{\varphi}$, the \emph{axiom name} of $\varphi$. For a finite
$\Gamma_0 \subseteq \Sen_{\instK}(\Sigma)$, the context $\instax{\Gamma_0}$ declares
$\instax{\varphi} : \varphi$ for each $\varphi \in \Gamma_0$, in any order. Because
every $\varphi$ is closed, no declaration depends on another, the order is
immaterial, and $\Sigma; \instax{\Gamma_0} \vdash$ is a well-formed context.

\begin{definition}[Kernel entailment]\label{def:inst-kernel-ent}
For $\Gamma \cup \{\varphi\} \subseteq \Sen_{\instK}(\Sigma)$,
$\Gamma \vdash_\Sigma \varphi$ iff there are a finite $\Gamma_0 \subseteq \Gamma$ and a
term $t$ with $\Sigma; \instax{\Gamma_0} \vdash t : \varphi$.
\end{definition}

\begin{definition}[Kernel witnesses]\label{def:inst-kernel-proofs}
Let $T = (\Sigma, \Gamma)$ be a theory. A \emph{witness} in $T$ is a pair
$(t, \varphi)$ with $\varphi \in \Sen_{\instK}(\Sigma)$ and $t$ a term, taken up to
renaming of bound variables and \emph{not} up to definitional equality, such that
$\Sigma; \instax{\Gamma_0} \vdash t : \varphi$ for some finite $\Gamma_0 \subseteq
\Gamma$. $\proofs_{\instK}(T)$ is the set of witnesses in $T$, and
$\pi_T(t, \varphi) = \varphi$.

The proof-theoretic structure $P_{\instK}(T)$ is the multicategory whose basic objects
are the elements of $\Sen_{\instK}(\Sigma)$ and whose morphisms
$\varphi_1, \dots, \varphi_n \to \psi_1, \dots, \psi_m$ are tuples
$(t_1, \dots, t_m)$ with $\Sigma; \instax{\Gamma_0}, x_1 : \varphi_1, \dots, x_n :
\varphi_n \vdash t_j : \psi_j$ for some finite $\Gamma_0 \subseteq \Gamma$, the
variables $x_i$ named by position. Composition substitutes the components of the first
tuple for the positional variables of the second; the monoidal product is
concatenation with reindexing; identities are tuples of variables.
$\instPr(P_{\instK}(T))$ is the set of morphisms with empty domain and a single
codomain object, which is $\proofs_{\instK}(T)$.

For an axiom-preserving $H : (\Sigma, \Gamma) \to (\Sigma', \Gamma')$ with underlying
signature morphism $\sigma$, $P_{\instK}(H)$ sends a tuple $(t_j)$ to
$(\insthat{\sigma}(t_j)[\instax{H(\varphi)} / \instax{\varphi}]_{\varphi \in \Gamma})$,
renaming each axiom name of $\Gamma$ to the axiom name of its image in $\Gamma'$.
\end{definition}

The witness records the proposition it proves. A term may have several types up to
definitional equality and subsumption; $\pi$ returns the recorded one. This is the form
of Meseguer's Example~32, where a proof is a triple that records its endpoints.

\subsection{The five lemmas}
\label{sec:inst-five-lemmas}

\begin{lemma}[Reflexivity]\label{thm:inst-refl}
$\{\varphi\} \vdash_\Sigma \varphi$.
\end{lemma}
\begin{proof}
$\Sigma; \cdot \vdash \varphi : \Prop$, so~\ref{inst:T-var} gives
$\Sigma; \instax{\varphi} : \varphi \vdash \instax{\varphi} : \varphi$.
\end{proof}

\begin{lemma}[Monotonicity]\label{thm:inst-mono}
$\Gamma \vdash_\Sigma \varphi$ and $\Gamma \subseteq \Gamma'$ imply
$\Gamma' \vdash_\Sigma \varphi$.
\end{lemma}
\begin{proof}
A finite $\Gamma_0 \subseteq \Gamma$ witnessing $\Gamma \vdash_\Sigma \varphi$ is a
finite subset of $\Gamma'$.
\end{proof}

\begin{lemma}[Transitivity: cut is substitution]\label{thm:inst-trans}
If $\Gamma \vdash_\Sigma \varphi_i$ for all $i \in I$ and
$\Gamma \cup \{\varphi_i \mid i \in I\} \vdash_\Sigma \psi$, then
$\Gamma \vdash_\Sigma \psi$.
\end{lemma}
\begin{proof}
We have $\Sigma; \instax{\Delta_0}, \instax{\varphi_{1}} : \varphi_{1}, \dots,
\instax{\varphi_{k}} : \varphi_{k} \vdash s : \psi$ for a finite $\Delta_0 \subseteq
\Gamma$ and finitely many $\varphi_{1}, \dots, \varphi_{k}$ among the $\varphi_i$ not
in $\Delta_0$. For each $j \le k$ there are a finite $\Delta_j \subseteq \Gamma$ and
$t_j$ with $\Sigma; \instax{\Delta_j} \vdash t_j : \varphi_{j}$. Let
$\Delta = \Delta_0 \cup \dots \cup \Delta_k$, a finite subset of $\Gamma$. By
weakening~\ref{inst:T-weak}, each $t_j$ and $s$ are typed in contexts extending
$\instax{\Delta}$. Applying substitution~\ref{inst:T-subst} $k$ times yields
$\Sigma; \instax{\Delta} \vdash s[t_1/\instax{\varphi_1}, \dots,
t_k/\instax{\varphi_k}] : \psi$, since $\psi$ and all $\varphi_{j}$ are closed and
therefore unchanged by the substitution.
\end{proof}

\begin{lemma}[$\vdash$-translation]\label{thm:inst-ent-trans}
If $\Gamma \vdash_\Sigma \varphi$ and $\sigma : \Sigma \to \Sigma'$, then
$\insthat{\sigma}(\Gamma) \vdash_{\Sigma'} \insthat{\sigma}(\varphi)$.
\end{lemma}
\begin{proof}
From $\Sigma; \instax{\Gamma_0} \vdash t : \varphi$, Lemma~\ref{thm:inst-translation}
gives $\Sigma'; \insthat{\sigma}(\instax{\Gamma_0}) \vdash \insthat{\sigma}(t) :
\insthat{\sigma}(\varphi)$. The context declares variables $\instax{\varphi'}$ of type
$\insthat{\sigma}(\varphi')$; renaming each to $\instax{\insthat{\sigma}(\varphi')}$,
and identifying variables whose types coincide by substituting one for the other
(property~\ref{inst:T-subst} with $u$ a variable), gives a derivation in context
$\instax{\insthat{\sigma}(\Gamma_0)}$.
\end{proof}

\begin{lemma}[Naturality of $\pi$]\label{thm:inst-nat}
$P_{\instK}$ is a functor $\Thz \to \instMultCat$, and $\pi : \proofs_{\instK}
\Rightarrow \Sen_{\instK}$ is natural.
\end{lemma}
\begin{proof}
For axiom-preserving $H$, every axiom name $\instax{\varphi}$ with $\varphi \in \Gamma$
is renamed to $\instax{H(\varphi)}$ with $H(\varphi) \in \Gamma'$. By the argument of
Lemma~\ref{thm:inst-ent-trans}, the image of a morphism of $P_{\instK}(T)$ is a
morphism of $P_{\instK}(H(T))$ with translated domain and codomain. $P_{\instK}(H)$
preserves composition because $\insthat{\sigma}$ and renaming commute with
substitution, preserves concatenation and identities by construction, and sends basic
objects to basic objects. Functoriality in $H$ is the composition law of
$\Sign_{\instK}$. Naturality of $\pi$ reads
$\pi_{H(T)}(\proofs(H)(t, \varphi)) = \insthat{\sigma}(\varphi) = \Sen(H)(\pi_T(t,
\varphi))$, which holds by definition.
\end{proof}

\begin{theorem}[The kernel is a proof calculus]\label{thm:inst-kernel-pc}
$\instK = (\Sign_{\instK}, \Sen_{\instK}, \vdash, P_{\instK}, \instPr, \pi)$ is a
proof calculus in the sense of Definition~\ref{def:inst-proof-calculus}. Its
entailment system is compact, and $\vdash_\Sigma$ is a Tarskian consequence relation
for which each $\Sen_{\instK}(\sigma)$ is a logical translation.
\end{theorem}
\begin{proof}
Lemmas~\ref{thm:inst-refl}--\ref{thm:inst-ent-trans} are the four axioms of
Definition~\ref{def:inst-entailment}. Lemma~\ref{thm:inst-nat} gives the functor and
the natural transformation. The image of $\pi_T$ is $\instclo{\Gamma}$ because
$\Gamma \vdash_\Sigma \varphi$ holds exactly when a witness $(t, \varphi)$ in $T$
exists. Compactness holds by Definition~\ref{def:inst-kernel-ent}. Reflexivity,
monotonicity, and transitivity are the Tarskian conditions, and
Lemma~\ref{thm:inst-ent-trans} is the translation condition of Remark~1.7 of
\citet{bittencourt-rios2022}.
\end{proof}

\subsection{Proof irrelevance and the witness level}
\label{sec:inst-proof-irrelevance}

The kernel's $\Prop$ is proof-irrelevant by the axiom
$\irrel : \Pi P{:}\Prop.\,\Pi p{:}P.\,\Pi q{:}P.\,\Id\,P\,p\,q$
(Chapter~\ref{ch:mltt}, Section~\ref{sec:mltt-prop}): if $\Sigma$ declares
$\irrel$, $\Sigma; \Delta \vdash p : \varphi$, and $\Sigma; \Delta \vdash q : \varphi$
with $\varphi : \Prop$, then
$\Sigma; \Delta \vdash \irrel\,\varphi\,p\,q : \Id\,\varphi\,p\,q$, while
$p \defeq q$ does not hold. An extension of a signature that declares $\irrel$
declares it. Definition~\ref{def:inst-kernel-proofs} takes witnesses up
to renaming of bound variables and not up to provable identity. Quotienting by
provable identity collapses the witness level.

\begin{proposition}\label{thm:inst-collapse}
Let $\instK_{=}$ be $\instK$, over the signatures that declare $\irrel$, with
$\proofs(T)$ replaced by its quotient under
$(t, \varphi) \sim (t', \varphi)$ iff $\Sigma; \instax{\Gamma_0} \vdash e :
\Id\,\varphi\,t\,t'$ for some $e$ and some finite $\Gamma_0 \subseteq \Gamma$. Then each
$\pi_T$ is injective, and $\instK_{=}$ is isomorphic, by the unit of the adjunction
$\mathit{ent} \dashv (\ )^{\natural}$, to $\mathit{ent}(\instK)^{\natural}$ over the
same signatures.
\end{proposition}
\begin{proof}
Two witnesses with the same $\pi$-image $\varphi$ are typed in contexts
$\instax{\Gamma_0}$ and $\instax{\Gamma_1}$; by weakening both are typed in
$\instax{\Gamma_0 \cup \Gamma_1}$, and Rule~\textsc{PI} of Chapter~\ref{ch:mltt}, the
axiom $\irrel$ applied to $\varphi$ and the two terms, proves them identical there.
Hence $\pi_T$ is injective, and its image is $\instclo{\Gamma}$. The unit
of the adjunction of Proposition~34 of \citet{meseguer1989} at a proof calculus is
$(1_{\Sign}, 1_{\Sen}, \pi)$; with each $\pi_T$ a bijection onto
$\mathit{thm}(T)$ it is an isomorphism.
\end{proof}

Under this quotient, every theorem has at most one witness, so any two proofs of the
same proposition become indistinguishable. The quotient is the explicit witness-level
counterpart of what definitional proof irrelevance would do silently inside conversion
(Chapter~\ref{ch:mltt}, Section~\ref{sec:mltt-prop}): it is a forgetful projection that
may be applied when only provability matters, rather than an identification baked into
the kernel's typing judgment.

Keeping the unquotiented witness set $\proofs_{\instK}(T)$ is essential for the rest of
the architecture. Justification logic records which specific term justifies a
proposition (Chapter~\ref{ch:justification-logic}), and the system records which witness
established a claim in order to compute provenance and audit trails; both require
distinct witnesses of the same proposition to remain distinct objects. Because proof
irrelevance in the kernel is an axiom ($\irrel$) rather than a conversion rule, two
proofs of a proposition are provably equal---so no computation in $\Setu$ or $\Typeu_k$
can distinguish them---while the kernel still retains both proof terms as distinct
witnesses. At any point in time, the system stores a finite subset of
$\proofs_{\instK}(T)$.

\subsection{Axioms as sentences}
\label{sec:inst-axioms}

An axiom $a : \varphi$ with $\varphi : \Prop$ is a closed constant whose type the
kernel admits without a proof. Its warrant is \emph{declared}
(Chapter~\ref{ch:justification-logic}). Two placements are available: $a$ as a
constant of the signature, or $\varphi$ as an element of the theory's axiom set.

\begin{definition}[Theory of an environment]\label{def:inst-env-theory}
An environment $E$ is \emph{separable} if no declaration of $E$ whose type is not a
proposition mentions a $\Prop$-typed axiom constant. The theory of a separable
environment is $T(E) = (\Sigma_E, \Gamma_E)$, where $\Sigma_E$ consists of the
declarations of $E$ other than $\Prop$-typed axioms and $\Prop$-typed definitions, and
$\Gamma_E$ is the set of types of the $\Prop$-typed axioms. Each $\Prop$-typed
definition $c : \psi := t$ of $E$ is the witness $(t', \psi) \in \proofs_{\instK}(T(E))$,
where $t'$ results from $t$ by renaming each axiom constant $a : \varphi$ to
$\instax{\varphi}$ and unfolding earlier $\Prop$-typed definitions.
\end{definition}

Non-$\Prop$-typed opaque constants, such as named individuals, remain signature
constants. A non-separable environment becomes separable by abstracting each offending
declaration over the axiom hypotheses it uses; this is the relativization of
Section~\ref{sec:inst-policies}.

The placement in $\Gamma$ is forced by two observations.
\begin{enumerate}[label=(\roman*)]
\item If $a$ were a signature constant, a signature morphism would have to assign to
  $a$ a term $\sigma(a)$ with $\Sigma' \vdash \sigma(a) : \insthat{\sigma}(\varphi)$
  by~\ref{inst:S-type}, that is, a target proof of the translated axiom. A change of
  vocabulary would then require a proof, which an axiom does not have.
\item A signature constant used in a witness is indistinguishable from a proved lemma
  of the same type. With axioms in $\Gamma$, the witnesses that depend on an axiom are
  exactly those whose term contains its axiom name.
\end{enumerate}
In $P_{\instK}(T)$, an axiom $\varphi \in \Gamma$ contributes the witness
$(\instax{\varphi}, \varphi)$ and nothing else; $\pi(\instax{\varphi}, \varphi) =
\varphi$.

\subsection{Definitional equality and subsumption}
\label{sec:inst-defeq}

Two properties of the kernel theory interact with the five lemmas.

\paragraph{Definitional equality is joinability.}
Definitional equality of the kernel theory is joinability under a confluent reduction
(Chapter~\ref{ch:mltt}, Definition~\ref{def:mltt-defeq}); it is transitive, and the
kernel's conversion test decides it on every term whose reduction terminates
(Theorem~\ref{thm:checking-conv}). Theories
with proof irrelevance as a conversion rule, Lean's among them, have neither property:
the rule together with elimination of subsingleton propositions into types breaks
normalization~\citep{abel-coquand2020}, and the algorithmic conversion check of
Lean's kernel theory is not transitive \citep[Section~3.1]{carneiro2019};
Chapter~\ref{ch:mltt} records why the kernel theory does not adopt the rule. The
lemmas of Section~\ref{sec:inst-five-lemmas} use
only~\ref{inst:T-var}--\ref{inst:T-level}. A checker that is sound for the declarative
relation and complete on terminating terms accepts the subset of
$\proofs_{\instK}(T)$ on which it terminates. Closure of that subset under the cut of
Lemma~\ref{thm:inst-trans} is obtained without new conversion problems by forming the
cut as a redex.

\begin{proposition}\label{thm:inst-cut-redex}
Let $\Sigma; \instax{\Delta}, \instax{\varphi_1} : \varphi_1, \dots, \instax{\varphi_k}
: \varphi_k \vdash s : \psi$ and $\Sigma; \instax{\Delta} \vdash t_j : \varphi_j$.
Then
\[
  \Sigma; \instax{\Delta} \vdash
  (\lambda \instax{\varphi_1} {:} \varphi_1 . \cdots \lambda \instax{\varphi_k} {:}
  \varphi_k . \, s)\, t_1 \cdots t_k : \psi ,
\]
and a bidirectional checker that accepts $s$ and each $t_j$ in these contexts accepts
the redex with no conversion check other than syntactic identity of each
$\varphi_j$ with itself.
\end{proposition}
\begin{proof}
The $\Pi$-introduction rule gives the abstraction the type
$\Pi \instax{\varphi_1} {:} \varphi_1 \ldots \Pi \instax{\varphi_k} {:} \varphi_k .\,
\psi$, which is a non-dependent iterated implication because $\psi$ and the
$\varphi_j$ are closed. Each application checks $t_j$ against the domain $\varphi_j$,
which is the type at which $t_j$ was accepted, and the result type
$\psi[t_j/\instax{\varphi_j}]$ is $\psi$ syntactically.
\end{proof}

The checker properties assumed in Proposition~\ref{thm:inst-cut-redex} (acceptance in
a context depends on the subterms and the conversion checks between their types) hold for the
type checker of Chapter~\ref{ch:checking} (Proposition~\ref{prop:checking-compositional}).

\paragraph{Subsumption.}
The subsumption judgment $\Sigma; \Delta \vdash A \subty B$ of Chapter~\ref{ch:mltt}
enters typing through one rule: $a : A$ and $A \subty B$ give $a : B$, with the term
unchanged. Weakening and substitution hold for it because its base cases are
definitional equality (preserved by both), universe cumulativity (independent of the
context), and the base cases of Chapter~\ref{ch:knowledge-graphs}, whose premises are
judgments that weakening and substitution preserve, and its structural $\Pi$ rule is
schematic. Lemma~\ref{thm:inst-translation} covers it explicitly. The
algorithmic restriction of subsumption to checking positions affects which terms a
checker accepts and not the declarative relation. Subsumption reappears in the
sentence translation into Lean's theory, which has no cumulativity
(Section~\ref{sec:inst-real-models}).

\subsection{Witness kinds of the other logics}
\label{sec:inst-other-witnesses}

Each hosted logic is a proof calculus with its own witness kind. This section fixes the shape;
Volume~II gives the details.

\begin{center}
\small
\begin{tabularx}{\textwidth}{@{}l>{\raggedright\arraybackslash}X
  >{\raggedright\arraybackslash}X@{}}
\toprule
Logic & Sentences & Witness and $\pi$ \\
\midrule
Lean's kernel theory &
  closed propositions over a Lean environment &
  a proof accepted by a hosted checker for Lean's kernel theory under a fixed axiom
  set $A$; $\pi$ returns the stated proposition \\
Justification logic &
  $j : P$, the kernel proposition $\JB(j, P)$ &
  the kernel's decision that $P \in \Evid(j)$; $\pi$ returns $j : P$ \\
Statistics &
  claims relative to an analysis plan &
  a verdict with its analysis trace; $\pi$ returns the claim the verdict supports \\
\bottomrule
\end{tabularx}
\end{center}

\paragraph{Justification logic.}
The sentences of the justification logic are the propositions $\JB(j, P)$, written
$j : P$, where $j$ ranges over the term algebra $\JTerm$ and $P$ over kernel
propositions (Chapter~\ref{ch:justification-logic}). A witness is the kernel's
decision that $P \in \Evid(j)$, and the projection $\pi$ sends this witness to the
sentence $j : P$ rather than to $P$ itself. Indeed, a justification term is not a proof
of $P$: the judgment $j : P$ is non-factive, so the step from $j : P$ to $P$ is not a
witness transport. Because the kernel's decision is already a kernel witness of
$\JB(j, P)$, the inclusion $j : P \mapsto \JB(j, P)$ into the kernel calculus---with
$\gamma$ the identity on witnesses---satisfies the transport invariant by construction.
By contrast, the erasure map $j : P \mapsto P$ has no witness component $\gamma$,
because a decision that $P \in \Evid(j)$ is not a proof term of type $P$.

\paragraph{Statistics.}
Whether statistical inference satisfies reflexivity, monotonicity, and transitivity of
Definition~\ref{def:inst-entailment} depends on the verdict semantics: composing
verdicts compounds error rates, which directly affects transitivity. These axioms are
obligations of the statistics institution treated in Volume~II and remain open here.

\section{Comorphisms: Witness Transport and its Consequences}
\label{sec:inst-comorphisms}

A translation between hosted logics consists of a signature translation, a sentence
translation, and a witness transport, subject to the requirement that every transported
witness prove the translated sentence. Definition~\ref{def:inst-pc-map} formalizes that
requirement, and Volume~II constructs the concrete translations between hosted logics.
Because a comorphism preserves provability of identities rather than definitional
conversion, the witnesses of a source logic---together with the justifications that
Chapter~\ref{ch:justification-logic} attaches to them---must arrive in the target as
inspectable objects. In a target institution with definitional proof irrelevance, a
witness-preserving comorphism must therefore route witnesses around the target's $\Prop$
into a data type, so the translation is no longer a direct map between proposition
universes; Lean's kernel theory is such a target (Section~\ref{sec:inst-real-models}).

\begin{lemma}[The transport invariant gives a map of proof calculi]
\label{thm:inst-invariant}
Let $\instP, \instP'$ be proof calculi, $\Psi : \Sign \to \Thz'$ a functor,
$\alpha : \Sen \Rightarrow \Sen' \circ \instsign' \circ \Psi$ a natural transformation
(where $\instsign' : \Thz' \to \Sign'$ is the signature projection), and $\Phi$ the
$\alpha$-extension of $\Psi$ to theories. Let $\gamma : \proofs \Rightarrow \proofs'
\circ \Phi$ be natural with
\[
  \pi'_{\Phi T}(\gamma_T(p)) = \alpha_T(\pi_T(p))
  \qquad\text{for all } T, p \qquad\text{(transport invariant).}
\]
Then $(\Phi, \alpha, \gamma)$ is a map of proof calculi.
\end{lemma}
\begin{proof}
$\Phi$ is $\alpha$-simple, hence $\alpha$-sensible. It remains to show the condition of
Definition~\ref{def:inst-ent-map}. Let $T = (\Sigma, \Gamma)$ and $\Gamma \vdash_\Sigma
\varphi$. Since the image of $\pi_T$ is $\instclo{\Gamma}$, there is $p \in \proofs(T)$
with $\pi_T(p) = \varphi$. Then $\gamma_T(p) \in \proofs'(\Phi T)$, and the image of
$\pi'_{\Phi T}$ is $\instclo{(\Gamma')}$ with $\Gamma' = \emptyset' \cup
\alpha_\Sigma(\Gamma)$; the invariant gives $\alpha_\Sigma(\varphi) \in
\instclo{(\emptyset' \cup \alpha_\Sigma(\Gamma))}$, which is
$\alpha_\Sigma(\Gamma) \vdash'_{\Phi(\Sigma, \emptyset)} \alpha_\Sigma(\varphi)$. The
equation of cells in Definition~\ref{def:inst-pc-map} is the invariant.
\end{proof}

Provability preservation is a consequence of witness transport and is not a separate
obligation. A translation that transports witnesses and checks each transported
witness establishes the invariant pointwise; naturality of $\gamma$ is the requirement
that transport commute with change of vocabulary.

\subsection{Axiom policies}
\label{sec:inst-policies}

A witness in $T = (\Sigma, \Gamma)$ may contain axiom names. Because an axiom has no
proof, a translation has no intrinsic evidence to transport for it. Meseguer's two
categories of theories ($\Thz$ and $\Th$) give two policies for handling source axioms,
and the deduction theorem gives a third; each translation declares one of them as its
\emph{axiom policy}. Whereas adoption operates uniformly on all theories in $\Thz$,
discharge is relative to a category of theories equipped with chosen target witnesses for
the translated axioms, and relativization operates on theories equipped with an ordered
list of distinct axioms ($\instThl$) so that the iterated implication and
$\lambda$-abstraction are uniquely determined.

\begin{center}
\small
\begin{tabularx}{\textwidth}{@{}l>{\raggedright\arraybackslash}p{0.22\textwidth}
  >{\raggedright\arraybackslash}X>{\raggedright\arraybackslash}p{0.2\textwidth}@{}}
\toprule
Policy & Theory map & $\gamma$ on an axiom name $\instax{\varphi}$ &
  Transported warrant \\
\midrule
Adopt & $\Thz$: $\alpha(\Gamma) \subseteq \Gamma'$ &
  $\instax{\alpha(\varphi)}$; the target assumes $\alpha(\varphi)$ &
  declared \\
Discharge & $\Th$: $\Gamma' \vdash' \alpha(\varphi)$ &
  a target witness of $\alpha(\varphi)$, supplied in the target &
  that of the target witness \\
Relativize & $\Gamma \vdash \varphi$ becomes
  $\vdash' \alpha(\Gamma) \instimp \alpha(\varphi)$ &
  a bound variable; the witness is abstracted over the axioms &
  verified, conditional on the stated hypotheses \\
\bottomrule
\end{tabularx}
\end{center}

\begin{proposition}[Adopt]\label{thm:inst-adopt}
Let $(\Phi, \alpha, \gamma)$ be as in Lemma~\ref{thm:inst-invariant} with $\Phi$
$\alpha$-plain, so that $\Phi(\Sigma, \Gamma) = (\Phi^{\instdia}\Sigma,
\alpha_\Sigma(\Gamma))$. Then $\Phi$ maps $\Thz$ to $\Thz'$ with $\alpha(\Gamma)$ as
the target axioms, and $(\Phi, \alpha, \gamma)$ is a map of proof calculi. If the
target has canonical axiom names, the assignment $\gamma_T(\instax{\varphi}, \varphi)
= (\instax{\alpha(\varphi)}, \alpha(\varphi))$ on axiom witnesses satisfies the
transport invariant.
\end{proposition}
\begin{proof}
The first two claims are Lemma~\ref{thm:inst-invariant} for an $\alpha$-plain
extension. For the third, $\alpha(\varphi) \in \alpha(\Gamma)$ is an axiom of
$\Phi(T)$, so $(\instax{\alpha(\varphi)}, \alpha(\varphi))$ is a witness in $\Phi(T)$
by reflexivity, and $\pi'$ maps it to $\alpha(\pi_T(\instax{\varphi}, \varphi))$.
\end{proof}

For a translation into the kernel calculus defined by structural recursion on terms,
the axiom names occurring in $\gamma_T(t)$ are the images of those occurring in $t$.
The identity translation along an axiom-preserving theory morphism, $P_{\instK}(H)$ of
Definition~\ref{def:inst-kernel-proofs}, is the basic instance of adoption.

\begin{proposition}[Discharge]\label{thm:inst-discharge}
Let $(\Phi, \alpha, \gamma)$ be an adopting map into the kernel calculus as in
Proposition~\ref{thm:inst-adopt}, $T = (\Sigma, \Gamma)$, and $T' = (\Sigma',
\Gamma'')$ a target theory with $\Sigma' = \Phi^{\instdia}\Sigma$. A \emph{discharge}
for $T$ in $T'$ is a family $d = (d_\chi)_{\chi \in \alpha(\Gamma)}$ of witnesses
$d_\chi \in \proofs_{\instK}(T')$ with $\pi'(d_\chi) = \chi$, indexed by the translated
axioms, so that two axioms with one translation have one discharge. Then
\[
  \gamma^d_T(p) = \gamma_T(p)[\,t_{d_\chi} / \instax{\chi}\,]_{\chi \in \alpha(\Gamma)},
  \qquad\text{where } d_\chi = (t_{d_\chi}, \chi),
\]
defines $\gamma^d_T : \proofs(T) \to \proofs_{\instK}(T')$ with
$\pi' \circ \gamma^d_T = \alpha \circ \pi_T$. On any subcategory of $\Thz$ on which an
assignment $T \mapsto (T', d)$ is functorial and $d$ natural, $(T \mapsto T', \alpha,
\gamma^d)$ satisfies Definition~\ref{def:inst-pc-map} on the subcategory, with the
entailment condition taken in the theory-relative form of Meseguer's Lemma~24, as in
Proposition~\ref{thm:inst-relativise}.
\end{proposition}
\begin{proof}
$\gamma_T(p)$ is typed in a context of axiom names $\instax{\chi}$ for finitely many
$\chi \in \alpha(\Gamma)$; each $t_{d_\chi}$ is typed in a context of axiom names for
$\Gamma''$. Lemma~\ref{thm:inst-trans} (weakening, then substitution) yields a witness in
$T'$ of the same, closed, proposition. $\gamma^d$ is natural because $\gamma$ and $d$ are
and $\insthat{\sigma}$ commutes with substitution. For the entailment condition, let
$\Delta \vdash_T \psi$; since $(\Phi, \alpha, \gamma)$ is a map of proof calculi,
$\alpha(\Gamma) \cup \alpha(\Delta) \vdash' \alpha(\psi)$ over $\Sigma'$, each sentence of
$\alpha(\Gamma)$ has the witness $d_\chi$ in $T'$, and transitivity gives
$\Gamma'' \cup \alpha(\Delta) \vdash' \alpha(\psi)$.
\end{proof}

Discharge is Meseguer's $\Th$: $H : (\Sigma, \Gamma) \to (\Sigma', \Gamma'')$ is a
theory morphism because each axiom is entailed in the target. The witness of an
entailed axiom is not determined by the theory morphism, which is the indeterminacy
that restricts $P$ to $\Thz$; the discharge $d$ supplies it.

Relativization requires an implication in the target. For a finite list
$\vec \psi = (\psi_1, \dots, \psi_k)$ of target sentences and a target sentence
$\chi$, write $\vec \psi \instimp \chi$ for $\psi_1 \to \cdots \to \psi_k \to \chi$.
For a natural $\alpha$, let $\instThl$ be the category whose objects are theories
$(\Sigma, \vec \varphi)$ with a finite list of axioms such that $\alpha(\varphi_1),
\dots, \alpha(\varphi_k)$ are pairwise distinct, and whose morphisms
$(\Sigma, \vec\varphi) \to (\Sigma', \vec\varphi')$ are signature morphisms $\sigma$
between lists of equal length with $\insthat{\sigma}(\varphi_i) = \varphi'_i$ for each
position $i$. Duplicates are removed from an axiom list before relativization, so the
distinctness condition loses no theory.

\begin{proposition}[Relativize]\label{thm:inst-relativise}
Let $(\Phi, \alpha, \gamma)$ be an adopting map into the kernel calculus as in
Proposition~\ref{thm:inst-adopt}. For $T = (\Sigma, \vec\varphi) \in \instThl$ define
\begin{gather*}
  \Phi_{\mathrm{rel}}(T) = (\Phi^{\instdia}\Sigma, \emptyset), \qquad
  \alpha_{\mathrm{rel}, T}(\psi) = \alpha(\vec\varphi) \instimp \alpha(\psi),
  \\
  \rho_T(t, \psi) = \big(\lambda h_1 {:} \alpha(\varphi_1) \ldots \lambda h_k {:}
  \alpha(\varphi_k) .\, \gamma_T(t)[h_i / \instax{\alpha(\varphi_i)}]_{i \le k},\
  \alpha_{\mathrm{rel}, T}(\psi)\big),
\end{gather*}
where $\gamma_T(t)$ denotes the term of the witness $\gamma_T(t, \psi)$. Then:
\begin{enumerate}[label=(\alph*)]
\item $\rho_T(t, \psi) \in \proofs_{\instK}(\Phi^{\instdia}\Sigma, \emptyset)$, a
  witness in the axiom-free fragment;
\item $\pi' \circ \rho_T = \alpha_{\mathrm{rel}, T} \circ \pi_T$;
\item $\alpha_{\mathrm{rel}}$ and $\rho$ are natural on $\instThl$;
\item $\Delta \vdash_{T} \psi$ implies
  $\alpha_{\mathrm{rel}, T}(\Delta) \vdash'_{\Phi_{\mathrm{rel}} T}
  \alpha_{\mathrm{rel}, T}(\psi)$.
\end{enumerate}
Hence $(\Phi_{\mathrm{rel}}, \alpha_{\mathrm{rel}}, \rho)$ satisfies
Definition~\ref{def:inst-pc-map} with $\Thz$ replaced by $\instThl$ and the
entailment condition taken in the theory-relative form of Meseguer's Lemma~24.
\end{proposition}
\begin{proof}
(a) The transported term $\gamma_T(t)$ is typed in a context of axiom names
$\instax{\chi}$ with $\chi \in \alpha(\vec\varphi)$. Renaming each
$\instax{\alpha(\varphi_i)}$ to $h_i$---which is unambiguous because the translated
axioms $\alpha(\varphi_1), \dots, \alpha(\varphi_k)$ are pairwise distinct---and
applying $\Pi$-introduction $k$ times yields a closed term of type
$\alpha(\vec\varphi) \instimp \alpha(\psi)$; any variable $h_i$ that does not occur in
$\gamma_T(t)$ is abstracted vacuously.

(b) Immediate from the definition of $\rho_T$.

(c) Every morphism of $\instThl$ is an axiom-preserving theory morphism, so $\gamma$ is
natural along it. Because the morphism preserves the axiom lists positionwise and both
lists have pairwise distinct $\alpha$-images, the renaming
$\instax{\alpha(\varphi_i)} \mapsto h_i$ commutes with translation; moreover,
$\insthat{\sigma}$ commutes with $\lambda$-abstraction and with $\instimp$, and $\alpha$
is natural.

(d) By compactness, we may assume $\Delta = \{\delta_1, \dots, \delta_m\}$ is finite,
with $\Delta \cup \vec\varphi \vdash \psi$ witnessed by a term $s$ in the theory
$(\Sigma, \Delta \cup \vec\varphi)$ on which $\gamma$ is defined. In the context
$g_1 : \alpha(\vec\varphi) \instimp \alpha(\delta_1), \dots, g_m : \alpha(\vec\varphi)
\instimp \alpha(\delta_m)$, the term
\[
  \lambda \vec h .\, \gamma(s)[h_i / \instax{\alpha(\varphi_i)}]_{i \le k}[g_l\,
  \vec h / \instax{\alpha(\delta_l)}]_{l \le m},
\]
with the first substitution taking precedence whenever $\alpha(\delta_l) =
\alpha(\varphi_i)$, has type $\alpha(\vec\varphi) \instimp \alpha(\psi)$.
\end{proof}

Note that $\alpha_{\mathrm{rel}, T}$ depends on the axioms of the theory $T$, not only
on its signature $\Sigma$. By Lemma~22 of \citet{meseguer1989}, it therefore cannot be a
natural transformation on all of $\Thz$, so restricting the domain from $\Thz$ to
$\instThl$ is essential: $\instThl$ has no morphism from $(\Sigma, ())$ to
$(\Sigma, \vec\varphi)$ when $\vec\varphi$ is non-empty, which is the morphism that the
proof of Lemma~22 exploits.

\paragraph{Relativization requires implication introduction in the target.}
Proposition~\ref{thm:inst-relativise} uses $\Pi$-introduction over $\Prop$ in the
target. The kernel has it. The justification-term algebra has no implication
introduction (Chapter~\ref{ch:justification-logic}): there is no term former that
abstracts a justification over a hypothesis. A bridge $A \to B$ therefore enters a
justification term as a leaf, $\jver$ when a checked proof of $A \to B$ exists and
$\jdecl$ otherwise. Relativization into justification terms is the composite of
relativization into the kernel with the formation of a $\jver$ leaf for the resulting
closed implication.

\paragraph{Default policy.}
Under relativization the transported witness proves an implication from the translated
axioms and holds unconditionally in the target. Adoption adds the translated axioms to the
target as assumptions; discharge requires a target witness for each of them; relativization
requires neither. It is the default for targets that
restrict their axioms (Section~\ref{sec:inst-real-models}). Adoption is used between
logics that share an axiom set.

\paragraph{Warrant transport.}
A transported claim has a warrant no stronger than that of the source witness. The
case analysis by policy is: under adoption, every transported witness depending on an
adopted axiom is \emph{declared} in that axiom; under discharge, the warrant is that of
the target witnesses $d_\chi$ combined with that of the source witness; under
relativization, the warrant of the implication is \emph{verified}, and the antecedents
are stated hypotheses. In justification terms, a translated claim with source
justification $j$ receives $\japp(\jdecl(c), j)$, for the translation $c$, unless the
satisfaction of $c$ is proved, in which case it receives $\japp(\jver(c), j)$.
Chapter~\ref{ch:justification-logic} states the grade algebra in which this ordering is
expressed.

\subsection{Forgetting witnesses}
\label{sec:inst-forgetting}

Meseguer's $\Phi$ acts on theories; a $\pi$-institution comorphism acts on signatures.
The construction below makes theories the signatures of a $\pi$-institution, so that a
map of entailment systems becomes a $\pi$-institution comorphism without restriction,
and identifies the case in which the signature level suffices.

\begin{definition}\label{def:inst-pi-of-ent}
Let $\instE = (\Sign, \Sen, \vdash)$ be an entailment system.
\begin{enumerate}[label=(\alph*)]
\item $\instJ(\instE) = \langle \Sign, \Sen, C \rangle$ with $C_\Sigma(\Gamma) =
  \instclo{\Gamma}$.
\item $\instJTh(\instE) = \langle \Thz, \Sen \circ \instsign, C^{\mathrm{th}} \rangle$
  with $C^{\mathrm{th}}_{(\Sigma, \Gamma)}(\Delta) = \instclo{(\Gamma \cup \Delta)}$.
\end{enumerate}
\end{definition}

\begin{lemma}\label{thm:inst-pi-of-ent}
$\instJ(\instE)$ and $\instJTh(\instE)$ are $\pi$-institutions, and
$\langle \instF, 1 \rangle : \instJ(\instE) \to \instJTh(\instE)$, with
$\instF(\Sigma) = (\Sigma, \emptyset)$, is a $\pi$-institution comorphism with
$C^{\mathrm{th}}_{\instF\Sigma} = C_\Sigma$.
\end{lemma}
\begin{proof}
$\Gamma \mapsto \instclo{\Gamma}$ is a closure operator by reflexivity, monotonicity,
and transitivity. The condition of Definition~\ref{def:inst-pi-institution} for
$\instJ(\instE)$ is $\vdash$-translation: $\varphi \in \instclo{\Gamma}$ gives
$H(\Gamma) \vdash H(\varphi)$. For $\instJTh(\instE)$, $\Delta \mapsto
\instclo{(\Gamma \cup \Delta)}$ is extensive and monotone, and idempotent because
$\Gamma \subseteq \instclo{(\Gamma \cup \Delta)}$ gives
$\instclo{(\Gamma \cup \instclo{(\Gamma \cup \Delta)})} =
\instclo{(\instclo{(\Gamma \cup \Delta)})} = \instclo{(\Gamma \cup \Delta)}$. For
axiom-preserving $H : (\Sigma, \Gamma) \to (\Sigma', \Gamma')$,
$H(\instclo{(\Gamma \cup \Delta)}) \subseteq \instclo{(H(\Gamma) \cup H(\Delta))}
\subseteq \instclo{(\Gamma' \cup H(\Delta))}$ by $\vdash$-translation and
$H(\Gamma) \subseteq \Gamma'$. The last claim is $\instclo{(\emptyset \cup \Delta)} =
\instclo{\Delta}$.
\end{proof}

\begin{theorem}[Forgetting witnesses]\label{thm:inst-forgetting}
The assignments $\instP \mapsto \instJTh(\mathit{ent}(\instP))$ and
$(\Phi, \alpha, \gamma) \mapsto \langle \Phi, \alpha \rangle$ define a functor
$\instU : \instPC \to \piIns^{\instco}$, which factors through
$\mathit{ent} : \instPC \to \Ent$.
\end{theorem}
\begin{proof}
Let $(\Phi, \alpha) : \instE \to \instE'$ be a map of entailment systems, $T = (\Sigma,
\Gamma)$, $\Phi(T) = (\Sigma', \Gamma')$, and $\Phi(\Sigma, \emptyset) = (\Sigma',
\emptyset')$; by condition~(1) of $\alpha$-sensibility the signature of both is
$\Phi^{\instdia}\Sigma$, and by Lemma~22 of \citet{meseguer1989} $\alpha_T =
\alpha_{(\Sigma, \emptyset)} = \alpha_\Sigma$. Let $\varphi \in
C^{\mathrm{th}}_T(\Delta)$, that is $\Gamma \cup \Delta \vdash_\Sigma \varphi$. By
Definition~\ref{def:inst-ent-map}, $\emptyset' \cup \alpha_\Sigma(\Gamma) \cup
\alpha_\Sigma(\Delta) \vdash'_{\Sigma'} \alpha_\Sigma(\varphi)$. Condition~(2) gives
$\emptyset' \cup \alpha_\Sigma(\Gamma) \subseteq \instclo{(\Gamma')}$, so by
transitivity $\Gamma' \cup \alpha_\Sigma(\Delta) \vdash'_{\Sigma'}
\alpha_\Sigma(\varphi)$, that is $\alpha_T(\varphi) \in
C'^{\mathrm{th}}_{\Phi T}(\alpha_T(\Delta))$. Hence $\langle \Phi, \alpha \rangle$ is a
comorphism $\instJTh(\instE) \to \instJTh(\instE')$. (This is the theory-relative
form, Lemma~24 of \citet{meseguer1989}.) Composition in $\Ent$ is $(\Phi_2\Phi_1,
\alpha_2\Phi_1 \cdot \alpha_1)$ and in $\piIns^{\instco}$ is $\langle \Phi_2\Phi_1,
\alpha_2\Phi_1 \cdot \alpha_1 \rangle$; identities are $(1, 1)$ in both. The
component $\gamma$ is discarded, so the functor factors through $\mathit{ent}$.
\end{proof}

\begin{corollary}[Signature level]\label{thm:inst-forgetting-sig}
Let $(\Phi, \alpha)$ be a map of entailment systems such that $\emptyset'_\Sigma
\subseteq \instclo{\emptyset}$ in $\Phi^{\instdia}\Sigma$ for every $\Sigma$; in
particular, let $\Phi$ be $\alpha$-plain. Then $\langle \Phi^{\instdia}, \alpha
\rangle : \instJ(\instE) \to \instJ(\instE')$ is a $\pi$-institution comorphism.
$\alpha$-plain maps are closed under composition, and on them the assignment is a
functor into $\piIns^{\instco}$.
\end{corollary}
\begin{proof}
From $\Gamma \vdash_\Sigma \varphi$, Definition~\ref{def:inst-ent-map} gives
$\alpha_\Sigma(\varphi) \in \instclo{(\emptyset' \cup \alpha_\Sigma(\Gamma))}$, and
$\emptyset' \subseteq \instclo{\emptyset} \subseteq
\instclo{(\alpha_\Sigma(\Gamma))}$ gives $\instclo{(\emptyset' \cup
\alpha_\Sigma(\Gamma))} = \instclo{(\alpha_\Sigma(\Gamma))}$. If $\Phi_1, \Phi_2$ are
plain, $\Phi_2\Phi_1(\Sigma, \Gamma) = (\Phi_2^{\instdia}\Phi_1^{\instdia}\Sigma,
\alpha_2\alpha_1(\Gamma))$, which is plain for $\alpha_2\Phi_1 \cdot \alpha_1$.
\end{proof}

A translation whose $\Phi(\Sigma, \emptyset)$ contains axioms not entailed by the empty
theory is theory-level: it maps a signature to a theory with new axioms, as
in Meseguer's Example~18, where predicates are encoded as characteristic functions
together with axioms forcing a two-element Boolean algebra. For such a translation,
Theorem~\ref{thm:inst-forgetting} applies and Corollary~\ref{thm:inst-forgetting-sig}
does not. The adopt and discharge policies yield $\alpha$-plain maps when $\Phi$
introduces no axioms of its own; relativization lands in axiom-free target theories.

\subsection{Composition}
\label{sec:inst-composition}

Maps of proof calculi compose (\citet{meseguer1989}, after Definition~33), and so do
$\pi$-institution comorphisms (Definition~\ref{def:inst-pi-comorphism}). The hosted
logics with their translations therefore form a subcategory of $\instPC$, and
Theorem~\ref{thm:inst-forgetting} maps it functorially into $\piIns^{\instco}$. The
transport invariant composes: if $\pi'\Phi_1 \cdot \gamma_1 = \alpha_1 \cdot \pi$ and
$\pi''\Phi_2 \cdot \gamma_2 = \alpha_2 \cdot \pi'$, then
$\pi''\Phi_2\Phi_1 \cdot \gamma_2\Phi_1 \cdot \gamma_1 = \alpha_2\Phi_1 \cdot
\pi'\Phi_1 \cdot \gamma_1 = \alpha_2\Phi_1 \cdot \alpha_1 \cdot \pi$. A chain of
translations whose links each satisfy the invariant satisfies it end to end, with no
check at intermediate logics beyond those of the links. Relativized maps compose only
when the second map is defined on the axiom-free image of the first, which holds
because $\emptyset$ is a list of length zero.

\section{The Model Level}
\label{sec:inst-models}

\begin{definition}[The functor $\instG$]\label{def:inst-G}
For a $\pi$-institution $\instJ = \langle \Sign, \Sen, C \rangle$,
$\instG(\instJ) = (\Sign, \Sen, \Mod^{\instJ}, \models^{\instJ})$ with
\begin{itemize}
\item $\Mod^{\instJ}(\Sigma) = \{ C_\Sigma(\Gamma) \mid \Gamma \subseteq \Sen(\Sigma)
  \}$, the closed sets of $\Sigma$-sentences, as a codiscrete category (exactly one
  morphism between any two objects);
\item $\Mod^{\instJ}(f) = \Sen(f)^{-1}$ for $f : \Sigma_1 \to \Sigma_2$;
\item $m \models^{\instJ}_\Sigma \varphi$ iff $\varphi \in m$.
\end{itemize}
For a comorphism $\langle \Phi, \alpha \rangle$, $\instG\langle \Phi, \alpha \rangle =
\langle \Phi, \alpha, \alpha^{-1} \rangle$, where $(\alpha^{-1})_\Sigma(m') =
\alpha_\Sigma^{-1}(m')$.
\end{definition}

\begin{theorem}[{\citealp{mariano-pinto2017}; \citealp[Theorem~1.8]{bittencourt-rios2022}}]
\label{thm:inst-mariano-pinto}
$\instG : \piIns^{\instco} \to \Ins^{\instco}$ and the functor
$\instF : \Ins^{\instco} \to \piIns^{\instco}$, which sends an institution to the
$\pi$-institution of its semantic consequence closure, form an adjunction
$\instG \dashv \instF$. Moreover $\instF \circ \instG = \mathrm{Id}$, and the unit is
the identity, so $\piIns^{\instco}$ is a full coreflective subcategory of
$\Ins^{\instco}$.
\end{theorem}

Theorem~\ref{thm:inst-mariano-pinto} is due to \citet{mariano-pinto2017}, as recalled
by \citet{bittencourt-rios2022}, who refer to the original for the proofs.
Propositions~\ref{thm:inst-G-models}--\ref{thm:inst-completeness} give direct proofs of
the three consequences used below.

\begin{proposition}[Well-definedness]\label{thm:inst-G-models}
For $f : \Sigma_1 \to \Sigma_2$ and a closed $m \in \Mod^{\instJ}(\Sigma_2)$,
$\Sen(f)^{-1}(m)$ is closed. For a comorphism $\langle \Phi, \alpha \rangle$ and a
closed $m' \in \Mod^{\instJ'}(\Phi\Sigma)$, $\alpha_\Sigma^{-1}(m')$ is closed, and
$\alpha^{-1}$ is natural.
\end{proposition}
\begin{proof}
Let $\Gamma = \alpha_\Sigma^{-1}(m')$. By the comorphism condition and monotonicity of
$C'$, $\alpha_\Sigma(C_\Sigma(\Gamma)) \subseteq C'_{\Phi\Sigma}(\alpha_\Sigma(\Gamma))
\subseteq C'_{\Phi\Sigma}(m') = m'$, so $C_\Sigma(\Gamma) \subseteq \Gamma$. The first
claim is the same argument with the translation condition of
Definition~\ref{def:inst-pi-institution} in place of the comorphism condition.
Naturality: for $f : \Sigma_1 \to \Sigma_2$, naturality of $\alpha$ gives
$\alpha_{\Sigma_2} \circ \Sen(f) = \Sen'(\Phi f) \circ \alpha_{\Sigma_1}$; taking
preimages, $\Sen(f)^{-1} \circ \alpha_{\Sigma_2}^{-1} = \alpha_{\Sigma_1}^{-1} \circ
\Sen'(\Phi f)^{-1}$. Functors between codiscrete categories are determined by their
object maps.
\end{proof}

\begin{proposition}[Satisfaction condition]\label{thm:inst-satisfaction}
For every $\pi$-institution comorphism $\langle \Phi, \alpha \rangle : \instJ \to
\instJ'$, $\Sigma$, $m' \in \Mod^{\instJ'}(\Phi\Sigma)$, and $\varphi \in \Sen(\Sigma)$,
\[
  m' \models'_{\Phi\Sigma} \alpha_\Sigma(\varphi) \iff
  \alpha_\Sigma(\varphi) \in m' \iff
  \varphi \in \alpha_\Sigma^{-1}(m') \iff
  \alpha_\Sigma^{-1}(m') \models_\Sigma \varphi .
\]
\end{proposition}
\begin{proof}
Each equivalence is a definition, and $\alpha_\Sigma^{-1}(m')$ is a model by
Proposition~\ref{thm:inst-G-models}. The satisfaction condition of
Definition~\ref{def:inst-institution} within $\instG(\instJ)$ is the case $\Phi =
\mathrm{Id}$, $\alpha = \Sen(f)$.
\end{proof}

\begin{proposition}[Completeness]\label{thm:inst-completeness}
In $\instG(\instJ)$, $\Gamma \models_\Sigma \varphi$ iff $\varphi \in C_\Sigma(\Gamma)$.
For an entailment system $\instE$, $(\Sign, \Sen, \Mod^{\instJ(\instE)}, \vdash,
\models^{\instJ(\instE)})$ is a complete logic in the sense of
Definition~\ref{def:inst-logic}.
\end{proposition}
\begin{proof}
If $\varphi \in C_\Sigma(\Gamma)$ and $m \supseteq \Gamma$ is closed, then
$C_\Sigma(\Gamma) \subseteq C_\Sigma(m) = m$, so $\varphi \in m$. Conversely,
$C_\Sigma(\Gamma)$ is a model containing $\Gamma$; if it satisfies $\varphi$, then
$\varphi \in C_\Sigma(\Gamma)$. For $\instJ(\instE)$, $C_\Sigma(\Gamma) =
\instclo{\Gamma}$, so $\vdash$ and $\models$ coincide.
\end{proof}

\begin{corollary}\label{thm:inst-every-comorphism}
Every translation between hosted logics that is a map of proof calculi yields, by
Theorem~\ref{thm:inst-forgetting} and $\instG$, an institution comorphism
$\langle \Phi, \alpha, \alpha^{-1} \rangle : \instG\instJTh(\instE) \to
\instG\instJTh(\instE')$ satisfying the satisfaction condition; when the map is
$\alpha$-plain, also $\langle \Phi^{\instdia}, \alpha, \alpha^{-1} \rangle :
\instG\instJ(\instE) \to \instG\instJ(\instE')$. In $\instG\instJTh(\instE)$ the models
over a theory $(\Sigma, \Gamma)$ are the closed theories over $\Sigma$ that contain
$\Gamma$.
\end{corollary}

\paragraph{Reading.}
At this level the models of each hosted logic are its closed theories, and $\vdash$
and $\models$ coincide; the system stores no other models. No interpretation of types
in a semantic universe is involved. The distinction between universes \`a la Russell
and \`a la Tarski, which matters for the coherence of a semantic interpretation, does
not arise; Chapter~\ref{ch:mltt} presents universes \`a la Russell.

\subsection{Meseguer's \texorpdfstring{$\instE^\dagger$}{E-dagger}}
\label{sec:inst-dagger}

\begin{proposition}[Meseguer, Proposition~9]\label{thm:inst-dagger}
Every entailment system $\instE$ underlies a logic $\instE^\dagger$ with
$\Mod(\Sigma) = (\Sigma, \emptyset)/\Th$, the slice category of theory morphisms out of
$(\Sigma, \emptyset)$, and $G \models_\Sigma \varphi$ iff $\Gamma' \vdash_{\Sigma'}
G(\varphi)$ for $G : (\Sigma, \emptyset) \to (\Sigma', \Gamma')$. $\instE^\dagger$ is
complete, exact, and admits initial models; if $\Th$ has pushouts, it is liberal.
\end{proposition}

The models of $\instG\instJ(\instE)$ and of $\instE^\dagger$ over one signature are
related by a satisfaction-preserving retraction.

\begin{proposition}\label{thm:inst-dagger-G}
For any morphism $G : (\Sigma, \emptyset) \to (\Sigma', \Gamma')$ in $\Mod_{\instE^\dagger}(\Sigma)$,
let $\mathrm{th}(G) = \Sen(G)^{-1}(\instclo{(\Gamma')})$. Then $\mathrm{th}(G)$ is
closed, $G \models \varphi$ iff $\varphi \in \mathrm{th}(G)$, and
$\mathrm{th}(1 : (\Sigma, \emptyset) \to (\Sigma, m)) = m$ for every closed $m$.
\end{proposition}
\begin{proof}
If $\Delta = \mathrm{th}(G)$ and $\Delta \vdash_\Sigma \varphi$, then $G(\Delta)
\vdash_{\Sigma'} G(\varphi)$ by $\vdash$-translation, and $G(\Delta) \subseteq
\instclo{(\Gamma')}$, so $G(\varphi) \in \instclo{(\Gamma')}$ by transitivity. The
second claim is the definition of satisfaction in $\instE^\dagger$. The third holds
because $m = \instclo{m}$.
\end{proof}

Meseguer's $\instE^\dagger$ is not used as the model level in Eigenius for two reasons:
\begin{enumerate}[label=(\roman*)]
\item \emph{Direction of model maps.} It does not extend functorially to the comorphisms
  of Eigenius. As Meseguer notes in the remark following Proposition~31, the construction
  does not yield a functor $(\ )^\dagger : \Ent \to \Log$: for a map $(\Phi, \alpha)$,
  the induced transformation sends a model $H : T \to T'$ of $\instE^\dagger$ forward to
  $\Phi(H) : \Phi(T) \to \Phi(T')$, whereas an institution comorphism (and Meseguer's map
  of logics, Definitions~27 and~29) requires a model map in the backward direction. By
  contrast, $\instG$ takes closed theories over the same signature as its models, and
  the preimage $\alpha^{-1}$ provides the required backward model map.
\item \emph{Unused categorical structure.} The extra categorical properties of
  $\instE^\dagger$---exactness and liberality---are not used by any result in this book
  (liberality of $\instE^\dagger$ relies on pushouts in $\Th$, whose existence for the
  kernel remains open; see Section~\ref{sec:inst-status}). Moreover, by
  Proposition~\ref{thm:inst-dagger-G}, every $\instE^\dagger$-model already determines a
  $\instG$-model satisfying the exact same sentences.
\end{enumerate}

\section{Limits of the Three Levels}
\label{sec:inst-limits}

\paragraph{Semantic models.}
The models of Section~\ref{sec:inst-models} are syntactic. For a hosted logic with its
own semantics, such as the kernel theory and Lean's kernel theory with their
set-theoretic models, soundness and the satisfaction condition with respect to
\emph{those} models are separate statements, treated in
Section~\ref{sec:inst-real-models}.

\paragraph{Proof relevance beyond $\gamma$.}
$\instU$ discards $\gamma$, and $\instG$ acts on what remains. Every statement about
which witness is transported, including provenance and warrant grades, is a statement
at the witness level and has no counterpart at the provability or model level.

\paragraph{Heterogeneous specification.}
A Grothendieck construction over the indexed category of hosted logics yields a single
institution in which sentences of all hosted logics and their translations coexist
\citep{diaconescu2008,diaconescu2025}; \citet[Section~3]{bittencourt-rios2022} obtain the
adjunctions of Theorem~\ref{thm:inst-mariano-pinto} from a corresponding construction
on rooms, and \citet{mossakowski2007hets} implement heterogeneous specification over a
Grothendieck logic. Eigenius requires translations between pairs of logics and their
composites, which Section~\ref{sec:inst-composition} supplies; the Grothendieck
construction is not used.

\section{Logics with Semantic Models: Soundness per Hosted Checker}
\label{sec:inst-real-models}

At the level of Section~\ref{sec:inst-models}, soundness is immediate from
Proposition~\ref{thm:inst-completeness}. It becomes an obligation where a hosted logic
also has models that are not its theories.

\begin{center}
\small
\begin{tabularx}{\textwidth}{@{}>{\raggedright\arraybackslash}p{0.16\textwidth}
  >{\raggedright\arraybackslash}X>{\raggedright\arraybackslash}X
  >{\raggedright\arraybackslash}X@{}}
\toprule
Logic & Semantic models & Soundness of $\vdash$ & Translation into it at the model
  level \\
\midrule
Kernel theory &
  set-theoretic models (Chapter~\ref{ch:mltt}) &
  proved, relative to the inaccessible cardinals of Chapter~\ref{ch:mltt}
  (Theorem~\ref{thm:inst-kernel-logic}) &
  --- \\
Lean's kernel theory &
  set-theoretic model \citep{carneiro2019} &
  cited; relative to correctness of the hosted checker and to the axiom set $A$ &
  requires a model map back to the kernel theory;
  Conjecture~\ref{thm:inst-initiality} \\
Justification logic &
  basic (Mkrtychev) models and Fitting models &
  cited \citep{artemov-fitting2019} & Chapter~\ref{ch:justification-logic}; Volume~II \\
Statistics &
  possibly none & per analysis plan & Volume~II \\
\bottomrule
\end{tabularx}
\end{center}

\paragraph{The kernel theory.}
Chapter~\ref{ch:mltt} constructs a set-theoretic model of the kernel theory in ZFC with
an increasing sequence of inaccessible cardinals, in which the value of a closed term lies
in the value of its type. Its models of the environments are semantic models of the
kernel theory, and the kernel calculus of Section~\ref{sec:inst-witness} is sound for
them.

\begin{theorem}[The kernel theory is a logic]\label{thm:inst-kernel-logic}
In ZFC with the inaccessible cardinals of Chapter~\ref{ch:mltt}, let a model $M$ of a
kernel signature $\Sigma$ be a signature morphism $\tau : \Sigma \to \Sigma''$
(Definition~\ref{def:inst-kernel-sign}) into a signature $\Sigma''$ that the kernel checks,
together with a valuation of the constants of $\Sigma''$ under which $\Sigma''$ has the
model of Chapter~\ref{ch:mltt}, and let $M \models_\Sigma \varphi$ when
$\llbracket \insthat{\tau}(\varphi) \rrbracket = \{\emptyset\}$ under that valuation. The
reduct of a model $(\tau', \dots)$ of $\Sigma'$ along $\sigma : \Sigma \to \Sigma'$ is
$(\tau' \circ \sigma, \dots)$, with the same valuation. The kernel entailment system on
$\Sign_{\instK}$ and these models form a logic (Definition~\ref{def:inst-logic}): the
satisfaction condition holds, and $\Gamma \vdash_\Sigma \varphi$ implies
$\Gamma \models_\Sigma \varphi$. In particular, a theory with a model does not entail
$\tttFalse$. A valuation under which $\Sigma$, or an extension of $\Sigma$ that the kernel
checks, has the model of Chapter~\ref{ch:mltt} is the model whose morphism is the
inclusion, and it satisfies $\varphi$ iff $\llbracket \varphi \rrbracket = \{\emptyset\}$
over $\Sigma$.
\end{theorem}

\begin{proof}
Translation is functorial, $\insthat{\tau' \circ \sigma} = \insthat{\tau'} \circ
\insthat{\sigma}$, so a model of $\Sigma'$ satisfies $\insthat{\sigma}(\varphi)$ iff its
reduct satisfies $\varphi$; this is the satisfaction condition. For soundness, let
$\Gamma \vdash_\Sigma \varphi$ and let $M = (\tau, \dots)$ satisfy $\Gamma$.
Lemma~\ref{thm:inst-ent-trans} along $\tau$ gives $\insthat{\tau}(\Gamma)
\vdash_{\Sigma''} \insthat{\tau}(\varphi)$, so it suffices to show soundness over
$\Sigma''$ under its valuation. Let $\Sigma''; \instax{\Gamma_0} \vdash t : \psi$ with
$\Gamma_0 \subseteq \insthat{\tau}(\Gamma)$ finite. Relativization
(Proposition~\ref{thm:inst-relativise}) gives a closed proof $\lambda \Gamma_0.\, t$ of
$\Gamma_0 \instimp \psi$, whose value lies in the value of $\Gamma_0 \instimp \psi$ by
the proof of Theorem~\ref{thm:mltt-declarative-consistency}. That value is a proposition
with an element, so it is $\{\emptyset\}$. An implication $\chi \instimp \chi'$ with
$\llbracket \chi \rrbracket = \{\emptyset\}$ denotes $\{\emptyset\}$ only if $\chi'$
does, so discharging the sentences of $\Gamma_0$ one at a time gives
$\llbracket \psi \rrbracket = \{\emptyset\}$. $\tttFalse$ denotes $\emptyset$, and so
does its translation, which is $\tttFalse$. For the last claim, a sentence of $\Sigma$
has the same value over $\Sigma$ and over an extension: by induction on its derivation,
the value of a term typed in an environment does not change when the environment is
extended and the valuation of its constants is kept, since the sorts that the
interpretation reads off derivations are the same in the extension.
\end{proof}

A model $(\tau, \dots)$ of $\Sigma$ assigns to each constant $c$ of $\Sigma$ the value
$\llbracket \tau(c) \rrbracket$. That induced valuation on $\Sigma$ need not itself be a
model of $\Sigma$ in the direct sense of Chapter~\ref{ch:mltt}, where each inductive
type denotes its own least fixed point and its recursor denotes recursion on that fixed
point. For example, let $\mathsf{Unit} : \Setu$ be the one-constructor inductive type
with constructor $\mathsf{tt} : \mathsf{Unit}$. The signature morphism that sends
$\tttBool$ to $\mathsf{Option}\,\mathsf{Unit}$, $\mathsf{true}$ to $\mathsf{none}$,
$\mathsf{false}$ to $\mathsf{some}\,\mathsf{tt}$, and $\rec_{\tttBool}$ to the recursor
of $\mathsf{Option}$ composed with that of $\mathsf{Unit}$ satisfies
conditions~\ref{inst:S-type}--\ref{inst:S-iota}, and its induced valuation interprets
$\tttBool$ by the set interpreting $\mathsf{Option}\,\mathsf{Unit}$. More generally, any
class of models defined purely as valuations on $\Sigma$ and closed under pullback along
signature morphisms must allow an inductive type to be interpreted by an arbitrary set
equipped with operations satisfying its computation rules; soundness for such a broader
class of models is not needed here. Constructing a semantic model map between the kernel
theory and Lean's theory is the subject of Conjecture~\ref{thm:inst-initiality} below.

\paragraph{Lean's kernel theory.}
Lean's kernel theory has a set-theoretic model in ZFC with inaccessible cardinals
\citep{carneiro2019}, and its typing relation and type checker have been partially
formalized and verified \citep{carneiro2024}. Soundness of a Lean witness accepted by a
hosted checker is therefore relative to two assumptions: that the hosted checker accepts
only declaratively well-typed terms, and that the axioms in the fixed set $A$ hold in
the model. When translating sentences from the Eigenius kernel theory into Lean's
theory, each use of cumulative subsumption (Chapter~\ref{ch:mltt}) must be replaced by
an explicit universe lift, because Lean's universes are not cumulative (Volume~II).
Furthermore, because Lean's proof irrelevance is a definitional conversion rule, a
kernel witness transported as a proof inside Lean's $\Prop$ would become definitionally
equal to every other proof of the same proposition; consequently, witness transport must
route witnesses around Lean's $\Prop$ into a data type, as noted in
Section~\ref{sec:inst-comorphisms}.

\paragraph{The axiom set is part of the target logic.}
The hosted checker admits a fixed set $A$ of axioms; write $\mathbb{L}_A$ for the proof
calculus of Lean's kernel theory with axiom set $A$. A translation into Lean that
\emph{adopts} a source axiom $\varphi$ requires the checker to admit $\alpha(\varphi)$,
and its target is $\mathbb{L}_{A \cup \{\alpha(\varphi)\}}$, a different logic. A
verdict of the hosted checker is a verdict relative to its axiom set, which is
therefore a component of the verdict. \emph{Relativization}
(Proposition~\ref{thm:inst-relativise}) leaves $A$ unchanged: the translated sentence
$\alpha(\vec\varphi) \instimp \alpha(\psi)$ is proved in $\mathbb{L}_A$. Relativization
is therefore the default policy for translations into Lean.

\paragraph{Model maps.}
A comorphism from the kernel theory into Lean's theory at the level of semantic models
requires a natural transformation $\beta$ from Lean models to kernel-theory models.
One route is categorical logic: if the kernel's logic is a categorical logic on some
category $\mathbb{T}$ of structured categories (\citet{meseguer1989}, Definition~40),
then by Theorem~42 of \citet{meseguer1989} it is complete, liberal, exact, and admits
initial models, and model maps are obtained by composing functors out of the
classifying object.

\begin{conjecture}[Initiality for the kernel theory]\label{thm:inst-initiality}
There is a category $\mathbb{T}$ of structured categories with pushouts and a faithful
functor $\mathbb{T} \to \Cat$ such that the term model of each kernel theory is initial
among its $\mathbb{T}$-models, and conditions~1--3 of Definition~40 of
\citet{meseguer1989} hold for the kernel's logic. The missing step is an initiality
theorem for the kernel theory with impredicative $\Prop$, proof irrelevance,
cumulative universes with subsumption, and inductive families.
\end{conjecture}

Without Conjecture~\ref{thm:inst-initiality}, soundness of a Lean verdict with respect
to Lean's semantics is stated as a cited assumption, and the satisfaction condition at
the semantic level is not claimed.

\paragraph{Justification logic.}
Basic models (Mkrtychev models) and Fitting models are the semantics of justification
logic, with soundness and completeness of the basic systems established in the
literature \citep{artemov2008justification} (and the monograph cited above). The
operational evidence function of Chapter~\ref{ch:justification-logic} is the minimal
evidence function of a basic model; the soundness argument is given there.

\paragraph{Statistics.}
A statistical logic may have no models beyond those of Section~\ref{sec:inst-models},
in which case only the witness and provability levels apply. Volume~II states the analysis-plan
semantics.

\paragraph{Trusted computing base.}
Admitting a hosted checker adds its soundness, relative to its axiom set, to the trust
assumptions of every claim whose witness it produced.

\section{Status of the Obligations}
\label{sec:inst-status}

\begin{center}
\small
\begin{longtable}{@{}>{\raggedright\arraybackslash}p{0.5\textwidth}
  >{\raggedright\arraybackslash}p{0.13\textwidth}
  >{\raggedright\arraybackslash}p{0.28\textwidth}@{}}
\toprule
Obligation & Status & Where \\
\midrule
\endhead
Kernel signature morphisms preserve all judgments & proved &
  Lemma~\ref{thm:inst-translation} \\
Reflexivity, monotonicity, transitivity, $\vdash$-translation & proved &
  Lemmas~\ref{thm:inst-refl}--\ref{thm:inst-ent-trans} \\
$P_{\instK}$ functorial on $\Thz$; $\pi$ natural & proved &
  Lemma~\ref{thm:inst-nat} \\
The kernel is a compact proof calculus with Tarskian $\vdash$ & proved &
  Theorem~\ref{thm:inst-kernel-pc} \\
Weakening, substitution, level instantiation & cited &
  Chapter~\ref{ch:mltt} \\
Quotienting witnesses by provable identity collapses to $\instE^\natural$ & proved &
  Proposition~\ref{thm:inst-collapse} \\
Cut realizable as a redex without new conversion checks & proved &
  Proposition~\ref{thm:inst-cut-redex}; Proposition~\ref{prop:checking-compositional} \\
Definitional equality is joinability: transitive, decided on terminating terms & proved &
  Lemma~\ref{lem:mltt-confluence}; Theorem~\ref{thm:checking-conv} \\
Transport invariant implies map of proof calculi & proved &
  Lemma~\ref{thm:inst-invariant} \\
Adopt is a map of proof calculi & proved &
  Proposition~\ref{thm:inst-adopt} \\
Discharge is a map of proof calculi on subcategories with natural discharge &
  proved & Proposition~\ref{thm:inst-discharge} \\
Relativize is a map of proof calculi over $\instThl$, into the axiom-free fragment &
  proved & Proposition~\ref{thm:inst-relativise} \\
Forgetting witnesses is a functor $\instPC \to \piIns^{\instco}$ & proved &
  Theorem~\ref{thm:inst-forgetting} \\
Signature-level comorphism for plain maps & proved &
  Corollary~\ref{thm:inst-forgetting-sig} \\
$\instG \dashv \instF$, $\instF\instG = \mathrm{Id}$, coreflectivity & cited &
  Theorem~\ref{thm:inst-mariano-pinto} \\
Closure of $\alpha^{-1}(m')$; naturality of $\alpha^{-1}$ & proved &
  Proposition~\ref{thm:inst-G-models} \\
Satisfaction condition for every comorphism & proved &
  Proposition~\ref{thm:inst-satisfaction} \\
Completeness at the model level & proved &
  Proposition~\ref{thm:inst-completeness} \\
The kernel theory is a logic with its set-theoretic models: satisfaction condition along
  signature morphisms, soundness & proved, relative to the inaccessible cardinals of
  Chapter~\ref{ch:mltt} & Theorem~\ref{thm:inst-kernel-logic} \\
$\instE^\dagger$ complete, exact, initial models & cited &
  Proposition~\ref{thm:inst-dagger} \\
$\instE^\dagger$ models retract onto $\instG$ models & proved &
  Proposition~\ref{thm:inst-dagger-G} \\
Pushouts in $\Th$ for the kernel (liberality of $\instE^\dagger$) & open, not pursued & Volume~IV
for merges \\
Grade algebra and warrant monotonicity along translations & stated elsewhere &
  Chapter~\ref{ch:justification-logic} \\
Non-factivity of $j : P$ & stated elsewhere &
  Chapter~\ref{ch:justification-logic} \\
Structural axioms for the statistics logic & open & Volume~II \\
Soundness of Lean's theory for its set-theoretic model & cited &
  Section~\ref{sec:inst-real-models} \\
Soundness and completeness of justification logic for basic models & cited &
  Section~\ref{sec:inst-real-models} \\
Initiality for the kernel theory; semantic model maps into Lean & open &
  Conjecture~\ref{thm:inst-initiality} \\
\bottomrule
\end{longtable}
\end{center}

\chapter{Representing Knowledge Graphs}
\label{ch:knowledge-graphs}

\providecommand{\kgIRI}{\mathsf{IRI}}
\providecommand{\kgBool}{\mathsf{Bool}}
\providecommand{\kgtrue}{\mathsf{true}}
\providecommand{\kgfalse}{\mathsf{false}}
\providecommand{\kgeqb}{\mathsf{eqb}}
\providecommand{\kgProps}{\mathcal{P}}
\providecommand{\kgClasses}{\mathcal{C}}
\providecommand{\kgRes}{\mathcal{R}}
\providecommand{\kgG}{\mathcal{G}}
\providecommand{\kgReq}{\mathrm{Req}}
\providecommand{\kgReqz}{\mathrm{Req}_0}
\providecommand{\kgRecm}{\mathrm{Rec}}
\providecommand{\kgPar}{\mathrm{Par}}
\providecommand{\kgdom}{\mathrm{dom}}
\providecommand{\kganc}{\sqsubseteq^{*}}
\providecommand{\kgeqv}{\approx}
\providecommand{\kgty}[1]{\tau_{#1}}
\providecommand{\kgrho}[1]{\rho_{#1}}
\providecommand{\kgS}[1]{S_{#1}}
\providecommand{\kglabels}{\mathrm{labels}}
\providecommand{\kgFL}{\mathrm{FL}}
\providecommand{\kgcanon}{\mathrm{canon}}
\providecommand{\kgrcd}[1]{\langle #1\rangle}
\providecommand{\kgtel}{\mathrm{tel}}
\providecommand{\kgRecType}{\mathsf{RecType}}
\providecommand{\kgEn}{\mathrm{En}}
\providecommand{\kgsingE}{\mathsf{singE}}
\providecommand{\kgjoin}{\mathsf{join}}
\providecommand{\kgsing}{\mathsf{sing}}
\providecommand{\kgjcase}{\mathsf{jcase}}
\providecommand{\kgown}[1]{C_{#1}}
\providecommand{\kgresof}[1]{\ulcorner #1 \urcorner}
\providecommand{\kgM}{\mathcal{M}}
\providecommand{\kgPhi}{\Phi}
\providecommand{\kgApprop}{\mathrm{Approp}}
\providecommand{\kgIntro}{\mathrm{Intro}}
\providecommand{\kgFeat}{\mathrm{Feat}}
\providecommand{\kgTyp}{\mathrm{Typ}}
\providecommand{\kgfsub}{\sqsubseteq_{\mathrm{fs}}}
\providecommand{\kgunify}{\sqcup}
\providecommand{\kgtr}{\mathrm{tr}}
\providecommand{\kgchi}{\chi}
\providecommand{\kgAt}{\mathcal{A}}
\providecommand{\kgisA}{\mathrm{isA}}
\providecommand{\kgreqA}{\mathrm{requires}}
\providecommand{\kgsrc}[1]{\textup{[#1]}}
\providecommand{\tttFalse}{\mathsf{False}}

This chapter gives knowledge graphs a type-theoretic semantics. The guiding framework is
Cooper's Type Theory with Records (TTR) \citep{cooper2023perception}, embedded in the
dependent type theory of Chapter~\ref{ch:mltt}. In this reading, a class denotes a named
record type, a resource denotes a record, and the statement that a resource belongs to a
class is a judgment at two levels (declared and structural) and a proposition at a third.
TTR also serves as the foundation for the unification-based grammar of Volume~III.
Accordingly, Section~\ref{sec:kg-ttr} restates the TTR definitions used across the book,
and Section~\ref{sec:kg-tfs} relates TTR records to typed feature structures.
Sections~\ref{sec:kg-embedding}--\ref{sec:kg-model} then develop the type-theoretic
embedding, class hierarchies, resource membership, joins and singletons, propositional
membership, and the consistency of the extended theory; finally,
Appendix~\ref{sec:kg-atomic-data} maps Atomic Data \citep{atomicdata} into this theory.
The realization of these constructions in the system implementation is treated in
Volume~II.

\section{Knowledge graphs as typed data}
\label{sec:kg-typed-data}

A knowledge graph comprises three kinds of entities: resources, properties, and classes.
All three are first-class citizens named by IRIs. A resource carries properties; a
property has a declared value type; and a class declares required properties, recommended
properties, and parent classes. Because properties and classes are themselves resources,
the schema vocabulary and the instance data inhabit a single graph.

\begin{definition}[Knowledge graph]
\label{def:kg-graph}
Fix a countably infinite set $\kgIRI$ of IRIs. A \emph{knowledge graph} $\kgG$
consists of finite sets $\kgProps, \kgClasses \subseteq \kgRes \subseteq \kgIRI$
of \emph{properties}, \emph{classes}, and \emph{resources}, together with:
\begin{enumerate}[nosep]
\item for each property $p \in \kgProps$, a \emph{field type} $\kgty{p}$
  (Section~\ref{sec:kg-classes}) and an optional \emph{domain}
  $\kgdom(p) \in \kgClasses$;
\item for each class $C \in \kgClasses$, a set $\kgReqz(C) \subseteq \kgProps$ of
  \emph{own required} properties, a set $\kgRecm(C) \subseteq \kgProps$ of
  \emph{recommended} properties, and a set $\kgPar(C) \subseteq \kgClasses$ of
  \emph{declared parents};
\item for each resource $r \in \kgRes$, a finite map $\kgrho{r}$ from IRIs to
  values, the \emph{record} of $r$, and a finite set
  $\kgS{r} \subseteq \kgClasses$ of \emph{declared classes}.
\end{enumerate}
A \emph{value} is a literal of a basic type, an IRI used as a reference, a finite
list of values, or an embedded record without an IRI.
\end{definition}

In the vocabulary of the data model, $\kgReqz$, $\kgRecm$, and $\kgPar$ correspond to the
properties \code{requires}, \code{recommends}, and \code{subclass\_of}; $\kgS{r}$ is the
value of \code{is\_a}; and $\kgdom$ is \code{domain}. Field types are built from
\code{data\_type}, \code{class\_types}, and \code{allows\_only}
(Section~\ref{sec:kg-joins}).

\paragraph{Open world, bounded.} A resource may carry any declared property beyond those
mentioned by its classes. However, it may not carry an IRI that is not a declared
property, nor may it carry a property whose declared \code{domain} it fails to satisfy
(Definition~\ref{def:kg-admissible}). In TTR terms, membership is width-inclusive, while
the admissible width is bounded by the vocabulary and its domain constraints.

\paragraph{Lineage.} An RDF graph is a set of triples $(s, p, o)$ \citep{rdf}. Grouping
the triples with subject $s$ by predicate yields the record $\kgrho{s}$; in our data
model, a resource has at most one value per property, and a multi-valued property is
assigned a list type. Atomic Data \citep{atomicdata} enriches RDF with typed,
dereferenceable properties (Appendix~\ref{sec:kg-atomic-data}). The type-theoretic
reading adds two further ingredients: a type for every property value, and classes as
genuine types---whereas \code{rdf:type} in RDF is merely a tag.

\section{TTR: the definitions this book uses}
\label{sec:kg-ttr}

TTR models types and their witnesses as set-theoretic objects. Records, record
types, ptypes, and function types are \emph{labeled sets}: sets of
\emph{fields} $\langle \ell, a\rangle$ in which no label occurs twice
\citep[App.~A1]{cooper2023perception}. The twelve definitions below are those the
book uses. Each is stated in the book's notation; the bracketed source is the
section and clause of the formal appendix of \citet{cooper2023perception}
(``App.'') or the section and page of its Chapter~1 (``Ch.~1''). Each carries a
usage tag: \textsf{KG} for this chapter, \textsf{P} for the parsing work of Volume~III.
Section~\ref{sec:kg-embedding} maps each into the dependent type theory
of Chapter~\ref{ch:mltt}.

\begin{definition}[Type system; witness judgment]
\label{def:kg-ttr-system}
\kgsrc{App.~A2.1, cl.~1--4; App.~A3.2; Ch.~1 \S1.3, p.~12} \textsf{KG, P}.
A \emph{system of basic types} is a pair $\langle \mathit{Type}, A\rangle$ in
which $\mathit{Type}$ is a non-empty set, $A$ is a function with domain
$\mathit{Type}$, $A(T)$ is disjoint from $\mathit{Type}$ for every
$T \in \mathit{Type}$, and $a : T$ iff $a \in A(T)$. A \emph{system of complex
types} is a quadruple
$\langle \mathit{Type}, \mathit{BType}, \langle \mathit{PType}, \mathit{Pred},
\mathit{ArgIndices}, \mathit{Arity}\rangle, \langle A, F\rangle\rangle$ that
extends a system of basic types $\langle \mathit{BType}, A\rangle$ with ptypes
(Definition~\ref{def:kg-ttr-ptype}); the pair $\langle A, F\rangle$ is its
\emph{model}. The judgment $a : T$ reads ``$a$ is a witness of $T$''. Types are
objects in their own right, not the sets of their witnesses.
\end{definition}

\begin{definition}[Basic types]
\label{def:kg-ttr-basic}
\kgsrc{App.~A2.1; App.~A3.2, cl.~1--3} \textsf{KG, P}.
A \emph{basic type} is a member of $\mathit{BType}$. Its witnesses are given
directly by $A$, and every witness of a basic type in
$\langle \mathit{BType}, A\rangle$ is a witness of it in the complex system.
\end{definition}

In the knowledge graph the basic types are the literal datatypes (strings,
integers, floats, booleans, dates, timestamps).

\begin{definition}[Ptypes]
\label{def:kg-ttr-ptype}
\kgsrc{App.~A3.1; App.~A3.2, cl.~5--8} \textsf{P; KG}.
A \emph{predicate signature} is a triple
$\langle \mathit{Pred}, \mathit{ArgIndices}, \mathit{Arity}\rangle$ with
$\mathit{Arity}$ mapping each predicate to a finite sequence of argument indices
(to a set of such sequences, for a polymorphic signature). Then
$P(a_1, \ldots, a_n) \in \mathit{PType}$ iff $P \in \mathit{Pred}$,
$\mathit{Arity}(P) = \langle T_1, \ldots, T_n\rangle$, and $a_i : T_i$ for each $i$.
For $T \in \mathit{PType}$, $F(T)$ is a set disjoint from $\mathit{Type}$, and
$a : T$ iff $a \in F(T)$. The witnesses of a ptype are situations or evidence for
the predication.
\end{definition}

\begin{definition}[Record types]
\label{def:kg-ttr-rectype}
\kgsrc{App.~A1; App.~A11.2, cl.~1, 3, 5, 7} \textsf{KG, P}.
Fix a countably infinite set $\mathcal{L}$ of labels. Record types are labeled
sets carrying the flavor $\mathrm{RT}$, which distinguishes them from records
with the same fields. The set $\mathit{RType}$ of record types contains the type
$\mathit{Rec}$ of all records and the type $\mathit{ERec}$ of the empty record
(cl.~1, 3); the type $\{\mathrm{RT}, \langle \ell, T\rangle\}$ for
$\ell \in \mathcal{L}$ and $T \in \mathit{Type}$ (cl.~5); and
$R \cup \{\langle \ell, T\rangle\}$ for $R \in \mathit{RType} - \{\mathit{Rec},
\mathit{ERec}\}$ whose labels do not include $\ell$ (cl.~7). A record type is
written $\rtype{\ell_1 : T_1, \ldots, \ell_n : T_n}$. Labels are unique, and the
fields are unordered.
\end{definition}

\begin{definition}[Records; membership]
\label{def:kg-ttr-record}
\kgsrc{App.~A11.1; App.~A11.2, cl.~2, 4, 6, 8; Ch.~1 \S1.4.3, p.~23} \textsf{KG, P}.
A \emph{record} is a finite labeled set whose labels lie in $\mathcal{L}$ and
outside the reserved labels, and each of whose values is a witness of some type.
The value of $\ell$ in $r$ is $r.\ell$. Membership is defined by: $r : \mathit{Rec}$
iff $r$ is a record (cl.~2); $r : \mathit{ERec}$ iff $r = \emptyset$ (cl.~4);
$r : \{\mathrm{RT}, \langle\ell, T\rangle\}$ iff $r : \mathit{Rec}$,
$\langle \ell, a\rangle \in r$, and $a : T$ (cl.~6);
$r : R \cup \{\langle \ell, T\rangle\}$ iff $r : R$, $\langle \ell, a\rangle \in r$,
and $a : T$ (cl.~8). The clauses constrain only the labels of the type, so a
record of a record type may carry further fields.
\end{definition}

\begin{definition}[Dependent fields; dependency families]
\label{def:kg-ttr-dependent}
\kgsrc{App.~A11.2, dependent record types, cl.~1--2; App.~A11.6} \textsf{KG, P}.
A \emph{dependent field} is $\langle \ell, \langle \mathcal{F}, \langle \pi_1,
\ldots, \pi_m\rangle\rangle\rangle$, where $\pi_1, \ldots, \pi_m$ are paths in the
record type and $\mathcal{F} : (T_1 \to \cdots \to (T_m \to \mathit{Type}^n))$.
Then $r : R \cup \{\langle \ell, \langle \mathcal{F}, \langle \pi_1, \ldots,
\pi_m\rangle\rangle\rangle\}$ iff $r : R$, $\langle \ell, a\rangle$ is a field of
$r$, $r.\pi_i : T_i$ for each $i$, and $a : \mathcal{F}(r.\pi_1)\cdots(r.\pi_m)$.
A path $\pi_2$ \emph{depends on} $\pi_1$ if the type at $\pi_2$ refers to
$\pi_1$; the \emph{dependency family} $\mathit{paths}_\pi(T)$ of $\pi$ is the
least set of paths containing $\pi$ and closed under dependency in both
directions.
\end{definition}

\begin{definition}[Singleton types; manifest fields]
\label{def:kg-ttr-singleton}
\kgsrc{App.~A6, cl.~1--2; App.~A11.2} \textsf{KG, P}.
For $T, T' \in \mathit{Type}$ and $a : T'$, the \emph{singleton type} $\sing{a}$
(formed over $T$) is a type, and $b : \sing{a}$ iff $b : T$ and $a = b$. A field
$\langle \ell, \sing{a}\rangle$ is \emph{manifest} and is written
$\ell = a : T$.
\end{definition}

\begin{definition}[Meet and join]
\label{def:kg-ttr-meet-join}
\kgsrc{App.~A7, cl.~1--2; App.~A8, cl.~1--2; App.~A11.3} \textsf{KG, P}.
For types $T_1, T_2$, $a : T_1 \ttrmeet T_2$ iff $a : T_1$ and $a : T_2$, and
$a : T_1 \ttrjoin T_2$ iff $a : T_1$ or $a : T_2$. The generalized forms
$\bigwedge \mathcal{T}$ (for non-empty finite $\mathcal{T}$) and
$\bigvee \mathcal{T}$ (for finite $\mathcal{T}$) are witnessed by $a$ iff $a$
witnesses every, respectively some, member of $\mathcal{T}$. The \emph{merge}
$\mu(T_1 \ttrmeet T_2)$ of two record types is the record type whose fields are
the union of the fields of $T_1$ and $T_2$, with the meet of the two field types
at a shared label (App.~A11.3, clause $\mu'$~1).
\end{definition}

\begin{definition}[Subtyping]
\label{def:kg-ttr-subtyping}
\kgsrc{App.~A9, restrictive modal notions, cl.~2; Ch.~1 \S1.4.3.5, (55)--(57),
pp.~36--38} \textsf{KG, P}.
In a modal system of complex types, whose possibilities $M$ assign witnesses to
basic types and ptypes, $T_1 \sqsubseteq T_2$ iff for every possibility $M$,
$\{a \mid a :_M T_1\} \subseteq \{a \mid a :_M T_2\}$. For record types,
$T_2 \subseteq T_1$ as sets of fields implies $T_1 \sqsubseteq T_2$ (Ch.~1, rule
(57)).
\end{definition}

Rule (57) is width subtyping. Depth subtyping, field-wise subtyping at shared
labels, is not stated as a rule by \citet{cooper2023perception}; for
non-dependent fields it follows from clauses~6 and~8 of App.~A11.2 and the
definition of $\sqsubseteq$. Section~\ref{sec:kg-embedding} states both as rules
and proves them sound.

\begin{definition}[Function and list types]
\label{def:kg-ttr-function-list}
\kgsrc{App.~A4, cl.~1--2; App.~A10; App.~A12, cl.~1--2} \textsf{KG, P}.
$f : (T_1 \to T_2)$ iff $f$ is a function whose domain is
$\{a \mid a : T_1\}$ and whose range is included in $\{a \mid a : T_2\}$.
$f : ((a : T) \to \mathcal{T}(a))$ iff $f$ is a function with domain
$\{a \mid a : T\}$ and $f(a) : \mathcal{T}(a)$ for each such $a$. For the list
type $\mathit{list}(T)$: $[\,] : \mathit{list}(T)$, and
$a \mid L : \mathit{list}(T)$ iff $a : T$ and $L : \mathit{list}(T)$.
\end{definition}

\begin{definition}[Intensionality]
\label{def:kg-ttr-intensional}
\kgsrc{Ch.~1 \S1.3, p.~12; App.~A10} \textsf{KG}.
Types are not determined by their witnesses: two distinct types may have the same
set of witnesses in every possibility. An \emph{intensional system of complex
types} is a family of systems of complex types indexed by $n \in \Nat$, with
$\mathit{Type}^n \in \mathit{Type}^{n+1}$, in which a type of order $n-1$ is
a witness of $\mathit{Type}^{n}$.
\end{definition}

\begin{definition}[Records and $\Sigma$-types]
\label{def:kg-ttr-sigma}
\kgsrc{Ch.~1 \S1.4.3, pp.~22--26} \textsf{KG, P}.
A record type is a set of labeled fields; a $\Sigma$-type is positional. The
types $\Sigma x{:}T_1.\,\Sigma y{:}T_2.\,T_3$ and
$\Sigma y{:}T_2.\,\Sigma x{:}T_1.\,T_3$ are distinct and have distinct sets of
witnesses ($\langle a, \langle b, s\rangle\rangle$ against
$\langle b, \langle a, s\rangle\rangle$); the record types
$\rtype{x : T_1, y : T_2, e : T_3}$ and $\rtype{y : T_2, x : T_1, e : T_3}$ are
one type. Labels give access to components by name. TTR uses record types in
place of both $\Sigma$-types and contexts. The same comparison is made in
\citet[\S2.3]{cooper2021records}.
\end{definition}

\begin{table}[ht]
\centering\small
\caption{The TTR definitions used in the book, with their sources in
\citet{cooper2023perception} and their uses.}
\label{tab:kg-ttr-defs}
\begin{tabularx}{\textwidth}{@{}rlXl@{}}
\toprule
\# & Definition & Source & Used in \\
\midrule
1 & type system; witness judgment & App.~A2.1 cl.~1--4; A3.2; Ch.~1 \S1.3 & KG, P \\
2 & basic types & App.~A2.1; A3.2 cl.~1--3 & KG, P \\
3 & ptypes & App.~A3.1; A3.2 cl.~5--8 & P; KG (\S\ref{sec:kg-propositions}) \\
4 & record types & App.~A1; A11.2 cl.~1, 3, 5, 7 & KG, P \\
5 & records; membership & App.~A11.1; A11.2 cl.~2, 4, 6, 8 & KG, P \\
6 & dependent fields; dependency families & App.~A11.2 (dependent) cl.~1--2; A11.6 & KG, P \\
7 & singleton types; manifest fields & App.~A6 cl.~1--2; A11.2 & KG, P \\
8 & meet and join; merge & App.~A7, A8 cl.~1--2; A11.3 & KG, P \\
9 & subtyping & App.~A9 (restrictive) cl.~2; Ch.~1 \S1.4.3.5 & KG, P \\
10 & function and list types & App.~A4 cl.~1--2; A10; A12 cl.~1--2 & KG, P \\
11 & intensionality & Ch.~1 \S1.3; App.~A10 & KG \\
12 & records and $\Sigma$-types & Ch.~1 \S1.4.3 & KG, P \\
\bottomrule
\end{tabularx}
\end{table}

\section{TTR and typed feature structures}
\label{sec:kg-tfs}

Typed feature structures are the data structures of unification-based grammar,
among them Head-Driven Phrase Structure Grammar
(HPSG)~\citep{pollard-sag1994}; unification formalisms are surveyed
in~\citet{shieber1986}. Their logic is given
in~\citet{carpenter1992}. TTR record types model feature structures, with labels in
record types corresponding to attributes \citep[Ch.~1 \S1.4.3,
p.~25]{cooper2023perception}. The definitions below follow Carpenter's
presentation; the order symbol $\kgfsub$ is written with the more general element
on the left.

\begin{definition}[Typed feature structures]
\label{def:kg-tfs}
Let $\kgFeat$ be a finite set of features.
\begin{enumerate}[nosep]
\item A \emph{type hierarchy} is a finite bounded-complete partial order
  $\langle \kgTyp, \kgfsub\rangle$; $s \kgfsub t$ means that $s$ is more general
  than $t$.
\item An \emph{appropriateness specification} is a partial function
  $\kgApprop : \kgFeat \times \kgTyp \rightharpoonup \kgTyp$.
\item \emph{Feature introduction}: each feature $f$ has a most general type
  $\kgIntro(f)$ at which $\kgApprop(f, \kgIntro(f))$ is defined, and
  $\kgApprop(f, t)$ is defined iff $\kgIntro(f) \kgfsub t$.
  \emph{Upward closure}: if $\kgApprop(f, s)$ is defined and $s \kgfsub t$,
  then $\kgApprop(f, s) \kgfsub \kgApprop(f, t)$.
\item A \emph{feature structure} is a tuple $F = \langle Q, \bar q, \theta,
  \delta\rangle$ of a finite set $Q$ of nodes, a root $\bar q \in Q$, a typing
  $\theta : Q \to \kgTyp$, and a partial transition function
  $\delta : \kgFeat \times Q \rightharpoonup Q$, with every node reachable from
  $\bar q$. $\delta$ extends to paths.
\item $F$ is \emph{well-typed} if, whenever $\delta(f, q)$ is defined,
  $\kgApprop(f, \theta(q))$ is defined and
  $\kgApprop(f, \theta(q)) \kgfsub \theta(\delta(f, q))$. It is \emph{totally
  well-typed} if it is well-typed and $\kgApprop(f, \theta(q))$ defined implies
  $\delta(f, q)$ defined.
\item $F$ \emph{subsumes} $F'$, $F \kgfsub F'$, if there is $h : Q \to Q'$ with
  $h(\bar q) = \bar q'$, $\theta(q) \kgfsub \theta'(h(q))$, and
  $h(\delta(f, q)) = \delta'(f, h(q))$ whenever $\delta(f,q)$ is defined. The
  \emph{unification} $F \kgunify F'$ is the least upper bound of $F$ and $F'$
  under $\kgfsub$, when it exists.
\item $F$ is \emph{reentrant} at $q$ if $q = \delta(\pi, \bar q) =
  \delta(\pi', \bar q)$ for distinct paths $\pi \neq \pi'$.
\end{enumerate}
\end{definition}

\begin{table}[ht]
\centering\small
\caption{Typed feature structures, TTR, and the knowledge graph.}
\label{tab:kg-tfs}
\begin{tabularx}{\textwidth}{@{}>{\raggedright\arraybackslash}X
  >{\raggedright\arraybackslash}X >{\raggedright\arraybackslash}X@{}}
\toprule
Typed feature structures & TTR & Knowledge graph \\
\midrule
type hierarchy (bounded-complete partial order) & subtyping between named types &
  class hierarchy $\kganc$ (a preorder) \\
appropriateness $\kgApprop(f, t)$ & the record type of a named type &
  $\RC{C}$: required properties and their field types \\
feature introduction & width inheritance along $\sqsubseteq$ &
  requirement inheritance (Definition~\ref{def:kg-req}) \\
well-typed; totally well-typed & $r : T$ for record types &
  structural membership; totality: all required fields present \\
subsumption $F \kgfsub F'$, the more general structure on the left &
  subtyping $T' \sqsubseteq T$, the more specific type on the left &
  width subtyping \\
unification; failure & meet of record types; uninhabited meet &
  refinement (Definition~\ref{def:kg-refinement}) \\
reentrancy (structure sharing by tag) & dependent and manifest fields &
  reference: two fields hold the IRI of one resource \\
closed: no feature outside $\kgApprop$ & membership admits extra fields &
  open, bounded by declarations and domains \\
\bottomrule
\end{tabularx}
\end{table}

\paragraph{Naming direction.} Subsumption of feature structures and subtyping of
TTR types order the same pairs in opposite notation. A feature structure with
fewer constraints subsumes one with more: $F \kgfsub F'$. A record type with more
fields is the subtype: $T' \sqsubseteq T$. The translation below sends
$s \kgfsub t$ to $t \subty s$.

\paragraph{Differences.} TTR types are first-class and intensional
(Definition~\ref{def:kg-ttr-intensional}), and ptypes represent propositions whose
witnesses are evidence, whereas feature structures are purely data structures.
Moreover, feature-structure grammars are closed-world (well-typedness forbids any feature
outside $\kgApprop$), whereas knowledge graphs are open-world (membership admits any
declared property permitted by the domain bound). Finally, feature introduction in
Carpenter's sense allows depth refinement, since upward closure permits
$\kgApprop(f, t)$ to be strictly more specific than $\kgApprop(f, \kgIntro(f))$; by
contrast, class record types fix each property's field type globally and differ only in
width (Remark~\ref{rem:kg-no-depth}).

\paragraph{Reentrancy as reference.} In a feature structure, reentrancy is explicit
structure sharing: two paths reach the same node, indicated by a shared tag. In a
knowledge graph, reentrancy is expressed by reference: a property value can be the IRI of
a resource, so two fields---either within a single record or across different
records---can hold the same IRI. Because resource identity provides the sharing, the data
naturally form a graph rather than a tree. By contrast, values that are not referenced
resources (such as literals or embedded records without an IRI) are strictly tree-shaped,
with each embedded value occurring at a single path. Two such values at distinct paths
$\pi \neq \pi'$ can be provably \emph{equal}, $\Id(T, r.\pi, r.\pi')$, but not
\emph{shared}. To share an embedded value, one must either assign it an IRI and reference
it from both paths, or state the equality via a dependent manifest field
$\pi' : \sing{r.\pi}$, which is TTR's path alignment $T_{\pi'=\pi}$
\citep[App.~A11.8]{cooper2023perception}.

\paragraph{The correspondence.} The \emph{reference fragment} consists of those feature
structures in which every reentrant node carries an IRI. (In particular, any node on a
cycle is reachable by two distinct paths and is therefore reentrant and named.) Given a
type hierarchy equipped with appropriateness and feature introduction, each
$t \in \kgTyp$ induces a named type $t$ with record type
\[
  \RC{t} \;=\; \rtype{f : \kgApprop(f, t) \mid \kgApprop(f, t) \text{ defined}},
\]
and named subsumption $t \subty s$ whenever $s \kgfsub t$. A feature structure $F$ in the
reference fragment translates to a graph $\kgtr(F)$ as follows: each node $q$ carries the
record $\kgrcd{f = \kgtr(\delta(f, q))}_{\delta(f, q)\text{ defined}}$ and the declared
class set $\{\theta(q)\}$. A named node becomes a resource and appears as its IRI at
every path that reaches it, whereas an unnamed node is embedded at its unique path (and,
for checking membership, is treated as an anonymous resource with that record and
declared class set).

\begin{lemma}[Feature introduction is requirement inheritance]
\label{lem:kg-tfs-inheritance}
If $s \kgfsub t$, then $\RC{t} \subty \RC{s}$ by width and depth subsumption.
\end{lemma}

\begin{proof}
If $\kgApprop(f, s)$ is defined, then $\kgIntro(f) \kgfsub s \kgfsub t$, so
$\kgApprop(f, t)$ is defined: $\kglabels(\RC{s}) \subseteq \kglabels(\RC{t})$.
By upward closure, $\kgApprop(f, s) \kgfsub \kgApprop(f, t)$, so the field type
of $f$ in $\RC{t}$ is a subtype of that in $\RC{s}$. Rule~\textsc{Rec-Sub} of
Section~\ref{sec:kg-embedding} applies.
\end{proof}

\begin{proposition}[Membership correspondence]
\label{prop:kg-tfs-membership}
Let $F$ be in the reference fragment. $F$ is totally well-typed iff for every node
$q$: (i) $\kgtr(q)$ is a member of $\theta(q)$ in the sense of
Definition~\ref{def:kg-membership}, and (ii)
$\kglabels(\kgtr(q)) \subseteq \kglabels(\RC{\theta(q)})$.
\end{proposition}

\begin{proof}[Proof sketch]
Condition~(ii) expresses the first clause of well-typedness: every feature present on $q$
is appropriate for $\theta(q)$. Conversely, totality requires
$\kglabels(\RC{\theta(q)}) \subseteq \kglabels(\kgtr(q))$, which is the field-presence
half of structural membership. The field-typing half of structural membership requires
$\kgtr(\delta(f, q))$ to inhabit $\kgApprop(f, \theta(q))$: declared membership holds
because $\theta(\delta(f,q)) \subty \kgApprop(f, \theta(q))$ translates the second clause
of well-typedness, while structural membership of the child is condition~(i) at
$\delta(f,q)$. Because reentrant nodes carry IRIs, their membership is a single condition
on one resource shared across all incoming paths, whereas unnamed nodes occur at a unique
path. Finally, because membership is defined as a greatest fixed point
(Definition~\ref{def:kg-membership}), cycles through named nodes require no
well-foundedness assumption. Every step of the correspondence is reversible.
\end{proof}

Each node $q$ of a feature structure in the reference fragment also determines a
\emph{characteristic record type} $\kgchi(q)$: the record type
$\rtype{f : \kgchi(\delta(f,q))}$ at an unnamed node, and the manifest type
$\sing{i}$ at a node named $i$. Write $\kgchi(F) = \kgchi(\bar q)$.

\begin{conjecture}[Subsumption and unification]
\label{conj:kg-tfs-unification}
On the reference fragment, $F \kgfsub F'$ iff $\kgchi(F') \sqsubseteq \kgchi(F)$,
and $F \kgunify F'$ exists iff the merge $\mu(\kgchi(F) \ttrmeet \kgchi(F'))$ is
inhabited in some possibility; when it exists, $\kgchi(F \kgunify F')$ and
$\mu(\kgchi(F) \ttrmeet \kgchi(F'))$ have the same witnesses.
\end{conjecture}

The grammar of Volume~III uses the same correspondence: grammatical categories have
features that combine by meet, which is unification under
Conjecture~\ref{conj:kg-tfs-unification}.

\section{Embedding TTR in the dependent type theory}
\label{sec:kg-embedding}

TTR is set-theoretic; the type theory of Chapter~\ref{ch:mltt} is syntactic. The
embedding sends record types to labeled dependent telescopes, records to
inhabitants, the witness judgment to typing, ptypes to $\Prop$-valued
predicates, and TTR subtyping to two new rules of the subsumption judgment
$\Gamma \vdash A \subty B$ of Chapter~\ref{ch:mltt}. Labels are property IRIs.

\begin{definition}[Field sets and canonical order]
\label{def:kg-canonical}
A \emph{field set} is a finite set $\Phi = \{(\ell_1, A_1), \ldots, (\ell_n,
A_n)\}$ of pairs of distinct labels and type expressions in which labels occur as
variables. Let $\kgFL(A)$ be the set of labels free in $A$; require
$\kgFL(A_i) \subseteq \kglabels(\Phi) \setminus \{\ell_i\}$. The
\emph{dependency relation} is $\ell_j \prec \ell_i$ iff $\ell_j \in
\kgFL(A_i)$. Fix a total order $<$ on labels (for IRIs, lexicographic order). The
\emph{canonical order} $\kgcanon(\Phi)$ is the sequence obtained by repeatedly
selecting, among the fields not yet selected all of whose $\prec$-predecessors
are selected, the one with the $<$-least label.
\end{definition}

\begin{theorem}[Canonical order]
\label{thm:kg-canonical-order}
\begin{enumerate}[nosep]
\item $\kgcanon(\Phi)$ is defined iff $\prec$ is acyclic.
\item If defined, $\kgcanon(\Phi)$ is a telescope: the type at position $i$
  mentions only labels at positions $< i$.
\item $\kgcanon(\Phi)$ is a function of the set $\Phi$. Two record types with
  equal field sets have syntactically identical canonical telescopes, and are
  therefore definitionally equal by reflexivity.
\end{enumerate}
\end{theorem}

\begin{proof}
(1) The selection step is Kahn's topological-sort step. If $\prec$ is acyclic,
some unselected field has all predecessors selected at every stage, since
otherwise following unselected predecessors would produce an infinite descending
chain in a finite set, hence a cycle; the procedure therefore selects all $n$
fields. If $\prec$ has a cycle, no field of the cycle is ever selectable, since
each waits on its predecessor in the cycle. (2) A field is selected only after
all labels free in its type, which are its $\prec$-predecessors. (3) Each step is
determined by $\Phi$, the set of already selected labels, and the fixed order
$<$; by induction on the number of steps, the whole sequence is determined by
$\Phi$. Equal sets yield the same sequence, so the same syntactic telescope.
\end{proof}

A record type $\rtype{\ell_1 : A_1, \ldots, \ell_n : A_n}$ denotes the canonical
telescope of its field set. This realizes Definition~\ref{def:kg-ttr-sigma}:
fields are unordered at the level of the type, and the order of the underlying
telescope is determined by the field set.
Write $\kgtel(R)$ for the context $\ell_1 : A_1, \ldots, \ell_n : A_n$ in
canonical order, and $A_\ell$ for the type of $\ell$ in $R$.

\paragraph{Rules.} Record formation, introduction, projection, and equality are
the following. Introduction is width-inclusive, realizing
Definition~\ref{def:kg-ttr-record}: a record literal with fields indexed by $K$
inhabits every record type whose labels lie in $K$. A record type has a
\emph{level} $w \ge 1$, which is part of the type and which the notation leaves
implicit; in \textsc{Rec-I}, \textsc{Width}, \textsc{Depth}, and \textsc{Rec-Sub},
$w$ is the level of $R$ and $w'$ that of $R'$.
\begin{mathpar}
\inferrule*[right=Rec-F]
  {\Gamma, \ell_1 : A_1, \ldots, \ell_{i-1} : A_{i-1} \vdash A_i : \Srt\,u_i
   \quad (1 \le i \le n,\ \text{canonical order}) \\
   u_1, \ldots, u_n \le w}
  {\Gamma \vdash \rtype{\ell_1 : A_1, \ldots, \ell_n : A_n} : \Srt\,(w+1)}

\inferrule*[right=Rec-I]
  {\kglabels(R) \subseteq K \\
   \Gamma \vdash a_\ell : A_\ell[a_{\ell'}/\ell']_{\ell' \prec^{+} \ell}
     \quad (\ell \in \kglabels(R)) \\
   \Gamma \vdash a_k : B_k \quad \Gamma \vdash B_k : \Srt\,w_k \quad w_k \le w
     \quad (k \in K \setminus \kglabels(R))}
  {\Gamma \vdash \kgrcd{k = a_k}_{k \in K} : R}

\inferrule*[right=Rec-E]
  {\Gamma \vdash t : R \\ \ell \in \kglabels(R)}
  {\Gamma \vdash t.\ell : A_\ell[t.\ell'/\ell']_{\ell' \prec^{+} \ell}}

\inferrule*[right=Rec-$\beta$]
  {k \in K}
  {\kgrcd{k = a_k}_{k \in K}.k \tttstep a_k}
\end{mathpar}
\textsc{Rec-$\beta$} is a computation rule of the definitional equality of
Chapter~\ref{ch:mltt} (Definition~\ref{def:mltt-defeq}). By contrast, there is no
$\eta$-rule for records: two records that agree on the fields of $R$ but differ on
additional fields remain distinct terms at type $R$, matching the kernel's conversion
check, which compares canonical records syntactically. It is width-inclusive typing
(\textsc{Rec-I} and \textsc{Width}), rather than definitional equality, that allows both
records to inhabit $R$.

Notice also that a record type of level $w$ lives in $\Srt\,(w+1)$, one universe above
its field types: because its inhabitants may carry arbitrary extra fields whose types lie
in $\Srt\,w$ (\textsc{Rec-I}), the record type implicitly quantifies over $\Srt\,w$, just
as a $\Pi$-type over $\Srt\,w$ does. In particular, the record types of a knowledge graph
have level~1, since all of their field types live in $\Setu = \Srt\,1$
(Section~\ref{sec:kg-classes}).

\paragraph{The witness judgment and ptypes.} The TTR witness judgment $a : T$ is mapped to
the typing judgment $\Gamma \vdash a : T$. A predicate $P$ with arity
$\langle T_1, \ldots, T_n\rangle$ is declared as a constant
$P : T_1 \to \cdots \to T_n \to \Prop$; the ptype $P(a_1, \ldots, a_n)$ is then the
proposition $P\,a_1 \cdots a_n$, and its witnesses are its proofs
(Definition~\ref{def:kg-ttr-ptype}). Records whose fields are proofs---such as
$\rtype{x : \Entity, e : P\,x}$, where $\Entity$ is the type of resources introduced in
Section~\ref{sec:kg-classes}---realize TTR's record types with ptype fields.

\paragraph{A type system relative to a theory.} The basic types, predicates, and
classes available at a point are the constants declared in a theory in the sense
of Chapter~\ref{ch:institutions}. A TTR system of complex types corresponds to
one such theory; the model $\langle A, F\rangle$ corresponds to the closed terms
the theory admits at each type.

\paragraph{Subsumption.} Two rules extend the subsumption judgment of
Chapter~\ref{ch:mltt}. \textsc{Width} drops fields; \textsc{Depth} refines field
types; \textsc{Rec-Sub} combines them.
\begin{mathpar}
\inferrule*[right=Width]
  {\kglabels(R') \subseteq \kglabels(R) \\
   \Gamma, \kgtel(R) \vdash A_\ell \defeq A'_\ell \quad (\ell \in \kglabels(R')) \\
   w \le w'}
  {\Gamma \vdash R \subty R'}

\inferrule*[right=Depth]
  {\kglabels(R') = \kglabels(R) \\
   \Gamma, \kgtel(R) \vdash A_\ell \subty A'_\ell \quad (\ell \in \kglabels(R)) \\
   w \le w'}
  {\Gamma \vdash R \subty R'}

\inferrule*[right=Rec-Sub]
  {\kglabels(R') \subseteq \kglabels(R) \\
   \Gamma, \kgtel(R) \vdash A_\ell \subty A'_\ell \quad (\ell \in \kglabels(R')) \\
   w \le w'}
  {\Gamma \vdash R \subty R'}
\end{mathpar}
The premises are checked in the context $\kgtel(R)$, which binds every label that
a field type of $R'$ can mention, since $\kgFL(A'_\ell) \subseteq
\kglabels(R') \subseteq \kglabels(R)$. \textsc{Width} and \textsc{Depth} are the
instances of \textsc{Rec-Sub} with equal field types and with equal label sets.
The level premise is needed because a field of $R$ that $R'$ does not name is a
further field at $R'$, whose type must lie in $\Srt\,w'$. The empty record type
$\rtype{}$ of level $w'$ is a supertype of every record type of level at most
$w'$; it is TTR's $\mathit{Rec}$, not $\mathit{ERec}$.

\begin{theorem}[Soundness of width and depth: identity on terms]
\label{thm:kg-subsumption-sound}
Assume that subsumption is stable under substitution and that Chapter~\ref{ch:mltt}'s
subsumption rule at checking holds. Let $\Gamma \vdash R \subty R'$ by
\textsc{Rec-Sub}.
\begin{enumerate}[nosep]
\item If $\Gamma \vdash \kgrcd{k = a_k}_{k \in K} : R$ by \textsc{Rec-I}, then
  $\Gamma \vdash \kgrcd{k = a_k}_{k \in K} : R'$ by \textsc{Rec-I}, with the same
  term.
\item If $\Gamma \vdash t : R$ and $\ell \in \kglabels(R')$, the type that
  \textsc{Rec-E} assigns to $t.\ell$ at $R$ is a subtype of the type it assigns
  at $R'$.
\end{enumerate}
Hence admitting $t : R'$ from $t : R$ adds no term former: the coercion is the
identity, and every elimination at $R'$ is an elimination at $R$ followed by
subsumption at the field type.
\end{theorem}

\begin{proof}
Write $\sigma_a = [a_{\ell'}/\ell']$ and $\sigma_t = [t.\ell'/\ell']$.

(1) $\kglabels(R') \subseteq \kglabels(R) \subseteq K$. For $\ell \in
\kglabels(R')$, $\Gamma \vdash a_\ell : A_\ell\sigma_a$ by the premise at $R$, and
$\Gamma \vdash A_\ell\sigma_a \subty A'_\ell\sigma_a$ by stability of
$\Gamma, \kgtel(R) \vdash A_\ell \subty A'_\ell$ under the substitution
$\sigma_a$, which is well-typed because each $a_{\ell'}$ inhabits
$A_{\ell'}\sigma_a$ (induction along the canonical order of $R$). The subsumption
rule gives $\Gamma \vdash a_\ell : A'_\ell\sigma_a$. Labels in
$\kglabels(R) \setminus \kglabels(R')$ now fall under the third premise of
\textsc{Rec-I} at $R'$, with $B_k = A_k\sigma_a$, a type of $\Srt\,u_k$ with
$u_k \le w \le w'$ by \textsc{Rec-F} at $R$ and the level premise of
\textsc{Rec-Sub}. All premises of \textsc{Rec-I} at $R'$ hold, and the term is
unchanged.

(2) At $R$, $\Gamma \vdash t.\ell : A_\ell\sigma_t$; at $R'$ the type is
$A'_\ell\sigma_t$. By \textsc{Rec-E} at $R$, $\sigma_t$ is a well-typed
instance of $\kgtel(R)$, so stability under substitution gives
$\Gamma \vdash A_\ell\sigma_t \subty A'_\ell\sigma_t$.

\end{proof}

\begin{corollary}
\label{cor:kg-ttr-subtyping}
If $\Gamma \vdash R \subty R'$ by \textsc{Rec-Sub}, then $R \sqsubseteq R'$ in the
sense of Definition~\ref{def:kg-ttr-subtyping}, for the interpretation that
assigns to each type the closed terms the theory admits at it.
\end{corollary}

\begin{proof}
By Theorem~\ref{thm:kg-subsumption-sound}(1) for record literals and the
subsumption rule for all other terms, every closed term of $R$ is a closed term
of $R'$.
\end{proof}

\begin{definition}[Named type]
\label{def:kg-named-type}
A \emph{named type} is a pair $(N, R_N)$ of a constant $N$ declared in the theory
and a record type $R_N$. Its rules are formation, $\vdash N : \Setu$; and
projection: from $\Gamma \vdash x : N$ and $\ell \in \kglabels(R_N)$ derive
$\Gamma \vdash x.\ell : A_\ell[x.\ell'/\ell']$. $N$ is not definitionally equal
to $R_N$, and two named types $N \neq N'$ are distinct constants even when
$R_N = R_{N'}$.
\end{definition}

A class is a named type (Section~\ref{sec:kg-classes}). The distinctness of
$N$ and $N'$ with equal record types is TTR intensionality
(Definition~\ref{def:kg-ttr-intensional}): two types with the same witnesses in
every possibility are distinct types.

\section{Classes}
\label{sec:kg-classes}

A class $C$ denotes the named type $(C, \RC{C})$, where $\RC{C}$ is the record
type over the required properties of $C$ and of its ancestors.

\paragraph{Field types.} A property $p$ determines its field type $\kgty{p}$,
which is one of: a basic type (a literal datatype, possibly refined by facets
such as bounds, lengths, or a pattern); a class type $D$, whose values are
references to members of $D$; a list type $\List\,T$ of a field type $T$
(Definition~\ref{def:kg-ttr-function-list}); or a join of class types or of
singleton types (Section~\ref{sec:kg-joins}). Every field type is a type of
$\Setu$.

\begin{definition}[Declared subclassing and inherited requirements]
\label{def:kg-req}
$C \sqsubclass D$ (\emph{declared subclassing}) holds iff $D \in \kgPar(C)$, or
$C$ and $D$ are listed together in an \code{EquivalentClasses} declaration (which
contributes both $C \sqsubclass D$ and $D \sqsubclass C$). $\kganc$ is the
reflexive--transitive closure of $\sqsubclass$; it is a preorder. The
\emph{requirements} of $C$ are
\[
  \kgReq(C) \;=\; \bigcup \{\kgReqz(D) \mid C \kganc D\},
\]
and the \emph{record type of $C$} is $\RC{C} = \rtype{p : \kgty{p} \mid p \in
\kgReq(C)}$, of level~1, in canonical order. $C \kgeqv D$ (\emph{equivalence}) holds iff
$C \kganc D$ and $D \kganc C$.
\end{definition}

The rules for classes extend those for named types
(Definition~\ref{def:kg-named-type}) with the inclusion of every class in
$\Entity$ and with declared subclassing, along its reflexive--transitive closure
$\kganc$, as a base case of subsumption, as Chapter~\ref{ch:mltt} anticipates:
\begin{mathpar}
\inferrule*[right=Class-F]
  {C \in \kgClasses}
  {\vdash C : \Setu}

\inferrule*[right=Class-Ent]
  {C \in \kgClasses}
  {\vdash C \subty \Entity}

\inferrule*[right=Class-Sub]
  {C \kganc D}
  {\vdash C \subty D}

\inferrule*[right=Class-E]
  {\Gamma \vdash x : C \\ p \in \kgReq(C)}
  {\Gamma \vdash x.p : \kgty{p}}
\end{mathpar}
\textsc{Class-E} needs no substitution of earlier fields, because field types of
classes depend on no other field; conditional requirements, which do, are
treated in Section~\ref{sec:kg-joins}. \textsc{Class-Sub} takes the closure
$\kganc$, so that it is one rule in the declarative system and in
$\mathcal{K}_1$, whose subsumption has no rule of transitivity
(Section~\ref{sec:kg-model}). $\Entity$ is not a class.

\begin{theorem}[Coherence]
\label{thm:kg-coherence}
If $C \kganc D$, then $\vdash \RC{C} \subty \RC{D}$ by \textsc{Width}.
\end{theorem}

\begin{proof}
Let $p \in \kgReq(D)$, so $p \in \kgReqz(E)$ for some $E$ with $D \kganc E$. By
transitivity $C \kganc E$, so $p \in \kgReq(C)$. Hence $\kglabels(\RC{D})
\subseteq \kglabels(\RC{C})$. The field type of $p$ is $\kgty{p}$ in both record
types, so the second premise of \textsc{Width} holds by reflexivity of
$\defeq$, and both record types have level~1.
\end{proof}

Coherence is a theorem and is not imposed as a condition on declarations. A subclass
cannot omit a requirement of its parent, because requirements are computed by
inheritance; this is feature introduction (Lemma~\ref{lem:kg-tfs-inheritance})
with all features introduced at the class that requires them.

\begin{corollary}[Equivalent classes]
\label{cor:kg-equivalence}
If $C \kgeqv D$, then $\RC{C}$ and $\RC{D}$ are syntactically identical, and
$C$ and $D$ are distinct named types.
\end{corollary}

\begin{proof}
Theorem~\ref{thm:kg-coherence} in both directions gives $\kgReq(C) =
\kgReq(D)$, so the field sets are equal and Theorem~\ref{thm:kg-canonical-order}
gives identical telescopes. $C$ and $D$ are distinct constants
(Definition~\ref{def:kg-named-type}).
\end{proof}

Equivalence is mutual subsumption, not identity. Theorem~\ref{thm:kg-coherence}
applies to it directly: requirements are inherited along $\kganc$, hence along
both directions of $\kgeqv$. Equivalent classes have equal record types and are
distinct classes, as are two unrelated classes whose required properties
coincide; both cases are instances of Definition~\ref{def:kg-ttr-intensional}.

\begin{remark}[Recommended properties]
\label{rem:kg-recommends}
$\kgRecm(C)$ does not enter $\RC{C}$. A recommended property describes what may
be present; width-inclusive membership admits it already. Recommendations have
no typing consequence.
\end{remark}

\begin{remark}[No depth refinement per class]
\label{rem:kg-no-depth}
The field type of $p$ is $\kgty{p}$ in every class that requires $p$. If
$p \in \kglabels(\RC{C}) \cap \kglabels(\RC{D})$, the two field types are
identical, so class record types differ only in width. In feature-structure
terms, $\kgApprop(f, t) = \kgApprop(f, \kgIntro(f))$ for every $t$ with
$\kgIntro(f) \kgfsub t$: a feature's value type is fixed where the feature is
introduced and is not specialized by subtypes.
\end{remark}

\begin{remark}[Inductive types]
\label{rem:kg-inductive}
Inductive types ($\Nat$, $\List\,A$, and the families of Chapter~\ref{ch:mltt})
are types of a different kind: closed data generated by constructors, eliminated
by recursors. They are not record types, not subtypes of $\Entity$, and not
classes.
\end{remark}

\section{Resources and membership}
\label{sec:kg-membership}

A resource $r$ denotes the record $\kgrho{r}$ of all its properties, together
with its set $\kgS{r}$ of declared classes. Membership of $r$ in $C$ has two
conditions: a declaration and a structure.

Field values that are references make membership recursive: the value of a
class-typed field must itself be a member of that class. Graphs contain
reference cycles, so membership is defined as a greatest fixed point.

\begin{definition}[Value typing relative to a membership relation]
\label{def:kg-value-typing}
Let $\kgM \subseteq \kgRes \times \kgClasses$. The relation $v :_{\kgM} T$
between values and field types is: for a basic type, $v$ is a literal of that
type satisfying its facets; for a class type $D$, $v$ is an IRI with
$(v, D) \in \kgM$; for $\List\,T$, $v$ is a list each of whose elements has
$:_{\kgM} T$; for a join $T_1 \ttrjoin \cdots \ttrjoin T_n$, $v :_{\kgM} T_i$ for
some $i$; for a singleton $\sing{a}$, $v = a$; for $\Entity$, $v \in \kgRes$. For a record $\rho$
and a record
type $R$, $\rho :_{\kgM} R$ iff for every $\ell \in \kglabels(R)$,
$\ell \in \kglabels(\rho)$ and $\rho.\ell :_{\kgM} A_\ell$.
\end{definition}

\begin{definition}[Membership]
\label{def:kg-membership}
Define $\kgPhi$ on relations $\kgM \subseteq \kgRes \times \kgClasses$ by
\[
  \kgPhi(\kgM) \;=\; \{(r, C) \mid \exists D \in \kgS{r}.\; D \kganc C
  \;\text{ and }\; \kgrho{r} :_{\kgM} \RC{C}\}.
\]
$r$ is a \emph{member} of $C$, written $r : C$, iff $(r, C) \in \nu\kgPhi$, the
greatest fixed point of $\kgPhi$. The first conjunct is \emph{declared}
membership; the second is \emph{structural} membership, width-inclusive by
Definition~\ref{def:kg-ttr-record}.
\end{definition}

\begin{proposition}[Membership is well-defined and decidable]
\label{prop:kg-membership-decidable}
$\nu\kgPhi$ exists, and $r : C$ iff (declared) some $D \in \kgS{r}$ has
$D \kganc C$ and (structural) $\kgrho{r} :_{\nu\kgPhi} \RC{C}$. For a finite
graph, membership is decidable.
\end{proposition}

\begin{proof}
Class types occur positively in value typing: enlarging $\kgM$ enlarges
$:_{\kgM}$ at every field type, since lists, joins, and records are monotone in
their components and the other cases do not mention $\kgM$. Hence $\kgPhi$ is
monotone on the complete lattice of subsets of $\kgRes \times \kgClasses$, and
$\nu\kgPhi$ exists by the Knaster--Tarski theorem and is a fixed point, which is
the stated equivalence. The lattice is finite, so the descending sequence
$\kgRes \times \kgClasses \supseteq \kgPhi(\kgRes \times \kgClasses) \supseteq
\cdots$ stabilizes after at most $|\kgRes| \cdot |\kgClasses|$ steps at
$\nu\kgPhi$; each step evaluates finitely many decidable conditions.
\end{proof}

A least fixed point would reject every resource on a reference cycle whose
classes require the cycle's fields; the greatest fixed point accepts a cycle when
every resource on it meets its conditions.

\begin{lemma}[Named subsumption is sound]
\label{lem:kg-named-sound}
If $r : C$ and $C \kganc D$, then $r : D$.
\end{lemma}

\begin{proof}
Let $\kgM' = \nu\kgPhi \cup \{(r, D) \mid (r, C) \in \nu\kgPhi,\ C \kganc D\}$.
We show $\kgM' \subseteq \kgPhi(\kgM')$; then $\kgM' \subseteq \nu\kgPhi$ by
coinduction. Elements of $\nu\kgPhi$ lie in $\kgPhi(\nu\kgPhi) \subseteq
\kgPhi(\kgM')$ by monotonicity. For a new pair $(r, D)$ from $(r, C)$: declared,
some $E \in \kgS{r}$ has $E \kganc C \kganc D$; structural,
$\kgrho{r} :_{\nu\kgPhi} \RC{C}$, and $\kglabels(\RC{D}) \subseteq
\kglabels(\RC{C})$ with equal field types (Theorem~\ref{thm:kg-coherence}), so
$\kgrho{r} :_{\nu\kgPhi} \RC{D}$ and, by monotonicity,
$\kgrho{r} :_{\kgM'} \RC{D}$.
\end{proof}

Lemma~\ref{lem:kg-named-sound} justifies \textsc{Class-Sub}: a member of $C$
is a member of every ancestor of $C$, with the same IRI and the same record.

\begin{lemma}[Value typing respects subsumption]
\label{lem:kg-value-sub}
Let $\vdash T \subty T'$ be derived for field types by reflexivity,
transitivity, \textsc{Class-Sub}, \textsc{Class-Ent}, covariance of $\List$,
\textsc{Rec-Sub}, and the join and singleton rules of
Section~\ref{sec:kg-joins}, with the premise $a : T$ of \textsc{Sing-Sub} read as
$a :_{\nu\kgPhi} T$. If $v :_{\nu\kgPhi} T$, then $v :_{\nu\kgPhi} T'$.
\end{lemma}

\begin{proof}
Induction on the derivation. Reflexivity and transitivity are immediate.
\textsc{Class-Sub}: Lemma~\ref{lem:kg-named-sound}. \textsc{Class-Ent}:
$\nu\kgPhi \subseteq \kgRes \times \kgClasses$. $\List$: elementwise.
\textsc{Rec-Sub}: the labels of $R'$ are labels of $R$, and the induction
hypothesis applies to each field type. \textsc{Join-I}: by the clause for joins
in Definition~\ref{def:kg-value-typing}. \textsc{Join-L}: $v :_{\nu\kgPhi} T_i$
for some $i$, and the induction hypothesis applies to $T_i \subty U$.
\textsc{Sing-Sub}: $v = a$, the premise gives $a :_{\nu\kgPhi} T$, and the
induction hypothesis applies to $T \subty U$.
\end{proof}

\begin{definition}[Refinement]
\label{def:kg-refinement}
For a record type $R$ and a finite set $S$ of classes, the \emph{refinement}
$R \ttrmeet \bigwedge S$ is the meet (Definition~\ref{def:kg-ttr-meet-join}):
$r : R \ttrmeet \bigwedge S$ iff $\kgrho{r} :_{\nu\kgPhi} R$ and $r : D$ for
every $D \in S$. The \emph{own record type} of $r$ is
$R_r = \rtype{p : \kgty{p} \mid p \in \kglabels(\kgrho{r})}$, and the \emph{type
of $r$} is $R_r \ttrmeet \bigwedge \kgS{r}$.
\end{definition}

\begin{mathpar}
\inferrule*[right=Ref-Sub]
  {\Gamma \vdash R \subty R' \\
   \forall D' \in S'.\ \exists D \in S.\ D \kganc D'}
  {\Gamma \vdash R \ttrmeet \textstyle\bigwedge S \subty R' \ttrmeet \bigwedge S'}
\end{mathpar}

Here the first premise covers the structural record component, while the second premise
requires entailment of declared classes. Mere structural coverage of a target class
($R \subty \RC{D'}$) is not sufficient on its own, because membership in $D'$ also
requires a class declaration (Definition~\ref{def:kg-membership}). Refinements and the
proofs of $\isa$ (Section~\ref{sec:kg-propositions}) are first-class witnesses retained
by the store: two records refined by different proofs remain distinct terms, and
subsumption---not conversion---relates their types (Chapter~\ref{ch:mltt},
Section~\ref{sec:mltt-prop}).

\begin{theorem}[Soundness of refinement subsumption]
\label{thm:kg-refinement-sound}
If $r : R \ttrmeet \bigwedge S$ and the premises of \textsc{Ref-Sub} hold, then
$r : R' \ttrmeet \bigwedge S'$, with the same resource and record.
\end{theorem}

\begin{proof}
The record component follows from Lemma~\ref{lem:kg-value-sub} at the
record type $R$. For $D' \in S'$, pick $D \in S$ with $D \kganc D'$; then
$r : D$ by assumption and $r : D'$ by Lemma~\ref{lem:kg-named-sound}.
\end{proof}

\textsc{Ref-Sub} is a sound sufficient condition that defines the syntactic subsumption
judgment on refinements. It is deliberately incomplete with respect to the semantic
inclusion order of Definition~\ref{def:kg-ttr-subtyping}: an uninhabited refinement is
semantically a subtype of every type, whereas \textsc{Ref-Sub} does not derive that.

\begin{definition}[Admissible width; validity]
\label{def:kg-admissible}
The record of $r$ is \emph{admissible} iff every $p \in \kglabels(\kgrho{r})$ is a
declared property, $p \in \kgProps$, and $r : \kgdom(p)$ whenever $\kgdom(p)$ is
declared. A resource $r$ is \emph{valid} iff its record is admissible,
$\kgrho{r}.p :_{\nu\kgPhi} \kgty{p}$ for every $p \in \kglabels(\kgrho{r})$, and
$r : D$ for every $D \in \kgS{r}$. A graph is valid iff each of its resources is.
\end{definition}

Domain restrictions bound the admissible width of a record: membership remains
width-inclusive, but any extra field carried by a resource must be a declared property
whose domain the resource satisfies. A property without a declared domain may appear on
any resource, whereas an advisory domain---such as schema.org's \code{domainIncludes}
\citep{schemaorg}---is recorded as \code{recommends} on the domain class
(Remark~\ref{rem:kg-recommends}) and imposes no typing constraint. Notice that
Definition~\ref{def:kg-admissible} involves no circularity: admissibility refers to
membership $\nu\kgPhi$, which is defined independently of admissibility.

\begin{proposition}[Validity gives the type of a resource]
\label{prop:kg-resource-type}
If $r$ is valid, then $r : R_r \ttrmeet \bigwedge \kgS{r}$. If $C \in \kgS{r}$ and
$\kgrho{r} :_{\nu\kgPhi} \RC{C}$, then $r : \RC{C} \ttrmeet C$.
\end{proposition}

\begin{proof}
The first claim restates the second and third clauses of validity against
Definition~\ref{def:kg-refinement}. For the second, declared membership holds
with $D = C$, so $r : C$ by Proposition~\ref{prop:kg-membership-decidable}, and
the record component is the hypothesis.
\end{proof}

The second clause governs construction: a record built against $\RC{C}$ and declared in
$C$ inhabits the refinement $\RC{C} \ttrmeet C$.

\paragraph{Why both conditions are needed.} Structure alone cannot distinguish two
classes that happen to have identical required properties (for instance, two classes that
each require only a name). Conversely, a class declaration alone does not guarantee that
the required fields are actually present and well-typed on the resource.
Definition~\ref{def:kg-membership} therefore requires both conditions.

\section{Joins, singletons, and dependent fields}
\label{sec:kg-joins}

Union-typed and enumerated properties are joins
(Definition~\ref{def:kg-ttr-meet-join}): of class types for unions, of singleton
types for enumerations. Conditional requirements are dependent fields over such
joins.

\paragraph{Formers.} The join and the singleton are formed over types of a
universe above $\Prop$, so that their inhabitants are data. The formers and their
eliminators are constants of the signature (Definition~\ref{def:kg-theory}), at a
level parameter $u$:
\begin{align*}
  \kgjoin &: \Pi (T_1\,T_2 : \Srt\,(u+1)).\ \Srt\,(u+1), \\
  \kgsing &: \Pi (T : \Srt\,(u+1))\,(a : T).\ \Srt\,(u+1), \\
  \kgjcase &: \Pi (T_1\,T_2 : \Srt\,(u+1))\,(P : T_1 \ttrjoin T_2 \to \Prop)\,
    (t : T_1 \ttrjoin T_2). \\
  &\qquad (\Pi x{:}T_1.\,P\,x) \to (\Pi x{:}T_2.\,P\,x) \to P\,t, \\
  \kgsingE &: \Pi (T : \Srt\,(u+1))\,(a : T)\,(x : \sing{a}).\ \Id(\sing{a}, x, a).
\end{align*}
$T_1 \ttrjoin T_2$ is $\kgjoin\,T_1\,T_2$, $\sing{a}$ is $\kgsing\,T\,a$,
$\kgsingE(x)$ is $\kgsingE\,T\,a\,x$, and $\kgjcase(t;\, x.h_1, x.h_2)$ is
\[
  \kgjcase\,T_1\,T_2\,P\,t\,(\tttlam{x}{T_1} h_1)\,(\tttlam{x}{T_2} h_2).
\]
The theory gains no term former:
\textsc{Join-E} and \textsc{Sing-E} below are Rule~\tttrule{Const} followed by
applications, \textsc{Sing-I} is a typing rule, and the other rules are base
cases of subsumption. The joins are binary. For $n \ge 2$,
$T_1 \ttrjoin \cdots \ttrjoin T_n$ is the nested join
$T_1 \ttrjoin (T_2 \ttrjoin \cdots \ttrjoin T_n)$, and for $n = 1$ it is $T_1$.

\begin{definition}[Graph type]
\label{def:kg-graph-type}
A \emph{graph type} is a type definitionally equal to $\Entity$, a class, a record
type, a refinement, a join, a singleton, or a basic type, that is, to a type former
of this chapter applied to arguments.
\end{definition}

\paragraph{Rules.} Joins and singletons enter the type theory by the following
rules. \textsc{Join-L} and \textsc{Sing-Sub} lead into graph types, so that no join
or singleton lies below a universe, a $\Pi$-type, or a $\Sigma$-type, and above a
graph type lie only graph types.
\begin{mathpar}
\inferrule*[right=Join-I]
  {\Gamma \vdash T_1 : \Srt\,(u+1) \\ \Gamma \vdash T_2 : \Srt\,(u+1) \\
   i \in \{1, 2\}}
  {\Gamma \vdash T_i \subty T_1 \ttrjoin T_2}

\inferrule*[right=Join-L]
  {\Gamma \vdash T_1 : \Srt\,(u+1) \\ \Gamma \vdash T_2 : \Srt\,(u+1) \\\\
   \Gamma \vdash T_1 \subty U \\ \Gamma \vdash T_2 \subty U \\
   U \text{ a graph type}}
  {\Gamma \vdash T_1 \ttrjoin T_2 \subty U}

\inferrule*[right=Join-E]
  {\Gamma \vdash t : T_1 \ttrjoin T_2 \\
   \Gamma, x : T_1 \ttrjoin T_2 \vdash P\,x : \Prop \\\\
   \Gamma, x : T_i \vdash h_i : P\,x \quad (i = 1, 2)}
  {\Gamma \vdash \kgjcase(t;\, x.h_1, x.h_2) : P\,t}
\end{mathpar}
\begin{mathpar}
\inferrule*[right=Sing-I]
  {\Gamma \vdash T : \Srt\,(u+1) \\ \Gamma \vdash a : T}
  {\Gamma \vdash a : \sing{a}}

\inferrule*[right=Sing-Sub]
  {\Gamma \vdash T : \Srt\,(u+1) \\ \Gamma \vdash a : T \\
   \Gamma \vdash T \subty U \\ U \text{ a graph type}}
  {\Gamma \vdash \sing{a} \subty U}

\inferrule*[right=Sing-E]
  {\Gamma \vdash x : \sing{a}}
  {\Gamma \vdash \kgsingE(x) : \Id(\sing{a}, x, a)}
\end{mathpar}
The rules for $n$ disjuncts hold for nested joins: $T_i \subty T_1 \ttrjoin \cdots
\ttrjoin T_n$ by \textsc{Join-I} and transitivity, and \textsc{Join-L} and
\textsc{Join-E} by recursion on the nesting. A join is a graph type, so
\textsc{Join-Incl} is derivable from \textsc{Join-I}, \textsc{Join-L}, and
transitivity:
\begin{mathpar}
\inferrule*[right=Join-Incl]
  {\forall i.\ \exists j.\ \vdash T_i \subty T'_j}
  {\vdash \textstyle\bigvee_i T_i \subty \bigvee_j T'_j}
\end{mathpar}
At a graph type $T$, \textsc{Sing-Sub} gives $\sing{a} \subty T$. \textsc{Sing-E}
states the identity at the singleton, the type of both $x$ and $a$; where
$\sing{a} \subty T$, transport of $\refl_x : \Id(T, x, x)$ along $\kgsingE(x)$
with $\Jelim$ and motive $\lambda z{:}\sing{a}.\, \Id(T, x, z)$ gives
$\Id(T, x, a)$.

The join $T_1 \ttrjoin T_2$ is untagged: its inhabitants are the inhabitants of $T_1$ and
$T_2$ directly, without injection constructors. Consequently, \textsc{Join-E} is
restricted to $\Prop$-valued motives: because an inhabitant of an untagged join carries no
constructor tag indicating which disjunct it came from, no $\iota$-reduction rule could
select a branch to compute data in $\Setu$ or $\Typeu_k$, whereas proofs of propositions
in $\Prop$ require no computation rule (Chapter~\ref{ch:mltt}).

\paragraph{Unions of classes.} A property with \code{class\_types}
$[C_1, \ldots, C_n]$ has field type $C_1 \ttrjoin \cdots \ttrjoin C_n$. Its
inhabitants are the resources that are members of some $C_i$
(Definition~\ref{def:kg-value-typing}). A resource that is a member of $C_1$ and
of $C_2$ is one inhabitant.

\paragraph{Enumerations.} A property with \code{allows\_only}
$[v_1, \ldots, v_n]$ has field type
\[
  \kgEn(V) \;=\; \sing{v_1} \ttrjoin \cdots \ttrjoin \sing{v_n}, \qquad
  V = \{v_1, \ldots, v_n\},
\]
where each singleton is formed over $\kgIRI$. This is the same former as for
unions, applied to values instead of classes.

\begin{proposition}[Enumerations as subset types]
\label{prop:kg-enum-subset}
\begin{enumerate}[nosep]
\item In TTR, $a : \kgEn(V)$ iff $a : \kgIRI$ and $a \in V$: $\kgEn(V)$ has the
  witnesses of the subset type $\{x : \kgIRI \mid x = v_1 \vee \cdots \vee x =
  v_n\}$, and these are the values themselves.
\item $\vdash \kgEn(V) \subty \kgIRI$, with identity coercion.
\item If $V \subseteq V'$, then $\vdash \kgEn(V) \subty \kgEn(V')$.
\end{enumerate}
\end{proposition}

\begin{proof}
(1) By the generalized join clause of Definition~\ref{def:kg-ttr-meet-join},
$a : \kgEn(V)$ iff $a : \sing{v_i}$ for some $i$, and by
Definition~\ref{def:kg-ttr-singleton} iff $a : \kgIRI$ and $a = v_i$. (2)
\textsc{Join-L} with premises \textsc{Sing-Sub}, both into the graph type
$\kgIRI$. (3) \textsc{Join-Incl}: each
$\sing{v_i}$ with $v_i \in V \subseteq V'$ is a disjunct of $\kgEn(V')$.
\end{proof}

A value that occurs in two enumerations is one value, of both types.

\paragraph{Distinctness.} The theory makes no unique-names assumption for
resources: two IRIs may denote one resource, and $\Id(\Entity, i, j)$ for distinct
IRIs $i, j$ is consistent and may be postulated. Distinctness of enumeration
values therefore requires an assumption about the type of the values, not about
what they denote.

\begin{definition}[Tags]
\label{def:kg-tags}
$\kgIRI$ is a basic type whose closed canonical terms are IRI literals. It carries
a constant $\kgeqb : \kgIRI \to \kgIRI \to \kgBool$ with the computation rules
$\kgeqb\,v\,w \defeq \kgtrue$ if $v$ and $w$ are the same literal and
$\kgeqb\,v\,w \defeq \kgfalse$ otherwise. Enumeration values are \emph{tags}:
elements of $\kgIRI$, compared by $\kgeqb$.
\end{definition}

\begin{lemma}[Distinct tags are distinct]
\label{lem:kg-distinct}
For distinct IRI literals $v, w$, $\vdash \neg\,\Id(\kgIRI, v, w)$.
\end{lemma}

\begin{proof}
Assume $e : \Id(\kgIRI, v, w)$. We transport
$\refl : \Id(\kgBool, \kgeqb\,v\,v, \kgtrue)$, well-typed since
$\kgeqb\,v\,v \defeq \kgtrue$, along $e$ with $\Jelim$ and motive
$\lambda z{:}\kgIRI.\, \Id(\kgBool, \kgeqb\,v\,z, \kgtrue)$, obtaining
$\Id(\kgBool, \kgeqb\,v\,w, \kgtrue) \defeq \Id(\kgBool, \kgfalse, \kgtrue)$.
Transport along this identity with the motive
$\lambda b{:}\kgBool.\, \mathsf{if}\ b\ \mathsf{then}\ \Empty\ \mathsf{else}\ \Unit$,
defined by large elimination of the inductive type $\kgBool$
(Chapter~\ref{ch:mltt}), sends the inhabitant of the motive at $\kgfalse$,
which is $\Unit$, to an inhabitant of the motive at $\kgtrue$, which is
$\Empty$.
\end{proof}

\begin{remark}[Identifiers and referents]
\label{rem:kg-no-entity-iri}
An IRI literal $i \in \kgRes$ can be viewed both as a syntactic identifier of type
$\kgIRI$ and as a resource of type $\Entity$. Whereas $\Id(\kgIRI, i, j)$ expresses
syntactic identity of identifiers and is decided by $\kgeqb$, $\Id(\Entity, i, j)$
expresses extensional identity of the underlying referents and is not decided by syntax.
Accordingly, the theory includes no subsumption rule $\Entity \subty \kgIRI$: if such a
rule were present, $\Jelim$ would transport a consistent aliasing postulate
$\Id(\Entity, i, j)$ between distinct IRIs into $\Id(\kgIRI, i, j)$, contradicting
Lemma~\ref{lem:kg-distinct}. A property that ranges over resources---where aliasing must
remain possible---therefore takes as its field type a join of singletons formed over
$\Entity$ (or a join of classes), rather than an enumeration formed over $\kgIRI$.
\end{remark}

\begin{theorem}[Exhaustiveness]
\label{thm:kg-exhaustive}
$x : \kgEn(V) \vdash \Id(\kgIRI, x, v_1) \vee \cdots \vee \Id(\kgIRI, x, v_n)$.
\end{theorem}

\begin{proof}
The motive $P\,x = \bigvee_i \Id(\kgIRI, x, v_i)$ is a proposition. For each
$i$, in context $x : \sing{v_i}$, \textsc{Sing-E} gives $\kgsingE(x) :
\Id(\sing{v_i}, x, v_i)$, which $\Jelim$ transports to $\Id(\kgIRI, x, v_i)$ since
$\sing{v_i} \subty \kgIRI$, and the $i$-th disjunction introduction gives
$h_i : P\,x$. \textsc{Join-E}, for the nested join, applies.
\end{proof}

Elimination over an enumeration is by equality tests with $\kgeqb$, completed by
Theorem~\ref{thm:kg-exhaustive}: the tests cover every inhabitant.

\paragraph{The labeled sum.} The alternative representation of an enumeration is
the inductive type $E_V$ with nullary constructors $c_1, \ldots, c_n$
(Chapter~\ref{ch:mltt}). Define
\begin{align*}
  f &: \kgIRI \to E_V, &
  f\,x &= \mathsf{if}\ \kgeqb\,x\,v_1\ \mathsf{then}\ c_1\ \mathsf{else}\ \cdots
         \ \mathsf{if}\ \kgeqb\,x\,v_{n-1}\ \mathsf{then}\ c_{n-1}\
         \mathsf{else}\ c_n, \\
  g &: E_V \to \kgEn(V), &
  g\,c_i &= v_i,
\end{align*}
where $f$ applies to $x : \kgEn(V)$ by subsumption, since $\kgEn(V) \subty
\kgIRI$ (Proposition~\ref{prop:kg-enum-subset}), and $g$ is defined by the recursor of
$E_V$ with $v_i : \kgEn(V)$ by \textsc{Sing-I} and \textsc{Join-I}.

\begin{theorem}[Join and labeled sum]
\label{thm:kg-join-sum}
Let $v_1, \ldots, v_n$ be distinct IRI literals. Then
$c : E_V \vdash \Id(E_V, f(g\,c), c)$ and
$x : \kgEn(V) \vdash \Id(\kgIRI, g(f\,x), x)$. $\kgEn(V)$ and $E_V$ are
isomorphic as value sets.
\end{theorem}

\begin{proof}
First equation: by induction on $c$. For $c = c_i$, $g\,c_i \defeq v_i$. By
distinctness of the literals, $\kgeqb\,v_i\,v_j \defeq \kgfalse$ for $j < i$, and
$\kgeqb\,v_i\,v_i \defeq \kgtrue$ (for $i = n$, the final branch is taken). So
$f(g\,c_i) \defeq c_i$ and $\refl$ proves the case.

Second equation: the goal is a proposition, so the disjunction of
Theorem~\ref{thm:kg-exhaustive} may be eliminated. In the case
$e : \Id(\kgIRI, x, v_i)$, $g(f\,v_i) \defeq g\,c_i \defeq v_i$ by the
computation just shown, so $\refl : \Id(\kgIRI, g(f\,v_i), v_i)$; transport along
the inverse of $e$ with motive $\lambda z{:}\kgIRI.\, \Id(\kgIRI, g(f\,z), z)$ gives
$\Id(\kgIRI, g(f\,x), x)$.
\end{proof}

The two types have isomorphic sets of values and are distinct types;
Table~\ref{tab:kg-join-sum} lists their differences.

\begin{table}[ht]
\centering\small
\caption{Join of singletons and labeled sum, for an enumeration $V$.}
\label{tab:kg-join-sum}
\begin{tabularx}{\textwidth}{@{}l>{\raggedright\arraybackslash}X
  >{\raggedright\arraybackslash}X@{}}
\toprule
 & join of singletons $\kgEn(V)$ & labeled sum $E_V$ (inductive) \\
\midrule
values & the IRIs themselves & new constructors $c_i$ \\
relation to $\kgIRI$ & subtype, identity coercion & separate type; $g$ computes \\
$V \subseteq V'$ & $\kgEn(V) \subty \kgEn(V')$ & $E_V$, $E_{V'}$ unrelated \\
distinctness & Lemma~\ref{lem:kg-distinct}, from $\kgeqb$ & built in (no
  confusion) \\
elimination & $\kgeqb$ tests and Theorem~\ref{thm:kg-exhaustive} & recursor,
  \code{match} \\
\bottomrule
\end{tabularx}
\end{table}

Treating $E_V$ as a subtype of $\kgIRI$ requires the computing coercion $g$,
that is, coercive subtyping proper \citep{chatzikyriakidis-luo-2020}. The join lies in
the identity-coercion fragment that the rest of the theory uses
(Chapter~\ref{ch:mltt}).

\paragraph{Conditional requirements.} A conditional requirement states: if
property $k$ holds a value among $v_1, \ldots, v_n$, then the properties in $Q$
are required.

\begin{definition}[Conditional requirement]
\label{def:kg-conditional}
A \emph{conditional requirement} on a class $C$ is a tuple $\kappa = (k, V, Q,
Q_{\mathrm{rec}})$ of a property $k$, a finite set $V$ of tags, a set
$Q \subseteq \kgProps$ of conditionally required properties, and a set
$Q_{\mathrm{rec}}$ of conditionally recommended ones. It determines the
dependent family $\Phi_\kappa : \kgty{k} \to \kgRecType$,
\[
  \Phi_\kappa(x) \;=\;
  \begin{cases}
    \rtype{q : \kgty{q} \mid q \in Q} & \text{if } \kgeqb\,x\,v \defeq \kgtrue
      \text{ for some } v \in V,\\
    \rtype{} & \text{otherwise.}
  \end{cases}
\]
Conditional requirements are inherited along $\kganc$ as required properties are.
Structural membership of $r$ in $C$ (Definition~\ref{def:kg-membership})
additionally requires $\kgrho{r} :_{\kgM} \Phi_\kappa(\kgrho{r}.k)$ for each
inherited $\kappa$ with $k \in \kglabels(\kgrho{r})$.
\end{definition}

In TTR terms, each $q \in Q$ is a dependent field whose type is a function of
the value at path $k$ (Definition~\ref{def:kg-ttr-dependent}); $\{k\} \cup Q$ is
a dependency family; and the case distinction is decided over the join of
singletons $\kgEn(V)$ (Definitions~\ref{def:kg-ttr-singleton}
and~\ref{def:kg-ttr-meet-join}). $Q_{\mathrm{rec}}$ has no typing consequence
(Remark~\ref{rem:kg-recommends}). The proofs of
Lemma~\ref{lem:kg-named-sound} and Theorem~\ref{thm:kg-coherence} extend to conditional
requirements:
the inherited conditional requirements of an ancestor are among those of the
descendant, and the added conjunct is monotone in $\kgM$.

\section{Membership as a proposition}
\label{sec:kg-propositions}

Membership $r : C$ (Definition~\ref{def:kg-membership}) is a typing judgment. In the
signature of a knowledge graph (Definition~\ref{def:kg-theory}), each resource $r$ is
declared as a constant of its own class $\kgown{r}$, which is declared as a subclass of
every class of which $r$ is a member: $\kgown{r} \sqsubclass C$ for all
$(r, C) \in \nu\kgPhi$. The membership rule \textsc{Mem} is then derivable from
$r : \kgown{r}$ by \textsc{Class-Sub} and subsumption:
\begin{mathpar}
\inferrule*[right=Mem]
  {(r, C) \in \nu\kgPhi}
  {\vdash r : C}
\end{mathpar}
Conversely, for any class $C$ of the graph, $\kgown{r} \kganc C$ holds only if
$(r, C) \in \nu\kgPhi$, because membership is upward closed along $\kganc$
(Lemma~\ref{lem:kg-named-sound}).

Because a typing judgment cannot be hypothesized, negated, or quantified over inside the
object language, we also introduce two $\Prop$-valued predicates over resources so that
membership and subclassing can be stated as propositions.

\begin{definition}[The type of classes]
\label{def:kg-class-type}
$\Class$ is the type of class types, $\Class := \{C : \Setu \mid C \subty
\Entity\}$, given by the rules
\begin{mathpar}
\inferrule*{ }{\vdash \Class : \Typeu_1}

\inferrule*{C \in \kgClasses}{\vdash C : \Class}

\inferrule*{\Gamma \vdash X : \Class}{\Gamma \vdash X : \Setu}

\inferrule*{\Gamma \vdash X : \Class}{\Gamma \vdash X \subty \Entity}
\end{mathpar}
Basic types and inductive types are not classes.
\end{definition}

\begin{definition}[$\isa$ and $\subclassof$]
\label{def:kg-isa}
For $\isa : \Entity \to \Class \to \Prop$ and $\subclassof : \Class \to \Class \to \Prop$,
\[
  \isa(x, C) := \exists y{:}C.\; \Id(\Entity, x, y), \qquad
  \subclassof(C, D) := \Pi x{:}C.\; \isa(x, D).
\]
\end{definition}

$\Id(\Entity, x, y)$ is well-typed because $y : C \subty \Entity$. Both
definitions are $\Prop$-valued: $\exists$ is the existential of $\Prop$, and
$\Pi$ into $\Prop$ is in $\Prop$ by impredicativity. The identity type compares
inhabitants of classes, which are entities, so it is always taken at $\Entity$. A
class argument is the type $C$, an inhabitant of $\Class$, and not the resource
$\kgresof{C}$ that the IRI $C$ names (Definition~\ref{def:kg-theory}). Read as
predicates with arities $\langle \Entity, \Class\rangle$
and $\langle \Class, \Class\rangle$, $\isa$ and $\subclassof$ form ptypes in the
sense of Definition~\ref{def:kg-ttr-ptype}, whose witnesses are proofs.

\begin{theorem}[Declaration implies proposition]
\label{thm:kg-decl-prop}
\begin{enumerate}[nosep]
\item If $r : C$ (Definition~\ref{def:kg-membership}), then
  $\vdash \kgrcd{r, \refl_r} : \isa(r, C)$.
\item If $C \kganc D$, then $\vdash \lambda x{:}C.\, \kgrcd{x, \refl_x} :
  \subclassof(C, D)$.
\end{enumerate}
\end{theorem}

\begin{proof}
(1) \textsc{Mem} gives $\vdash r : C$, and $r : \Entity$ by \textsc{Class-Ent}.
Then $\refl_r : \Id(\Entity, r, r)$, and $\exists$-introduction with witness
$y = r$ gives the claim. (2) By \textsc{Class-Sub},
$\vdash C \subty D$. For $x : C$, subsumption gives $x : D$, and
$\kgrcd{x, \refl_x} : \isa(x, D)$ as in (1). $\lambda$-abstraction over $x : C$
gives the claim. The proof uses only typing; Lemma~\ref{lem:kg-named-sound}
ensures that \textsc{Class-Sub} is sound for membership.
\end{proof}

\begin{theorem}[Inheritance, reflexivity, transitivity]
\label{thm:kg-isa-laws}
For $x : \Entity$ and $C, D, E : \Class$:
\begin{enumerate}[nosep]
\item $\isa(x, C) \to \subclassof(C, D) \to \isa(x, D)$;
\item $\subclassof(C, C)$;
\item $\subclassof(C, D) \to \subclassof(D, E) \to \subclassof(C, E)$.
\end{enumerate}
\end{theorem}

\begin{proof}
(1) Let $u : \isa(x, C)$ and $s : \subclassof(C, D)$. The goal is a proposition,
so $u$ may be eliminated: $y : C$ and $e : \Id(\Entity, x, y)$. Then
$s\,y : \isa(y, D)$; eliminating it gives $z : D$ and $e' : \Id(\Entity, y, z)$.
Transitivity of $\Id$, by $\Jelim$, gives $\Id(\Entity, x, z)$, and
$\kgrcd{z, e \cdot e'} : \isa(x, D)$. (2) $\lambda x{:}C.\, \kgrcd{x, \refl_x}$.
(3) Given $s : \subclassof(C, D)$ and $s' : \subclassof(D, E)$, abstract over
$x : C$ and apply the proof of (1) to $s\,x : \isa(x, D)$ and $s'$.
\end{proof}

\begin{proposition}[One way]
\label{prop:kg-one-way}
The converse of Theorem~\ref{thm:kg-decl-prop}(1) does not hold: there are
theories in which $\isa(r, C)$ is provable and $r : C$ is not derivable.
\end{proposition}

\begin{proof}
Suppose $s : C$ holds by \textsc{Mem}, no $D \in \kgS{r}$ satisfies $D \kganc C$, and
the theory includes an identity axiom $a : \Id(\Entity, r, s)$. Then
$\kgrcd{s, a} : \isa(r, C)$ is derivable. On the other hand, declared membership of $r$
in $C$ fails, so $(r, C) \notin \nu\kgPhi$ and $\kgown{r} \kganc C$ does not hold.
Because the constant $r$ can be assigned type $C$ only via Rule~\tttrule{Const} at its
declared type $\kgown{r}$ followed by a subsumption $\kgown{r} \subty C$---which between
classes reduces to $\kgown{r} \kganc C$---the typing judgment $r : C$ is not derivable.
(Postulating an axiom $a' : \isa(r, C)$ directly yields the same separation.)
\end{proof}

Whereas class declarations are intensional, the propositions $\isa$ and $\subclassof$ are
extensional: $\isa(x, C)$ holds of any entity provably equal to a member of $C$. A
membership proposition that holds without a corresponding declaration is a substantive
claim; Chapter~\ref{ch:justification-logic} treats the justification terms that ground
such claims.

\begin{table}[ht]
\centering\small
\caption{Three levels of membership.}
\label{tab:kg-levels}
\begin{tabularx}{\textwidth}{@{}l>{\raggedright\arraybackslash}X
  >{\raggedright\arraybackslash}X@{}}
\toprule
Level & Judgment or proposition & Character \\
\midrule
declared & $\exists D \in \kgS{r}.\ D \kganc C$ & intensional; decidable \\
structural & $\kgrho{r} :_{\nu\kgPhi} \RC{C}$ & extensional over fields;
  decidable (Proposition~\ref{prop:kg-membership-decidable}) \\
propositional & $\isa(r, C)$, $\subclassof(C, D)$ & a claim: can be hypothesized,
  proved, grounded \\
\bottomrule
\end{tabularx}
\end{table}

The membership judgment $r : C$ is the conjunction of the first two levels.
Theorem~\ref{thm:kg-decl-prop} maps it to the third;
Proposition~\ref{prop:kg-one-way} shows the map is not onto.

\paragraph{Domain: resources only.} Membership and subclassing are relations
among resources. They are not stated for basic or inductive types, such as the
integer datatype or $\Nat$; there the question ``is $n$ a natural number'' is the
typing judgment $n : \Nat$, not a claim about the graph, and $\Nat$ is not an
inhabitant of $\Class$.

\section{The ontology as a theory}
\label{sec:kg-theory}

An ontology is a theory in the sense of Chapter~\ref{ch:institutions}: a
signature, a set of axioms, and well-formedness conditions on both.

\begin{definition}[Theory of a graph]
\label{def:kg-theory}
The theory $\mathit{Th}(\kgG) = (\Sigma_\kgG, \mathit{Ax}_\kgG)$ of a knowledge
graph $\kgG$ has:
\begin{enumerate}[nosep]
\item \emph{signature} $\Sigma_\kgG$:
  \begin{itemize}[nosep]
  \item the type $\Entity : \Setu$, which is not a class;
  \item a class $C : \Setu$ with record type $\RC{C}$ for each $C \in \kgClasses$,
    with declared subclassing $\sqsubclass$ (Definition~\ref{def:kg-req});
  \item for each resource $r \in \kgRes$, its \emph{own class} $\kgown{r} : \Setu$,
    a class with no own requirements, declared below the classes of which $r$ is
    a member, $\kgown{r} \sqsubclass C$ for $(r, C) \in \nu\kgPhi$, and a
    constant $r : \kgown{r}$, written $\kgresof{r}$ where $r$ is also a class,
    apart from the type $r$;
  \item a label $p$ with field type $\kgty{p}$ for each $p \in \kgProps$;
  \item the formers and eliminators of joins and singletons
    (Section~\ref{sec:kg-joins}), the basic types, and the constants $\isa$,
    $\subclassof$ of Definition~\ref{def:kg-isa};
  \end{itemize}
\item \emph{axioms} $\mathit{Ax}_\kgG$: declarations $a : P$ with $P : \Prop$,
  such as hypothesized memberships $\isa(r, C)$ and identities
  $\Id(\Entity, r, s)$.
\end{enumerate}
\end{definition}

\paragraph{The class system.} The constants of $\Sigma_\kgG$ other than the
definitions $\isa$ and $\subclassof$ extend an environment of
Chapter~\ref{ch:mltt} (Definition~\ref{def:mltt-environment}) as one
declaration, the \emph{class system} of $\kgG$, which binds them together as an
inductive declaration binds a family with its constructors and recursor:
\begin{mathpar}
\inferrule*[right=\tttrule{E-KG}]
  {\tttEnv \tttwf \\
   \text{the names of } \Sigma_\kgG \text{ are distinct and not bound in } \tttEnv \\
   \text{the type of each resource is a class} \\
   \text{the field types are well formed (W2)}}
  {\tttEnv, \Sigma_\kgG \tttwf}
\end{mathpar}
\textsc{Class-F} is Rule~\tttrule{Const} at the declared type $\Setu$ of a class,
and a resource infers its own class by Rule~\tttrule{Const}; \textsc{Mem} and
$r : \Entity$ follow by \textsc{Class-Sub} and \textsc{Class-Ent}.

\begin{definition}[Well-formed ontology]
\label{def:kg-wellformed}
$\kgG$ is \emph{well-formed} iff:
\begin{enumerate}[label=(W\arabic*), nosep]
\item every property named in $\kgReqz(C)$, $\kgRecm(C)$, or a conditional
  requirement is declared, and every class named in $\kgPar(C)$, an
  \code{EquivalentClasses} declaration, or a \code{domain} is declared;
\item every field type is well-formed: its class types name declared classes,
  and the values of its singleton types are well-typed at their base type;
\item for a property with \code{class\_types} $[E]$ and \code{allows\_only} $V$,
  with field type $E \ttrmeet \kgEn(V)$, every $v \in V$ is a member of $E$;
\item for each conditional requirement $(k, V, Q, Q_{\mathrm{rec}})$, every
  $v \in V$ is a well-typed value of $\kgty{k}$;
\item every resource is valid (Definition~\ref{def:kg-admissible}).
\end{enumerate}
\end{definition}

The coherence conditions of Sections~\ref{sec:kg-classes}
and~\ref{sec:kg-joins} are not among the conditions: requirement inheritance
makes them theorems (Theorem~\ref{thm:kg-coherence},
Corollary~\ref{cor:kg-equivalence}, Lemma~\ref{lem:kg-named-sound}).
Well-formedness of a finite graph is decidable by
Proposition~\ref{prop:kg-membership-decidable}. The judgments of the theory
are relative to $\mathit{Th}(\kgG)$; this is the sense in which a TTR type
system is relative to a theory (Section~\ref{sec:kg-embedding}).

Volume~IV treats the change of theories over time by layers, branches, and merges.

\section{Consistency of the extended theory}
\label{sec:kg-model}

The rules of this chapter extend the fragment $\mathcal{K}_1$ of
Chapter~\ref{ch:mltt} (Definition~\ref{def:mltt-k0}) by the formers of
records, classes, refinements, joins, singletons, and the basic types. The
formers are constants of the class system (Section~\ref{sec:kg-theory}), applied
to their type arguments, which the notation of this chapter leaves implicit, and
the rules are base cases of subsumption and checking rules, so the term syntax
and the inference rules of Chapter~\ref{ch:mltt} stay as they are. The
model of Theorem~\ref{thm:mltt-k1-consistency} extends to them.

\paragraph{The rules in $\mathcal{K}_1$.} In the bidirectional system $\mathcal{K}_1$,
the added rules inspect types up to definitional equality $\defeq_K$, just as the rules
of Chapter~\ref{ch:mltt} do: for instance, \textsc{Class-Sub} relates any types
$A \defeq_K C$ and $B \defeq_K D$ with $C \kganc D$, and \textsc{Join-L} relates any type
$A \defeq_K T_1 \ttrjoin T_2$ to a graph type $U$, while checking the well-formedness of
the components ($T_i \chk \Srt\,(u+1)$ and $a \chk T$). Moreover, \textsc{Join-I} in
$\mathcal{K}_1$ builds transitivity into its premise---deriving
$A \subty_K T_1 \ttrjoin T_2$ from $A \subty_K T_i$, just as \textsc{Join-L} and
\textsc{Sing-Sub} do in both systems---so that full transitivity of $\subty_K$ remains
admissible (Lemma~\ref{lem:mltt-k1-trans}) by induction on the right-hand derivation.
Finally, both \textsc{Sing-I} ($a \chk C$ for $C \defeq_K \sing{a}$ from $a \chk T$) and
\textsc{Join-I} ($a \chk C$ for $C \defeq_K T_1 \ttrjoin T_2$ from $a \chk T_i$) are
included as checking rules in $\mathcal{K}_1$, because annotated reduction erases
type annotations in checking positions, leaving $\lambda$-abstractions and pairs checked
directly against joins.

\begin{theorem}[Model of a knowledge graph]\label{thm:kg-model}
Let $\kgG$ be a well-formed knowledge graph (Definition~\ref{def:kg-wellformed})
and let $\tttEnv$ be a well-formed environment extended by the signature
$\Sigma_\kgG$ of Definition~\ref{def:kg-theory}, with the rules of
Sections~\ref{sec:kg-embedding}--\ref{sec:kg-propositions} added to
$\mathcal{K}_1$. Under the hypotheses of
Theorem~\ref{thm:mltt-k1-consistency}, the model extends to the new formers
so that every added rule holds, every added subsumption rule denotes an
inclusion, and every constant of $\Sigma_\kgG$ denotes an element of its type,
as Definition~\ref{def:mltt-obligation} asks of an axiom. In particular no closed
term of the
theory $(\Sigma_\kgG, \emptyset)$ inhabits $\tttFalse$, and the same holds for
$(\Sigma_\kgG, \mathit{Ax}_\kgG)$ whenever every axiom in $\mathit{Ax}_\kgG$
denotes $\{\emptyset\}$.
\end{theorem}

\begin{proof}
Semantic types are added as follows.
\begin{itemize}[nosep]
\item $\kgIRI$ denotes the set of IRI literals, $\kgBool$ the set of its two
  constructors, and $\kgeqb$ the equality test on literals; the computation
  rules of Definition~\ref{def:kg-tags} hold because distinct literals are
  distinct sets. The other basic types denote the sets of their literals
  restricted by their facets.
\item $\Entity$ denotes $\kgRes$, and the constant $r$ for a resource, written
  $\kgresof{r}$ where $r$ is also a class, denotes $r$.
\item A record type $\rtype{\ell_1 : A_1, \ldots, \ell_n : A_n}$ of level $w$
  denotes the set of finite functions $m$ from labels to $V_{w-1}$ whose domain
  contains $\ell_1, \ldots, \ell_n$ and that satisfy
  $m(\ell_i) \in \mathrm{El}\,\llbracket A_i \rrbracket[\ell_j \mapsto
  m(\ell_j)]_{j < i}$ in canonical order. The labels form a set in $V_0$ and
  $V_{w-1} \in V_w$, so this set lies in $V_w$, which $\Srt\,(w+1)$ denotes. A
  record literal denotes the function $k \mapsto \llbracket a_k \rrbracket$, and
  $t.\ell$ denotes $\llbracket t \rrbracket(\ell)$. \textsc{Rec-F},
  \textsc{Rec-I}, \textsc{Rec-E}, and \textsc{Rec-$\beta$} hold by
  construction; \textsc{Rec-I} is width-inclusive because the domain of $m$ may
  exceed the labels of the type, and its further fields, of types of
  $\Srt\,w$, have values in $V_{w-1}$.
\item A class $C$ denotes the set of the resources $r$ with $\kgown{r} \kganc C$.
  For a class of the graph these are its members
  (Definition~\ref{def:kg-membership}), since $\kgown{r} \sqsubclass D$ for the
  classes $D$ of which $r$ is a member and membership is closed upward along
  $\kganc$ (Lemma~\ref{lem:kg-named-sound}); the own class $\kgown{r}$ denotes
  $\{r\}$. $x.p$ for $x : C$ denotes the value in the graph of the property $p$
  of the resource $\llbracket x \rrbracket$. That value lies in
  $\mathrm{El}\,\llbracket \kgty{p} \rrbracket$ because the graph is valid
  (W5) and value typing relative to $\nu\kgPhi$ agrees with membership in the
  model, by induction on the field type.
\item A refinement $R \ttrmeet \bigwedge S$ denotes
  $\{r \in \kgRes \mid \kgrho{r} \in \mathrm{El}\,\llbracket R \rrbracket
  \text{ and } r : D \text{ for all } D \in S\}$.
\item A join $T_1 \ttrjoin T_2$ denotes
  $\mathrm{El}\,\llbracket T_1 \rrbracket \cup \mathrm{El}\,\llbracket T_2 \rrbracket$
  and a singleton $\sing{a}$ denotes $\{\llbracket a \rrbracket\}$: $\kgjoin$ at
  level $u$ denotes the function $A, B \mapsto A \cup B$ on $V_u$ and $\kgsing$
  the function $A, a \mapsto \{a\}$, which lie in their types because a
  Grothendieck universe is closed under binary unions and singletons. The types
  of $\kgjcase$ and $\kgsingE$ are propositions, and both constants denote
  $\emptyset$. $\Class$ denotes the set of denotations of the classes of the
  graph, and $\isa$ and $\subclassof$ are definitions.
\end{itemize}
\emph{Subsumption.} \textsc{Width}, \textsc{Depth}, and \textsc{Rec-Sub}
denote inclusions when the field subsumptions do, since an element of the
subtype is a finite function whose values at the labels of the supertype lie
in the supertype's field sets and whose values lie in
$V_{w-1} \subseteq V_{w'-1}$, by the level premise; a field subsumption by the
$\Pi$-case of $\mathcal{K}_1$ is an inclusion by
Theorem~\ref{thm:mltt-k1-consistency}. \textsc{Class-Ent} is the inclusion of a
set of resources in $\kgRes$, \textsc{Class-Sub} is an inclusion since $\kganc$
is transitive, and \textsc{Mem} is \textsc{Class-Sub} from $\kgown{r}$;
\textsc{Ref-Sub} is Theorem~\ref{thm:kg-refinement-sound} read in the model.
\textsc{Join-I} and \textsc{Join-L} are the inclusions of a union,
\textsc{Sing-Sub} is
$\{\llbracket a \rrbracket\} \subseteq \mathrm{El}\,\llbracket T \rrbracket
\subseteq \mathrm{El}\,\llbracket U \rrbracket$, and \textsc{Sing-I} and
\textsc{Join-I} as checking rules are memberships. The rule
$X \subty \Entity$ for $X : \Class$ is \textsc{Class-Ent} at every class. No
subsumption changes a value, so an element lying in two disjuncts of a join has
one value.

\emph{Elimination.} The value $\emptyset$ of $\kgjcase$ lies in its type: for
values of the arguments, $\llbracket t \rrbracket$ lies in
$\mathrm{El}\,\llbracket T_1 \rrbracket$ or in
$\mathrm{El}\,\llbracket T_2 \rrbracket$, and $h_1$ or $h_2$ inhabits the
motive there, so $\llbracket P\,t \rrbracket = \{\emptyset\}$. The value
$\emptyset$ of $\kgsingE$ lies in its type because an element of
$\{\llbracket a \rrbracket\}$ equals $\llbracket a \rrbracket$, so
$\llbracket \Id(\sing{a}, x, a) \rrbracket = \{\emptyset\}$.

\emph{Obligation.} Each constant of $\Sigma_\kgG$ has an element of its type:
a resource in its own class, a class in $V_0$, a class of the graph also in the
denotation of $\Class$, a property label through its field type, the formers and
eliminators of joins and singletons as above, and the basic types in $V_0$.
An axiom $a : P$ with $\llbracket P \rrbracket = \{\emptyset\}$ denotes
$\emptyset$ and meets the obligation.

\emph{The extended system.} Theorem~\ref{thm:mltt-k1-consistency} therefore extends to
the enriched system, whose conversion is joinability under the computation rules of the
added formers: the metatheoretic and semantic inductions of Chapter~\ref{ch:mltt} gain
one case per added rule, with the sorts of the stratified model read off derivations of
the extended system and the new computation rules preserving the values of typed redexes.
Because all added subsumption rules have a graph type on the right, any subsumption into
a non-graph type---such as a universe, a $\Pi$-type, a $\Sigma$-type, or an identity
type---has the exact same inversions as in Chapter~\ref{ch:mltt}. Conversely, because a
type may lie below a join of which it is a component, lemmas that invert a subsumption
with a universe or a $\Pi$-type on the left proceed by induction on the subsumption
derivation. Similarly, a term checks against a type that is neither a join nor a
singleton only via the checking rules of Chapter~\ref{ch:mltt}, preserving the inversion
needed for subject reduction at $\beta$- and $\iota$-redexes. Sort agreement
(Lemma~\ref{lem:mltt-k1-sorts}) likewise extends by one case per rule, since every added
subsumption relates data types in universes above $\Prop$.
\end{proof}

Two remarks apply when the theory includes propositional axioms in $\mathit{Ax}_\kgG$:
\begin{itemize}[nosep]
\item \emph{Identity axioms.} An identity postulate $\Id(\Entity, r, s)$ between distinct
  resources $r \neq s$ denotes $\emptyset$ in the base model above, where distinct
  resources denote distinct IRIs. Such axioms are satisfied by modifying the model so
  that $\Entity$ denotes the quotient of $\kgRes$ by the equivalence relation generated by
  the postulated identities, and each class $C$ denotes the image of its members in that
  quotient.
\item \emph{Membership axioms.} A postulated membership $\isa(r, C)$ for a resource $r$
  that is not a declared and structural member of $C$ (Proposition~\ref{prop:kg-one-way})
  is satisfied in that same quotient model whenever the axioms identify $r$ with some
  member $s$ of $C$. By contrast, no model in which $r : C$ holds directly can validate
  such a resource unless $r$ actually carries every required property of $C$, as
  \textsc{Class-E} demands. Which propositional axioms a theory postulates is a choice of
  the theory rather than of the base model.
\end{itemize}
\section{Comparisons}
\label{sec:kg-comparisons}

Table~\ref{tab:kg-comparisons} states, for each related framework, what this
chapter shares with it and the deltas. The sources are: for TTR,
\citet{cooper2023perception}; for typed feature structures and HPSG,
\citet{carpenter1992} and
\citet{pollard-sag1994}; for Luo's program, common nouns as
types \citep{luo2012cnt}, coercive subtyping and modern-type-theoretic
semantics \citep{chatzikyriakidis-luo-2020}, and dependent record
types~\citep{luo2009records}; for the semantic web and
description logic, \citet{rdf}, \citet{owl}, and \citet{baader2003dlh}; for
SHACL, \citet{shacl}; for ontologies on dependent record
types, \citet{barlatier-dapoigny2012}.

\begin{table}[ht]
\centering\small
\caption{Related frameworks: shared constructions and deltas.}
\label{tab:kg-comparisons}
\begin{tabularx}{\textwidth}{@{}>{\raggedright\arraybackslash}p{0.2\textwidth}
  >{\raggedright\arraybackslash}X >{\raggedright\arraybackslash}X@{}}
\toprule
Framework & Shared & Deltas \\
\midrule
TTR & records, record types, singletons, meets,
  joins, dependent fields, intensional types (Section~\ref{sec:kg-ttr}) &
  embedded in a dependent type theory with $\Prop$; record types as canonical
  telescopes; width and depth as subsumption rules; membership as a greatest
  fixed point over a graph \\
Typed feature structures; HPSG & appropriateness,
  feature introduction, total well-typedness, unification as meet
  (Section~\ref{sec:kg-tfs}) & reversed naming of the order; open, not closed;
  no depth refinement per class; reentrancy only by reference \\
Luo: common nouns as types; coercive subtyping; dependent record types & classes as types;
subsumption between
  classes; $\Sigma$-like records & identity-coercion fragment only; no
  computing coercions \\
RDF, RDFS, OWL; description logic & $\sqsubclass$
  as concept inclusion; \code{EquivalentClasses} as concept equivalence; a
  required property $p$ as $C \sqsubseteq \exists p.\kgty{p}$; open world as
  width membership & \code{domain} rejects a field instead of inferring
  membership; membership requires a declaration; no property-level reasoning
  (Appendix~\ref{sec:kg-atomic-data}) \\
SHACL & required properties as minimum cardinality one; value
  types; validation that rejects & shapes are separate from classes; here the
  class is the shape \\
Barlatier--Dapoigny & ontologies on dependent record types & classes are named
  (intensional) types; membership has a declared component \\
\bottomrule
\end{tabularx}
\end{table}

\begin{subappendices}
\section{Mapping Atomic Data}
\label{sec:kg-atomic-data}

Atomic Data \citep{atomicdata} is RDF with IRI identity in which properties are
first-class, dereferenceable resources. Schemas are resources: a class and a
property are described by the same means as any other datum. There are no blank
nodes and no separation of vocabulary from data. The mapping of this appendix
sends an Atomic Data schema to a theory (Section~\ref{sec:kg-theory}) and a
resource valid under its classes to a record of their meet.

\begin{table}[ht]
\centering\small
\caption{Atomic Data, the data model of this chapter, and the theory.}
\label{tab:kg-atomic}
\begin{tabularx}{\textwidth}{@{}>{\raggedright\arraybackslash}X
  >{\raggedright\arraybackslash}p{0.26\textwidth}
  >{\raggedright\arraybackslash}p{0.27\textwidth}@{}}
\toprule
Atomic Data & Data model & Theory \\
\midrule
resource (subject URL) & resource (\code{urn:} IRI) & record
  (Section~\ref{sec:kg-membership}) \\
\code{Property} (requires \code{shortname}, \code{datatype},
  \code{description}; recommends \code{classtype}, \code{allows-only}, \ldots)
  & property declaration & a label with its field type \\
\code{datatype}: slug, atomicURL, URI, string, markdown, integer, float,
  boolean, date, timestamp, resourceArray, json, ydoc & \code{data\_type} (with
  \code{format}, \code{content\_type}) & basic types
  (Definition~\ref{def:kg-ttr-basic}) \\
\code{resourceArray} & list-valued property & list type
  (Definition~\ref{def:kg-ttr-function-list}) \\
\code{classtype}: one class, for atomicURL or resourceArray values &
  \code{class\_types}: one or several classes & a class type; several give a
  join (Section~\ref{sec:kg-joins}) \\
\code{allows-only}: a resourceArray of allowed values & \code{allows\_only} &
  join of singletons; values compared by IRI, as tags
  (Section~\ref{sec:kg-joins}) \\
\code{Class} (requires \code{shortname}, \code{description}; recommends
  \code{requires}, \code{recommends}) & class & named record type:
  \code{requires} gives fields, \code{recommends} is admitted by width
  (Section~\ref{sec:kg-classes}) \\
\code{isA}: several classes per resource & \code{is\_a} & declared classes
  $\kgS{r}$; the type is a meet (Section~\ref{sec:kg-membership}) \\
\code{parent}: navigational and authorization hierarchy, inverse of
  \code{children} & --- & not subclassing; no counterpart \\
Commit: changes one resource, signed by an Agent, ordered by
  \code{previousCommit} & layer: a commit of several resources & outside this chapter (Volume~IV) \\
\bottomrule
\end{tabularx}
\end{table}

The \code{allows-only} property is specified as turning a property into an enum
\citep{atomicdata}; the theory reads that enum as a join of singletons. The
Atomic Data documentation lists \code{description} as optional on
\code{Property}; the published \code{Property} class requires it.\footnote{Checked against the
specification and the published class
definitions on 2026-10-28. The specification is versioned by date.}

\paragraph{Shared with Atomic Data.} Requirements are attached to the class
(\code{requires}, \code{recommends}) in both. A resource may have several classes
in both.

\paragraph{Divergences.}
\begin{itemize}[nosep]
\item \emph{Naming.} Atomic Data uses camelCase in URLs (\code{isA},
  \code{allowsOnly}, \code{classtype}) and kebab-case in shortnames
  (\code{allows-only}). The data model uses snake\_case throughout
  (\code{is\_a}, \code{allows\_only}, \code{class\_types}, \code{data\_type},
  \code{short\_name}).
\item \emph{Identifiers.} \code{urn:} identifiers need not dereference; Atomic
  Data URLs dereference.
\item \emph{Syntax.} No \code{@context} and no prefixes; \code{short\_name} is a
  value, not a document key.
\item \emph{Typing.} Literal typing is split into \code{data\_type},
  \code{format}, and \code{content\_type}.
\item \emph{Class types.} \code{class\_types} admits several classes, a join;
  \code{classtype} names one.
\item \emph{Commits.} An Atomic Data commit changes one resource; a layer
  commits several.
\end{itemize}

\paragraph{Additions for RDFS and OWL.} Table~\ref{tab:kg-rdfs} places the
constructs added for mapping RDFS, OWL, and schema.org \citep{rdf, owl, schemaorg,
shacl}.

\begin{table}[ht]
\centering\small
\caption{Constructs added for RDFS, OWL, and schema.org.}
\label{tab:kg-rdfs}
\begin{tabularx}{\textwidth}{@{}>{\raggedright\arraybackslash}p{0.26\textwidth}
  >{\raggedright\arraybackslash}p{0.28\textwidth}
  >{\raggedright\arraybackslash}X@{}}
\toprule
RDFS / OWL / schema.org & Data model & Theory \\
\midrule
\code{rdfs:subClassOf} & \code{core:subclass\_of} (absent in Atomic Data) &
  declared subsumption between named types; requirement inheritance
  (Section~\ref{sec:kg-classes}) \\
\code{owl:equivalentClass} & \codebrk{core:EquivalentClasses}, an $n$-ary resource
  listing \code{core:classes} & mutual subsumption; equal record types, distinct
  classes (Corollary~\ref{cor:kg-equivalence}) \\
\code{rdfs:domain} (restrictive) & \code{core:domain} & bound on width
  (Definition~\ref{def:kg-admissible}) \\
schema.org \code{domainIncludes} (advisory) & \code{core:recommends} on each
  domain class & admitted by width; not in the type \\
\code{rdfs:range}, \code{rangeIncludes} over classes & \code{class\_types} &
  class type, or join of classes \\
\code{owl:oneOf}; schema.org enumeration members & \code{allows\_only}, with
  \code{class\_types} $= [E]$ for an enumeration class $E$ & join of singletons
  met with the class type \\
conditional constraints & \code{conditional\_requires} (\code{when\_property},
  \code{has\_value}, \code{then\_requires}, \code{then\_recommends}) & dependent
  fields over a join of singletons (Definition~\ref{def:kg-conditional}) \\
XSD facets; SHACL value constraints & \code{min\_value}, \code{max\_value},
  \code{min\_length}, \code{max\_length}, \code{pattern}, \code{format} &
  refinements of basic types \\
\bottomrule
\end{tabularx}
\end{table}

\paragraph{Not mapped.} The constructs \codebrk{rdfs:subPropertyOf},
\code{owl:inverseOf}, \codebrk{owl:equivalentProperty}, \code{supersededBy}, and
the Role pattern are recorded as data and given no semantics. The theory has no property-level
reasoning: properties are labels, and labels stand in no order and no inverse
relation.

\paragraph{Other additions.} Class membership is also a claim
(Section~\ref{sec:kg-propositions}); property values may be type-theoretic terms
(Chapter~\ref{ch:mltt}); and the open world is bounded, with \code{domain} as
the explicit bound (Definition~\ref{def:kg-admissible}). The Atomic Data
documentation does not state whether a resource may carry properties beyond
those of its classes.

\paragraph{The mapping.} An Atomic Data store $\mathcal{S}$ is a set of resources,
each with a subject URL, a set $\kgisA(r)$ of classes, and a map from property
URLs to values; its \code{Property} and \code{Class} resources form its schema.

\begin{definition}[Atomic validity]
\label{def:kg-atomic-valid}
A resource $r$ of $\mathcal{S}$ is \emph{Atomic-valid} iff:
\begin{enumerate}[label=(A\arabic*), nosep]
\item every property key of $r$ is the URL of a \code{Property} resource of
  $\mathcal{S}$;
\item for every $K \in \kgisA(r)$, every property in $\kgreqA(K)$ is present on
  $r$;
\item every value conforms to the \code{datatype} of its property;
\item if the property has \code{classtype} $K'$, every referenced resource $u$
  has $K' \in \kgisA(u)$;
\item if the property has \code{allows-only} $V$, every value lies in $V$.
\end{enumerate}
$\mathcal{S}$ is \emph{closed} iff every URL in an atomicURL or resourceArray
value is the subject of a resource of $\mathcal{S}$.
\end{definition}

\begin{definition}[The mapping $\kgAt$]
\label{def:kg-atomic-map}
$\kgAt(\mathcal{S})$ is the graph with: $\kgProps$ the \code{Property}
resources; $\kgClasses$ the \code{Class} resources; $\kgReqz(K) =
\kgreqA(K)$, $\kgRecm(K)$ the \code{recommends} of $K$, $\kgPar(K) =
\emptyset$; no domains; $\kgS{r} = \kgisA(r)$ and $\kgrho{r}$ the property map of
$r$. The field type $\kgty{p}$ is: the basic type of the \code{datatype} for
literal datatypes; $K'$ for atomicURL with \code{classtype} $K'$, else
$\Entity$; $\List$ of that type for resourceArray; met with $\kgEn(V)$, at the
element type for resourceArray, when \code{allows-only} $V$ is present.
\end{definition}

\begin{theorem}[Soundness of the Atomic Data mapping]
\label{thm:kg-atomic-sound}
Let $\mathcal{S}$ be closed and let every resource of $\mathcal{S}$ be
Atomic-valid. Then $\kgAt(\mathcal{S})$ is a valid graph, and for every resource
$r$ and every $K \in \kgisA(r)$, $r : K$; hence
$r : R_r \ttrmeet \bigwedge \kgisA(r)$.
\end{theorem}

\begin{proof}
Let $\kgM_0 = \{(r, K) \mid K \in \kgisA(r)\}$. We show $\kgM_0 \subseteq
\kgPhi(\kgM_0)$; coinduction then gives $\kgM_0 \subseteq \nu\kgPhi$. Let
$(r, K) \in \kgM_0$. Declared: $K \in \kgS{r}$ and $K \kganc K$. Structural:
$\kgPar = \emptyset$, so $\kgReq(K) = \kgreqA(K)$, and every $p \in \kgReq(K)$
is present by (A2). Its value $v$ satisfies $v :_{\kgM_0} \kgty{p}$, by cases on
$\kgty{p}$: a basic type by (A3); a class type $K'$ by (A4), since $K' \in
\kgisA(u)$ gives $(u, K') \in \kgM_0$; $\Entity$ by closedness; a list
elementwise; a meet with $\kgEn(V)$ additionally by (A5) and
Proposition~\ref{prop:kg-enum-subset}(1). Hence $r : K$. Validity
(Definition~\ref{def:kg-admissible}): admissibility by (A1), with no domains to
check; well-typed values for all present properties, required or not, by the
same case analysis; and $r : D$ for $D \in \kgS{r}$ as shown. The last claim is
Proposition~\ref{prop:kg-resource-type}.
\end{proof}

Atomic Data has no subclassing, so its schemas map into the fragment with
$\kgPar = \emptyset$ and $\kganc$ the identity; Atomic Data has no domains, so
the width bound is the declaration of properties alone.

\begin{conjecture}[Faithfulness on the shared fragment]
\label{conj:kg-atomic-faithful}
On closed stores, if $\kgAt(\mathcal{S})$ is valid then every resource of
$\mathcal{S}$ is Atomic-valid; and $\kgAt$ is injective on schemas up to the
renaming of vocabulary listed under Divergences.
\end{conjecture}
\end{subappendices}

\chapter{Deciding the Kernel Theory}
\label{ch:checking}

\providecommand{\tttEnv}{\mathcal{S}}
\providecommand{\tttlam}[2]{\lambda #1{:}#2.\,}
\providecommand{\tttpair}[2]{\langle #1, #2\rangle}
\providecommand{\tttann}[2]{(#1 : #2)}
\providecommand{\tttinst}[2]{#1.\{#2\}}
\providecommand{\tttstep}{\longrightarrow}
\providecommand{\tttFV}{\mathrm{FV}}
\providecommand{\tttsim}{\simeq}
\providecommand{\tttrule}[1]{\textsc{#1}}
\providecommand{\tttFalse}{\mathsf{False}}
\providecommand{\tttAcc}{\mathsf{Acc}}
\providecommand{\kganc}{\sqsubseteq^{*}}
\providecommand{\kgsing}{\mathsf{sing}}
\providecommand{\kglabels}{\mathrm{labels}}
\providecommand{\kgtel}{\mathrm{tel}}
\providecommand{\instax}[1]{\mathsf{a}_{#1}}
\providecommand{\chkwh}{\longrightarrow_{\mathsf{wh}}}     
\providecommand{\chkred}{\searrow}                         
\providecommand{\chkconv}{\Leftrightarrow_{\mathsf{a}}}    
\providecommand{\chkinf}{\Rightarrow_{\mathsf{a}}}         
\providecommand{\chkchk}{\Leftarrow_{\mathsf{a}}}          
\providecommand{\chksub}{\leq_{\mathsf{a}}}                
\providecommand{\chkspine}{\Rightarrow_{\mathsf{s}}}       
\providecommand{\chkSN}{\mathrm{SN}}
\providecommand{\chkeval}[2]{\llbracket #1 \rrbracket_{#2}} 
\providecommand{\chkelim}{\mathsf{elim}}
\providecommand{\chkrb}{\mathsf{R}}
\providecommand{\chknbe}{\mathsf{nbe}}
\providecommand{\chkq}[1]{\ulcorner #1 \urcorner}           
\providecommand{\chkv}[1]{\hat{#1}}                         

Chapter~\ref{ch:mltt} states the kernel theory declaratively and isolates two bidirectional
fragments, $\mathcal{K}_0$ and $\mathcal{K}_1$ (Definition~\ref{def:mltt-k0}), whose rules are
syntax-directed and read every premise $C \defeq \Pi x{:}A.\,B$ through a reduct of $C$. This
chapter gives the procedures that decide them: weak-head reduction
(Section~\ref{sec:checking-whnf}), a conversion test that compares weak-head normal forms
(Section~\ref{sec:checking-conversion}), normalization by evaluation
(Section~\ref{sec:checking-nbe}), and a type checker that reads the rules of $\mathcal{K}_0$ and
$\mathcal{K}_1$ through weak-head normal forms (Section~\ref{sec:checking-bidir}).
Section~\ref{sec:checking-inversion} establishes the inversion lemmas on which the type checker
rests, Section~\ref{sec:checking-kg} extends it with the rules of
Chapter~\ref{ch:knowledge-graphs}, and Section~\ref{sec:checking-properties} derives the
properties of the checker that Chapters~\ref{ch:justification-logic} and~\ref{ch:institutions}
assume.

Every procedure is sound, complete on every input on which it terminates, and terminating on the
inputs that the kernel presents to it if the erasures of typable terms are strongly normalizing
(Hypothesis~\ref{hyp:checking-normalization}). Soundness and completeness do not use this
hypothesis. Whereas the hypothesis is expected, it is not proved here;
Section~\ref{sec:checking-termination} records what is known about it and which results depend on
it.

\section{Procedures and their correctness}
\label{sec:checking-procedures}

\paragraph{Erasures.} Definitional equality ignores annotations, so the procedures compare
terms without them. Write $|t|$ for the erasure of $t$, which deletes every annotation
(Section~\ref{sec:mltt-consistency}), and $|\tttEnv|$ for the environment $\tttEnv$ with each
body replaced by its erasure. Every term $a$ reduces to its erasure by \tttrule{Ann-$\tttstep$},
so $a \defeq |a|$. Moreover, definitional equality over $\tttEnv$ coincides with definitional
equality over $|\tttEnv|$ by the argument of Lemma~\ref{lem:mltt-env-annotation}(2), which uses
only that the bodies of the two environments have the same erasures. Since reduction over
$|\tttEnv|$ keeps a term annotation-free, $a \defeq b$ holds if and only if $|a|$ and $|b|$ are
joinable by reduction over $|\tttEnv|$ through annotation-free terms. Throughout this chapter,
$\tttstep$ denotes that reduction, and every term is annotation-free except the terms that the
type checker of Section~\ref{sec:checking-bidir} checks, whose annotations fix the checking and
inference modes.

\paragraph{Strong normalization.} A term $t$ is \emph{strongly normalizing}, written
$t \in \chkSN$, if there is no infinite sequence of $\tttstep$-steps from $t$. Because a term has
finitely many redexes, $\tttstep$ is finitely branching, and for $t \in \chkSN$ the length
$\nu(t)$ of the longest reduction sequence from $t$ is finite by K\H{o}nig's lemma. Every subterm
and every reduct of a term in $\chkSN$ is again in $\chkSN$, with $\nu$ no larger, and
$\nu(t') < \nu(t)$ whenever $t \tttstep^{+} t'$.

\begin{hypothesis}[Normalization]\label{hyp:checking-normalization}
For the environment $\tttEnv$: if $\ctx \vdash a : A$ in the declarative system of
Chapter~\ref{ch:mltt} over $\tttEnv$, then $|a| \in \chkSN$.
\end{hypothesis}

Every derivation of $\mathcal{K}_1$ in a context whose entries are types erases to a declarative
derivation of the same judgment (Section~\ref{sec:mltt-pi-sigma}), so under the hypothesis the
erasure of every term that infers or checks in $\mathcal{K}_0$ or $\mathcal{K}_1$, in such a
context, is strongly normalizing.

\paragraph{Correctness.} A procedure \emph{decides} a relation on a set of inputs if it
terminates on each of them and accepts exactly the related ones. Because the procedures of this
chapter are recursive procedures that may fail to terminate on untyped inputs, their correctness
is established in three parts. Each procedure is shown \emph{sound} (what it accepts stands in the
relation), \emph{complete relative to termination} (every related input on which it terminates is
accepted), and \emph{terminating} on a specified set of inputs under
Hypothesis~\ref{hyp:checking-normalization}. The first two properties together ensure that the
procedure decides the relation on every input on which it terminates.

\section{Weak-head reduction}
\label{sec:checking-whnf}

The procedures reduce a term only until its outermost form is determined, and compare the parts
of that form recursively.

\begin{definition}[Spines and weak-head reduction]\label{def:checking-wh}
An \emph{elimination} is an argument $a$, a projection $\fst$ or $\snd$, or $\Jelim\,a\,C\,d\,b$;
applied to a term $t$, it gives $t\,a$, $\fst\,t$, $\snd\,t$, or $\Jelim\,a\,C\,d\,b\,t$. Every term
is uniquely of the form $h\,e_1 \cdots e_n$, written $h\,\bar e$, of a \emph{head} $h$, which is
none of these four forms, and a \emph{spine} $\bar e$ of eliminations; the \emph{prefixes} of
$h\,\bar e$ are the terms $h\,e_1 \cdots e_i$ for $i \le n$. A \emph{redex} is an instance of the
left-hand side of a computation rule of Definition~\ref{def:mltt-defeq} other than
\tttrule{Ann-$\tttstep$}, and its \emph{contractum} is the corresponding instance of the
right-hand side. \emph{Weak-head reduction} $\chkwh$ is the least relation such that
\begin{enumerate}
\item $s\,\bar e \chkwh s'\,\bar e$ for every spine $\bar e$, if $s$ is a redex with contractum
  $s'$;
\item $\tttinst{c}{\bar\ell}\,a_1 \cdots a_{n-1}\,a\,\bar e \chkwh
  \tttinst{c}{\bar\ell}\,a_1 \cdots a_{n-1}\,a'\,\bar e$, if $c$ is a recursor whose type has
  $n$ premises and $a \chkwh a'$.
\end{enumerate}
A term is in \emph{weak-head normal form} if no step of $\chkwh$ applies to it, and
$t \chkred w$ means that $t \chkwh^{*} w$ with $w$ in weak-head normal form.
\end{definition}

The last premise of a recursor is its major premise (Definition~\ref{def:mltt-recursor}), and the
major premise of $\Jelim$ is the term to which the elimination $\Jelim\,a\,C\,d\,b$ applies, so
clause~(2) and clause~(1) at an elimination by $\Jelim$ reduce the major premise of a stuck
eliminator.

\begin{lemma}[Weak-head reduction]\label{lem:checking-wh}
\leavevmode
\begin{enumerate}
\item $\chkwh$ is contained in $\tttstep$, and a term has at most one $\chkwh$-step. Hence
  $t \chkred w$ for at most one $w$, the \emph{weak-head normal form} of $t$.
\item A term $h\,\bar e$ is in weak-head normal form if and only if $h$ is not a transparent
  constant, no prefix of $h\,\bar e$ is a redex, and, if $h$ is a recursor with $n$ premises
  and $\bar e$ begins with $n$ arguments, the $n$-th of them is in weak-head normal form.
\end{enumerate}
\end{lemma}

\begin{proof}
(1) A step by clause~(1) contracts a redex at a prefix, whereas a step by clause~(2) reduces
inside an argument, so both are steps of $\tttstep$. At most one prefix of a term is a redex:
the prefix one shorter than a redex is an abstraction, a pair, $\refl\,a$, or a recursor applied
to fewer arguments than it has premises, and no shorter prefix of any of these is a redex, while
a transparent constant is a redex only as the head. Moreover, clauses~(1) and~(2) cannot apply at
the same prefix: if the prefix $\tttinst{c}{\bar\ell}\,a_1 \cdots a_n$ of a recursor with $n$
premises is a redex, then $a_n$ is a constructor applied to arguments, whose head is not
transparent and none of whose prefixes is a redex, so $a_n$ is in weak-head normal form and
clause~(2) does not apply. Determinacy follows by induction on the size of the term.

(2) This restates the negation of the two clauses of Definition~\ref{def:checking-wh}.
\end{proof}

The \emph{shape} of a weak-head normal form $h\,\bar e$ consists of the form of its head---the
variable, the constant with its level instance, the universe with its level, or one of the
formers $\Pi$, $\Sigma$, $\lambda$, pairing, $\Id$, and $\refl$---and the forms of its
eliminations: argument, $\fst$, $\snd$, or $\Jelim$. Its \emph{components} are the immediate
subterms of the head and the terms in the eliminations. Two shapes are \emph{equivalent} if they
are equal up to the replacement of level expressions by equivalent ones.

\begin{lemma}[Stability of weak-head normal forms]\label{lem:checking-whnf-stable}
If $w$ is in weak-head normal form and $w \tttstep^{*} w'$, then $w'$ is in weak-head normal form,
$w'$ has the shape of $w$, and each component of $w$ reduces to the corresponding component of
$w'$.
\end{lemma}

\begin{proof}
It suffices to treat a single step $w \tttstep w'$, by induction on the size of $w$. Because no
prefix of $w$ is a redex and its head is not a transparent constant, the step must contract a
redex inside a component of $w$; thus the shape is unchanged, and that component takes a step. It
remains to check that no redex arises at a prefix of $w'$, and that the condition of
Lemma~\ref{lem:checking-wh}(2) on the major premise of a recursor persists. First, the left-hand
sides of $\beta$, of the projections, and of \tttrule{Id-$\iota$} match a prefix whose next
shorter prefix is an abstraction, a pair, or $\refl\,a$; because the step inside a component does
not change the form of that shorter prefix, none of these rules matches at $w'$. Second, the
left-hand side of $\iota$ matches the prefix $\tttinst{c}{\bar\ell}\,a_1 \cdots a_n$ of a
recursor with $n$ premises only when $a_n$ is a constructor applied to all its arguments. In $w$,
the argument $a_n$ is in weak-head normal form and is not such a constructor application; by the
induction hypothesis, its reduct in $w'$ remains in weak-head normal form with the same shape, so
it is not such a constructor application either.
\end{proof}

\begin{corollary}[Joinable weak-head normal forms]\label{cor:checking-whnf-join}
If $w_1$ and $w_2$ are in weak-head normal form and $w_1 \defeq w_2$, then their shapes are
equivalent and their corresponding components are joinable. In particular, a weak-head normal
form joinable with a $\Pi$-type is a $\Pi$-type, and likewise for every former and every head.
\end{corollary}

\begin{proof}
By Lemma~\ref{lem:mltt-confluence}, $w_1 \tttstep^{*} u_1$ and $w_2 \tttstep^{*} u_2$ for terms
$u_1$ and $u_2$ equal up to the replacement of levels by equivalent ones. By
Lemma~\ref{lem:checking-whnf-stable}, each $u_i$ has the shape of $w_i$ and components that are
reducts of those of $w_i$; $u_1$ and $u_2$ have equivalent shapes and corresponding components
equal up to equivalent levels.
\end{proof}

\section{The conversion test}
\label{sec:checking-conversion}

\begin{definition}[Conversion test]\label{def:checking-conv}
$a \chkconv b$ is the least relation on annotation-free terms such that $a \chkconv b$
whenever $a \chkred w_1$, $b \chkred w_2$, the shapes of $w_1$ and $w_2$ are equivalent, and
each component of $w_1$ is related by $\chkconv$ to the corresponding component of $w_2$,
components under a binder with the bound variables identified. The \emph{conversion test} on
$a$ and $b$ computes $w_1$ and $w_2$, compares their shapes, deciding the equivalence of levels
by Proposition~\ref{prop:mltt-levels}, and runs the test on the pairs of components in order; it
accepts if every comparison accepts. On terms with annotations, the test is the test on their
erasures.
\end{definition}

\begin{theorem}[Conversion]\label{thm:checking-conv}
For terms $a$ and $b$:
\begin{enumerate}
\item (Soundness) if the test accepts $a$ and $b$, then $a \defeq b$;
\item (Completeness) if $a \defeq b$ and the test on $a$ and $b$ terminates, it accepts;
\item (Termination) if $|a|, |b| \in \chkSN$, the test on $a$ and $b$ terminates.
\end{enumerate}
Hence the test decides definitional equality on pairs of terms whose erasures are strongly
normalizing, and, under Hypothesis~\ref{hyp:checking-normalization}, on pairs of terms typable in
the declarative system.
\end{theorem}

\begin{proof}
Erasure does not change definitional equality, so let $a$ and $b$ be annotation-free.
(1) By induction on the derivation of $a \chkconv b$. The components are joinable by the
induction hypothesis, and $\defeq$ is closed under every term former and every elimination and
contains the equivalence of levels (Lemma~\ref{lem:mltt-confluence},
\tttrule{Sort-$\defeq$}), so $w_1 \defeq w_2$; and $a \tttstep^{*} w_1$, $b \tttstep^{*} w_2$.
(2) By induction on the run. From $a \tttstep^{*} w_1$ and $b \tttstep^{*} w_2$,
$w_1 \defeq w_2$; Corollary~\ref{cor:checking-whnf-join} gives equivalent shapes and joinable
components, on which the test accepts by the induction hypothesis. (3) Weak-head reduction of a
term in $\chkSN$ terminates, since $\chkwh \subseteq \tttstep$. Each component $c$ of $w_1$ has
$\nu(c) \le \nu(w_1) \le \nu(a)$, with $\nu(w_1) < \nu(a)$ if $a \ne w_1$, and is smaller than
$w_1$; likewise for $b$. So the pair $(\nu(a) + \nu(b), |a|_{\#} + |b|_{\#})$, with $|t|_{\#}$
the size of $t$, decreases lexicographically from each run to the runs on its components.
\end{proof}

\begin{remark}[Strategies]\label{rem:checking-strategies}
Soundness uses only that the test reduces its inputs, completeness that it compares weak-head
normal forms, and termination that reduction from a term in $\chkSN$ is finite. Any strategy that
reaches the weak-head normal form therefore decides the same relation, as does a test that
compares two terms before reducing them and reduces only when the comparison fails. In
particular, a test that immediately accepts two terms equal up to equivalent levels, or that
compares $\tttinst{c}{\bar\ell}\,\bar a$ with $\tttinst{c}{\bar\ell'}\,\bar b$ argument by
argument before unfolding a transparent constant $c$---as the lazy unfolding of definitions in
Lean's kernel does~\citep{type_checking_in_lean4}---is sound because each comparison it accepts
witnesses joinability, and complete because a failed comparison falls back to the weak-head
normal forms. When the heads of the two terms are different transparent constants, the choice of
which definition to unfold first affects the cost of the test rather than its result.
\end{remark}

The conversion test consults no type. Because definitional equality has no $\eta$-rule and proof
irrelevance is an axiom rather than a conversion rule (Section~\ref{sec:mltt-prop}), whether two
terms are definitionally equal does not depend on the type at which they are compared. Even the
$\eta$-rule for functions alone admits an untyped test, which compares an abstraction with the
application of the other term to a fresh variable~\citep{coquand1996}. By contrast, $\eta$ for
unit types and proof irrelevance as a conversion rule make the test type-directed: at a unit
type, or at a proposition, the test accepts any two terms without comparing
them~\citep{abel2012irrelevance}, and its correctness is proved by a logical
relation~\citep{abel-ohman-vezzosi2018}. Moreover, as shown in
Section~\ref{sec:mltt-subsumption}, combining $\eta$ with domain-contravariant subsumption makes
the answer of a type-directed test depend on the type at which the two terms are compared.

\section{Normalization by evaluation}
\label{sec:checking-nbe}

Normalization by evaluation computes the normal form of a term by evaluating it in a domain of
values, in which a closure stands for a function, and reading the value back into a
term~\citep{nbe,abel2013habil}. Definitional equality has no $\eta$-rule, so the readback
consults no type: the procedure is the untyped normalization by evaluation
of~\citet{aehlig-joachimski2004}, extended by the constants and the computation rules of the
kernel theory.

\begin{definition}[Values]\label{def:checking-values}
Values $v, w$, eliminations of values $\epsilon$, and closures $\kappa$ are given by
\[
\begin{array}{r@{\ }c@{\ }l}
v, w &::=& \Srt\,\ell \mid \Pi\,v\,\kappa \mid \Sigma\,v\,\kappa \mid \lambda\,v\,\kappa
  \mid \tttpair{v}{w} \mid \Id\,v\,w\,w' \mid \refl\,v
  \mid \chkv{x} \mid \tttinst{c}{\bar\ell} \mid v \cdot \epsilon, \\[2pt]
\epsilon &::=& w \mid \fst \mid \snd \mid \Jelim\,v_a\,v_C\,v_d\,v_b, \qquad
\kappa ::= \langle x.\,t, \rho\rangle,
\end{array}
\]
where $\chkv{x}$ is a variable as a value, $c$ ranges over the constants of $|\tttEnv|$ that are
not transparent definitions, and a closure $\langle x.\,t, \rho\rangle$ is a term $t$ with a bound
variable $x$ and an environment $\rho$, a finite map from the other free variables of $t$ to
values.
\end{definition}

\begin{definition}[Evaluation and readback]\label{def:checking-nbe}
Evaluation $\chkeval{t}{\rho}$, elimination $\chkelim(v, \epsilon)$, and readback $\chkrb(v)$ are
given by the following equations, read as a recursive procedure; $\chknbe(t)$ is
$\chkrb(\chkeval{t}{\rho_t})$, where $\rho_t$ sends each free variable $x$ of $t$ to $\chkv{x}$.
\begin{gather*}
\chkeval{x}{\rho} = \rho(x), \qquad
\chkeval{\Srt\,\ell}{\rho} = \Srt\,\ell, \\
\chkeval{\tttinst{c}{\bar\ell}}{\rho} =
  \begin{cases}
  \chkeval{t[\bar\ell/\bar u]}{\emptyset} & c \text{ transparent, with body } t,\\
  \tttinst{c}{\bar\ell} & \text{otherwise,}
  \end{cases} \\
\chkeval{\Pi x{:}A.\,B}{\rho} = \Pi\,\chkeval{A}{\rho}\,\langle x.\,B, \rho\rangle
  \quad (\text{likewise } \Sigma \text{ and } \lambda), \\
\chkeval{b\,a}{\rho} = \chkelim(\chkeval{b}{\rho}, \chkeval{a}{\rho}), \qquad
\chkeval{\fst\,a}{\rho} = \chkelim(\chkeval{a}{\rho}, \fst)
  \quad (\text{likewise } \snd), \\
\chkeval{\Jelim\,a\,C\,d\,b\,p}{\rho} =
  \chkelim(\chkeval{p}{\rho},
    \Jelim\,\chkeval{a}{\rho}\,\chkeval{C}{\rho}\,\chkeval{d}{\rho}\,\chkeval{b}{\rho}),
\end{gather*}
and homomorphically for pairs, $\Id$, and $\refl$;
\begin{gather*}
\chkelim(\lambda\,v\,\langle x.\,t, \rho\rangle, w) = \chkeval{t}{\rho[x \mapsto w]}, \\
\chkelim(\tttpair{v}{w}, \fst) = v, \qquad \chkelim(\tttpair{v}{w}, \snd) = w, \\
\chkelim(\refl\,v, \Jelim\,v_a\,v_C\,v_d\,v_b) = v_d, \qquad
\chkelim(\tttinst{c}{\bar\ell} \cdot w_1 \cdots w_{n-1}, w) = \chkeval{r}{\sigma},
\end{gather*}
where in the last equation $c$ is a recursor with $n$ premises, $w$ is a constructor applied to
all its arguments, $r$ is the right-hand side of Rule~\tttrule{$\iota$} for that constructor, and
$\sigma$ assigns to the pattern variables of $r$ the values that the rule reads off the redex;
in every other case $\chkelim(v, \epsilon) = v \cdot \epsilon$. Readback is
\begin{gather*}
\chkrb(\Pi\,v\,\langle x.\,t, \rho\rangle) =
  \Pi y{:}\chkrb(v).\,\chkrb(\chkeval{t}{\rho[x \mapsto \chkv{y}]})
  \quad (y \text{ fresh; likewise } \Sigma \text{ and } \lambda), \\
\chkrb(\chkv{x}) = x, \qquad
\chkrb(v \cdot w) = \chkrb(v)\,\chkrb(w), \\
\chkrb(v \cdot \Jelim\,v_a\,v_C\,v_d\,v_b) =
  \Jelim\,\chkrb(v_a)\,\chkrb(v_C)\,\chkrb(v_d)\,\chkrb(v_b)\,\chkrb(v),
\end{gather*}
and homomorphically for the other values and for $\fst$ and $\snd$.
\end{definition}

\begin{theorem}[Normalization by evaluation]\label{thm:checking-nbe}
For annotation-free terms:
\begin{enumerate}
\item (Soundness) if $\chknbe(t)$ is defined, then $t \tttstep^{*} \chknbe(t)$ and
  $\chknbe(t)$ is normal;
\item (Termination) if $t \in \chkSN$, then $\chknbe(t)$ is defined;
\item (Completeness) if $a, b \in \chkSN$, then $a \defeq b$ if and only if $\chknbe(a)$ and
  $\chknbe(b)$ are equal up to the replacement of levels by equivalent ones.
\end{enumerate}
\end{theorem}

\begin{proof}
A value stands for a term $\chkq{v}$: $\chkq{\Pi\,v\,\langle x.\,t, \rho\rangle} =
\Pi x{:}\chkq{v}.\,t[\chkq{\rho}]$, where $\chkq{\rho}$ substitutes $\chkq{\rho(y)}$ for each
$y$, and likewise for $\Sigma$ and $\lambda$; $\chkq{\chkv{x}} = x$; $\chkq{v \cdot \epsilon}$
is the elimination $\epsilon$, with its values quoted, applied to $\chkq{v}$; and $\chkq{\cdot}$
is homomorphic otherwise.

(1) By induction on the computation, we show three simultaneous claims: if
$\chkeval{t}{\rho} = v$, then $t[\chkq{\rho}] \tttstep^{*} \chkq{v}$; if
$\chkelim(v, \epsilon) = v'$, then $\chkq{v \cdot \epsilon} \tttstep^{*} \chkq{v'}$; and if
$\chkrb(v) = s$, then $\chkq{v} \tttstep^{*} s$ and $s$ is normal. In $\chkelim$, the equation
for an abstraction applied to a value corresponds to a $\beta$-step,
$\chkq{\lambda\,v\,\langle x.\,t, \rho\rangle}\,\chkq{w} \tttstep t[\chkq{\rho}][\chkq{w}/x]
= t[\chkq{\rho[x \mapsto w]}]$, while the equations for pairs, $\refl$, and recursors correspond
to steps of the projections, \tttrule{Id-$\iota$}, and $\iota$. In $\chkeval{\cdot}{\rho}$, the
equation for a transparent constant is a $\delta$-step, and the remaining equations are
congruences. In the readback $\chkrb$, an elimination $v \cdot \epsilon$ is a value only when no
equation of $\chkelim$ applies; since $\chkrb$ preserves the outermost former, variable, or
constant of a value and no value is a transparent constant, the readback of an elimination is
not a redex, and its subterms are normal by the induction hypothesis.

(2) Assign to each call the term it stands for: $t[\chkq{\rho}]$ to $\chkeval{t}{\rho}$,
$\chkq{v \cdot \epsilon}$ to $\chkelim(v, \epsilon)$, and $\chkq{v}$ to $\chkrb(v)$. By the
three claims of~(1), each of these terms is a reduct of a subterm of a reduct of $t$, and hence
belongs to $\chkSN$. Each recursive call decreases in the lexicographic order on
$(\nu, \text{size}, \text{mode})$, with evaluation ordered above elimination:
\begin{itemize}
\item a call on a proper subterm stands for a term of no larger $\nu$ and strictly smaller size;
\item the evaluation called by a reduction equation of $\chkelim$, or the evaluation of the body
  of a transparent constant, stands for a term of strictly smaller $\nu$, since the equation
  performs a reduction step;
\item the elimination called by the evaluation of an application, a projection, or $\Jelim$
  stands for a reduct of the term of that evaluation, which either has strictly smaller $\nu$ or
  is the same term at the lower mode; and
\item the readback of an evaluated closure body stands for a reduct of a proper subterm of the
  term of the call.
\end{itemize}
The terms therefore decrease along every chain of calls, so the computation is finite.

(3) By~(1), $\chknbe(a)$ and $\chknbe(b)$ are normal forms of $a$ and $b$. If $a \defeq b$, they
are joinable by Lemma~\ref{lem:mltt-confluence}, and a normal form reduces only to itself, so
they are equal up to equivalent levels. The converse holds since $\defeq$ contains the
equivalence of levels.
\end{proof}

The conversion test and normalization by evaluation decide the same relation on the same terms,
but they organize reduction differently. Whereas the conversion test reduces only to weak-head
normal form and stops at the first difference, normalization by evaluation computes full normal
forms and shares the evaluated value of a variable across all of its occurrences via the
environment. The typed normalization by evaluation of~\citet{nbe} reads a value back at its type
in order to compute $\eta$-long normal forms; without $\eta$, untyped readback suffices, and a
single procedure serves every type. In practice, the kernel compares values directly, evaluating
the bodies of two closures at a single fresh variable and thereby interleaving readback with the
comparison of Definition~\ref{def:checking-conv}; Volume~II relates that implementation to the
two procedures studied here.

\section{Inversion lemmas}
\label{sec:checking-inversion}

The type checker of Section~\ref{sec:checking-bidir} reads each premise $C \defeq \Pi x{:}A.\,B$
of $\mathcal{K}_0$ and $\mathcal{K}_1$ through the weak-head normal form of $C$. Its correctness
rests on the inversion and typed-reduct lemmas of this section. Here and in the remainder of the
chapter, $\tttEnv$ is an environment in which every declared type is a type of $\mathcal{K}_1$ in
the empty context and every definition body checks against its declared type over the preceding
declarations; this is the hypothesis of Proposition~\ref{prop:mltt-k1-complete}, and it holds of
every environment accepted by the kernel. A \emph{type in} $\ctx$ is a term $A$ such that
$\ctx \vdash A \infr D$ for some $D \defeq \Srt\,\ell$, and a context is a sequence of types,
each of which is a type in the context preceding it.

\begin{lemma}[Validity, uniqueness, and generation]\label{lem:checking-unique}
In $\mathcal{K}_0$ and in $\mathcal{K}_1$:
\begin{enumerate}
\item if $\ctx \vdash a \infr A$, then $A$ is a type in $\ctx$;
\item if $\ctx \vdash a \infr A$ and $\ctx \vdash a \infr A'$, then $A \defeq A'$; in particular
  the universe of a type is unique up to the equivalence of levels;
\item for each form of term, at most one rule of inference concludes a judgment
  $\ctx \vdash t \infr T$ with $t$ of that form, and its premises are determined by $t$ up to the
  reducts they name; and $\ctx \vdash t \chk C$ holds only by the checking rule for the form of
  $t$, if $t$ is an abstraction, a pair, or $\refl\,a$, or by \tttrule{Chk-Sub}.
\end{enumerate}
\end{lemma}

\begin{proof}
(1) By induction on the derivation. Each rule whose conclusion names a type not already carried by
a premise constructs it from types carried by its premises: \tttrule{Inf-Lam} uses
\tttrule{$\Pi$-F}; \tttrule{Inf-App} and \tttrule{Inf-Snd} substitute the annotated argument,
which is well-typed by Lemma~\ref{lem:mltt-k0-annotated}(2); \tttrule{Id-E} applies the motive;
and \tttrule{Const} instantiates the level parameters in the declared type from $\tttEnv$.

(2) As noted after Definition~\ref{def:mltt-k0}, each inference rule reads its types off the
term, off the context, or off reducts named in its premises; any two reducts named for the same
type are joinable, and the type formers are injective up to $\defeq$.

(3) By inspection of the rules.
\end{proof}

\begin{lemma}[Inversion of subsumption]\label{lem:checking-sub-inv}
In $\mathcal{K}_0$ and in $\mathcal{K}_1$, let $\ctx \vdash A \subty_K B$.
\begin{enumerate}
\item If $B \defeq \Srt\,n$, then $A \defeq \Srt\,m$ for some $m \cumle n$.
\item If $B \defeq \Pi x{:}B_1.\,B_2$ for a type $\Pi x{:}B_1.\,B_2$ in $\ctx$, then
  $A \defeq \Pi x{:}A_1.\,A_2$ with $A_1 \defeq B_1$ and
  $\ctx, x{:}B_1 \vdash A_2 \subty_K B_2$; in $\mathcal{K}_0$, $A \defeq B$.
\item If $B$ is joinable with a $\Sigma$-type, an identity type, or a weak-head normal form
  headed by a variable, then $A \defeq B$; without the rules of
  Chapter~\ref{ch:knowledge-graphs}, likewise if $B$ is joinable with a constant applied to
  arguments.
\end{enumerate}
\end{lemma}

\begin{proof}
By induction on the derivation of $\ctx \vdash A \subty_K B$. Each subsumption rule reads its two
sides through reducts, and types with different head formers---or universes at inequivalent
levels---are not joinable (Lemma~\ref{lem:mltt-confluence}). Consequently, a universe on the
right can be reached only by joinability or cumulativity, a $\Pi$-type only by joinability or
the $\Pi$-rule, and the remaining forms in~(3) only by joinability; when the rules of
Chapter~\ref{ch:knowledge-graphs} are present, they lead only into joins and into types headed by
constants of a class system (Section~\ref{sec:kg-model}).

In~(2), if the last rule is joinability, then $A_1 = B_1$ and $A_2 = B_2$. If the last rule is
the $\Pi$-rule, its codomain premise is derived in the context extended by the domain of the
reduct on the right, which is joinable with $B_1$; context conversion replaces that domain by
$B_1$, and the compatibility of $\subty_K$ with $\defeq$ on both sides (established in the proof
of Lemma~\ref{lem:mltt-k1-trans}) replaces the codomain by $B_2$.
\end{proof}

\begin{lemma}[Typed reducts]\label{lem:checking-typed-reducts}
In $\mathcal{K}_0$ and in $\mathcal{K}_1$, let $A$ be a type in $\ctx$ with $|A| \chkred w$.
There is a type $A'$ in $\ctx$, not an annotation, with $A' \defeq A$ and $|A'| = w$. The
subterms of $A'$ satisfy the premises of the inference rule for its form: if
$w = \Pi x{:}B.\,C$, then $A' = \Pi x{:}B'.\,C'$ with $B'$ a type in $\ctx$ and $C'$ a type in
$\ctx, x{:}B'$, and likewise for $\Sigma$; if $w = \Id\,B\,a\,b$, then
$A' = \Id\,B'\,a'\,b'$ with $B'$ a type in $\ctx$ and $\ctx \vdash a' \chk B'$,
$\ctx \vdash b' \chk B'$; and if $w$ is a variable or a constant applied to arguments, so is
$A'$, each argument checking against the domain of a reduct of the type that the preceding
prefix infers.
\end{lemma}

\begin{proof}
The reduction $|A| \tttstep^{*} w$ is a sequence of steps over $|\tttEnv|$ starting from the
erasure of a typable term. By Lemma~\ref{lem:mltt-k0-annotated}(4) (and
Lemma~\ref{lem:mltt-k1-annotated} for $\mathcal{K}_1$), each step lifts to annotated steps whose
result has the erasure $w$; by part~(3) of the same lemma, the resulting term infers a joinable
type, and annotated steps preserve definitional equality. Thus $A \tttstep_a^{*} A''$ for some
type $A''$ in $\ctx$ with $|A''| = w$ and $A'' \defeq A$.

If $A''$ is an annotation $\tttann{A_1}{U}$, then $U \defeq \Srt\,\ell$ and
$\ctx \vdash A_1 \chk U$. Because $|A_1| = w$ is a type, $A_1$ cannot be an abstraction, a pair,
or $\refl\,a$; by Lemma~\ref{lem:checking-unique}(3) it must be checked by \tttrule{Chk-Sub}, so
it infers a type below $U$, which is joinable with a universe by
Lemma~\ref{lem:checking-sub-inv}(1). Stripping any root annotations in this way yields $A'$. The
statement on the subterms of $A'$ then follows from Lemma~\ref{lem:checking-unique}(3) applied to
the inference rule for its head former, and by repeated application of \tttrule{Inf-App} along a
spine.
\end{proof}

\section{Deciding inference, checking, and subsumption}
\label{sec:checking-bidir}

The rules of $\mathcal{K}_0$ and $\mathcal{K}_1$ are syntax-directed
(Lemma~\ref{lem:checking-unique}(3)): the term and the mode determine the rule, except that a term
that infers may also be checked by
\tttrule{Chk-Sub}, and the rule names a reduct of each type it reads. The procedure replaces each
named reduct by a weak-head normal form, and it computes types without annotations, which do not
affect definitional equality.

\begin{figure}[tp]
\begin{mathpar}
\inferrule*[right=\tttrule{A-Var}]
  {(x{:}T) \in \ctx}
  {\ctx \vdash x \chkinf T}
\and
\inferrule*[right=\tttrule{A-Sort}]
  {\tttFV_{\mathrm{lvl}}(\ell) \subseteq \bar u}
  {\ctx \vdash \Srt\,\ell \chkinf \Srt\,(\lsucc\,\ell)}
\and
\inferrule*[right=\tttrule{A-Const}]
  {(\tttinst{c}{\bar u} : A) \in \tttEnv \\ |\bar\ell| = |\bar u| \\
   \tttFV_{\mathrm{lvl}}(\bar\ell) \subseteq \bar v}
  {\ctx \vdash \tttinst{c}{\bar\ell} \chkinf |A|[\bar\ell/\bar u]}
\\
\inferrule*[right=\tttrule{A-$\Pi$}]
  {\ctx \vdash A \chkinf C_1 \\ C_1 \chkred \Srt\,\ell_1 \\
   \ctx, x{:}|A| \vdash B \chkinf C_2 \\ C_2 \chkred \Srt\,\ell_2}
  {\ctx \vdash \Pi x{:}A.\,B \chkinf \Srt\,(\limax\,\ell_1\,\ell_2)}
\and
\inferrule*[right=\tttrule{A-Lam}]
  {\ctx \vdash A \chkinf C \\ C \chkred \Srt\,\ell \\ \ctx, x{:}|A| \vdash b \chkinf B}
  {\ctx \vdash \tttlam{x}{A} b \chkinf \Pi x{:}|A|.\,B}
\\
\inferrule*[right=\tttrule{A-App}]
  {\ctx \vdash f \chkinf C \\ C \chkred \Pi x{:}A.\,B \\ \ctx \vdash a \chkchk A}
  {\ctx \vdash f\,a \chkinf B[|a|/x]}
\and
\inferrule*[right=\tttrule{A-Fst}]
  {\ctx \vdash p \chkinf C \\ C \chkred \Sigma x{:}A.\,B}
  {\ctx \vdash \fst\,p \chkinf A}
\and
\inferrule*[right=\tttrule{A-Snd}]
  {\ctx \vdash p \chkinf C \\ C \chkred \Sigma x{:}A.\,B}
  {\ctx \vdash \snd\,p \chkinf B[\fst\,|p|/x]}
\\
\inferrule*[right=\tttrule{A-Id}]
  {\ctx \vdash A \chkinf C \\ C \chkred \Srt\,\ell \\
   \ctx \vdash a \chkchk |A| \\ \ctx \vdash b \chkchk |A|}
  {\ctx \vdash \Id\,A\,a\,b \chkinf \Prop}
\and
\inferrule*[right=\tttrule{A-Ann}]
  {\ctx \vdash A \chkinf C \\ C \chkred \Srt\,\ell \\ \ctx \vdash a \chkchk |A|}
  {\ctx \vdash \tttann{a}{A} \chkinf |A|}
\\
\inferrule*[right=\tttrule{A-J}]
  {\ctx \vdash C \chkinf T \\ T \chkred \Pi y{:}A.\,T_1 \\ T_1 \chkred \Pi z{:}T_2.\,T_3 \\\\
   T_2 \chkred \Id\,A'\,a'\,y' \\ T_3 \chkred \Srt\,v \\ A' \chkconv A \\ y' \chkconv y \\\\
   \ctx \vdash a \chkchk A \\ a' \chkconv |a| \\ \ctx \vdash b \chkchk A \\\\
   \ctx \vdash d \chkchk |C|\,|a|\,(\refl\,|a|) \\ \ctx \vdash p \chkchk \Id\,A\,|a|\,|b|}
  {\ctx \vdash \Jelim\,a\,C\,d\,b\,p \chkinf |C|\,|b|\,|p|}
\end{mathpar}
\caption{Inference in the algorithmic systems $\mathcal{A}_0$ and $\mathcal{A}_1$. The types of
the context and of the judgments are annotation-free. \tttrule{A-$\Sigma$} is
\tttrule{A-$\Pi$} with $\lmax$ in place of $\limax$. In \tttrule{A-J}, the premises compare the
type of the motive with $\Pi y{:}A.\,\Id\,A\,a\,y \to \Srt\,v$ for the base point $a$ of the
term, and they are evaluated from left to right. As in Chapter~\ref{ch:mltt}, $\bar u$ are the
level parameters in scope, and in \tttrule{A-Const} $\bar u$ are the parameters of the declaration
and $\bar v$ those in scope.}
\label{fig:checking-alg}
\end{figure}

\begin{figure}[tp]
\begin{mathpar}
\inferrule*[right=\tttrule{A-Chk-Lam}]
  {C \chkred \Pi x{:}A'.\,B \\ \ctx \vdash A \chkinf D \\ D \chkred \Srt\,\ell \\
   |A| \chkconv A' \\ \ctx, x{:}|A| \vdash b \chkchk B}
  {\ctx \vdash \tttlam{x}{A} b \chkchk C}
\and
\inferrule*[right=\tttrule{A-Chk-Pair}]
  {C \chkred \Sigma x{:}A.\,B \\ \ctx \vdash a \chkchk A \\ \ctx \vdash b \chkchk B[|a|/x]}
  {\ctx \vdash \tttpair{a}{b} \chkchk C}
\\
\inferrule*[right=\tttrule{A-Chk-Refl}]
  {C \chkred \Id\,A\,x\,y \\ \ctx \vdash a \chkchk A \\ x \chkconv |a| \\ |a| \chkconv y}
  {\ctx \vdash \refl\,a \chkchk C}
\and
\inferrule*[right=\tttrule{A-Chk-Sub}]
  {\ctx \vdash a \chkinf A \\ A \chksub C}
  {\ctx \vdash a \chkchk C}
\\
\inferrule*[right=\tttrule{A-Sub-Conv}]
  {A \chkconv B}
  {A \chksub B}
\and
\inferrule*[right=\tttrule{A-Sub-Cum}]
  {A \chkred \Srt\,m \\ B \chkred \Srt\,n \\ m \cumle n}
  {A \chksub B}
\and
\inferrule*[right=\tttrule{A-Sub-$\Pi$}]
  {A \chkred \Pi x{:}A_1.\,B_1 \\ B \chkred \Pi x{:}A_2.\,B_2 \\
   A_1 \chkconv A_2 \\ B_1 \chksub B_2}
  {A \chksub B}
\end{mathpar}
\caption{Checking and subsumption in the algorithmic system $\mathcal{A}_1$;
$\mathcal{A}_0$ omits \tttrule{A-Sub-$\Pi$}.}
\label{fig:checking-alg-chk}
\end{figure}

\begin{definition}[The algorithmic systems]\label{def:checking-alg}
$\mathcal{A}_1$ is given by the rules of Figures~\ref{fig:checking-alg}
and~\ref{fig:checking-alg-chk}, with the judgments
$\ctx \vdash a \chkinf T$, $\ctx \vdash a \chkchk T$, and $T \chksub T'$, over contexts of
annotation-free types; $\mathcal{A}_0$ omits \tttrule{A-Sub-$\Pi$}. Because the subsumption rules
of Figure~\ref{fig:checking-alg-chk} do not consult $\ctx$, the context is omitted from
$T \chksub T'$ in this section (it is written explicitly as $\ctx \vdash A \chksub B$ in
Section~\ref{sec:checking-kg}, where Rules~\tttrule{A-Class-Var} and~\tttrule{A-Rec-Sub} read
$\ctx$). A premise $C \chkred P$, for a pattern $P$, computes the weak-head normal form of $C$
and matches it against $P$; a premise $a \chkconv b$ runs the conversion test. Every type that a
rule takes from the term, from the context, or from the environment is erased, and a term
substituted into a type is erased.
\end{definition}

\paragraph{The procedure.} The procedure for $\mathcal{A}_i$ takes a context and a term, with an
expected type in checking mode, or two types in subsumption mode. It applies the rules whose
conclusion matches the input, in the order of the figure, evaluating the premises from left to
right, and tries the alternatives of a checking or subsumption judgment until one accepts:
the checking rule for the form of the term, if any, before \tttrule{A-Chk-Sub}, and
\tttrule{A-Sub-Conv} before \tttrule{A-Sub-Cum} and \tttrule{A-Sub-$\Pi$}. Inference is a partial
function: at most one rule matches a term, and its premises determine the type. The \emph{run}
of the procedure on an input is the tree of the premises it evaluates; the run terminates if the
tree is finite, and it then accepts or rejects. Since $\mathcal{K}_i$ reads all its types
through joinability, the procedure may use any type joinable with the one that $\mathcal{K}_i$
names, and the three theorems below are stated for such types.

\begin{theorem}[Soundness]\label{thm:checking-sound}
For $i \in \{0, 1\}$ and a context $\ctx$:
\begin{enumerate}
\item if $|\ctx| \vdash a \chkinf T$ in $\mathcal{A}_i$, then $\ctx \vdash a \infr A$ in
  $\mathcal{K}_i$ for some $A$ with $|A| = T$;
\item if $|\ctx| \vdash a \chkchk T$ in $\mathcal{A}_i$ and $A$ is a type in $\ctx$ with
  $A \defeq T$, then $\ctx \vdash a \chk A$ in $\mathcal{K}_i$;
\item if $T \chksub T'$ in $\mathcal{A}_i$ and $A$, $B$ are types in $\ctx$ with $A \defeq T$ and
  $B \defeq T'$, then $\ctx \vdash A \subty_K B$ in $\mathcal{K}_i$.
\end{enumerate}
\end{theorem}

\begin{proof}
By simultaneous induction on the derivations of $\mathcal{A}_i$. Each rule of $\mathcal{A}_i$ is a
rule of $\mathcal{K}_i$ with the reducts that $\mathcal{K}_i$ names replaced by weak-head normal
forms of erasures. For each such premise, Lemma~\ref{lem:checking-typed-reducts} yields a reduct
that is a type of $\mathcal{K}_i$, joinable with the type of $\mathcal{K}_i$, whose erasure is the
weak-head normal form computed by $\mathcal{A}_i$ and whose components are well-typed;
Theorem~\ref{thm:checking-conv}(1) turns each premise $a \chkconv b$ into $a \defeq b$; and
Lemma~\ref{lem:mltt-k0-annotated}(1) moves a checking judgment to a joinable type.

We detail the representative cases:
\begin{itemize}
\item For \tttrule{A-App}, the induction hypothesis gives $\ctx \vdash f \infr C'$ with $|C'| = C$.
  Applying Lemma~\ref{lem:checking-typed-reducts} to $C'$ (which is a type by
  Lemma~\ref{lem:checking-unique}(1)) gives a type $\Pi x{:}A''.\,B''$ joinable with $C'$ such that
  $|A''| = A$ and $|B''| = B$. The induction hypothesis gives $\ctx \vdash a \chk A''$, and
  \tttrule{Inf-App} yields $\ctx \vdash f\,a \infr B''[\tttann{a}{A''}/x]$, whose erasure is
  $B[|a|/x]$.
\item For \tttrule{A-Chk-Lam}, the typed reduct $\Pi x{:}A''.\,B''$ of the expected type satisfies
  $|B''| = B$ and $A'' \defeq |A| \defeq A$. Thus $B''$, which is a type in $\ctx, x{:}A''$, is also
  a type in $\ctx, x{:}A$ by context conversion, and the induction hypothesis checks $b$ against it.
\item For \tttrule{A-Chk-Sub}, the induction hypotheses give $\ctx \vdash a \infr A'$ with
  $|A'| = A$ and $\ctx \vdash A' \subty_K B$, so \tttrule{Chk-Sub} applies.
\item For \tttrule{A-Sub-$\Pi$}, the induction hypothesis for the codomains is applied in the
  context extended by the domain of the typed reduct of the right-hand side.
\item For \tttrule{A-J}, the reduct $\Pi y{:}A''.\,\Pi z{:}\Id\,A''\,a\,y.\,\Srt\,v$ named by the
  rule for $\Jelim$ uses the domain $A''$ of the typed reduct that
  Lemma~\ref{lem:checking-typed-reducts} supplies for $T$, together with the base point $a$ of the
  term; this reduct is a type because $a$ checks against $A''$, and it is joinable with the type of
  the motive by the typed reducts for $T_1$ and $T_2$ and the premises $A' \chkconv A$,
  $a' \chkconv |a|$, and $y' \chkconv y$.
\end{itemize}
The remaining rules are similar.
\end{proof}

\begin{theorem}[Completeness relative to termination]\label{thm:checking-complete}
For $i \in \{0, 1\}$, a context $\ctx$, and annotation-free terms $T$ and $T'$:
\begin{enumerate}
\item if $\ctx \vdash a \infr A$ in $\mathcal{K}_i$ and the run on $|\ctx|$ and $a$ in inference
  mode terminates, it returns a type $T \defeq A$;
\item if $\ctx \vdash a \chk A$ in $\mathcal{K}_i$, $T \defeq A$, and the run on $|\ctx|$, $a$,
  and $T$ in checking mode terminates, it accepts;
\item if $\ctx \vdash A \subty_K B$ in $\mathcal{K}_i$, $T \defeq A$, $T' \defeq B$, and the run on
  $T$ and $T'$ terminates, it accepts.
\end{enumerate}
\end{theorem}

\begin{proof}
By simultaneous induction on the derivations of $\mathcal{K}_i$; the runs on the premises
terminate because the run on the conclusion does. Each rule of $\mathcal{K}_i$ names a reduct of
each type it reads, and the procedure computes the weak-head normal form of a joinable type; by
Corollary~\ref{cor:checking-whnf-join}, that weak-head normal form has the same head as the reduct
and joinable components. Consequently, the corresponding rule of $\mathcal{A}_i$ matches, its
conversion premises accept by Theorem~\ref{thm:checking-conv}(2), and the induction hypothesis
applies to its remaining premises, whose types are joinable with those in $\mathcal{K}_i$.

In particular:
\begin{itemize}
\item For \tttrule{Inf-App}, the type $T_f$ returned for $f$ is joinable with the type that $f$
  infers, which in turn is joinable with the reduct $\Pi x{:}A.\,B$; hence
  $T_f \chkred \Pi x{:}A_0.\,B_0$ with $A_0 \defeq A$ and $B_0 \defeq B$, the argument $a$ is
  accepted against $A_0$, and the returned type $B_0[|a|/x]$ is joinable with
  $B[\tttann{a}{A}/x]$.
\item For \tttrule{Chk-Sub}, if the checking rule for the form of the term (when one exists)
  rejects, the procedure tries the alternative \tttrule{A-Chk-Sub}, and both of its premises accept
  by the induction hypothesis.
\item For the subsumption rules, joinability is accepted by \tttrule{A-Sub-Conv}, cumulativity by
  \tttrule{A-Sub-Cum} at the levels of the weak-head normal forms (which are equivalent to the named
  ones), and the $\Pi$-rule by \tttrule{A-Sub-$\Pi$}.
\item For $\Jelim\,a\,C\,d\,b\,p$, the reduct named for the type of the motive is
  $\Pi y{:}A.\,\allowbreak\Pi z{:}\Id\,A\,a\,y.\,\allowbreak\Srt\,v$, so the three weak-head normal
  forms in \tttrule{A-J} have components joinable with $A$, $a$, and $y$, its conversion premises
  accept, and its checking premises match those of the rule.
\end{itemize}
\end{proof}

\begin{theorem}[Termination]\label{thm:checking-terminates}
Under Hypothesis~\ref{hyp:checking-normalization}, for $i \in \{0, 1\}$ and a context $\ctx$, the
run of the procedure for $\mathcal{A}_i$ terminates on $|\ctx|$ and any term $a$ in inference
mode, on $|\ctx|$, any term $a$, and $|A|$ in checking mode for every type $A$ in $\ctx$, and on
$|A|$ and $|B|$ in subsumption mode for all types $A$ and $B$ in $\ctx$.
\end{theorem}

\begin{proof}
We first observe that every type or term that the procedure reduces or compares is the erasure of a
type or term of $\mathcal{K}_i$ in the current context, or a component of a weak-head normal form
of one:
\begin{itemize}
\item the entries of the context and the declared types in $\tttEnv$ are erasures of types by
  hypothesis;
\item any inferred type is the erasure of a type by Theorem~\ref{thm:checking-sound}(1) and
  Lemma~\ref{lem:checking-unique}(1), and the components of its weak-head normal form are components
  of that erasure;
\item any type formed from a subterm of the input---by substitution in \tttrule{A-App},
  \tttrule{A-Snd}, and \tttrule{A-Chk-Pair}, or by application in \tttrule{A-J}---is formed only
  after the premise that checks that subterm succeeds, so Theorem~\ref{thm:checking-sound}(2) types
  the subterm and the proof of Theorem~\ref{thm:checking-sound} exhibits the resulting type as an
  erasure; and
\item the terms compared by \tttrule{A-Chk-Refl} and \tttrule{A-J} are compared only after the
  premises that check them succeed.
\end{itemize}
By Hypothesis~\ref{hyp:checking-normalization} and the remark following it, all of these terms
belong to $\chkSN$, so every weak-head normalization and every conversion test performed during the
run terminates (Theorem~\ref{thm:checking-conv}(3)).

Furthermore, each recursive call of $\chksub$ is made on components of the weak-head normal forms of
its two input types, so $\chksub$ terminates by the measure of Theorem~\ref{thm:checking-conv}(3).
Finally, the recursive calls of inference and checking are made on proper subterms of the input
term, except for \tttrule{A-Chk-Sub}, which calls inference on the same term; ordered
lexicographically by the size of the term and the mode (with checking ordered above inference),
every chain of calls strictly decreases.
\end{proof}

\begin{corollary}[Decidability]\label{cor:checking-decidable}
Under Hypothesis~\ref{hyp:checking-normalization}, for $i \in \{0, 1\}$ and a context $\ctx$:
whether a term infers a type in $\mathcal{K}_i$, and which type, up to $\defeq$; whether a term
checks
against a type in $\ctx$; and whether $\ctx \vdash A \subty_K B$ for types $A$ and $B$ in $\ctx$
are decidable. The inferred type is unique up to $\defeq$ (Lemma~\ref{lem:checking-unique}(2)).
\end{corollary}

\begin{proof}
Theorems~\ref{thm:checking-sound}, \ref{thm:checking-complete},
and~\ref{thm:checking-terminates}, with $T = |A|$ in checking and subsumption mode.
\end{proof}

Corollary~\ref{cor:checking-decidable} decides the bidirectional judgments on annotated terms rather
than the declarative typability of an unannotated term: Proposition~\ref{prop:mltt-k1-complete}
inserts annotations (such as the type annotation on the head of a redex whose body only checks),
whereas the procedure checks a term using whatever annotations it carries. In practice, the kernel
checks terms with the annotations supplied by elaboration, and it checks an environment declaration
by declaration:
\begin{itemize}
\item an axiom is checked by inferring a universe for its declared type;
\item a definition is checked by inferring a universe for its declared type and checking its body
  against that type; and
\item an inductive declaration is checked by verifying the typing conditions of
  Definition~\ref{def:mltt-admissible} in $\mathcal{K}_1$ and deciding its syntactic conditions
  (strict positivity, the universe constraint, and non-dependence) by inspecting the declaration and
  comparing universe levels.
\end{itemize}
Each declaration accepted in this way preserves the environment invariant of
Section~\ref{sec:checking-inversion}.

\section{The rules of Chapter~\ref{ch:knowledge-graphs}}
\label{sec:checking-kg}

Chapter~\ref{ch:knowledge-graphs} extends $\mathcal{K}_1$ by the formers of records, classes,
refinements, joins, singletons, and basic types, which are constants of a class system applied to
arguments, by base cases of subsumption, and by the checking rules \textsc{Sing-I} and
\textsc{Join-I}, all reading their types through reducts (Section~\ref{sec:kg-model}). Its
computation rules, \textsc{Rec-$\beta$} and the tests of Definition~\ref{def:kg-tags}, are
computation rules of $\tttstep$ whose left-hand sides match a record literal or literals at fixed
argument positions of a constant. Weak-head reduction contracts them as it contracts $\iota$, with
those argument positions reduced, from left to right, as the major premise of a recursor is;
evaluation applies them as it applies $\iota$; and Lemmas~\ref{lem:checking-wh}
and~\ref{lem:checking-whnf-stable} and Theorem~\ref{thm:checking-nbe} hold with the same proofs.

\begin{figure}[t]
\begin{mathpar}
\inferrule*[right=\tttrule{A-Class-Ent}]
  {A \chkred C \\ C \text{ a class} \\ B \chkred \Entity}
  {\ctx \vdash A \chksub B}
\and
\inferrule*[right=\tttrule{A-Class-Sub}]
  {A \chkred C \\ B \chkred D \\ C \kganc D}
  {\ctx \vdash A \chksub B}
\\
\inferrule*[right=\tttrule{A-Class-Var}]
  {A \chkred n \\ \ctx \vdash n \chkspine U \\ U \chkred \Class \\ B \chkred \Entity}
  {\ctx \vdash A \chksub B}
\\
\inferrule*[right=\tttrule{A-Join-I}]
  {B \chkred T_1 \ttrjoin T_2 \\ \ctx \vdash A \chksub T_i}
  {\ctx \vdash A \chksub B}
\and
\inferrule*[right=\tttrule{A-Join-L}]
  {A \chkred T_1 \ttrjoin T_2 \\ B \chkred U \\ U \text{ a graph type} \\
   \ctx \vdash T_1 \chksub B \\ \ctx \vdash T_2 \chksub B}
  {\ctx \vdash A \chksub B}
\\
\inferrule*[right=\tttrule{A-Sing-Sub}]
  {A \chkred \kgsing\,T\,b \\ B \chkred U \\ U \text{ a graph type} \\ \ctx \vdash T \chksub B}
  {\ctx \vdash A \chksub B}
\\
\inferrule*[right=\tttrule{A-Rec-Sub}]
  {A \chkred R \\ B \chkred R' \\ \kglabels(R') \subseteq \kglabels(R) \\
   \ctx, \kgtel(R) \vdash A_\ell \chksub A'_\ell \quad (\ell \in \kglabels(R')) \\ w \le w'}
  {\ctx \vdash A \chksub B}
\\
\inferrule*[right=\tttrule{A-Ref-Sub}]
  {A \chkred R \ttrmeet \textstyle\bigwedge S \\ B \chkred R' \ttrmeet \textstyle\bigwedge S' \\
   \ctx \vdash R \chksub R' \\ \forall D' \in S'.\ \exists D \in S.\ D \kganc D'}
  {\ctx \vdash A \chksub B}
\\
\inferrule*[right=\tttrule{A-Chk-Sing}]
  {C \chkred \kgsing\,T\,b \\ \ctx \vdash a \chkchk T \\ |a| \chkconv b}
  {\ctx \vdash a \chkchk C}
\and
\inferrule*[right=\tttrule{A-Chk-Join}]
  {C \chkred T_1 \ttrjoin T_2 \\ \ctx \vdash a \chkchk T_i}
  {\ctx \vdash a \chkchk C}
\end{mathpar}
\caption{The rules that the procedure gains for Chapter~\ref{ch:knowledge-graphs}, with
$i \in \{1, 2\}$. $\kgsing\,T\,b$ is the singleton $\sing{b}$ at $T$. In
\tttrule{A-Rec-Sub}, $A_\ell$ and $A'_\ell$ are the field types of the record types $R$ and $R'$,
and $w$ and $w'$ their levels. A graph type is a weak-head normal form headed by a former of the
class system (Definition~\ref{def:kg-graph-type}). $\ctx \vdash n \chkspine U$ reads the type of a
neutral term off its spine.}
\label{fig:checking-alg-kg}
\end{figure}

Figure~\ref{fig:checking-alg-kg} gives the rules that the procedure gains. With them,
subsumption consults the context, which \tttrule{A-Sub-$\Pi$} extends by the domain of its right
side and \tttrule{A-Rec-Sub} by the telescope of $R$. Rule \tttrule{A-Class-Var} decides the rule
$X \subty \Entity$ for $X : \Class$ of Definition~\ref{def:kg-class-type} at a type whose
weak-head normal form $n$ is headed by a variable or by a constant that is not a former of the
class system. Its premise $\ctx \vdash n \chkspine U$ computes the type of $n$ from the type of
its head, the entry of the context or the declared type, by reading it along the spine as
\tttrule{A-App}, \tttrule{A-Fst}, \tttrule{A-Snd}, and \tttrule{A-J} read the type of their head,
without checking the arguments again: $n$ is the erasure of a typed reduct, whose arguments
check (Lemma~\ref{lem:checking-typed-reducts}). The alternatives of the procedure are tried in the
order of Figures~\ref{fig:checking-alg-chk} and~\ref{fig:checking-alg-kg}, with \tttrule{A-Chk-Sub}
last among the checking rules.

\begin{proposition}[The rules of Chapter~\ref{ch:knowledge-graphs}]\label{prop:checking-kg}
Let $\tttEnv$ contain a class system (Rule~\tttrule{E-KG}), and let $\mathcal{K}_1$ include the
rules of Sections~\ref{sec:kg-embedding} to~\ref{sec:kg-propositions}, read through reducts
(Section~\ref{sec:kg-model}). Theorems~\ref{thm:checking-sound}, \ref{thm:checking-complete},
and~\ref{thm:checking-terminates} and Corollary~\ref{cor:checking-decidable} hold for
$\mathcal{K}_1$ and for $\mathcal{A}_1$ with the rules of Figure~\ref{fig:checking-alg-kg}. The
side conditions of these rules are decided by inspecting the finite class system and by
Proposition~\ref{prop:mltt-levels}.
\end{proposition}

\begin{proof}
The rules of Figure~\ref{fig:checking-alg-kg} correspond one to one to the added rules of
$\mathcal{K}_1$, and the inductions of the three theorems gain one case for each, argued along the
same lines as \tttrule{A-Sub-$\Pi$} and \tttrule{A-Chk-Lam}: the procedure computes weak-head
normal forms with the heads of the reducts named by $\mathcal{K}_1$
(Corollary~\ref{cor:checking-whnf-join}), Lemma~\ref{lem:checking-typed-reducts} supplies typed
reducts with the erasures computed by the procedure, and the premises operate on their components.
Lemmas~\ref{lem:checking-sub-inv}(1) and~(2) continue to hold with the added subsumption rules,
which lead only into joins and graph types, as does Lemma~\ref{lem:checking-unique}(3) with
\textsc{Sing-I} and \textsc{Join-I} among the checking rules.

For completeness, the procedure tries every alternative in order: a type below a join by
\textsc{Join-I} is accepted by \tttrule{A-Join-I} at the component used by the derivation, a term
checked against a join by \textsc{Join-I} is accepted by \tttrule{A-Chk-Join}, and a term checked
by \tttrule{Chk-Sub} at a join or a singleton is accepted when the procedure falls back to
\tttrule{A-Chk-Sub}. In \tttrule{A-Class-Var}, the type $U$ read off the spine is the erasure of
the type inferred by the typed reduct of $A$, which is joinable with $\Class$; that reduct then
checks against $\Class$ by \tttrule{Chk-Sub}, giving soundness, and conversely any reduct that
checks against $\Class$ infers a type below $\Class$, which must be joinable with $\Class$ because
no rule other than joinability has $\Class$ on its right.

For termination, the added rules call subsumption and checking on components of weak-head normal
forms; in \tttrule{A-Chk-Sing} and \tttrule{A-Chk-Join}, checking keeps the same term while passing
to a component of the weak-head normal form of the expected type, so the recursive calls decrease in
the lexicographic order of Theorem~\ref{thm:checking-terminates} extended by $(\nu, \text{size})$ of
the expected type. Finally, the type $U$ computed by \tttrule{A-Class-Var} is the erasure of a type
of $\mathcal{K}_1$, and therefore belongs to $\chkSN$.
\end{proof}

\section{Properties of the checker used in Chapters~\ref{ch:justification-logic}
  and~\ref{ch:institutions}}
\label{sec:checking-properties}

Proposition~\ref{thm:inst-cut-redex} is stated for a checker whose acceptance of a term in a
context depends on its subterms and on the conversion checks between their types, and
Corollary~\ref{thm:jl-decidable} assumes decidable definitional equality, decidable checking of
instances, and weak-head normalization of the propositions in the range of the constant
specification. The procedure of Section~\ref{sec:checking-bidir} has these properties.

\begin{proposition}[Compositionality]\label{prop:checking-compositional}
\leavevmode
\begin{enumerate}
\item Whether the procedure for $\mathcal{A}_i$ accepts a term in a context is determined by its
  results on the immediate subterms of the term, in the contexts and at the types that the rule
  for the form of the term computes (and, when the rules of Figure~\ref{fig:checking-alg-kg} are
  included in $\mathcal{A}_1$, on the term itself against components of the weak-head normal form
  of the expected type in \tttrule{A-Chk-Sing} and \tttrule{A-Chk-Join}), together with the
  conversion and subsumption tests between the types that the rule reads.
\item The conversion test accepts two terms that are equal up to the replacement of levels by
  equivalent ones whenever their weak-head normalization terminates.
\item In the situation of Proposition~\ref{thm:inst-cut-redex}, if the procedure for
  $\mathcal{A}_1$ accepts $s$ against $\psi$ in the context
  $\instax{\Delta}, \instax{\varphi_1} : \varphi_1, \ldots, \instax{\varphi_k} : \varphi_k$ and
  each $t_j$ against $\varphi_j$ in $\instax{\Delta}$, then it accepts the redex of that
  proposition against $\psi$, with the abstraction annotated by
  $\Pi \instax{\varphi_1}{:}\varphi_1 \cdots \Pi \instax{\varphi_k}{:}\varphi_k.\,\psi$, and the
  conversion tests it adds compare each $\varphi_j$ with itself and $\psi$ with itself.
\end{enumerate}
\end{proposition}

\begin{proof}
(1) By inspection of Figures~\ref{fig:checking-alg}, \ref{fig:checking-alg-chk},
and~\ref{fig:checking-alg-kg}: the premises of each rule are judgments on immediate subterms (or,
in \tttrule{A-Chk-Sing} and \tttrule{A-Chk-Join}, on the same term at a component of the weak-head
normal form of the expected type), weak-head normalizations and conversion tests of types, and
subsumption tests. (2) By Theorem~\ref{thm:checking-conv}(2), since such terms are joinable.
(3) The annotation of the abstraction infers its type by \tttrule{A-Ann}, which checks
the abstraction against it by \tttrule{A-Chk-Lam}, comparing each domain $\varphi_j$ with itself
and checking $s$ against $\psi$, which is accepted; the type of the annotation is a type, since
the $\varphi_j$ and $\psi$ are closed propositions. Each application then reads a
$\Pi$-type, which is its own weak-head normal form, checks $t_j$ against $\varphi_j$, which is
accepted, and infers the codomain, in which the substitution is vacuous since $\psi$ and the
$\varphi_j$ are closed. The redex infers $\psi$, and \tttrule{A-Chk-Sub} compares $\psi$ with
itself.
\end{proof}

\begin{corollary}[Decidability of justification]\label{cor:checking-jl}
Under Hypothesis~\ref{hyp:checking-normalization} for the environment of the constant
specification, the assumptions of Corollary~\ref{thm:jl-decidable} hold, and
$\vdash \JB(j, P)$ is decidable.
\end{corollary}

\begin{proof}
The propositions that the evidence function compares, and the propositions in the range of the
constant specification, are erasures of types of $\mathcal{K}_1$, hence in $\chkSN$: the
conversion test decides definitional equality on them and their weak-head normalization
terminates (Theorem~\ref{thm:checking-conv}). Checking an instance against the domain of a
universal proposition is decided by Corollary~\ref{cor:checking-decidable}.
\end{proof}

\section{Termination}
\label{sec:checking-termination}

Hypothesis~\ref{hyp:checking-normalization} is the one statement of this chapter that is not
proved. The decidability results of the chapter rest on it: Theorem~\ref{thm:checking-conv}(3)
for typable terms, Theorem~\ref{thm:checking-nbe}(2), Theorem~\ref{thm:checking-terminates}, and
Corollaries~\ref{cor:checking-decidable} and~\ref{cor:checking-jl}, and through them
Corollary~\ref{thm:jl-decidable} and the decidability of the conversion checks in
Proposition~\ref{thm:inst-cut-redex}. The soundness and the completeness of the procedures do not
depend on it.

\paragraph{What is known.} The set-theoretic model of Section~\ref{sec:mltt-consistency} interprets
terms in sets and establishes consistency rather than normalization. Nevertheless, the theories for
which normalization has been proved cover the main components of the kernel theory: ECC, with its
cumulative universes and impredicative $\Prop$, is strongly normalizing
(Theorem~\ref{thm:mltt-ecc}), and conversion is decidable for predicative Martin-L\"of type theory
with a universe hierarchy over $\Pi$ and $\Nat$~\citep{abel-ohman-vezzosi2018}. Moreover, the
normalization failures recorded in Theorem~\ref{thm:mltt-failure} do not transfer to the kernel
theory. Both counterexamples rely on proof irrelevance as a conversion rule, under which $\Jelim$
or a recursor reduces on a proof that conversion identifies with $\refl$ or with a constructor
term; in the kernel theory, by contrast, proof irrelevance is the axiom $\irrel$, which has no
computation rule, and $\Jelim$ and the recursors reduce only on literal $\refl$ and constructor
terms. Subsumption and cumulativity add no computation rules, though they enlarge the set of
typable terms and thus make Hypothesis~\ref{hyp:checking-normalization} stronger than normalization
for the non-cumulative fragment. Finally, the reduction of~\citet{pujet-tabareau2023} is blocked by
the large elimination of $\tttAcc$, through which impredicativity contributes to computation
(Section~\ref{sec:mltt-metatheory}).

\paragraph{Routes.} One natural route to a proof is to interpret types as reducibility candidates in
the manner of Girard~\citep{girard1972}: the impredicative universe quantifies over all candidates,
inductive families are interpreted as least fixed points of monotone operators on candidates, and
the subsumption judgment is interpreted as candidate inclusion, mirroring the set-theoretic model of
Chapter~\ref{ch:mltt} in which subsumption is set inclusion. A second route is to construct a
Kripke logical relation for untyped conversion in the manner of~\citet{abel-ohman-vezzosi2018},
extended by the impredicative universe $\Prop$; here the absence of $\eta$-conversion removes the
type-directed component of their algorithmic equality.

\begin{table}[t]
\centering\small
\begin{tabular}{@{}p{0.52\textwidth}p{0.4\textwidth}@{}}
\toprule
Result & Status \\
\midrule
Weak-head reduction is deterministic; stability of weak-head normal forms & proved
  (Lemmas~\ref{lem:checking-wh}, \ref{lem:checking-whnf-stable}) \\
Conversion test sound, complete relative to termination & proved
  (Theorem~\ref{thm:checking-conv}(1), (2)) \\
Normalization by evaluation sound, complete on $\chkSN$ & proved
  (Theorem~\ref{thm:checking-nbe}) \\
Inversion lemmas & proved (Section~\ref{sec:checking-inversion}) \\
Type checker sound, complete relative to termination & proved
  (Theorems~\ref{thm:checking-sound}, \ref{thm:checking-complete};
  Proposition~\ref{prop:checking-kg}) \\
Properties of the checker for Chapters~\ref{ch:justification-logic}
  and~\ref{ch:institutions} & proved (Proposition~\ref{prop:checking-compositional}) \\
Termination of the conversion test, of normalization by evaluation, and of the type checker on
  kernel inputs & relative to Hypothesis~\ref{hyp:checking-normalization}
  (Theorems~\ref{thm:checking-conv}(3), \ref{thm:checking-nbe}(2),
  \ref{thm:checking-terminates}) \\
Hypothesis~\ref{hyp:checking-normalization} & open \\
\bottomrule
\end{tabular}
\caption{The results of the chapter and their status.}
\label{tab:checking-status}
\end{table}

\appendix
\clearpage
\chapter{Formalization}
\label{app:formalization}

The theory of this volume is formalized in Lean~4 over Mathlib, in the library
\texttt{EigeniusTheory} that accompanies the book, published at
\url{https://github.com/eigenius/theory}. This appendix maps the definitions and
results of each chapter to their Lean declarations. Every declaration listed is proved without
\texttt{sorry} and depends only on propositional extensionality, choice, and quotient
soundness, the standard axioms of Lean. A row whose status is \emph{definition} formalizes a
definition and the well-formedness obligations it carries, such as the laws of a category or
of a closure operator; a row whose status is \emph{proved} formalizes a stated result.

The library has one module per section of the book, and the docstring of each declaration
names the label it formalizes, so the mapping below can be checked in either direction. All
declarations are in the namespace \texttt{Eigenius}, which the tables omit; the module column
names the file within the directory given for the chapter. Where a formal definition differs
from the text in its presentation, the difference is stated with the chapter.

\section{Chapter~\ref{ch:mltt}}

Modules are in \texttt{EigeniusTheory/TypeTheory}. Variables are de Bruijn indices, and the
substitution of Definition~\ref{def:mltt-substitution} is the instantiation of an index by a
term; the level and term operations follow the expression layer of the lean4lean development.
The decidability of the orders on levels (Proposition~\ref{prop:mltt-levels}) is not stated.
Rule~\tttrule{S-Cum} of the declarative system, of $\mathcal{K}_0$ and $\mathcal{K}_1$, and of
$\mathcal{K}_0^{\mathrm{J}}$ compares levels in the order of cumulativity
\texttt{Level.CumLe}, so that $\Prop$ lies below no type universe
(\texttt{Level.not\_cumLe\_zero\_succ}); the inversion of subsumption at a universe returns a
level below in that order (\texttt{SubK.sort\_inv}, \texttt{Chk.inf\_sort\_le}), which the
translation of \tttrule{$\Pi$-F} and \tttrule{$\Sigma$-F} into $\mathcal{K}_1$ uses through
\texttt{CumLe.imax} and \texttt{CumLe.max}.

The term $\Jelim\,a\,C\,d\,b\,p$ carries its base point (\texttt{Term.jelim}), as in the chapter.
The rule for $\Jelim$ in $\mathcal{K}_0$ and $\mathcal{K}_1$ (\texttt{Bidir.infJ}) checks the base
point against the domain of the reduct that it names for the type of the motive, a premise that
the declarative rule leaves to the type of the motive; subject reduction for a step inside the
base point uses it, and the annotated $\Jelim$-rule annotates the minor premise at
$C\,a\,(\refl\,a)$, the type at which the rule checks it (\texttt{Inf.jRefl\_step}).

The computation rules that the global environment supplies, $\delta$ for transparent
definitions (Definition~\ref{def:mltt-environment}) and $\iota$ for recursors
(Definition~\ref{def:mltt-recursor}), enter one-step reduction through an interface,
\texttt{RuleSet}: a relation on terms whose redexes are a constant applied to as many arguments
as the constant's arity, which is deterministic and closed under lifting, substitution, and
level substitution. The
confluence proof is the complete-development argument for parallel reduction, and it asks one
more thing of the rule set, stated as \texttt{RuleSet.\allowbreak Stable}: when the arguments of
a redex reduce in parallel, the result is again a redex whose reduct is a parallel reduct of the
original reduct. This is the form in which the left-linearity and the non-overlap of the rules,
which the proof of Lemma~\ref{lem:mltt-confluence} invokes, are used. It is proved of
$\delta$, whose redexes carry no arguments, and of $\iota$ under the reading of the rule fixed
after Definition~\ref{def:mltt-recursor}, which matches the index arguments of the recursor and
the level instance and the parameter arguments of the constructor term as variables. The
redexes of $\beta$, the projections, annotation erasure, and \tttrule{Id-$\iota$} are headed by
an abstraction, a pair, an annotation, or $\Jelim$, so they do not overlap with a constant
application.

Definitional equality is joinability up to level equivalence, as Definition~\ref{def:mltt-defeq}
states it: \texttt{Term.LEquiv} is the congruence closure of level equivalence on terms, and
\texttt{RuleSet.\allowbreak DefEq} relates two terms that reduce to terms related by it. The
proof of Lemma~\ref{lem:mltt-confluence} takes level equivalence to commute with every rule; the
rule set states this as \texttt{RuleSet.\allowbreak LevelCompat}, which $\delta$ and $\iota$
satisfy because their right-hand sides are built from the levels and the arguments of their
left-hand sides. With it, level equivalence commutes with reduction, and confluence makes
$\defeq$ transitive. The remaining clauses of the lemma are proved: closure under the term
formers, under substitution, and under reduction; the injectivity of $\Pi$, $\Sigma$, $\Id$,
applied families, and abstractions; and separation, stated for the head former of a term, which
is a universe, $\Pi$, $\Sigma$, $\Id$, or a constant on which no rule fires. Applied families are
constant applications of the last kind.

The global environment of Definition~\ref{def:mltt-environment} is a sequence of declarations,
\texttt{Decl}: axioms and definitions with their level-parameter counts, types, and bodies, and
inductive declarations. An inductive declaration (Definition~\ref{def:mltt-inductive}) is the data
\texttt{IndDecl}: the names of the family and of its recursor, the number of level parameters, the
parameter and index telescopes, the result level, and the constructors, each with its name, its
argument telescope, whose entries are marked non-recursive or recursive, a recursive entry given by
its telescope $\Theta_j$ and its indices $\bar s_j$, and its result indices. Telescopes are
contexts, so each entry is typed under the entries before it. The declaration determines the types
of the family, of the constructors, of the motive, of the minor premises, and of the recursor as
Section~\ref{sec:mltt-inductive} states them. A declaration records whether its recursor eliminates
into every universe; if it does, the elimination level $v$ is the last level parameter of the
recursor, and otherwise $v$ is $\lzero$. Strict positivity (Definition~\ref{def:mltt-positivity}) is
a predicate on the data, read on the erasures of the argument types and index terms, since the
definition reads an annotation as the term it annotates. Rule~\tttrule{$\iota$} is the relation
\texttt{IndDecl.iotaR}: the right-hand side of the rule for a constructor is a template over the
parameters, the motive, the minor premises, and the fields as variables, which the rule instantiates
by the simultaneous substitution \texttt{Term.instRev} of the arguments matched, a field that is one
of the result indices of its constructor, at the type of that index, replaced by the corresponding
index argument of the recursor (\texttt{IndDecl.\allowbreak forcedIndex},
\texttt{IndDecl.\allowbreak forceFields}). The relation has the closure properties of a rule set,
since the template is closed over its pattern context, each redex has one reduct when the
declaration binds distinct names, and the relation is stable with respect to the parallel reduction
of any rule set in which the constructors carry no rules, which is the form in which the non-overlap
of $\iota$ with the other rules enters. The transparent definitions of an environment \texttt{Env}
give Rule~\tttrule{$\delta$}, and the inductive declarations in force give their $\iota$-rules;
together they are the rule set \texttt{Env.rules}, which is stable and commutes with level
equivalence, since level substitution is a congruence (\texttt{Level.\allowbreak
inst\_congr\_subst}), so confluence and the properties of definitional equality hold over every
environment. The rules of a declaration fire only when its bodies are closed and, for an inductive
declaration, when it is closed, binds distinct names, and is the declaration in force for its names,
which holds of every declaration of a well-formed environment.

The three judgment forms of Definition~\ref{def:mltt-judgements} are one inductive family,
\texttt{Derivable}, over the judgment shapes \texttt{Judgment}, so that an induction over
derivations is the induction of one family; \texttt{WfCtx}, \texttt{HasType}, and \texttt{Sub}
name the forms. The judgments are relative to the environment, to a rule set that
Rules~\tttrule{Conv} and \tttrule{S-Conv} consult through definitional equality, and to the
number of level parameters in scope. The subsumption rules carry the typing premises of
Figure~\ref{fig:mltt-subsumption}, so that both sides of every derivable subsumption are types
(\texttt{Sub.valid}), as Definition~\ref{def:mltt-subsumption} requires.
Rule~\tttrule{$\Pi$-Prop} is derived from \tttrule{$\Pi$-F} by \tttrule{Conv}. Of the
metatheory, the well-formedness of the context of every judgment,
Lemma~\ref{lem:mltt-env-weakening}, and the weakening and substitution clauses of
Lemma~\ref{lem:mltt-substitution} are proved. The two context changes of that lemma, the
insertion of a type at a depth and the instantiation of the variable at a depth, are relations
between contexts, since with de Bruijn indices the entries above the point of change are lifted
or instantiated as well. Both proofs assume the declared types of the environment closed.
Well-formedness of environments, Figure~\ref{fig:mltt-environment}, is the judgment
\texttt{Env.WF}, in which each declaration is checked in the environment of the declarations
before it, under that environment's rules. Rule~\tttrule{E-Ind} requires the names of the
declaration fresh and distinct and the declaration admissible,
\texttt{IndDecl.\allowbreak Admissible}: the conditions of
Definition~\ref{def:mltt-admissible}, and the elimination restriction of
Definition~\ref{def:mltt-recursor}, that the recursor eliminates into every universe only if
the result level is at least $\lsucc\,\lzero$ or the declaration is a subsingleton, with the
subsingleton criterion of Definition~\ref{def:mltt-subsingleton} the predicate
\texttt{IndDecl.\allowbreak Subsingleton}. The constructor condition~(2) is stated as the
kernel checks it, \texttt{IndDecl.\allowbreak CtorTyped}: over the environment extended by
the family alone, the argument telescope is well formed over the parameters, and the result
indices, and the indices of each recursive argument, are typed against the index telescope at
the parameters, each against its entry instantiated by the earlier ones
(\texttt{ArgsTyped}). The condition of the definition, that the constructor's type is a type,
follows (\texttt{CtorTyped.\allowbreak ctorType\_sort}). The converse reads the typing of
the indices off a derivation of that condition through the inversion of subsumption at
$\Pi$-types, one of the inversion lemmas of Chapter~\ref{ch:checking}, so the formal condition
is the stronger one. That the type of the recursor is a type over the declaration,
which the chapter leaves implicit, is proved from the typing of the family and of the
constructors (\texttt{IndDecl.\allowbreak recType\_sort}): the motive type is a type over
the parameters, since the family applied to the parameters and the indices is a type; each
minor premise is a type over the earlier ones, the motive, and the parameters, since the
constructor applied to the parameters and its fields has the family at its result indices,
and each induction hypothesis applies the motive to the lifted indices of a recursive argument
and to the argument applied to its telescope; and the conclusion applies the motive to the
index variables and the major premise. The lemmas this needs are those of telescopes: $\Pi$
over a telescope is a type exactly when the telescope is well formed and its body is a type;
weakening by a telescope inserted in the middle of a context lifts the entries above it at
their depths; a term of $\Pi$-telescope type applied to the variables of its telescope has
the body; the application to arguments typed against a telescope has the simultaneous
instance of the result; and substituting the variables of a telescope for the variables of a
term lifts it. A derivable judgment has its free variables
in its context, so the types and
bodies of a well-formed environment are closed, and so is every inductive declaration in it,
which discharges the hypotheses of the metatheory and the conditions under which the
$\iota$-rules fire.

The axiom $\irrel$ of Section~\ref{sec:mltt-prop} is the declaration \texttt{irrelDecl},
without level parameters; Rule~\tttrule{PI} is \tttrule{Const} followed by three
applications of \tttrule{$\Pi$-E} in every environment that declares it
(\texttt{HasType.irrel}), the declaration is well formed in every environment that does not
bind its name (\texttt{Env.WF.irrel}), and the function with value $\emptyset$ inhabits the
interpretation of its type in every model structure (\texttt{ModelStructure.\allowbreak
irrel\_mem}). The propositions of Sections~\ref{sec:mltt-subsumption} and
\ref{sec:mltt-identity} are
proved as derivations: Proposition~\ref{prop:mltt-spi-eta}, with the $\eta$-expansion typed
by \tttrule{Sub} at the variable and at the application; Proposition~\ref{prop:mltt-uip},
over an environment that declares $\irrel$, with uniqueness of identity proofs by
Rule~\tttrule{PI} and axiom~K as a term over any context
in which $A$ and $a$ are typed, with the motive's applications converted by
$\beta$-reduction; and
Proposition~\ref{prop:mltt-id-inductive}, with $\mathsf{Eq}$ an inductive declaration whose
recursor type is computed to be the type of $\Jelim$, whose $\iota$-rule rewrites the
recursor applied to a constructor term to the minor premise, which is a subsingleton, and
which is admissible, so that Rule~\tttrule{E-Ind} extends every well-formed environment in
which its names are fresh by it. The translation between the primitive and the inductive
presentation is not formalized. Propositions~\ref{prop:mltt-negative} and
\ref{prop:mltt-pi-large-elim} are proved over concrete environments, which the
well-formedness judgment rejects and which the typing rules, relative to any environment,
accept: the closed inhabitants of $\tttFalse$, the proposition without constructors, are the
chapter's terms, typed with the $\iota$-rules of the recursors through \tttrule{Conv}, the
environment of the second declaring $\irrel$, and
the term $g\,(\mathsf{mk}\,g)$ of the first reduces to itself in two steps.

The fragments $\mathcal{K}_0$ and $\mathcal{K}_1$ of Definition~\ref{def:mltt-k0} are the
family \texttt{Bidir} over the judgment shapes \texttt{KJudgment}, with \texttt{Inf},
\texttt{Chk}, and \texttt{SubK} naming inference, checking, and the relation $\subty_K$; a flag
selects the fragment, and the $\Pi$-case of subsumption, with joinable domains, is the rule it
enables. The rules are those of
Figure~\ref{fig:mltt-bidir} and the inferring rules for the remaining formers, with premises in
checking mode where the declarative rule has a premise of known type, and \tttrule{Chk-Refl}
with endpoints joinable with its argument. The conventions of the definition are premises: a reduct
named in a premise comes
with a derivation that it is a type, the expected type of a checking judgment comes with one,
and the type of an application is
$B[\tttann{a}{A}/x]$; the conventions that fix which derivation the model interprets are not
imposed. Rule~\tttrule{Inf-Var} has no context premise, and the soundness theorem assumes a
well-formed context. The statement of
Section~\ref{sec:mltt-pi-sigma} that every bidirectional derivation erases to a declarative one
is proved for both fragments, by induction over derivations with \tttrule{Conv} at the named
reducts, the annotated substitution read through \tttrule{Ann-$\tttstep$}, and the $\Pi$-case
of subsumption read through \tttrule{S-$\Pi$}; the proof uses generation lemmas for $\Pi$,
$\Sigma$, and $\Id$, and the codomain premise of \tttrule{S-$\Pi$}, formed under the domain of
the right side, is reached by narrowing along \tttrule{S-Conv}: a judgment in the context
$\ctx, x{:}A, \Delta$ holds in $\ctx, x{:}A', \Delta$ when $\ctx \vdash A' \subty A$.
Transitivity of $\subty_K$ (Lemma~\ref{lem:mltt-k1-trans}) is proved by induction on the first
subsumption with cases of the second (\texttt{SubK.trans}): a cumulativity meets a
subsumption from a universe, which leads to a universe (\texttt{SubK.\allowbreak
sort\_inv\_left}), and two $\Pi$-cases compose through the joinable domains, the codomain
premises moved between them by context conversion. An earlier version of $\mathcal{K}_1$ had a
$\Pi$-case contravariant in the domain and transitivity as a case of the definition, since the
composition of two such $\Pi$-cases narrows the codomain premise of the first, which no measure
on derivations bounds; with joinable domains transitivity is a lemma.

Level instantiation holds for judgments: a judgment with $\bar u$ in scope holds at levels
$\bar\ell$ over the parameters in scope, with the context, the term, and the type substituted,
and the environment unchanged, since its declarations carry their own parameters. This is the
form in which Rule~\tttrule{Const} uses a declaration. With it, validity holds: the type of a
typed term is a type, over an environment whose declared types are types, which a well-formed
environment is. Validity is what the conventions of Definition~\ref{def:mltt-k0} assume of
inferred types.

Of the ingredients of Proposition~\ref{prop:mltt-k1-complete}, the following are proved: the
erasure $|t|$, with a term definitionally equal to its erasure and two terms with one erasure
definitionally equal; that $\subty_K$ respects definitional equality on both sides;
Lemma~\ref{lem:mltt-k0-annotated}~(1), for an expected type that is a type; the closure of
checking under $\subty_K$ up to annotation, by \tttrule{Inf-Ann} and \tttrule{Chk-Sub}, and
without annotation (\texttt{Chk.sub}), by induction on the checking derivation through the
inversions of $\subty_K$ at a $\Pi$-type on either side (\texttt{SubK.\allowbreak pi\_inv\_left},
\texttt{SubK.\allowbreak pi\_inv\_right}); weakening of the bidirectional systems; the substitution
of a term that infers
the type of a variable, whose variable case inserts the derivation of the substituted term, with
Lemma~\ref{lem:mltt-k0-annotated}~(2) the instance for an annotated term; and context
conversion, that a judgment holds when an entry of its context is replaced by a definitionally
equal term, with the terms unchanged and an inferred type changed up to definitional equality.
In a well-formed context, the type inferred for a term is a type, with the applied motive of
\tttrule{Id-E} a universe (\texttt{Inf.valid}).

Proposition~\ref{prop:mltt-k1-complete} is proved. A type of $\mathcal{K}_1$ is a term that
infers a universe, \texttt{KType}, and a well-formed context of $\mathcal{K}_1$,
\texttt{KWfCtx}, has every entry a type under the entries before it; the systems have no
context judgment, so the predicate is external, and the type of a variable of such a context
is a type. Level instantiation holds for the bidirectional systems as for the declarative
one, which is how Rule~\tttrule{Const} reads a declaration whose type is a type of
$\mathcal{K}_1$ in the empty context, the hypothesis on the environment
(\texttt{Env.KTyped}). The translation is by induction on the declarative derivation, with
the statement quantified over every well-formed annotation of the context, so that the
premises of a rule are translated into one context; a context has such an annotation by the
same induction. A term checked against a universe is a type, by the inversion of $\subty_K$ at
a universe; the reduct that an elimination names is read off the
translated type of the head under its outermost annotations, which is a type by inversion of
\tttrule{Inf-Ann} and \tttrule{Chk-Sub}, and the head is annotated with its translated type,
so that no inversion of $\subty_K$ at a $\Pi$-type is used; Rule~\tttrule{S-$\Pi$} is
translated the same way, with the typing premises of Figure~\ref{fig:mltt-subsumption}
supplying the reducts, and Rule~\tttrule{S-Trans} by the transitivity of $\subty_K$. The levels that
the universes name are well formed by declarative
validity. The result for a closed term, with the annotated term and type definitionally equal
to the originals, is what Theorem~\ref{thm:mltt-declarative-consistency} uses.
Lemma~\ref{lem:mltt-env-annotation}, which brings a well-formed environment into the form the
proposition assumes, is formalized except for its model clause (\texttt{Env.WF.annotate}). An
environment annotates another when their
declarations do so in order, with the same names and levels and types and bodies of one
erasure; an inductive declaration annotates another when its telescopes, indices, and
constructor arguments have the same erasures, and then the types it derives and the
right-hand sides of its $\iota$-rules have the same erasures as those of the other. Clauses
(2) and (3) are proved for every pair of annotated environments that are well formed: the two
environments have one definitional equality, since a $\delta$-step of either unfolds a
constant to bodies of one erasure, and an $\iota$-step of either is an $\iota$-step of the
other, up to the reducts' erasure, the declaration in force in one being annotated by the one
in force in the other, which the freshness of names in a well-formed environment fixes; and
every judgment over the original holds over the annotation, with \tttrule{Conv} at each
constant, whose original type is typed over the annotation by level instantiation and
weakening. Clause (1), the construction of the annotation, is proved by induction on the
well-formedness derivation (\texttt{Env.WF.annotate}), each declaration
translated over the annotation of the declarations before it. The
typing of the recursor in $\mathcal{K}_1$ that the inductive case needs is proved
(\texttt{IndDecl.\allowbreak krecType\_sort}): for a declaration whose family type is a
type of $\mathcal{K}_1$ and whose constructors satisfy the kernel's condition in
$\mathcal{K}_1$ (\texttt{KCtorTyped}: the argument telescope well formed, the result and
recursive indices checked against the index telescope), the type of the recursor is a type
of $\mathcal{K}_1$ over the declaration, with the recursor's level parameters in scope. The
proof follows the declarative one with the inferred types tracked up to erasure, since
\tttrule{Inf-App} substitutes the annotated argument: at each application the inferred type
of the function is a type, its reduct under the outermost annotations is a $\Pi$-type, and
the argument is checked against the domain by conversion, the two erasures being equal. No
typing of the declared types is assumed; the types of the constructors are types of
$\mathcal{K}_1$ by the same argument, and the universe of elimination is well formed by the
choice of the recursor's level parameters. The annotation of a declaration is strictly
positive exactly when the declaration is
(\texttt{IndDecl.\allowbreak Annotates.\allowbreak positive\_iff}), since
Definition~\ref{def:mltt-positivity} reads an annotation as the term it annotates and an
annotation has the erasures of the original. The inductive case is proved
(\texttt{IndDecl.\allowbreak annotate\_ind}): the telescopes of the family and of each
recursive argument and the index terms are translated by
Proposition~\ref{prop:mltt-k1-complete}, each in the annotated context built from the
translations before it, with the family as the bare head of every recursive argument type
and result. The kernel's constructor condition holds in $\mathcal{K}_1$ for the annotated
constructors, and admissibility over the annotated environment follows by soundness of
$\mathcal{K}_1$ and conversion: the arguments are typed against the entries of the index
telescope because each instantiated entry is a type by validity of the application to the
earlier arguments (\texttt{ArgsTyped.\allowbreak of\_defEq}), and the subsingleton test
reads the result indices and the types on the erasure, as
Definition~\ref{def:mltt-subsingleton} does (\texttt{IndDecl.\allowbreak Annotates.\allowbreak
indexAtType}). For the universe constraint, the entries of a
recursive argument's telescope are translated at the levels that generation reads off the
original constraint (\texttt{HasType.\allowbreak pis\_inv\_levels}), and the result
universe of the family bounds the universe of its applications from below
(\texttt{HasType.\allowbreak const\_spine\_sort\_le}), so that the annotated argument
type is formed below the bound. The original declared types are types over the annotated
environment by the typing of the recursor over any environment typing the family and the
constructors at their declared types (\texttt{IndDecl.\allowbreak recType\_sort\_of}).
The syntactic part of
Theorem~\ref{thm:mltt-declarative-consistency}, that a closed typing judgment over a
well-formed environment translates into $\mathcal{K}_1$ over an annotation of the
environment, follows (\texttt{HasType.translate\_env}).

Annotated reduction (Lemma~\ref{lem:mltt-k0-annotated}~(3) to (5), and
Lemma~\ref{lem:mltt-k1-annotated}) is the relation \texttt{RedA}, indexed by the fragment whose
judgments its side conditions use, by the context, and by the position of the term, inferring or
checking, with the rules of the environment a parameter; one relation serves both fragments, with
the rule that moves the annotation of a head out of an application (\texttt{RedA.appAnn}) in both,
its side condition a subsumption from the type the head infers to the annotation. The rules with an
annotated head read the annotation through a reduct that is a type, and erasure in inference
position needs the inferred type joinable with the annotation, not subsumed under it: with
subsumption, the body of an abstraction could drop to a smaller universe, which no rule of
$\mathcal{K}_0$ absorbs (the example after the definition of $\tttstep_a$). Subject reduction (3) is
proved by induction on the derivation, for both fragments, with the inferred type of the reduct
joinable with the original and a checked term checked against every type above its inferred type
(\texttt{Bidir.srMotive}), relative to environment rules that preserve inference up to joinability
over a well-formed context. It rests on the inversions of $\subty_K$ at a $\Pi$-type on either side,
with joinable domains (\texttt{SubK.\allowbreak pi\_inv\_left}, \texttt{SubK.\allowbreak
pi\_inv\_right}), and at a type whose head is neither a universe nor $\Pi$ nor a constant with
rules, at which subsumption is joinability (\texttt{SubK.\allowbreak rigid\_left},
\texttt{SubK.\allowbreak rigid\_right}); on the closure of checking under subsumption
(\texttt{Chk.sub}), which an erasure in checking position uses; and, for a moved annotation, on the
typing of the application at the annotated argument (\texttt{Chk.\allowbreak app\_push}). The
annotated $\delta$-rule preserves inference for definitions whose bodies check in the fragment
(\texttt{Env.KTypedBodies}, \texttt{Env.DeltaRA.inf}), and so does the annotated $\iota$-rule
(\texttt{Env.IotaRA.inf}). For the latter, the arguments of the redex are typed against the
recursor's telescope by spine inversion (\texttt{Inf.\allowbreak const\_spine\_inv}) and the fields
of the major premise against the constructor's, the two agreeing at the parameters, the indices, and
the levels by the injectivity of the applied family; the pattern context, at the levels of the
redex, is instantiated by the arguments annotated with its entries (\texttt{Term.\allowbreak
annotateArgs}), which infer them (\texttt{KArgsDefEq.\allowbreak annotate\_inf}), a forced field's
index argument checking against the field's type by the condition on that type, which erases to the
type of the index at the earlier result indices (\texttt{Term.\allowbreak instRev\_instRev}); and
the template is typed once in the pattern context (\texttt{IndDecl.\allowbreak template\_inf}), the
minor premise applied to the field variables and to the induction hypothesis terms, each checked
against its hypothesis in the recursor's telescope (\texttt{IndDecl.\allowbreak ihTerm\_chk}), so
that the instance infers a type joinable with the type of the redex (\texttt{IndDecl.\allowbreak
iotaRA\_inf}). The environment version (\texttt{Inf.redA\_env}) takes the declarations in force well
formed as the kernel checks them (\texttt{Env.KIndWf}: the levels of the declared types among their
parameters, and the indices of each recursive field as many as the family's). The $\Jelim$-rule
preserves inference up to joinability, since $\refl\,a$ checks only against an identity whose
endpoints are joinable with $a$ (\texttt{Inf.jRefl\_step}). Every annotated step erases to at most
one step (\texttt{RedA.erase\_redStar}), and a step of the erasure of a typable term lifts to
annotated steps (\texttt{Bidir.liftMotive}): the annotations on the head of a $\beta$-redex move out
of the application down to the last, which the rule with an annotated head reads
(\texttt{Inf.lift\_beta}), and those above a pair under a projection erase in inference position,
subsumption at a $\Sigma$-type being joinability (\texttt{Inf.peel\_sigma}); for an environment
annotating a well-formed one, the $\delta$-steps lift by the correspondence of the bodies
(\texttt{Env.AnnotatedBy.\allowbreak deltaRA\_lift}) and the $\iota$-steps by moving the annotations
on the partial applications of the recursor out of its spine and out of the application to the major
premise (\texttt{Inf.\allowbreak spine\_peel}, \texttt{Inf.\allowbreak push\_all}), with subject
reduction typing the intermediate terms, and erasing those at the top of the major premise and on
the partial applications of the constructor in checking position
(\texttt{Env.AnnotatedBy.\allowbreak iotaRA\_lift}), so the environment version
(\texttt{Inf.lift\_env}) takes the declarations typed as the kernel checks them. Two terms are
joinable exactly when their erasures are, the erasures reducing through annotation-free terms
(\texttt{RuleSet.DefEq.erase\_iff}, clause (5)), and two annotated reducts of a term have joinable
erasures (clause (6), confluence up to erasure, \texttt{RedAStar.\allowbreak erase\_joinable\_env},
from the converse of (4) and \texttt{Env.rules\_confluent}). The stratified model uses clauses (3)
and (4): a step keeps a defined value (Lemma~\ref{lem:mltt-k0-strat}(2)), and joinable typed terms
have one value through the lifted reductions (Lemma~\ref{lem:mltt-k0-strat}(3)); clause (3) is the
subject reduction of $\mathcal{K}_0$ and $\mathcal{K}_1$ to which the remark on the typed-equality
route after Theorem~\ref{thm:mltt-k0-consistency} refers.

An earlier formalization of $\mathcal{K}_1$, with a contravariant $\Pi$-case, had a relation
of its own, with the argument of a $\beta$-redex at an annotated head annotated twice, erasure in
checking position only for terms that are not abstractions, and the codomain premises narrowed
from one domain to another; with joinable domains the relation of $\mathcal{K}_0$, with the rule
that moves an annotation, serves both fragments.

Lemma~\ref{lem:mltt-k0-sorts} (1) is proved for observations (\texttt{ObsSort}): paths of
applications to arguments checked against a type below the domain, first projections, and second
projections at the first projection of the term, ending at a type joinable with a universe. The
proofs of both clauses are stated for K$_0$ and K$_1$ alike, with the flag of the system as a
parameter, which gives Lemma~\ref{lem:mltt-k1-sorts}. The
induction over the derivation of the first term (\texttt{Bidir.sortsMotive}) is made under
simultaneous substitutions of joinable arguments inferring the entries of a telescope
(\texttt{KArgsInf}), with the second term's annotations peeled off (\texttt{ObsSort.peel}) and
the abstraction case extending the substitution by the argument, annotated by the type it checks
against and then by the domain (\texttt{SortsAgree.lam}). An observation passes along a
subsumption to the type below (\texttt{Bidir.obsSubMotive}, by induction on the length of the
path and, inside it, on the subsumption): at a $\Pi$-case of K$_1$ the argument keeps its place,
the domains being joinable, and the rest of the path follows the codomain premise instantiated
by the argument annotated by the domain of the right side. Agreement of the universes themselves is
the case of the empty path
(\texttt{Chk.sort\_zeq}, \texttt{Inf.sort\_zeq}). Observations, and not the types, are
compared: two terms of one erasure need not have joinable types, since a pair annotated by two
$\Sigma$-types whose second components agree only at its first component checks against both.
The case of joinable types (\texttt{Inf.sort\_zeq\_of\_defEq}) holds for rules compatible with
erasure, as those of every environment are (\texttt{Env.rules\_eraseCompat}), whose erased steps
lift to annotated steps preserving inference; the lifting of reductions of the erasure goes
through erased steps, a step followed by the erasure of the reduct (\texttt{RuleSet.ERed},
\texttt{RuleSet.RedStar.ered}, \texttt{Inf.lift\_ered}). Over an environment annotating a
well-formed one, with its definitions and inductive declarations typed as the kernel checks them
(\texttt{Env.KTypedBodies}, \texttt{Env.KIndWf}), the hypotheses are discharged by clauses (3)
and (4) (\texttt{Inf.sort\_zeq\_of\_defEq\_env}), in both fragments. The lifting is stated for
any
annotated reduction whose steps preserve inference and to which erased steps lift
(\texttt{Inf.lift\_ered}, \texttt{Inf.relStar}).

Lemma~\ref{lem:mltt-k0-sorts} (2) is proved the same way, for observations that end at a term
whose type is joinable with a type formed at a universe (\texttt{ObsProof}, with
\texttt{UnivOf}); two ends at joinable types are compared by the case of joinable types of (1),
taken as a hypothesis (\texttt{SortJoin}). The induction (\texttt{Bidir.proofsMotive}) differs
where an observation ends at an abstraction, a pair, or a reflexivity proof: the universe of the
$\Pi$-type is zero exactly where that of its codomain is, and the body is compared by the
induction hypothesis without the substitution, the universe of the codomain being unchanged by
it (\texttt{ProofsAgree.lam}); a pair is compared through its two components
(\texttt{Inf.sigma\_univ}, \texttt{UnivOf.subst\_ann}); and an identity type is a proposition
(\texttt{ObsProof.ident\_zeq}). An observation of a type ends at once at a universe
that is not $\Prop$ (\texttt{ObsProof.sort\_zeq}), which also covers the annotations that
change a type by cumulativity. An observation that ends at a $\Pi$-case of K$_1$ compares the
codomains under the domain of the right side, the codomain of the left side typed there by
context conversion (\texttt{Bidir.obsProofSubMotive}). The statement for checked terms is
\texttt{Chk.proof\_zeq}; over an annotation of a well-formed environment with kernel-checked
declarations, \texttt{Chk.proof\_zeq\_env} discharges \texttt{SortJoin} by
\texttt{Inf.sort\_zeq\_of\_defEq\_env}, in both fragments.

The model of a typed definitional equality of the remark after
Theorem~\ref{thm:mltt-k0-consistency}, for the fragment $\mathcal{K}_0^{\mathrm{J}}$ of
$\mathcal{K}_0$ whose joinability premises are typed equations, is formalized relative to an
abstract model structure, \texttt{ModelStructure}: a type of values with the codes $\mathrm{lam}$,
$\mathrm{app}$, and $\langle a, b \rangle$ and their projections, the elements $\mathrm{El}$
of a value, the predicates of being a semantic type and a proposition, the universes as a
sequence of values, and the operations $\widehat\Pi$, $\widehat\Sigma$, and $\Id$, with the
equations that the proofs use as the axioms of the structure:
$\mathrm{app}(\mathrm{lam}(f), x) = f(x)$ on the elements of the domain, two codes whose
functions agree on the domain equal, the elements of $\widehat\Pi(A, F)$ the codes on
$\mathrm{El}(A)$ with values in the codomains and those of $\widehat\Sigma(A, F)$ the pairs,
$\widehat\Pi$ and $\widehat\Sigma$ in the universes of Rules~\tttrule{$\Pi$-F} and
\tttrule{$\Sigma$-F}, the universes cumulative, every element of a universe a semantic type and
every element of $\Prop$ a proposition, which has at most one element, and $\Id$ a proposition
inhabited by $\emptyset$ exactly when its two endpoints are equal. A code need not record its
domain, so Aczel's trace coding is an instance; the set-theoretic instantiation, in ZFC with
inaccessibles, is not constructed. The interpretation is structural on terms, relative to an
environment, a function from indices to values, and to a valuation of the level parameters,
with the constants valued by the model of the environment. The substitution lemma, that
$t[a/x]$ has the value of $t$ at the environment that gives $x$ the value of $a$, is proved for
lifting and for substitution at every depth, with level instantiation moved into
the valuation, and two terms identical up to level equivalence, as a term and its erasure, have
one value. The fragment $\mathcal{K}_0^{\mathrm{J}}$ is the family \texttt{BidirJ}, over the shapes
of $\mathcal{K}_0$ together with the equation
$\ctx \vdash a \defeq b : A$ and the arguments of a redex checked against the telescope of a
$\Pi$-type, with the typing rules of $\mathcal{K}_0$, and with no equation between distinct
proofs. The model of an environment, \texttt{EnvModel}, is the data of
Definition~\ref{def:mltt-obligation} with the semantics of the declarations folded in: a
valuation of the constants under which each constant lies in the interpretation of its
declared type, a semantic type, and under which the rules $\delta$ and $\iota$ preserve the
value of a redex whose arguments lie in the telescope of the declared type of its head
(\texttt{SemArgs}); the $\delta$-part follows for a valuation that sends each transparent
definition to the value of its body (\texttt{EnvModel.\allowbreak delta\_sound}), and the
$\iota$-part, the semantics of inductive declarations, is a hypothesis.
The consistency of $\mathcal{K}_0^{\mathrm{J}}$ is one induction over its derivations
(\texttt{EnvModel.\allowbreak kj\_sound}), each case from the induction hypotheses alone. The
equivalence of untyped and typed conversion that the remark leaves open is, for $\mathcal{K}_0$,
the proposition \texttt{JoinBridge}, and the translation of $\mathcal{K}_0$ into
$\mathcal{K}_0^{\mathrm{J}}$ takes it as a hypothesis: every
$\mathcal{K}_0$ derivation in a context of types is a $\mathcal{K}_0^{\mathrm{J}}$ derivation of
the same judgment (\texttt{Bidir.toJ}), each joinability premise between types becoming an
equation at a universe that contains both, with the type inferred for a head a type by
validity in $\mathcal{K}_0$, and the endpoint premises of \tttrule{Chk-Refl} becoming
equations by the first clause of the equivalence.

The stratified model of Theorems~\ref{thm:mltt-k0-consistency}
and~\ref{thm:mltt-k1-consistency} is formalized relative to an abstract structure as well,
\texttt{SModel}, for both fragments, the flag of the system selecting the derivations from
which the sorts are read: codes $\mathrm{lam}(A, f)$ with application
defined exactly at the elements of the recorded domain, pairs and projections, the data and
the propositional $\widehat\Pi$ and $\widehat\Sigma$ (\texttt{piD}, \texttt{piP},
\texttt{sigmaD}, \texttt{sigmaP}), the universes of the levels below a bound, and $\Id$, with
every element of a proposition equal to $\emptyset$. The sorts are \texttt{KIsProof} and
\texttt{KIsProp} (\texttt{KStrata}): a term of the erasure up to level equivalence checks, in a
well-formed context joinable with the given one entry by entry (\texttt{CtxJoin}), against a type
formed at a universe that is $\Prop$ at the valuation. Lemma~\ref{lem:mltt-k0-sorts} makes every
derivation give one answer (\texttt{KIsProof.iff\_chk}, \texttt{KIsProp.iff\_inf}), from which the
commutation of the sorts with substitution, weakening, and annotated reduction follows; no term with
a type former at its core is a proof, typed or not (\texttt{KIsProof.not\_pi} and the like). The
interpretation \texttt{SModel.interp} takes values in \texttt{Option}, consults the sorts at
variables, constants, abstractions, applications, pairs, projections, and $\Jelim$, and chooses the
former of a $\Pi$- or $\Sigma$-type by \texttt{KIsProp}; it depends on the term through its erasure
up to level equivalence and on the context up to joinability (\texttt{SModel.interp\_congr}).
Lemma~\ref{lem:mltt-k0-strat} (1) and (2) are inductions on derivations (\texttt{Bidir.sWeak},
\texttt{Bidir.sSubst}, \texttt{Bidir.sInv}), the last following \texttt{Bidir.srMotive} case by
case, with the case of a proof uniform: a step keeps being a proof, and a proof denotes $\emptyset$
(\texttt{SModel.interp\_proof\_eq}). (3) lifts the two reductions to the common reduct
(\texttt{SModel.interp\_join\_env}). The interpretation of a typed term also commutes with level
instantiation (\texttt{Bidir.sInstL}, the sorts by \texttt{KIsProof.instL\_chk} and
\texttt{KIsProp.instL}), and the value of a closed typed term does not depend on the context
(\texttt{SModel.interp\_closed}). The model of an environment, \texttt{SEnvModel}, values each
constant, at each valuation of its level parameters, in the interpretation of its closed declared
type in the empty context, a proposition where the type is one, values each transparent definition
as its body, and assumes that the rules $\iota$ keep the defined values of typed redexes. The two
lemmas give the membership of a constant in its type at its levels in every context
(\texttt{SEnvModel.const\_mem'}) and the soundness of $\delta$ (\texttt{SEnvModel.delta\_sound}), so
that only $\iota$, the semantics of inductive declarations, remains assumed beyond the obligation of
the axioms. Theorems~\ref{thm:mltt-k0-consistency} and~\ref{thm:mltt-k1-consistency} are one
induction over the derivations of $\mathcal{K}_0$ or $\mathcal{K}_1$ (\texttt{Bidir.sValid},
\texttt{SEnvModel.sound}, \texttt{SEnvModel.consistency}), over an environment annotating a
well-formed one whose declared types, definitions, and inductive declarations are typed in the
fragment (\texttt{Env.KTyped}, \texttt{Env.KTypedBodies}, \texttt{Env.KIndWf}); the statement for an
inferring $\Pi$- or $\Sigma$-type carries the sorts of its components (\texttt{SEnvModel.CompOK}),
and the case of $\Jelim$ reads the universe of the two instances of the motive off
\texttt{Inf.jMotive\_app\_sort}. At the $\Pi$-case of $\mathcal{K}_1$ the domains denote one value
by the invariance under joinability, the two $\Pi$-types are propositions together
(\texttt{SubK.isProp\_iff}, by induction on the subsumption), and the codomains are included
pointwise under an environment extended by the domain, so a subsumption is an inclusion and no
coercion enters the model. Theorem~\ref{thm:mltt-declarative-consistency} is
\texttt{HasType.consistency}: a closed declarative judgment over a well-formed environment holds
over an annotation of it (\texttt{AnnotatedBy.\allowbreak transport}) and translates into
$\mathcal{K}_1$ with the type annotated (\texttt{HasType.\allowbreak translate\_closed}), and a type
and its annotation have one value (\texttt{SModel.interp\_congr}); the theorem takes the definitions
and inductive declarations of the annotation typed as the kernel checks them. An environment whose
declarations are checked in this way is an annotation of itself (\texttt{Env.AnnotatedBy.refl}, with
\texttt{Env.Annotates.refl}), so the theorem applies to it directly (\texttt{HasType.\allowbreak
consistency\_of\_kernel}).

The structure has an instance in Lean's sets (\texttt{SSet.model}), from sets
$V_0 \in V_1 \in \cdots$ that are Grothendieck universes (\texttt{IsGrothendieck}: transitive,
with $\emptyset$, and closed under pairs, power sets, and unions of families indexed by an
element), the sets $V_{\kappa_n}$ of the chapter. A semantic type is the set of its elements.
$\mathrm{lam}(A, f)$ is the graph of $f$ on $A$, a set of Kuratowski pairs from which $A$ is
read off, and $\mathrm{app}$ reads the graph; $\Prop$ is the set of subsets of $\{\emptyset\}$
and the universe $\mathcal{U}_n$ the set $V_n$; the data $\widehat\Pi(A, F)$ is the set of the
graphs of the dependent functions and the data $\widehat\Sigma(A, F)$ the set of the pairs. The
axioms of the structure ask for no tags, and the instance has none. The chain is a hypothesis,
as the inaccessible cardinals are in the chapter, and it has a universe at every level, as an
unbounded structure asks (\texttt{SModel.Unbounded}). With it the empty environment has a model
(\texttt{SSet.envModelNil}), and in the empty environment no closed declarative term has the
type $\Pi P{:}\Prop.\,P$ (\texttt{HasType.\allowbreak consistency\_nil}): its codomain at
$\emptyset \in \Prop$ denotes $\emptyset$. The environment that declares $\tttFalse$ alone
(\texttt{falseEnv}) is well formed and kernel-checked, the type of its recursor a type of
$\mathcal{K}_1$ by \texttt{IndDecl.\allowbreak krecType\_sort}, and has a model
(\texttt{SSet.\allowbreak falseModel}): $\tttFalse$ denotes $\emptyset$, and the recursor
$\emptyset$ where its type is a proposition and otherwise the code of the constant function
$\emptyset$ on the interpretation of the type of the motive, which lies in the interpretation of
the type since at every motive the codomain is a $\widehat\Pi$ over $\emptyset$; there is no
$\iota$-rule. So no closed declarative term has the type $\tttFalse$ in that environment
(\texttt{HasType.\allowbreak consistency\_false}). The interpretation of a term typed in an
environment is the same in every extension of it, under valuations of the constants that agree
on its constants (\texttt{Bidir.\allowbreak sEnvWeak}): on such a term an extension reads the
sorts as the environment does (\texttt{KIsProof.\allowbreak env\_iff},
\texttt{KIsProp.\allowbreak env\_iff}), from the term's own derivation lifted to the extension
and sort agreement there. The model of an environment is built declaration by declaration on
this.

The instance has inductive families. At values of its parameters, the family of a declaration is
the least fixed point of the operator of its constructors on families of sets indexed by the
values of the indices (\texttt{indFam}), a constructor application denoting the code of its
number and its arguments (\texttt{ctorCode}), and an inductive proposition $\{\emptyset\}$ or
$\emptyset$. The fibers of a data family lie in the universe of its level
(\texttt{indFam\_mem}), by a regular cardinal above the arities of the constructors
(\texttt{IsGrothendieck.\allowbreak lfp\_mem}) and the types of the constructors' arguments lying
in the universe below in the model in which the family denotes the singletons
(\texttt{argsIn\_of\_unit}). The recursor denotes the function defined by the least relation
closed under its $\iota$-equations (\texttt{recGraph}), which is functional on the fixed point
(\texttt{recGraph\_functional}): for a data family because constructor applications are
injective codes, and for a family of propositions eliminated into every universe because the
kernel's elimination restriction makes the declaration a subsingleton, whose argument values
its indices determine. Its value lies in its type (\texttt{RecSetting.\allowbreak recVal\_mem}).
The rules $\iota$ keep defined values at every environment, as \texttt{SEnvModel} asks
(\texttt{RecSetting.\allowbreak iota\_sound}). In a data family the major premise denotes the
code of a constructor application, which fixes the fields' values, and a forced field's index
argument has the same value. In a subsingleton as the kernel checks it
(\texttt{IndDecl.\allowbreak KForcedSubsingleton}), each field is a proof, which denotes
$\emptyset$, or forced: Rule~\tttrule{$\iota$} reads it off the index argument, whose value is
that of the field's result index at the argument values (\texttt{RecSetting.\allowbreak
index\_val}), as for the field $x$ of $\tttAcc$. Read off the constructor term, such a field
could have no value at an environment where the redex has one, since the major premise is a proof
and denotes $\emptyset$. A well-formed environment whose
declarations the kernel checks, each over the declarations before it (\texttt{Env.KChecked}), and
whose axioms have values in the model (\texttt{AxiomsInhabited}) has a model
(\texttt{envData\_nonempty}), built declaration by declaration: an axiom denotes its value, a
definition the value of its body, and an inductive declaration the family's code, the constructors'
values, and the recursor's value. The recursor's value reads its own telescope, which mentions only
the constants of the environment of the family with the constructors as axioms, where the recursor's
type is a type of $\mathcal{K}_1$ (\texttt{IndDecl.\allowbreak krecType\_sort\_of}), and which is
defined by validity in the model of that environment. So, relative to a chain of Grothendieck
universes with $\omega \in V_0$, no closed term of $\mathcal{K}_1$ or of the declarative system has
the type $\Pi P{:}\Prop.\,P$ in such an environment (\texttt{Chk.\allowbreak
consistency\_of\_model}, \texttt{HasType.\allowbreak consistency\_of\_model}); there, a closed
term of the declarative system has its value in the value of its type
(\texttt{HasType.\allowbreak sound\_of\_model}). For an environment
without axioms a chain of Grothendieck universes suffices, since its universes above the first
contain $\omega$ (\texttt{HasType.\allowbreak consistency\_of\_checked}). The environments of
$\tttAcc$ and of $\mathsf{Unit}$, $\tttBool$, and $\Nat$, declared one after the other, are well
formed and checked by the kernel, with the derivations written out (\texttt{SAcc},
\texttt{SUnitBoolNat}), so the theorem applies to them.

The hypothesis of the formal theorems, a Grothendieck universe at every level, is that of
Werner's soundness theorem for CIC, which assumes an increasing sequence of inaccessible
cardinals $(\lambda_i)_{i \in \mathbb{N}}$ and interprets the universes as the sets
$V_{\lambda_i}$; its corollary that no closed term has the type $\Pi \alpha{:}\Prop.\,\alpha$ is
\texttt{HasType.\allowbreak consistency\_nil}~\citep{werner1997}. \citet{lee-werner2011} assume
countably many inaccessible cardinals, and \citet{felicissimo2026acc} postulate Grothendieck
universes $V_0 \in V_1 \in \cdots$ in an intuitionistic set theory embedded in Rocq: there the
universes are axioms, here a premise of the theorem. Chapter~\ref{ch:mltt} states its theorems
with this hypothesis (Section~\ref{sec:mltt-consistency}, ``The model''). The sharper form
bounds each derivation. Werner proves that CIC with $n + 1$ universes, excluded middle, and
type-theoretic choice and description axioms is consistent if ZFC with $n$ inaccessible
cardinals is, and Carneiro's
soundness theorem for Lean holds as soon as the first $k$ of an increasing sequence of strong
limit cardinals are inaccessible, $k$ being the largest universe in the
derivation~\citep[Theorem~6.7]{carneiro2019}; Lean is then consistent relative to ZFC with $n$
inaccessible cardinals for each $n$, which is optimal. An increasing sequence of
inaccessible cardinals proves the consistency of ZFC with $n$ inaccessible cardinals for each
$n$, so the hypothesis of the formal theorems is the stronger one. The sharper form needs a bound
on each derivation, not on its judgment: Rule \tttrule{Inf-App} names a $\Pi$-type joinable with
the type of the head, so for $f : \Nat \to \Nat$ a derivation of $f\,a \infr \Nat$ may name
$\Pi x{:}((\tttlam{y}{\Srt\,99} \Nat)\,\Nat).\,\Nat$, which reduces to $\Nat \to \Nat$ and
mentions $\Srt\,99$. The formal theorems have this form relative to a structure whose universes
are those of the levels below a bound $N$ (\texttt{SModel.bound}). The copy \texttt{BidirB} of
$\mathcal{K}_0$ and $\mathcal{K}_1$ asks of the rules that name a universe, \tttrule{Sort},
\tttrule{$\Pi$-F}, \tttrule{$\Sigma$-F}, and cumulativity, that the level they name lie below $N$
at the valuation of the levels. These are the rules at which validity uses a universe of the
structure, so validity and the consistency theorems hold for its derivations
(\texttt{BidirB.\allowbreak sValid}, \texttt{SEnvModel.\allowbreak consistencyB}). Every derivation
is one with the bound $\top$ (\texttt{Bidir.\allowbreak toBidirB}), which gives the theorems for
every derivation in a structure with a universe at every level (\texttt{SModel.Unbounded}).
The instance in Lean's sets has the bounded form (\texttt{SSet.modelOf}): a chain whose sets
$V_n$ for the levels $n + 1$ below the bound are Grothendieck universes, each a member of the
next, satisfies the axioms for the levels below the bound, and gives the bounded theorems in the
empty environment and in the environment of $\tttFalse$ (\texttt{ChkB.\allowbreak
consistency\_nil}, \texttt{ChkB.\allowbreak consistency\_false}). For an inaccessible cardinal
$\kappa$ the set $V_\kappa$ is a Grothendieck universe (\texttt{isGrothendieck\_vonNeumann}):
its elements are the sets of rank below $\kappa$, which is a limit, and an element has fewer than
$\kappa$ elements, $\kappa$ being a strong limit, so a union of the family it indexes has rank
below $\kappa$ by regularity. Inaccessible cardinals $\kappa_0 < \cdots < \kappa_{N-1}$ thus give
the bounded theorems for the derivations whose universes lie below $N + 1$
(\texttt{ChkB.\allowbreak consistency\_nil\_\allowbreak of\_inaccessible}, \texttt{ChkB.\allowbreak
consistency\_false\_\allowbreak of\_inaccessible}), and Lean provides finitely many: the cardinality
of each
universe of Lean below $w$ is inaccessible among the cardinals of the universe~$w$
(\texttt{Cardinal.\allowbreak IsInaccessible.univ}). Those of the universes $0$, $1$, and $2$
give, with no hypothesis, that no closed term checks in $\mathcal{K}_0$ or $\mathcal{K}_1$
against $\Pi P{:}\Prop.\,P$ in the empty environment, or in $\mathcal{K}_1$ against $\tttFalse$ in
the environment that declares it, by a derivation whose universes lie below $4$, those of
$\Srt\,0$ to $\Srt\,3$ (\texttt{ChkB.\allowbreak consistency\_nil\_univ},
\texttt{ChkB.\allowbreak consistency\_false\_univ}). The bound is fixed in the statement, since
Lean does not index its universes by a natural number. For an environment with inductive
declarations the formal theorems keep the hypothesis of a universe at every level, the form of
Werner's soundness theorem and of the model of Felicissimo et al. Its model interprets each
declaration at every valuation of the declaration's levels, by validity in the model of the
declarations before it, and the derivations of a recursor name universes that grow with the
valuation, the motive's $\Srt\,v$. A bounded form needs the environment model only at the
instances of the constants within the bound, a bound on the levels of the constants that
$\delta$ and $\iota$ keep, and the validity of the bounded system for the derivations that the
construction derives; it is not formalized.

By Gödel's second incompleteness theorem, each of these consistency proofs runs in a theory of
greater consistency strength than the system it concerns, and Werner's two encodings make the
boundary exact: CIC with $n + 1$ universes is consistent relative to ZFC with $n$ inaccessible
cardinals, and ZFC with $n$ inaccessible cardinals is encoded in CIC with $n + 2$ universes,
excluded middle, and the description axiom, a result Werner states ahead of the formal
construction of the inaccessible cardinals in Coq. \citet[Section~1.2]{carneiro2019} gives the
same pair for Lean: ZFC with $n + 1$ inaccessible cardinals proves the consistency of Lean with
$n + 1$ universes, and Lean with $n + 2$ universes that of ZFC with $n$ inaccessible cardinals.
Lean, the metatheory of the formalization, proves the bounded statements one bound at a time,
with finitely many Grothendieck universes taken from its own universe levels. It does not prove
the full statement if $\mathcal{K}_1$ is as strong as CIC: an environment may declare the choice
axioms, which meet their obligation, and the inductive types of Werner's encoding, and if
$\mathcal{K}_1$ over such an environment encodes ZFC with $n$ inaccessible cardinals for each
$n$, as CIC does, then a proof in Lean, without a hypothesis, that every environment meeting its
obligation is consistent would prove the consistency of Lean.

The formalization differs from these works in three respects. Its system is the bidirectional
$\mathcal{K}_1$ with joinability as its conversion, carried to the declarative system by
completeness; Werner and \citet{miquel-werner2003} also convert untyped terms, while Lee and
Werner and Carneiro type their definitional equality. Its inductive types are declarations of
the environment under one schema, whose semantics the formal theorems take as the
$\iota$-condition of the environment model; Werner sketches the semantics of inductive types,
and Carneiro models eight primitive inductive types and translates the general schema into
them, leaving the translation of the recursors and the $\iota$-rules open. And it is mechanized
from the syntax to the set-theoretic instance, where Felicissimo et al.\ mechanize the
set-theoretic constructions and interpret the syntax on paper.

\begin{center}
\small
\setlength{\tabcolsep}{4pt}
\begin{longtable}{@{}>{\raggedright\arraybackslash}p{0.22\textwidth}
  >{\raggedright\arraybackslash}p{0.36\textwidth}
  >{\raggedright\arraybackslash}p{0.2\textwidth}
  >{\raggedright\arraybackslash}p{0.12\textwidth}@{}}
\toprule
Book & Lean & Module & Status \\
\midrule
\endhead
Definition~\ref{def:mltt-levels} &
  \texttt{Level}, \texttt{Level.eval}, the \texttt{Preorder} instance, the \texttt{Setoid}
  instance, the order of cumulativity \texttt{Level.CumLe} & \texttt{Level} & definition \\
Proposition~\ref{prop:mltt-levels} &
  \texttt{Level.\allowbreak succ\_congr}, \texttt{max\_congr}, \texttt{imax\_congr},
    \texttt{imax\_zero}, \texttt{imax\_succ}, \texttt{inst\_le\_inst},
  \texttt{inst\_congr}; for $\cumle$, \texttt{CumLe.refl}, \texttt{CumLe.trans},
  \texttt{CumLe.le}, \texttt{CumLe.of\_le}, \texttt{CumLe.inst}, \texttt{cumLe\_zero\_iff},
  \texttt{not\_cumLe\_zero\_succ} & \texttt{Level} & proved \\
Definition~\ref{def:mltt-terms} &
  \texttt{Term} & \texttt{Term} & definition \\
Definition~\ref{def:mltt-substitution} &
  \texttt{Term.liftN}, \texttt{Term.inst}, \texttt{Term.instL}, \texttt{Level.inst}, with
  the lifting and substitution lemmas & \texttt{Term} & definition \\
The rules $\delta$ and $\iota$ as an interface to reduction (Definitions~\ref{def:mltt-environment}
  and \ref{def:mltt-recursor}) &
  \texttt{RuleSet}: shape, determinism, closure under \texttt{liftN}, \texttt{inst}, and
  \texttt{instL} & \texttt{Reduction} & definition \\
Definition~\ref{def:mltt-defeq}, one-step reduction $\tttstep$ and $\tttstep^{*}$ &
  \texttt{RuleSet.\allowbreak Red}, \texttt{RuleSet.\allowbreak RedStar}; parallel reduction
  \texttt{RuleSet.\allowbreak Par} with \texttt{Par.liftN}, \texttt{Par.inst},
  \texttt{Red.toPar}, \texttt{Par.toRedStar} & \texttt{Reduction} & definition \\
Lemma~\ref{lem:mltt-confluence}, confluence of $\tttstep$ &
  \texttt{RuleSet.\allowbreak confluent}, for a \texttt{Stable} rule set, from
  \texttt{Par.diamond} by \texttt{Relation.\allowbreak church\_rosser} &
  \texttt{Confluence} & proved \\
Definition~\ref{def:mltt-defeq}, joinability $\defeq$, and Rule~\tttrule{Sort-$\defeq$} &
  \texttt{Term.LEquiv}, \texttt{RuleSet.\allowbreak LevelCompat},
  \texttt{RuleSet.\allowbreak DefEq}, \texttt{DefEq.sort} & \texttt{DefEq} & definition \\
Lemma~\ref{lem:mltt-confluence}, equivalence, closure, injectivity, and separation &
  \texttt{Red.lequiv}, \texttt{DefEq.\allowbreak equivalence}, the congruences
  \texttt{DefEq.pi} to \texttt{DefEq.mkApp}, \texttt{DefEq.of\_red}, \texttt{DefEq.pi\_inj},
  \texttt{sigma\_inj}, \texttt{ident\_inj}, \texttt{lam\_inj}, \texttt{mkApp\_const\_inj},
  \texttt{sort\_inj}, \texttt{head\_eq} & \texttt{DefEq} & proved \\
Lemma~\ref{lem:mltt-substitution}~(3) &
  \texttt{RedStar.inst}, \texttt{DefEq.inst}, \texttt{DefEq.liftN} & \texttt{DefEq} & proved \\
Definition~\ref{def:mltt-inductive}, inductive declarations, with the types of
  Definition~\ref{def:mltt-recursor} &
  \texttt{IndDecl}, \texttt{Ctor}, \texttt{CtorArg}, with \texttt{IndDecl.\allowbreak familyType},
  \texttt{ctorType}, \texttt{motiveType}, \texttt{minorType}, \texttt{recType} &
  \texttt{Inductive} & definition \\
Definition~\ref{def:mltt-positivity}, on the erasures &
  \texttt{IndDecl.Positive}, \texttt{IndDecl.\allowbreak Annotates.\allowbreak positive\_iff} &
  \texttt{Inductive}, \texttt{IndAnnotation} & definition, proved \\
The non-dependence condition of Definition~\ref{def:mltt-admissible}, on the erasures &
  \texttt{IndDecl.Nondep}, \texttt{IndDecl.\allowbreak argsNondep}, \texttt{Term.TeleFree},
  \texttt{IndDecl.\allowbreak Annotates.\allowbreak nondep\_iff} &
  \texttt{Inductive}, \texttt{IndAnnotation} & definition, proved \\
Definition~\ref{def:mltt-subsingleton} and the elimination restriction of
  Definition~\ref{def:mltt-recursor} &
  \texttt{IndDecl.\allowbreak Subsingleton}, \texttt{IndDecl.\allowbreak IndexAtType} (the
  argument is a result index at the type of that index, read on the erasure),
  \texttt{IndDecl.\allowbreak largeElim},
  \texttt{IndDecl.\allowbreak elimLevel}, \texttt{IndDecl.\allowbreak recLevels},
  \texttt{IndDecl.\allowbreak Admissible.elim} & \texttt{Inductive}, \texttt{Wellformed} &
  definition \\
Rule~\tttrule{$\iota$} (Definition~\ref{def:mltt-recursor}) &
  \texttt{IndDecl.iotaR}, with the templates \texttt{IndDecl.\allowbreak template}, the
  simultaneous substitution \texttt{Term.instRev}, and the forced fields
  \texttt{IndDecl.\allowbreak forceFields}; \texttt{iotaR\_det}, \texttt{iotaR\_liftN},
  \texttt{iotaR\_inst}, \texttt{iotaR\_instL}, \texttt{iotaR\_stable\_wrt},
  \texttt{iotaR\_levelCompat} & \texttt{InstRev}, \texttt{Iota} & definition, proved \\
Definition~\ref{def:mltt-environment} &
  \texttt{Decl}, \texttt{Env}, \texttt{Env.find?}, \texttt{Env.type?}, \texttt{Env.body?},
  \texttt{Env.InForce} & \texttt{Environment} & definition \\
Rules~\tttrule{$\delta$} and \tttrule{$\iota$} of an environment
  (Figure~\ref{fig:mltt-environment}) &
  \texttt{Env.rules}, with \texttt{Env.\allowbreak rules\_stable},
  \texttt{Env.\allowbreak rules\_levelCompat}, \texttt{Env.\allowbreak rules\_confluent},
  \texttt{Env.\allowbreak rules\_defEq\_equivalence} & \texttt{Environment} & proved \\
Definitions~\ref{def:mltt-context} and \ref{def:mltt-judgements}, with the typing rules of
  Sections~\ref{sec:mltt-judgements}--\ref{sec:mltt-identity}, \ref{sec:mltt-environment},
  and \ref{sec:mltt-prop} &
  \texttt{Ctx}, \texttt{Judgment}, \texttt{Derivable}, with \texttt{WfCtx}, \texttt{HasType},
  and \texttt{Sub}; one constructor per rule & \texttt{Typing} & definition \\
Rule~\tttrule{$\Pi$-Prop} & \texttt{HasType.piProp} & \texttt{Typing} & proved \\
Lemma~\ref{lem:mltt-env-weakening} &
  \texttt{Derivable.mono}, \texttt{Derivable.\allowbreak env\_weakening}, with
  \texttt{DefEq.mono} & \texttt{Typing} & proved \\
Lemma~\ref{lem:mltt-substitution}~(1) and (2) &
  \texttt{Ctx.InsertAt}, \texttt{Ctx.SubstAt}, \texttt{Derivable.\allowbreak weakenAt},
  \texttt{Derivable.\allowbreak substAt}, with the instances
  \texttt{HasType.weaken\textsubscript{0}} and \texttt{HasType.subst\textsubscript{0}} at the
  top of the context & \texttt{Substitution} & proved \\
Figure~\ref{fig:mltt-environment}, Rules~\tttrule{E-Empty} to \tttrule{E-Ind}, with
  Definition~\ref{def:mltt-admissible} &
  \texttt{Env.WF}, \texttt{IndDecl.\allowbreak Admissible},
  \texttt{IndDecl.\allowbreak CtorTyped}, \texttt{IndDecl.\allowbreak ArgsRecurTyped},
  \texttt{IndDecl.\allowbreak ArgsUniverseOk}, \texttt{IndDecl.\allowbreak withFamily} &
  \texttt{Wellformed} & definition \\
Definition~\ref{def:mltt-admissible}~(2) from \texttt{CtorTyped} &
  \texttt{IndDecl.\allowbreak CtorTyped.\allowbreak ctorType\_sort},
  \texttt{IndDecl.\allowbreak family\_app\_params} & \texttt{Wellformed} & proved \\
Telescopes (Definition~\ref{def:mltt-recursor}) &
  \texttt{HasType.\allowbreak pis\_inv}, \texttt{HasType.\allowbreak pis\_sort},
  \texttt{Derivable.\allowbreak weakenMid}, \texttt{HasType.\allowbreak app\_vars},
  \texttt{HasType.\allowbreak app\_vars\_mid}, \texttt{ArgsTyped},
  \texttt{HasType.\allowbreak mkApp\_of\_argsTyped},
  \texttt{ArgsTyped.\allowbreak vars\_liftTele}, \texttt{Term.instRevAt},
  \texttt{Term.\allowbreak instRevAt\_varsFrom}, \texttt{Derivable.monoU},
  \texttt{Derivable.levelWF}, \texttt{Term.\allowbreak instL\_params} &
  \texttt{InstRev}, \texttt{Telescopes} & proved \\
The type of the recursor is a type (Definition~\ref{def:mltt-recursor}) &
  \texttt{IndDecl.\allowbreak recType\_sort}, with \texttt{IndDecl.\allowbreak motive\_app},
  \texttt{IndDecl.\allowbreak ih\_sort}, \texttt{IndDecl.\allowbreak ihs\_wf},
  \texttt{IndDecl.\allowbreak minorType\_sort} & \texttt{Recursor} & proved \\
Well-formedness, consequences (Section~\ref{sec:mltt-environment}) &
  \texttt{Derivable.\allowbreak closedMotive} (the free variables of a judgment lie in its
  context), \texttt{Env.WF.\allowbreak closedTypes}, \texttt{Env.WF.\allowbreak closedBodies},
  \texttt{IndDecl.\allowbreak Admissible.closed}, \texttt{IndDecl.\allowbreak recType\_closed} &
  \texttt{Wellformed}, \texttt{Iota} & proved \\
Generation for $\Pi$, $\Sigma$, and $\Id$ (used in Section~\ref{sec:mltt-consistency}) &
  \texttt{HasType.pi\_inv}, \texttt{HasType.\allowbreak sigma\_inv},
  \texttt{HasType.\allowbreak ident\_inv} & \texttt{Typing} & proved \\
Narrowing (the codomain premise of the $\Pi$-case under the domain of the right side) &
  \texttt{Derivable.narrow}, with \texttt{HasType.\allowbreak narrow\textsubscript{0}} &
  \texttt{Narrowing} & proved \\
Definition~\ref{def:mltt-subsumption}, both sides of a subsumption are types &
  \texttt{Sub.valid} & \texttt{Validity} & proved \\
Proposition~\ref{prop:mltt-spi-eta} &
  \texttt{HasType.\allowbreak eta\_of\_sub} & \texttt{Subsumption} & proved \\
Rule~\tttrule{PI}, the axiom $\irrel$ (Section~\ref{sec:mltt-prop}) &
  \texttt{irrelDecl}, \texttt{irrelTm}, \texttt{HasType.irrel}, \texttt{irrelType\_typed},
  \texttt{Env.WF.irrel}, \texttt{Env.type?\_irrel}, \texttt{ModelStructure.\allowbreak
  irrel\_mem} & \texttt{Irrel} & proved \\
Proposition~\ref{prop:mltt-uip} &
  \texttt{HasType.uip}, \texttt{kProof}, \texttt{kTerm}, \texttt{HasType.\allowbreak axiomK} &
  \texttt{Identity} & proved \\
Proposition~\ref{prop:mltt-id-inductive} &
  \texttt{eqDecl}, \texttt{eqDecl\_recType}, \texttt{eqDecl\_iotaR},
  \texttt{eqDecl\_subsingleton}, \texttt{eqDecl\_admissible}, \texttt{eqDecl\_wf}; the
  translation between the presentations is not formalized & \texttt{Identity} & proved \\
Proposition~\ref{prop:mltt-negative} &
  \texttt{badEnv\_false}, \texttt{badEnv\_cycle} & \texttt{Inconsistency} & proved \\
Proposition~\ref{prop:mltt-pi-large-elim} &
  \texttt{largeElimEnv\_false} & \texttt{Inconsistency} & proved \\
Definition~\ref{def:mltt-k0}, the fragments $\mathcal{K}_0$ and $\mathcal{K}_1$ &
  \texttt{KJudgment}, \texttt{Bidir}, with \texttt{Inf}, \texttt{Chk}, and \texttt{SubK}; the
  $\Pi$-case \texttt{Bidir.subPi}, with joinable domains, is enabled by the flag &
  \texttt{Bidirectional} & definition \\
Every $\mathcal{K}_0$ or $\mathcal{K}_1$ derivation erases to a declarative one
  (Section~\ref{sec:mltt-pi-sigma}) &
  \texttt{Inf.sound}, \texttt{Chk.sound}, \texttt{SubK.sound}, \texttt{SubK.sound\_sub} &
  \texttt{Bidirectional} & proved \\
Level instantiation (Section~\ref{sec:mltt-universes}, universe polymorphism) &
  \texttt{Red.instL}, \texttt{DefEq.instL}, \texttt{Derivable.instL} & \texttt{Validity} &
  proved \\
Validity: the type of a typed term is a type (Section~\ref{sec:mltt-consistency}) &
  \texttt{HasType.valid}, \texttt{HasType.\allowbreak valid\_of\_wf}, with
  \texttt{Env.WF.\allowbreak typedDecls}, \texttt{WfCtx.\allowbreak lookup\_type},
  \texttt{HasType.\allowbreak sort\_inv} & \texttt{Validity} & proved \\
The erasure $|t|$ (Section~\ref{sec:mltt-consistency}) &
  \texttt{Term.erase}, \texttt{RedStar.erase}, \texttt{DefEq.erase},
  \texttt{DefEq.\allowbreak of\_erase\_eq} & \texttt{Erasure} & proved \\
Lemma~\ref{lem:mltt-k0-annotated}~(1) &
  \texttt{Chk.conv}, with \texttt{Chk.\allowbreak expected\_wf},
  \texttt{SubK.defEq} & \texttt{Bidirectional} & proved \\
Closure of checking under $\subty_K$ up to annotation
  (Proposition~\ref{prop:mltt-k1-complete}) &
  \texttt{Chk.sub\_ann} & \texttt{Bidirectional} & proved \\
Weakening of $\mathcal{K}_0$ and $\mathcal{K}_1$ &
  \texttt{Bidir.weakenAt}, \texttt{Inf.\allowbreak weakenMany} &
  \texttt{Bidir\allowbreak Substitution} & proved \\
Lemma~\ref{lem:mltt-k0-annotated}~(2) &
  \texttt{Bidir.substAt}, with \texttt{Chk.inf\_ann}, \texttt{Chk.subst\_ann},
  \texttt{Inf.subst\textsubscript{0}}, \texttt{Chk.subst\textsubscript{0}}, and
  \texttt{SubK.subst\textsubscript{0}} & \texttt{Bidir\allowbreak Substitution} & proved \\
Context conversion in $\mathcal{K}_0$ and $\mathcal{K}_1$ &
  \texttt{Bidir.ctxConv}, with \texttt{Inf.pi\_sort}, \texttt{Inf.\allowbreak sigma\_sort},
  \texttt{Inf.\allowbreak ident\_sort}; every entry, \texttt{Inf.\allowbreak ctxConv\_suffix} &
  \texttt{BidirConv}, \texttt{KTypes} & proved \\
Types and well-formed contexts of $\mathcal{K}_1$ (the conventions of
  Definition~\ref{def:mltt-k0}) &
  \texttt{KType}, \texttt{KWfCtx}, \texttt{KWfCtx.lookup}, \texttt{Env.KTyped},
  \texttt{KType.const} & \texttt{KTypes} & definition, proved \\
Level instantiation in $\mathcal{K}_0$ and $\mathcal{K}_1$ &
  \texttt{Bidir.instL} & \texttt{KTypes} & proved \\
Inversion of $\subty_K$ at a universe &
  \texttt{SubK.sort\_inv}, \texttt{SubK.\allowbreak sort\_inv\_left},
  \texttt{Chk.\allowbreak ktype\_of\_sort} & \texttt{KTypes} & proved \\
Lemma~\ref{lem:mltt-k1-trans} &
  \texttt{SubK.trans}, with \texttt{Bidir.\allowbreak transMotive}; \texttt{Chk.sub}, with
  \texttt{Bidir.\allowbreak chkSubMotive}, \texttt{SubK.\allowbreak pi\_inv\_left},
  \texttt{SubK.\allowbreak pi\_inv\_right} & \texttt{KTypes}, \texttt{Annotated\allowbreak
  Reduction} & proved \\
The reduct under the outermost annotations &
  \texttt{Term.stripAnn}, \texttt{Inf.\allowbreak pi\_of\_stripAnn},
  \texttt{Inf.\allowbreak sigma\_of\_stripAnn}, \texttt{Inf.\allowbreak ident\_of\_stripAnn} &
  \texttt{Erasure}, \texttt{KTypes} & proved \\
Proposition~\ref{prop:mltt-k1-complete} &
  \texttt{Derivable.transMotive}, with \texttt{HasType.translate},
  \texttt{HasType.\allowbreak translate\_at}, \texttt{HasType.\allowbreak translate\_closed},
  \texttt{Sub.translate\_at}, \texttt{WfCtx.translate} & \texttt{Completeness} & proved \\
Validity in $\mathcal{K}_0$ and $\mathcal{K}_1$ (the convention of Definition~\ref{def:mltt-k0}
  on inferred types) &
  \texttt{Inf.valid}, with \texttt{Inf.\allowbreak jMotive\_app}, \texttt{Inf.sort\_inv},
  \texttt{Inf.\allowbreak pi\_sort\_wf} & \texttt{KValidity} & proved \\
Annotated declarations and environments (Lemma~\ref{lem:mltt-env-annotation}) &
  \texttt{IndDecl.\allowbreak Annotates}, with \texttt{familyType\_erase},
  \texttt{ctorType\_erase}, \texttt{recType\_erase}, \texttt{template\_erase},
  \texttt{iotaR\_corr}; \texttt{Decl.Annotates}, \texttt{Env.\allowbreak Annotates} with
  \texttt{find?\_corr}, \texttt{find?\_pos}, \texttt{inForce\_corr};
  \texttt{Env.FreshNames}, \texttt{Env.WF.\allowbreak freshNames} &
  \texttt{IndAnnotation}, \texttt{EnvAnnotation} & definition, proved \\
Lemma~\ref{lem:mltt-env-annotation} (2) and (3), for annotated well-formed environments &
  \texttt{Env.\allowbreak AnnotatedBy}, \texttt{AnnotatedBy.\allowbreak defEq},
  \texttt{AnnotatedBy.\allowbreak transport}, \texttt{Bidir.mono},
  \texttt{DefEq.iff\_of\_rules} & \texttt{EnvAnnotation} & proved \\
Lemma~\ref{lem:mltt-env-annotation} (1), the cases of axioms and definitions &
  \texttt{Env.AnnotatedBy.\allowbreak translate\_type},
  \texttt{Env.AnnotatedBy.\allowbreak snoc\_ax}, \texttt{Env.AnnotatedBy.\allowbreak snoc\_defn} &
  \texttt{EnvAnnotation}, \texttt{IndTranslation} & proved \\
Telescopes in $\mathcal{K}_1$: weakening by a telescope inserted in the middle of a
  context, application to the variables of a telescope, arguments checked against a
  telescope &
  \texttt{Bidir.weakenMid}, \texttt{KWfCtx.\allowbreak weakenMid}, \texttt{Inf.app\_vars},
  \texttt{Inf.\allowbreak app\_vars\_mid}, \texttt{KArgsTyped},
  \texttt{Inf.\allowbreak mkApp\_of\_kargsTyped}, \texttt{KType.pis\_sort},
  \texttt{KType.pis\_inv}, \texttt{Bidir.monoU} & \texttt{KTelescopes} & proved \\
The type of the recursor is a type of $\mathcal{K}_1$ over the declaration, from the typing
  of the family in $\mathcal{K}_1$ and the kernel's check of the constructors &
  \texttt{IndDecl.\allowbreak krecType\_sort}, with \texttt{IndDecl.\allowbreak KCtorTyped},
  \texttt{KCtorTyped.sound}, \texttt{KCtorTyped.\allowbreak ctorType\_ktype},
  \texttt{IndDecl.\allowbreak kmotive\_app}, \texttt{IndDecl.\allowbreak kih\_sort},
  \texttt{IndDecl.\allowbreak kminorType\_sort} & \texttt{KRecursor} & proved \\
Generation with universe bounds: subsumption raises the universe at the end of a
  $\Pi$-telescope, $\Pi$-types and constant applications at a universe, and arguments typed
  up to definitional equality &
  \texttt{Sub.\allowbreak codomainSort\_le}, \texttt{HasType.\allowbreak pi\_inv\_le},
  \texttt{HasType.\allowbreak pis\_inv\_levels},
  \texttt{HasType.\allowbreak const\_spine\_sort\_le}, \texttt{ArgsTyped.\allowbreak of\_defEq} &
  \texttt{Generation} & proved \\
Lemma~\ref{lem:mltt-env-annotation} (1), the inductive case, and the lemma for every
  well-formed environment &
  \texttt{IndDecl.\allowbreak annotate\_ind}, with \texttt{translate\_tele},
  \texttt{translate\_index\_args}, \texttt{IndDecl.\allowbreak translate\_args},
  \texttt{IndDecl.\allowbreak translate\_ctor}; \texttt{Env.WF.annotate},
  \texttt{HasType.\allowbreak translate\_env} & \texttt{IndTranslation} & proved \\
Annotated reduction $\tttstep_a$ (Section~\ref{sec:mltt-consistency}) &
  \texttt{RedA}, for both fragments, with \texttt{RedA.appAnn} (the moved annotation),
  \texttt{RedA.\allowbreak mode\_irrel}, \texttt{RedA.defEq};
  \texttt{Env.RulesA} with \texttt{Env.DeltaRA}, \texttt{IndDecl.\allowbreak iotaRA} (the
  arguments annotated by \texttt{Term.\allowbreak annotateArgs}), \texttt{Env.IotaRA},
  \texttt{Env.RulesA.defEq}, \texttt{Env.RulesA.shape} &
  \texttt{Annotated\allowbreak Reduction} & definition \\
Lemma~\ref{lem:mltt-k0-annotated}~(3) &
  \texttt{Bidir.srMotive}, \texttt{Inf.redA}, \texttt{Chk.redA}, \texttt{RedAStar.inf}, with
  \texttt{SubK.\allowbreak pi\_inv\_left}, \texttt{SubK.\allowbreak pi\_inv\_right},
  \texttt{Chk.sub}, \texttt{Chk.\allowbreak app\_push}; the $\delta$-case
  \texttt{Env.DeltaRA.inf}; the $\Jelim$-case \texttt{Inf.\allowbreak jRefl\_step} &
  \texttt{Annotated\allowbreak Reduction} & proved \\
Lemma~\ref{lem:mltt-k0-annotated}~(3), the $\iota$-case &
  \texttt{IndDecl.\allowbreak iotaRA\_inf}, with spine inversion
  \texttt{Inf.\allowbreak const\_spine\_inv} and \texttt{Inf.\allowbreak mkApp\_of\_defEq},
  the annotated arguments \texttt{Term.\allowbreak annotateArgs} and
  \texttt{KArgsDefEq.\allowbreak annotate\_inf}, instantiation of a telescope
  \texttt{Bidir.\allowbreak instRevTele}, the pattern context \texttt{IndDecl.\allowbreak patTele},
  the induction hypotheses \texttt{IndDecl.\allowbreak ihTerm\_chk}, and the template
  \texttt{IndDecl.\allowbreak template\_inf}; stated for $\mathcal{K}_0$ and $\mathcal{K}_1$,
  with \texttt{SubK.\allowbreak rigid\_left} at the major premise & \texttt{KSpine},
  \texttt{KTeleInst}, \texttt{KIota} & proved \\
Lemma~\ref{lem:mltt-k0-annotated}~(3) for the rules of an environment &
  \texttt{Inf.redA\_env}, \texttt{Chk.redA\_env}, \texttt{RedAStar.inf\_env}, with
  \texttt{Env.KTypedBodies}, \texttt{Env.KIndWf}, \texttt{Env.RulesA.inf}, and the
  $\iota$-case \texttt{Env.IotaRA.inf} & \texttt{Annotated\allowbreak Reduction} & proved \\
Lemma~\ref{lem:mltt-k0-annotated}~(4), the converse &
  \texttt{RedA.\allowbreak erase\_redStar}, \texttt{RedA.\allowbreak erase\_redStar\_env} &
  \texttt{Annotated\allowbreak Reduction} & proved \\
Lemma~\ref{lem:mltt-k0-annotated}~(4), the forward direction &
  \texttt{Bidir.liftMotive}, \texttt{Inf.lift}, \texttt{Chk.lift}, with
  \texttt{Inf.\allowbreak lift\_beta}, \texttt{Inf.\allowbreak peel\_sigma},
  \texttt{Chk.\allowbreak pair\_sigma\_reduct}, \texttt{RedAStar.\allowbreak annChk\_stripAnn},
  \texttt{RedAStar.\allowbreak stripAnn\_lift}; for an environment annotating a well-formed
  one \texttt{Inf.lift\_env}, \texttt{Chk.lift\_env}, with
  \texttt{Env.AnnotatedBy.\allowbreak deltaRA\_lift} and
  \texttt{Env.AnnotatedBy.\allowbreak iotaRA\_lift} (the annotations on the partial
  applications of the recursor moved out of the redex, \texttt{Inf.\allowbreak spine\_peel},
  \texttt{Inf.\allowbreak push\_all}) & \texttt{Annotated\allowbreak Reduction} & proved \\
Lemma~\ref{lem:mltt-k0-annotated}~(5) &
  \texttt{RuleSet.DefEq.\allowbreak erase\_iff}, with \texttt{Term.AnnFree},
  \texttt{AnnFree.red} & \texttt{Annotated\allowbreak Reduction} & proved \\
Lemma~\ref{lem:mltt-k0-annotated}~(6), confluence up to erasure &
  \texttt{RedAStar.\allowbreak erase\_joinable\_env}, from
  \texttt{RedAStar.\allowbreak erase\_redStar\_env} and \texttt{Env.rules\_confluent} &
  \texttt{Annotated\allowbreak Reduction} & proved \\
Lemma~\ref{lem:mltt-k1-annotated} &
  the declarations of Lemma~\ref{lem:mltt-k0-annotated} with the flag \texttt{k1}, the moved
  annotation \texttt{RedA.appAnn} with \texttt{Chk.\allowbreak app\_push},
  \texttt{Inf.\allowbreak lift\_beta}, and \texttt{Inf.\allowbreak push\_all} &
  \texttt{Annotated\allowbreak Reduction} & proved \\
Lemma~\ref{lem:mltt-k0-sorts}~(1), for types of one erasure up to level equivalence &
  \texttt{ObsSort} with \texttt{ObsSort.zeq}, \texttt{ObsSort.congr}, \texttt{ObsSort.sub},
  \texttt{ObsSort.peel}; \texttt{SortsAgree}, \texttt{Bidir.sortsMotive},
  \texttt{SortsAgree.lam}; \texttt{Chk.sort\_zeq}, \texttt{Inf.sort\_zeq} &
  \texttt{KSorts} & proved \\
Lemma~\ref{lem:mltt-k0-sorts}~(1), for joinable types &
  \texttt{RuleSet.EraseCompat}, \texttt{Env.rules\_eraseCompat}, \texttt{RuleSet.ERed},
  \texttt{RuleSet.RedStar.ered}, \texttt{Inf.lift\_ered}, \texttt{Inf.sort\_zeq\_of\_defEq};
  over an annotation of a well-formed environment with kernel-checked declarations
  \texttt{Inf.sort\_zeq\_of\_defEq\_env} & \texttt{KSorts} & proved \\
Lemma~\ref{lem:mltt-k0-sorts}~(2) &
  \texttt{UnivOf}, \texttt{SortJoin}, \texttt{ObsProof} with \texttt{ObsProof.zeq},
  \texttt{ObsProof.sub}, \texttt{ObsProof.peel}; \texttt{ProofsAgree},
  \texttt{Bidir.proofsMotive}, \texttt{ProofsAgree.lam}; \texttt{Chk.proof\_zeq}; over an
  annotation of a well-formed environment with kernel-checked declarations
  \texttt{Chk.proof\_zeq\_env} & \texttt{KProofs} & proved \\
Lemma~\ref{lem:mltt-k1-sorts} &
  the same declarations with the flag \texttt{k1}; at the $\Pi$-case
  \texttt{KJudgment.\allowbreak ObsSubMotive}, \texttt{Bidir.\allowbreak obsSubMotive},
  \texttt{KJudgment.\allowbreak ObsProofSubMotive},
  \texttt{Bidir.\allowbreak obsProofSubMotive}; for joinable types \texttt{Inf.relStar},
  \texttt{Inf.lift\_ered}, \texttt{Inf.\allowbreak sort\_zeq\_of\_defEq\_env};
  \texttt{Chk.\allowbreak proof\_zeq\_env}; the sorts \texttt{SortJoin.env},
  \texttt{Inf.\allowbreak redA\_chk\_env}, \texttt{KIsProof.\allowbreak redA\_inf\_env},
  \texttt{KIsProof.\allowbreak redA\_chk\_env}, \texttt{KIsProof.\allowbreak redA\_inf\_chk\_env},
  \texttt{KIsProp.\allowbreak redA\_env}; a subsumption keeps whether a type is a proposition,
  \texttt{SubK.\allowbreak isProp\_iff} & \texttt{KSorts}, \texttt{KProofs}, \texttt{KStrata} &
  proved \\
The model structure of a typed equality (the remark after
  Theorem~\ref{thm:mltt-k0-consistency}): codes, elements, semantic types, propositions,
  universes, $\widehat\Pi$, $\widehat\Sigma$, $\Id$, with the equations the proofs use &
  \texttt{ModelStructure} (\texttt{app\_lam}, \texttt{lam\_ext}, \texttt{El\_pi},
  \texttt{El\_sigma}, \texttt{El\_ident}, \texttt{pi\_mem}, \texttt{sigma\_mem},
  \texttt{sort\_mem}, \texttt{El\_sort\_mono}, \texttt{prop\_subsingleton}),
  \texttt{mem\_pi\_iff}, \texttt{mem\_sigma\_iff}, \texttt{app\_mem},
  \texttt{eq\_of\_mem\_ident} & \texttt{Model\allowbreak Structure} & interface \\
The interpretation of terms in that structure, relative to an environment and a valuation of
  the level parameters, with the constants valued by the environment model &
  \texttt{ModelStructure.\allowbreak interp}, \texttt{levelsVal}, \texttt{SemEnv.push},
  \texttt{liftEnv}, \texttt{instEnv}, \texttt{dropEnv} & \texttt{Interpretation} &
  definition \\
Its substitution lemma, for lifting and substitution at every depth, with level
  instantiation, level equivalence, and erasure &
  \texttt{ModelStructure.\allowbreak interp\_liftN}, \texttt{interp\_inst},
  \texttt{interp\_instL}, \texttt{interp\_lequiv}, \texttt{interp\_erase} &
  \texttt{Interpretation} & proved \\
The fragment $\mathcal{K}_0^{\mathrm{J}}$ with a typed equality &
  \texttt{JJudgment}, \texttt{BidirJ} & \texttt{KJudgmental} & definition \\
The model of an environment for $\mathcal{K}_0^{\mathrm{J}}$:
  Definition~\ref{def:mltt-obligation} as the valuation of the
  constants, each in its declared type, with the rules $\delta$ and $\iota$ sound on redexes
  whose arguments lie in the telescope; the $\delta$-part from the bodies &
  \texttt{ModelStructure.\allowbreak SemArgs}, \texttt{EnvModel} (\texttt{const\_mem},
  \texttt{rules\_sound}), \texttt{EnvModel.\allowbreak delta\_sound} & \texttt{KJModel} &
  interface \\
Environments of a context &
  \texttt{EnvModel.EnvOf}, \texttt{envOf\_nil}, \texttt{envOf\_push} & \texttt{KJModel} &
  definition \\
The consistency of $\mathcal{K}_0^{\mathrm{J}}$, relative to an environment model, as one
  induction over its derivations &
  \texttt{EnvModel.\allowbreak JMotive}, \texttt{EnvModel.\allowbreak kj\_sound},
  \texttt{interp\_mem\_of\_chk}, \texttt{interp\_mem\_of\_inf}, \texttt{interp\_eq\_of\_eq},
  \texttt{not\_chk\_of\_El\_eq\_empty} & \texttt{KJModel} & proved \\
The equivalence of untyped and typed conversion for $\mathcal{K}_0$, as a hypothesis &
  \texttt{JoinBridge} & \texttt{KJBridge} & definition \\
Uniqueness of inferred types and of universes in $\mathcal{K}_0$ and $\mathcal{K}_1$
  (after Definition~\ref{def:mltt-k0}), and the invariance of being a proposition under checking
  against a universe &
  \texttt{Ctx.Lookup.unique}, \texttt{Inf.unique}, \texttt{Inf.level\_unique},
  \texttt{Chk.zero\_iff} & \texttt{KUnique} & proved \\
The translation of $\mathcal{K}_0$ into $\mathcal{K}_0^{\mathrm{J}}$, relative to that
  equivalence and an environment model &
  \texttt{JoinBridge.types}, \texttt{KJudgment.ToJ}, \texttt{Bidir.toJ},
  \texttt{BidirJ.ident\_inv}, \texttt{EnvModel.\allowbreak k0\_consistency},
  \texttt{EnvModel.\allowbreak k0\_not\_chk\_of\_El\_eq\_empty} & \texttt{KJBridge} &
  proved \\
The sorts of the stratified model (Section~\ref{sec:mltt-consistency}, ``The stratified
  model'') &
  \texttt{CtxJoin}, \texttt{Chk.ctxJoin}, \texttt{Inf.ctxJoin}, \texttt{KIsProof},
  \texttt{KIsProp}, \texttt{KIsProof.iff\_chk}, \texttt{KIsProp.iff\_inf},
  \texttt{KIsProof.iff\_isProp}, \texttt{KIsProof.app\_iff}, \texttt{KIsProof.not\_pi};
  \texttt{KIsProof.subst\_chk}, \texttt{KIsProof.weaken\_chk},
  \texttt{KIsProof.\allowbreak redA\_inf\_env} & \texttt{KStrata} & proved \\
The stratified model structure and the interpretation &
  \texttt{SModel}, \texttt{SModel.interp}, \texttt{SModel.\allowbreak interp\_congr} &
  \texttt{SModel} & interface \\
Lemma~\ref{lem:mltt-k0-strat}~(1), substitution and weakening &
  \texttt{SModel.interp\_proof}, \texttt{Bidir.sWeak}, \texttt{SModel.\allowbreak interp\_liftMany},
  \texttt{Bidir.sSubst} & \texttt{SSubst} & proved \\
Lemma~\ref{lem:mltt-k0-strat}~(1), level instantiation, and closed terms &
  \texttt{KIsProof.instL\_chk}, \texttt{KIsProp.instL}, \texttt{KIsProp.closed},
  \texttt{Bidir.sInstL}, \texttt{SModel.\allowbreak interp\_closed} & \texttt{SInstL} &
  proved \\
Lemma~\ref{lem:mltt-k0-strat}~(2) and~(3) &
  \texttt{SModel.\allowbreak interp\_proof\_eq}, \texttt{Bidir.sInv}, \texttt{RedAStar.sInv},
  \texttt{SModel.\allowbreak interp\_join\_env} & \texttt{SInvariance} & proved \\
Theorems~\ref{thm:mltt-k0-consistency} and~\ref{thm:mltt-k1-consistency}, relative to a
  stratified model structure and an environment model whose rules $\iota$ keep defined
  values &
  \texttt{SEnvModel}, \texttt{SEnvModel.\allowbreak const\_mem'},
  \texttt{SEnvModel.\allowbreak delta\_sound}, \texttt{SEnvModel.\allowbreak rules\_sound},
  \texttt{SEnvModel.EnvOf}, \texttt{envOf\_push}, \texttt{SEnvModel.CompOK},
  \texttt{Bidir.sValid}, \texttt{SEnvModel.sound}, \texttt{SEnvModel.\allowbreak consistency} &
  \texttt{SValidity} & proved \\
Theorems~\ref{thm:mltt-k0-consistency} and~\ref{thm:mltt-k1-consistency} for the derivations
  whose universes lie below the bound of the structure &
  \texttt{SModel.bound}, \texttt{SModel.Unbounded}, \texttt{BidirB}, \texttt{BidirB.toBidir},
  \texttt{Bidir.toBidirB}, \texttt{BidirB.sValid}, \texttt{SEnvModel.soundB},
  \texttt{SEnvModel.\allowbreak consistencyB} & \texttt{SModel}, \texttt{KBounded},
  \texttt{SValidity} & proved \\
Theorem~\ref{thm:mltt-declarative-consistency}, over an annotation of the environment whose
  definitions and inductive declarations are kernel-checked, relative to a model of it &
  \texttt{HasType.\allowbreak consistency}; for an environment with kernel-checked
  declarations, its own annotation, \texttt{HasType.\allowbreak consistency\_of\_kernel}, with
  \texttt{Env.AnnotatedBy.refl} & \texttt{SValidity}, \texttt{EnvAnnotation} & proved \\
The stratified model structure in set theory, relative to a chain of Grothendieck universes, and
  Theorem~\ref{thm:mltt-declarative-consistency} in the empty environment &
  \texttt{IsGrothendieck}, \texttt{SSet.model}, \texttt{SSet.\allowbreak envModelNil},
  \texttt{HasType.\allowbreak consistency\_nil} & \texttt{SSet} & proved \\
The environment of $\tttFalse$ and its model; Theorem~\ref{thm:mltt-declarative-consistency}
  there, relative to the chain &
  \texttt{falseDecl}, \texttt{falseEnv\_wf}, \texttt{falseEnv\_ktyped},
  \texttt{SSet.\allowbreak falseModel}, \texttt{HasType.\allowbreak consistency\_false} &
  \texttt{FalseDecl}, \texttt{SFalse} & proved \\
Environment weakening of the stratified interpretation &
  \texttt{KIsProof.\allowbreak env\_iff}, \texttt{KIsProp.\allowbreak env\_iff},
  \texttt{Bidir.\allowbreak sEnvWeak}, \texttt{SModel.\allowbreak interp\_cv\_congr},
  \texttt{SModel.\allowbreak interp\_closedN\_congr} & \texttt{SEnvWeak} & proved \\
Semantic telescopes, and least fixed points of operators on families of sets in a
  Grothendieck universe &
  \texttt{SModel.TeleVals}, \texttt{SModel.lamTele}, \texttt{SModel.\allowbreak
  appVals\_lamTele\_inv}, \texttt{SModel.\allowbreak interp\_mkApp}, \texttt{SModel.\allowbreak
  pis\_mem\_lamTele}, \texttt{IsGrothendieck.\allowbreak lfp\_mem} & \texttt{STele},
  \texttt{SFix} & proved \\
The family of an inductive declaration as the least fixed point of the operator of its
  constructors, and the size of its fibers &
  \texttt{ctorCode}, \texttt{ArgVals}, \texttt{indOp}, \texttt{indFam},
  \texttt{indFam\_fix}, \texttt{indFam\_mem} & \texttt{SInd}, \texttt{SIndSize} & proved \\
The family's code and the constructors' values in their types &
  \texttt{famCode}, \texttt{ctorVal}, \texttt{interp\_famApp}, \texttt{teleVals\_of\_argVals},
  \texttt{famCode\_mem}, \texttt{ctorVal\_mem}, \texttt{argsIn\_of\_unit} &
  \texttt{SIndInterp} & proved \\
The recursor: the graph of its $\iota$-equations, functional and total on the fixed point, and
  its value in its type under the kernel's elimination restriction &
  \texttt{recGraph}, \texttt{recGraph\_functional}, \texttt{recFun\_spec}, \texttt{RecSetting},
  \texttt{recVal}, \texttt{IndDecl.KSubsingleton}, \texttt{RecSetting.\allowbreak
  det\_of\_subsingleton}, \texttt{RecSetting.\allowbreak recVal\_mem} & \texttt{SIndRec},
  \texttt{SIndRecType} & proved \\
The rules $\iota$ keep defined values, the forced fields reading the index arguments &
  \texttt{SModel.\allowbreak interp\_instRev}, \texttt{RecSetting.\allowbreak template\_val},
  \texttt{RecSetting.\allowbreak graph\_eq}, \texttt{IndDecl.\allowbreak KForcedSubsingleton},
  \texttt{RecSetting.\allowbreak reduct\_val}, \texttt{RecSetting.\allowbreak field\_proof},
  \texttt{RecSetting.\allowbreak index\_val}, \texttt{RecSetting.\allowbreak iota\_sound} &
  \texttt{SIota}, \texttt{SIotaSound} & proved \\
The model of a well-formed environment whose declarations the kernel checks over their
  prefixes and whose axioms have values &
  \texttt{IndDecl.KChecked}, \texttt{Env.KChecked}, \texttt{SSet.EnvData},
  \texttt{EnvData.ax}, \texttt{EnvData.defn}, \texttt{famData}, \texttt{ctorData},
  \texttt{IndDecl.\allowbreak krecType\_sort\_of}, \texttt{IndDecl.\allowbreak recTele\_free},
  \texttt{recSetting}, \texttt{EnvData.ind}, \texttt{envData\_nonempty} &
  \texttt{SEnvData}, \texttt{SEnvInd}, \texttt{SEnvRec}, \texttt{KRecursor} & proved \\
Theorems~\ref{thm:mltt-k1-consistency} and~\ref{thm:mltt-declarative-consistency} for such an
  environment, relative to a chain of Grothendieck universes &
  \texttt{SSet.\allowbreak omega\_mem\_succ}, \texttt{Chk.\allowbreak consistency\_of\_model},
  \texttt{HasType.\allowbreak consistency\_of\_model}, \texttt{HasType.\allowbreak
  consistency\_of\_checked}, \texttt{HasType.\allowbreak sound\_of\_model} &
  \texttt{SConsistency} & proved \\
The structure in set theory for the levels below a bound, and
  Theorems~\ref{thm:mltt-k0-consistency} and~\ref{thm:mltt-k1-consistency} for the derivations
  whose universes lie below it, in the empty environment and in that of $\tttFalse$ &
  \texttt{SSet.modelOf}, \texttt{SSet.\allowbreak envModelNilOf}, \texttt{ChkB.\allowbreak
  consistency\_nil}, \texttt{ChkB.\allowbreak consistency\_false} & \texttt{SSet},
  \texttt{SFalse} & proved \\
Grothendieck universes from inaccessible cardinals, and the bounded theorems relative to finitely
  many of them and, from Lean's universes, for the derivations whose universes lie below $4$ &
  \texttt{isGrothendieck\_vonNeumann}, \texttt{SSet.\allowbreak
  exists\_chain\_\allowbreak of\_inaccessible}, \texttt{ChkB.\allowbreak
  consistency\_nil\_\allowbreak of\_inaccessible}, \texttt{ChkB.\allowbreak
  consistency\_false\_\allowbreak of\_inaccessible}, \texttt{SSet.univCard},
  \texttt{ChkB.\allowbreak
  consistency\_nil\_univ}, \texttt{ChkB.\allowbreak consistency\_false\_univ} &
  \texttt{SUniverses} & proved \\
The environment of $\tttAcc$, eliminating into every universe: well formed, checked by the
  kernel, and consistent relative to the chain &
  \texttt{accDecl}, \texttt{accDecl\_admissible}, \texttt{accEnv\_wf},
  \texttt{accDecl\_kchecked}, \texttt{accEnv\_kchecked}, \texttt{HasType.\allowbreak
  consistency\_acc} & \texttt{SAcc} & proved \\
The environment of $\mathsf{Unit}$, $\tttBool$, and $\Nat$, data families eliminating into every
  universe, declared one after the other: well formed, checked by the kernel, and consistent
  relative to the chain &
  \texttt{unitDecl}, \texttt{natDecl}, \texttt{dataEnv\_wf}, \texttt{dataEnv\_kchecked},
  \texttt{HasType.\allowbreak consistency\_data} & \texttt{SUnitBoolNat} & proved \\
\bottomrule
\end{longtable}
\end{center}

Not formalized are the chain of Grothendieck universes at every level that the instance of the
stratified model takes as a hypothesis, which Lean, with finitely many inaccessible cardinals, does
not provide; the bounded theorems for environments with inductive declarations; the model clause
of Lemma~\ref{lem:mltt-env-annotation}; and the equivalence of untyped and typed conversion of the
remark after Theorem~\ref{thm:mltt-k0-consistency}, which is open and which no consistency theorem
uses. The kernel's
checks of the declarations (\texttt{Env.KTyped}, \texttt{Env.KTypedBodies}, \texttt{Env.KIndWf}) are
hypotheses of the consistency theorems, as they are of the chapter's, and hold of every environment
a kernel has accepted; the theorems of the instance take them over the prefixes
(\texttt{Env.KChecked}), from which the global ones follow. The obligation of
Definition~\ref{def:mltt-obligation} is taken as the data of the model, a valuation of the constants
with each in the interpretation of its type, rather than as the non-emptiness condition that such a
valuation witnesses; in the instance, each axiom's type has an element in every model of the
declarations before it (\texttt{AxiomsInhabited}). Inductive
families are declarations of the global environment with their constructors, recursors, and
$\iota$-rules, with admissibility a predicate on declarations and the metatheory proved relative to
the environment, which is how Theorem~\ref{thm:mltt-k0-consistency} is stated; the encodings of
inductive families by well-founded trees or by a universe of descriptions are not used, since the
first needs extensionality principles the theory lacks and the second changes the theory's syntax.

\section{Chapter~\ref{ch:justification-logic}}

Modules are in \texttt{EigeniusTheory/Justification}. The kernel enters through an interface,
\texttt{PropositionSystem}: a type of propositions whose equality stands for definitional
equality, which is the chapter's convention that propositions are taken up to $\defeq$; the
implication and the universal proposition read off a weak-head normal form; typing of instance
terms; and provability, closed under application and instantiation, which is what the factivity
theorem uses. The rule Conv of Definition~\ref{def:jl-rules} is therefore absorbed, and the
hypotheses of Theorem~\ref{thm:jl-decision}, a transitive definitional equality with injective
$\Pi$-types, hold in the interface by construction. The interface has an instance from every
well-formed environment of Chapter~\ref{ch:mltt} (\texttt{KernelInstance}): the closed
propositions up to definitional equality as a quotient, the implications and universal
propositions read off a representative, typing and provability of closed terms, and classical
decidability; with Theorem~\ref{thm:mltt-declarative-consistency}, no verified justification
justifies $\tttFalse$ in an environment without axioms that the kernel checks. Of
Corollary~\ref{thm:jl-decidable} the
decidability is proved and the complexity bound is not stated. The basic systems of
Section~\ref{sec:jl-basic} have their own syntax in \texttt{Basic}, with a tautology read as a
formula true under every valuation of the letters and of the justification assertions; soundness
for Mkrtychev models is proved, completeness is cited, and
Proposition~\ref{thm:jl-minimal-characterizes} is proved with completeness as a hypothesis. The
provenance semirings are commutative semirings in Mathlib's sense, with
$\jlNX$ as \texttt{MvPolynomial}. A chain state carries its derived resources with their
terms and propositions, $\jlWF$ is Mathlib's least fixed point of a monotone map on sets, and
the edge check is acyclicity, equivalent on the finite set $D$ to well-foundedness of the
reference relation. A signature morphism of Section~\ref{sec:jl-logic} acts on the interface:
on propositions, terms, and types, preserving the destructors and typing.

\begin{center}
\small
\setlength{\tabcolsep}{4pt}
\begin{longtable}{@{}>{\raggedright\arraybackslash}p{0.22\textwidth}
  >{\raggedright\arraybackslash}p{0.36\textwidth}
  >{\raggedright\arraybackslash}p{0.2\textwidth}
  >{\raggedright\arraybackslash}p{0.12\textwidth}@{}}
\toprule
Book & Lean & Module & Status \\
\midrule
\endhead
Definition~\ref{def:jl-jterm} &
  \texttt{JTerm}, \texttt{Leaf}, \texttt{JTerm.leaves} & \texttt{Terms} & definition \\
Conventions of Section~\ref{sec:jl-evidence} &
  \texttt{PropositionSystem} & \texttt{Kernel} & interface \\
Definition~\ref{def:jl-leaf-cs} &
  \texttt{ConstantSpec}, \texttt{ConstantSpec.\allowbreak Admitted} & \texttt{Kernel} &
  definition \\
Definition~\ref{def:jl-evidence} &
  \texttt{ConstantSpec.\allowbreak evid} & \texttt{Evidence} & definition \\
Lemma~\ref{thm:jl-finite} &
  \texttt{ConstantSpec.\allowbreak card\_evid\_le} & \texttt{Evidence} & proved \\
Definition~\ref{def:jl-rules} &
  \texttt{ConstantSpec.JB} & \texttt{Evidence} & definition \\
Theorem~\ref{thm:jl-decision} &
  \texttt{ConstantSpec.\allowbreak jb\_iff\_mem\_evid} & \texttt{Evidence} & proved \\
Corollary~\ref{thm:jl-decidable} &
  \texttt{ConstantSpec.\allowbreak decidableJB} & \texttt{Evidence} & proved \\
Proposition~\ref{thm:jl-minimal} &
  \texttt{ConstantSpec.\allowbreak Admissible}, \texttt{admissible\_evid},
  \texttt{evid\_subset\_of\_admissible} & \texttt{Evidence} & proved \\
Definition~\ref{def:jl-support} &
  \texttt{ConstantSpec.\allowbreak supp} & \texttt{Support} & definition \\
Lemma~\ref{thm:jl-supp-nonempty} &
  \texttt{supp\_nonempty\_iff}, \texttt{admitted\_mem\_of\_mem\_supp} & \texttt{Support} &
  proved \\
Definition~\ref{def:jl-term-support} &
  \texttt{ConstantSpec.\allowbreak termSupp} & \texttt{Support} & definition \\
Proposition~\ref{thm:jl-refines} &
  \texttt{supp\_subset\_termSupp} & \texttt{Support} & proved \\
Theorem~\ref{thm:jl-phantom} &
  \texttt{supp\_sum\_of\_notMem} & \texttt{Support} & proved \\
Definition~\ref{def:jl-queries} &
  \texttt{fv}, \texttt{sw}, \texttt{lo}, \texttt{cited}, \texttt{claims};
  \texttt{lo\_subset\_cited} & \texttt{Support} & definition \\
Definition~\ref{def:jl-valuation} &
  \texttt{hv} & \texttt{Support} & definition \\
Theorem~\ref{thm:jl-lattice-hom} &
  \texttt{hv\_union}, \texttt{hv\_image$_2$}, \texttt{hv\_minimals}, \texttt{fv\_iff\_hv},
  \texttt{sw\_iff\_hv}, \texttt{grade} & \texttt{Support} & proved \\
Theorem~\ref{thm:jl-verified-factive} &
  \texttt{provable\_of\_mem\_supp}, \texttt{provable\_of\_fv} & \texttt{Support} & proved \\
Proposition~\ref{thm:jl-grade-loses} &
  \texttt{grade\_loses} & \texttt{Support} & proved \\
Definition~\ref{def:jl-semirings} &
  \texttt{Why}, \texttt{Pos}, \texttt{Lin}, each a \texttt{CommSemiring} & \texttt{Semiring} &
  definition \\
Definition~\ref{def:jl-prov} &
  \texttt{ConstantSpec.\allowbreak prov}, \texttt{Why.ofPoly} & \texttt{Semiring} & definition \\
Theorem~\ref{thm:jl-supp-hom} &
  \texttt{suppW\_eq\_ofPoly\_prov}, \texttt{suppW\_sum}, \texttt{suppW\_app} &
  \texttt{Semiring} & proved \\
Lemma~\ref{thm:jl-min-hom} &
  \texttt{Why.toPos}, \texttt{toPos\_surjective}, \texttt{Why.factors\_iff} & \texttt{Semiring} &
  proved \\
Proposition~\ref{thm:jl-lineage} &
  \texttt{Why.lin}, \texttt{lin\_not\_factors}, \texttt{lin\_suppW} & \texttt{Semiring} & proved \\
Definition~\ref{def:jl-chain-state} &
  \texttt{ChainState} & \texttt{WellFounded} & definition \\
Definition~\ref{def:jl-wellfounded} &
  \texttt{ChainState.phi}, \texttt{WF}, \texttt{WellFounded}, \texttt{suppWF} & \texttt{WellFounded}
  &
  definition \\
Lemma~\ref{thm:jl-wf-cycle} &
  \texttt{exists\_leaf\_notMem\_WF}, \texttt{exists\_cycle} & \texttt{WellFounded} & proved \\
Definition~\ref{def:jl-edge-check} &
  \texttt{ChainState.Ref}, \texttt{EdgeCheck}; \texttt{edgeCheck\_iff\_\allowbreak wellFounded} &
  \texttt{WellFounded} & definition \\
Theorem~\ref{thm:jl-wf-sum} &
  \texttt{wellFounded\_of\_edgeCheck}, \texttt{wellFounded\_sumAt},
  \texttt{Example.\allowbreak exists\_\allowbreak wellFounded\_\allowbreak not\_\allowbreak
  edgeCheck} &
  \texttt{WellFounded} & proved \\
Definition~\ref{def:jl-calculus} &
  \texttt{PropositionSystem.Hom}, \texttt{ConstantSpec.Hom}, \texttt{JTerm.map},
  \texttt{JSentence}, \texttt{closedTheory} & \texttt{Logic} & definition \\
Proposition~\ref{thm:jl-translation} &
  \texttt{ConstantSpec.Hom.\allowbreak jb\_map}, \texttt{mem\_evid\_map},
  \texttt{mapsTo\_closedTheory} & \texttt{Logic} & proved \\
Theorem~\ref{thm:jl-monotone} &
  \texttt{supp\_app\_bridge}, \texttt{hv\_app\_bridge}, \texttt{grade\_app\_bridge\_eq\_iff},
  \texttt{fv\_app\_bridge\_iff}, \texttt{not\_sw\_app\_bridge} & \texttt{Logic} & proved \\
Definition~\ref{def:jl-church} &
  \texttt{ChurchFamily} & \texttt{Church} & definition \\
Proposition~\ref{thm:jl-sum-splits} &
  \texttt{ChurchFamily.\allowbreak sum\_splits} & \texttt{Church} & proved \\
Proposition~\ref{thm:jl-pair-simulation} &
  \texttt{PairData}, \texttt{mem\_evid\_pairTerm}, \texttt{suppW\_pairTerm}, \texttt{hv\_pairTerm},
  \texttt{fv\_pairTerm\_iff}, \texttt{sw\_pairTerm\_iff} & \texttt{Church} & proved \\
Definitions~\ref{def:jl-language}, \ref{def:jl-axioms}, \ref{def:jl-hierarchy}, \ref{def:jl-cs} &
  \texttt{BTerm}, \texttt{BFormula}, \texttt{BasicSystem}, \texttt{IsAxiom},
  \texttt{ConstantSpecification}, \texttt{BDerivable} & \texttt{Basic} & definition \\
Definitions~\ref{def:jl-evidence-function}, \ref{def:jl-mkrtychev}, \ref{def:jl-fitting} &
  \texttt{EvidenceFunction}, \texttt{MkrtychevModel}, \texttt{FittingModel} & \texttt{Basic} &
  definition \\
Theorem~\ref{thm:jl-completeness-cited}, soundness &
  \texttt{MkrtychevModel.\allowbreak sat\_of\_derivable} & \texttt{Basic} & proved \\
Lemma~\ref{thm:jl-minimal-evidence} &
  \texttt{EvidenceFunction.\allowbreak minimal}, \texttt{minimal\_respects}, \texttt{minimal\_le} &
  \texttt{Basic} & proved \\
Proposition~\ref{thm:jl-minimal-characterizes} &
  \texttt{derivable\_\allowbreak just\_\allowbreak iff\_\allowbreak mem\_\allowbreak minimal} &
  \texttt{Basic} &
  proved \\
The interface from the kernel of Chapter~\ref{ch:mltt} &
  \texttt{KSen.setoid}, \texttt{KPrp}, \texttt{KUniv}, \texttt{kImpl?}, \texttt{kPi?},
  \texttt{KProvable}, \texttt{kProvable\_app}, \texttt{kProvable\_inst}, \texttt{kernelPS} &
  \texttt{KernelInstance} & proved \\
Theorem~\ref{thm:jl-verified-factive} with Theorem~\ref{thm:mltt-declarative-consistency},
  relative to a chain of Grothendieck universes &
  \texttt{kernelPS\_not\_\allowbreak provable\_false}, \texttt{kernelPS\_not\_fv\_false} &
  \texttt{KernelInstance} & proved \\
Proposition~\ref{thm:jl-conservative}, the replacement of the new constants by an
  interpretation, and the consistency of the extension, relative to a chain of Grothendieck
  universes &
  \texttt{Env.FreshFor}, \texttt{Env.\allowbreak rules\_mapC}, \texttt{Env.\allowbreak
  rules\_append\_ax}, \texttt{interpMap}, \texttt{KSigMor.\allowbreak interpAx},
  \texttt{HasType.\allowbreak conservative}, \texttt{HasType.\allowbreak
  consistency\_conservative} & \texttt{Conservativity}, \texttt{KernelSign} & proved \\
Theorem~\ref{thm:jl-nonfactive}, over the extension of
  Proposition~\ref{thm:jl-conservative} &
  \texttt{HasType.\allowbreak not\_ext\_of\_not}, \texttt{HasType.\allowbreak nonfactive},
  \texttt{HasType.\allowbreak nonfactive\_pi} & \texttt{Conservativity} & proved \\
Definitions~\ref{def:jl-code} and~\ref{def:jl-jterm} as kernel inductive types &
  \texttt{codeDecl}, \texttt{jtermDecl}, \texttt{jEnv}, \texttt{jEnv\_wf},
  \texttt{jEnv\_\allowbreak kchecked} & \texttt{KernelJ} & proved \\
Proposition~\ref{thm:jl-conservative} and Theorem~\ref{thm:jl-nonfactive} over $K$, the
  construction of $J$ &
  \texttt{eqNat}, \texttt{eqCode}, \texttt{enc}, \texttt{jJ}, \texttt{eqCode\_eval},
  \texttt{enc\_eval}, \texttt{jJ\_eval}, \texttt{jEnv\_\allowbreak freshFor},
  \texttt{jb\_hval}, \texttt{HasType.\allowbreak jb\_conservative},
  \texttt{HasType.\allowbreak jb\_consistent}, \texttt{HasType.\allowbreak jb\_nonfactive} &
  \texttt{KernelJ} & proved \\
\bottomrule
\end{longtable}
\end{center}

Not formalized are Proposition~\ref{thm:jl-prov-derivations}, whose derivations would be a
type-valued inductive family; Proposition~\ref{thm:jl-poly}, a complexity bound; the former
$\jpair$ of Definition~\ref{def:jl-pair} and the checker extension of
Section~\ref{sec:jl-checker} with Proposition~\ref{thm:jl-j4}, which add term formers to
$\JTerm$; Proposition~\ref{thm:jl-hyperintensional};
Proposition~\ref{thm:jl-relative-internalization}; Corollary~\ref{thm:jl-j-embedding}; and the
cited internalization and self-reference results of Section~\ref{sec:jl-basic}. The
conservativity of Section~\ref{sec:jl-open} is formalized in two steps. In
\texttt{Conservativity}, over any environment, the extension by a constant and finitely many
axioms whose names are fresh for the environment, interpreted by closed terms of the environment
at the types the axioms give, is conservative for the judgments that mention no new name, since
the interpretation is a signature morphism of Chapter~\ref{ch:institutions} and fixes the rules
of the environment; with Theorem~\ref{thm:mltt-declarative-consistency}, the extension of an
environment without axioms that the kernel checks is consistent, and
Theorem~\ref{thm:jl-nonfactive} follows, since a proof of $\JB(\lambda, P_0) \jlimp P_0$ applied
to the decision axiom proves $P_0$. In \texttt{KernelJ}, $K$ is the environment of $\tttBool$,
$\Nat$, $\jlCode$, and $\JTerm$, with resource names in $\Nat$, which is well formed and checked
by the kernel, and $J$ is the Boolean decision of the proof of
Proposition~\ref{thm:jl-conservative}: $\mathsf{eq}$, $\mathsf{enc}$, and $J$ are closed terms
of $K$, and on constructor forms $J(j, c)$ reduces to $\Pi P{:}\Prop.\, P \to P$ for a positive
pair and to $\tttFalse$ otherwise, by $\beta$- and $\iota$-steps at closed arguments; the
conversions of minor premises to the types the recursor expects are decided by
$\beta$-normalization. The decision rules are taken at constructor forms, with distinct fresh
names and no pair decided both ways. With this interpretation the conservativity, the
consistency of $K_{\JB}$, and Theorem~\ref{thm:jl-nonfactive} hold over $K$ without hypotheses
on $J$.
In Lemma~\ref{thm:jl-wf-cycle} the cycle is the
eventual periodicity of the iteration of a chosen successor function on $D \setminus \jlWF$. In
Proposition~\ref{thm:jl-pair-simulation} the leaf $\kappa$ enters through the interface as the
two instantiations and the two implications of $\Pi (A\, B : \Prop).\, A \jlimp B \jlimp A
\wedge B$, and the comparison with $\jpair(s, t)$ is stated as the values of $\jlfv$ and $\jlsw$
for the two summands.

\section{Chapter~\ref{ch:institutions}}

Modules are in \texttt{EigeniusTheory/Institutions}. A map of entailment systems is
formalized in the normal form of Lemma~22 of \citet{meseguer1989}: the theory translation is
a functor on signatures together with the axiom sets of the translated theories, and $\alpha$
is natural on signatures, so condition~(1) of $\alpha$-sensibility holds by construction. The
categories $\Ent$, $\instPC$, and $\piIns^{\instco}$ have their objects, morphisms, and
composition defined; their unit and associativity laws are not stated. The kernel calculus of
Section~\ref{sec:inst-witness} is formalized in \texttt{Kernel} over the kernel of
Chapter~\ref{ch:mltt}, with the inclusions of environments, extension at the end, as the
signature morphisms. Sentences are the closed propositions, the axiom names of a finite
$\Gamma_0$ are the entries of a context listing its sentences, a witness records its axiom
context, its term, and its sentence, and the five lemmas are the variable rule, monotonicity,
weakening followed by substitution, and environment weakening. The general signature morphisms
of Definition~\ref{def:inst-kernel-sign} and Lemma~\ref{thm:inst-translation} are in
\texttt{KernelSign}: the assignment is indexed by the levels of each constant, $\sigma\,c\,\bar\ell$
for $\sigma(c)[\bar\ell/\bar u]$, natural in the levels, so that identities and composition
satisfy the laws of a category exactly; a signature morphism fixes the constants of the class
systems. The sentence functor, the entailment system, the proof calculus, and the models of
Theorem~\ref{thm:inst-kernel-logic} extend to $\Sign_{\instK}$; along a signature morphism
$\sigma$ a witness keeps the form of \texttt{Kernel}, with its axiom context and sentence
translated by $\Sen_{\instK}(\sigma)$ and its term by $\insthat{\sigma}$, which is the identity
along an inclusion. The axiom policies into the kernel calculus over $\Sign_{\instK}$ are in
\texttt{KernelPolicies}. Discharge and relativization rest on one simultaneous substitution of
terms over a common axiom context for the variables of another, which commutes with
$\insthat{\sigma}$; a discharge is indexed by the translated axioms. Discharge on a category of
theories with functorial target theories and a natural discharge, and relativization on
$\instThl$, are maps of proof calculi in the theory-relative form of Meseguer's Lemma~24
(\texttt{ProofCalculusMapOn}), of which a map of proof calculi is the case $\Thz$. The
relativizing transport renames each axiom variable to the variable of its first position in the
translated list, and is natural along morphisms injective on the list, which the distinctness of
$\instThl$ provides; the entailment condition~(d) follows from modus ponens for iterated
implications and the deduction theorem for kernel entailment.
Lemma~\ref{thm:inst-invariant} and Meseguer's $\instE^{\natural}$ are in \texttt{Transport},
$\instE^\dagger$ in \texttt{Dagger}.
Logics are an entailment system extended by models, satisfaction, the satisfaction condition,
and soundness, so that the two share their sentences (\texttt{Logic}). The kernel theory as a
logic, Theorem~\ref{thm:inst-kernel-logic}, is in \texttt{KernelModels} for the inclusions,
relative to a chain of Grothendieck universes: a model of a signature is a valuation of the
constants under which the signature, or a well-formed extension of it that the kernel checks,
has the model of Chapter~\ref{ch:mltt}, so that the reduct along an inclusion keeps the
valuation. A sentence has one value over a signature and over an extension, through its
translation into $\mathcal{K}_1$ and the environment weakening of the interpretation. Soundness
takes the value of the relativized proof in the model and discharges its hypotheses one at a
time. Over $\Sign_{\instK}$ the theorem is in \texttt{KernelModelsC}: a model is a signature
morphism into a signature that the kernel checks with a model of Chapter~\ref{ch:mltt} of the
target, the reduct composes the morphisms, the satisfaction condition is the functoriality of
the translation, and soundness translates the entailment along the morphism and applies the
soundness of \texttt{KernelModels} in the target. A model of \texttt{KernelModels} is the case
of an inclusion, with the same satisfaction.

\begin{center}
\small
\setlength{\tabcolsep}{4pt}
\begin{longtable}{@{}>{\raggedright\arraybackslash}p{0.22\textwidth}
  >{\raggedright\arraybackslash}p{0.36\textwidth}
  >{\raggedright\arraybackslash}p{0.2\textwidth}
  >{\raggedright\arraybackslash}p{0.12\textwidth}@{}}
\toprule
Book & Lean & Module & Status \\
\midrule
\endhead
Definition~\ref{def:inst-institution} &
  \texttt{Institution} & \texttt{Basic} & definition \\
Definition~\ref{def:inst-comorphism}(b) &
  \texttt{Institution.\allowbreak Comorphism} & \texttt{Basic} & definition \\
Semantic consequence, after Definition~\ref{def:inst-institution} &
  \texttt{Institution.\allowbreak SemCons} & \texttt{Basic} & definition \\
Definition~\ref{def:inst-entailment} &
  \texttt{EntailmentSystem} & \texttt{Entailment} & definition \\
Closure function, Remark~2 after Definition~\ref{def:inst-entailment} &
  \texttt{EntailmentSystem.\allowbreak clo} and its laws & \texttt{Entailment} & proved \\
Definition~\ref{def:inst-logic} &
  \texttt{Logic}, \texttt{Logic.\allowbreak toInstitution}, \texttt{Logic.semCons},
  \texttt{Logic.Complete} & \texttt{Logic} & definition \\
Definition~\ref{def:inst-theories} &
  \texttt{Theory}, \texttt{Theory.Hom}, the instance \texttt{Category (Theory Sen)},
  \texttt{Theory.sign} & \texttt{Theories} & definition \\
Definition~\ref{def:inst-proof-calculus} &
  \texttt{ProofCalculus} & \texttt{ProofCalculus} & definition \\
Definition~\ref{def:inst-ent-map} &
  \texttt{EntailmentMap}, \texttt{EntailmentMap.\allowbreak natTrans},
  \texttt{EntailmentMap.\allowbreak theoryFunctor} & \texttt{EntailmentMap} & definition \\
Definition~\ref{def:inst-pc-map} &
  \texttt{ProofCalculusMap} & \texttt{ProofCalculus} & definition \\
Definition~\ref{def:inst-pi-institution} &
  \texttt{PiInstitution} & \texttt{PiInstitution} & definition \\
Definition~\ref{def:inst-pi-comorphism} &
  \texttt{PiInstitution.\allowbreak Comorphism} & \texttt{PiInstitution} & definition \\
Definition~\ref{def:inst-pi-of-ent}(a), (b) &
  \texttt{PiInstitution.\allowbreak ofEntailment},
  \texttt{PiInstitution.\allowbreak ofEntailmentTh} & \texttt{PiInstitution} & proved \\
Lemma~\ref{thm:inst-pi-of-ent} &
  the two above and \texttt{PiInstitution.\allowbreak ofEntailmentToTh} &
  \texttt{PiInstitution} & proved \\
Theorem~\ref{thm:inst-forgetting} &
  \texttt{EntailmentMap.\allowbreak comorphismTh}, \texttt{ProofCalculusMap.\allowbreak forget} &
  \texttt{EntailmentMap}, \texttt{ProofCalculus} & proved \\
Corollary~\ref{thm:inst-forgetting-sig} &
  \texttt{EntailmentMap.\allowbreak comorphismSig} & \texttt{EntailmentMap} & proved \\
Composition, Section~\ref{sec:inst-composition} &
  \texttt{EntailmentMap.\allowbreak comp}, \texttt{ProofCalculusMap.\allowbreak comp} &
  \texttt{EntailmentMap}, \texttt{ProofCalculus} & proved \\
Definition~\ref{def:inst-G} &
  \texttt{PiInstitution.\allowbreak Closed}, \texttt{PiInstitution.\allowbreak modelFunctor},
  \texttt{PiInstitution.\allowbreak toInstitution},
  \texttt{PiInstitution.\allowbreak Comorphism.\allowbreak toInstitution} &
  \texttt{Models} & definition \\
Proposition~\ref{thm:inst-G-models} &
  \texttt{PiInstitution.\allowbreak closed\_preimage}; naturality of $\alpha^{-1}$ in
  \texttt{Comorphism.\allowbreak toInstitution} & \texttt{Models} & proved \\
Proposition~\ref{thm:inst-satisfaction} &
  the fields \texttt{satisfaction} of \texttt{toInstitution} and \texttt{compat} of
  \texttt{Comorphism.\allowbreak toInstitution} & \texttt{Models} & proved \\
Proposition~\ref{thm:inst-completeness} &
  \texttt{PiInstitution.\allowbreak semCons\_iff},
  \texttt{PiInstitution.\allowbreak ofEntailment\_\allowbreak semCons\_iff} &
  \texttt{Models} & proved \\
Corollary~\ref{thm:inst-every-comorphism} &
  \texttt{ProofCalculusMap.\allowbreak toInstitution}, \texttt{toInstitutionSig},
  \texttt{ofTransportPlain\_\allowbreak ax\_empty},
  \texttt{PiInstitution.\allowbreak ofEntailmentTh\_\allowbreak closed\_iff} &
  \texttt{Models} & proved \\
Definition~\ref{def:inst-theories}, theory morphisms &
  \texttt{Theory.ThHom}, the instance \texttt{Category (Th E)}, \texttt{Th.ofHom},
  \texttt{Th.ofSign} & \texttt{Dagger} & definition \\
Lemma~\ref{thm:inst-invariant} &
  \texttt{EntailmentMap.\allowbreak alphaExtension}, \texttt{EntailmentMap.\allowbreak ofSimple},
  \texttt{ProofCalculusMap.\allowbreak ofTransport} & \texttt{Transport} & proved \\
Proposition~\ref{thm:inst-adopt} &
  \texttt{alphaExtension\_plain\_ax}, \texttt{ProofCalculusMap.\allowbreak ofTransportPlain};
  \texttt{KWitness.ofAxiom}, \texttt{KWitness.}$\pi$\texttt{\_ofAxiom};
  \texttt{KWitnessC.ofAxiom}, \texttt{sen\_ofAxiom}, \texttt{ofTransportPlain\_ax} &
  \texttt{Transport}, \texttt{Kernel}, \texttt{KernelPolicies} & proved \\
$\instE^{\natural}$ and the unit of $\mathit{ent} \dashv (\ )^{\natural}$, after
Definition~\ref{def:inst-pc-map} &
  \texttt{EntailmentSystem.\allowbreak thm}, \texttt{EntailmentSystem.\allowbreak natural},
  \texttt{EntailmentMap.id}, \texttt{ProofCalculus.unit} & \texttt{Transport} & proved \\
Definition~\ref{def:inst-kernel-sign}, inclusions &
  \texttt{KSign} with its \texttt{Preorder} & \texttt{Kernel} & definition \\
Definition~\ref{def:inst-kernel-sen} &
  \texttt{KSen}, \texttt{KSenF} & \texttt{Kernel} & definition \\
Definition~\ref{def:inst-kernel-ent} &
  \texttt{axCtx}, \texttt{KEnt} & \texttt{Kernel} & definition \\
Definition~\ref{def:inst-kernel-proofs} &
  \texttt{KWitness}, \texttt{KWitness.map}, \texttt{KProofs}, \texttt{KPi} & \texttt{Kernel} &
  definition \\
Lemmas~\ref{thm:inst-refl} to~\ref{thm:inst-ent-trans} &
  \texttt{KEnt.refl}, \texttt{KEnt.mono}, \texttt{KEnt.cut}, \texttt{KEnt.trans},
  \texttt{KEnt.\allowbreak translation}, \texttt{KEntSystem} & \texttt{Kernel} & proved \\
Lemma~\ref{thm:inst-nat} &
  \texttt{KWitness.map\_id}, \texttt{KWitness.map\_comp}, the naturality of \texttt{KPi} &
  \texttt{Kernel} & proved \\
Theorem~\ref{thm:inst-kernel-pc} for the inclusions &
  \texttt{KCalc}, \texttt{KEnt.compact} & \texttt{Kernel} & proved \\
Theorem~\ref{thm:inst-kernel-pc}, with Definition~\ref{def:inst-kernel-proofs} and
  Lemma~\ref{thm:inst-nat}, over $\Sign_{\instK}$ &
  \texttt{KWitnessC}, \texttt{KWitnessC.map}, \texttt{map\_id}, \texttt{map\_comp},
  \texttt{term\_map\_\allowbreak ofPrefix}, \texttt{KProofsC}, \texttt{KPiC}, \texttt{KCalcC},
  \texttt{KCalcC.compact} & \texttt{KernelCalcC} & proved \\
Proposition~\ref{thm:inst-collapse}, for signatures declaring $\irrel$ &
  \texttt{KWitness.\allowbreak proofIrrel}, \texttt{KWitness.Equiv}, \texttt{equiv\_iff},
  \texttt{KWitness.\allowbreak quotEquiv}; \texttt{ProofCalculus.\allowbreak
  unit\_}$\gamma$\texttt{\_bijective} & \texttt{Kernel}, \texttt{Transport} & proved \\
Proposition~\ref{thm:inst-discharge}, for the inclusions &
  \texttt{KWitness.\allowbreak discharge} & \texttt{Kernel} & proved \\
Proposition~\ref{thm:inst-relativise}(a), (b), for the inclusions &
  \texttt{KSign.HasType.lams}, \texttt{KSign.HasType.\allowbreak pis\_prop}, \texttt{KSen.pis},
  \texttt{KWitness.\allowbreak relativize}, \texttt{sen\_relativize} & \texttt{Kernel} & proved \\
Maps of proof calculi in the theory-relative form &
  \texttt{ProofCalculusMapOn}, \texttt{ProofCalculusMap.toOn} & \texttt{ProofCalculus} &
  definition \\
Proposition~\ref{thm:inst-discharge} over $\Sign_{\instK}$ &
  \texttt{Term.substAx}, \texttt{mapC\_substAx}, \texttt{KSign.HasType.\allowbreak substAx},
  \texttt{KWitnessC.\allowbreak dischargeTerms}, \texttt{KWitnessC.\allowbreak discharge},
  \texttt{map\_discharge}, \texttt{ProofCalculusMap.\allowbreak dischargeOn} &
  \texttt{KernelPolicies} & proved \\
Proposition~\ref{thm:inst-relativise} over $\Sign_{\instK}$, (a)--(d) &
  \texttt{ThLst}, \texttt{ThLst.}$\iota$, \texttt{KWitnessC.\allowbreak relativizeTo},
  \texttt{map\_relativizeTo}, \texttt{KEnt.mp}, \texttt{KEnt.\allowbreak deduction},
  \texttt{KEnt.\allowbreak relativize}, \texttt{ProofCalculusMap.\allowbreak relativizeOn} &
  \texttt{KernelPolicies} & proved \\
Definition~\ref{def:inst-kernel-sign} with its category &
  \texttt{Term.mapC}, \texttt{KSigMor}, \texttt{KSigMor.id}, \texttt{KSigMor.comp},
  \texttt{KSigMor.\allowbreak ofPrefix}, \texttt{KSigCat}, \texttt{KSign.\allowbreak toSigCat} &
  \texttt{KernelSign} & definition \\
Lemma~\ref{thm:inst-translation} &
  \texttt{Term.\allowbreak mapC\_liftN}, \texttt{mapC\_inst}, \texttt{mapC\_instL},
  \texttt{LEquiv.mapC}, \texttt{KSigMor.\allowbreak defEq\_of\_red}, \texttt{KSigMor.defEq},
  \texttt{KSigMor.lookup}, \texttt{Derivable.mapC} & \texttt{KernelSign} & proved \\
Lemma~\ref{thm:inst-ent-trans} along signature morphisms &
  \texttt{KSen.mapC}, \texttt{KSenC}, \texttt{KEnt.mapC}, \texttt{KEntSystemC} &
  \texttt{KernelSign} & proved \\
Theorem~\ref{thm:inst-kernel-logic} for the inclusions, relative to a chain of Grothendieck
  universes &
  \texttt{KModel}, \texttt{KModel.reduct}, \texttt{KModF}, \texttt{KSat},
  \texttt{KSen.\allowbreak interp\_prefix}, \texttt{KSat.\allowbreak prefix\_iff},
  \texttt{KSen.\allowbreak interp\_cons}, \texttt{KSat.of\_imp}, \texttt{KSat.of\_pis},
  \texttt{KEnt.sound}, \texttt{KEnt.\allowbreak consistent}, \texttt{KLogic} &
  \texttt{KernelModels} & proved \\
Theorem~\ref{thm:inst-kernel-logic} along signature morphisms, relative to a chain of
  Grothendieck universes &
  \texttt{KModelC}, \texttt{KModelC.\allowbreak reduct}, \texttt{KModFC}, \texttt{KSatC},
  \texttt{KSatC.\allowbreak mapC\_iff}, \texttt{KEnt.soundC}, \texttt{KEnt.\allowbreak consistentC},
  \texttt{KModel.toC}, \texttt{KModel.\allowbreak toC\_sat}, \texttt{KLogicC} &
  \texttt{KernelModelsC} & proved \\
Proposition~\ref{thm:inst-dagger} &
  \texttt{EntailmentSystem.\allowbreak dagger}, \texttt{dagger\_semCons\_iff} & \texttt{Dagger} &
  proved \\
Proposition~\ref{thm:inst-dagger-G} &
  \texttt{daggerTh}, \texttt{daggerSat\_iff\_mem\_daggerTh}, \texttt{clo\_daggerTh},
  \texttt{daggerTh\_daggerModel} & \texttt{Dagger} & proved \\
\bottomrule
\end{longtable}
\end{center}

Not formalized are Definition~\ref{def:inst-env-theory}; Proposition~\ref{thm:inst-cut-redex},
of which the abstraction over the axiom context is \texttt{KSign.HasType.lams}; exactness,
initial models, and liberality in Proposition~\ref{thm:inst-dagger}; the cited
Theorem~\ref{thm:inst-mariano-pinto}, of which the three consequences used in the text are
proved directly; and the category laws of $\Ent$, $\instPC$, and $\piIns^{\instco}$. The
isomorphism of Proposition~\ref{thm:inst-collapse} is stated componentwise: the witnesses of $T$
modulo provable identity are in bijection with the theorems of $T$, and the unit of
$\mathit{ent} \dashv (\ )^{\natural}$ has bijective components when each $\pi_T$ is injective.

\section{Chapter~\ref{ch:knowledge-graphs}}

Modules are in \texttt{EigeniusTheory/KnowledgeGraphs}, and the extension of the kernel of
Chapter~\ref{ch:mltt} is in \texttt{EigeniusTheory/TypeTheory}. The chapter is formalized at
two levels. The first is TTR's witness judgment over a graph, the level at which its
definitions and results are stated. A knowledge graph is a
structure of finite sets of IRIs with the field type of each property, the own requirements,
declared parents, equivalence declarations, and conditional requirements of each class, and the
record and declared classes of each resource; the basic types enter through their witness
relation on literals. Values are literals, references, lists, and embedded records; field types
are the formers of Sections~\ref{sec:kg-classes} and~\ref{sec:kg-joins}, with record types as
labels and a field type at each label, so that membership is width-inclusive by construction
(Definition~\ref{def:kg-ttr-record}). Value typing is an inductive predicate relative to a
membership relation, membership is Mathlib's greatest fixed point of the monotone operator
$\Phi$, and the coinductive proofs of the chapter are instances of \texttt{OrderHom.le\_gfp}.
Decidability for a finite graph is the stabilization of the descending iteration from
$\mathcal{R} \times \mathcal{C}$ within $|\mathcal{R}| \cdot |\mathcal{C}|$ steps. The canonical
order is Kahn's selection as a function on field sets with an abstract notion of free labels.
The propositional level is read in the model: a class denotes its members and the identity of
entities is an equivalence on IRIs.

The second level is the rules that extend the kernel. The formers of the chapter are constants
that a declaration of the environment binds, so that the syntax of terms and the inference rules
of Chapter~\ref{ch:mltt} stay as they are, and the rules of the chapter are new base cases. The
class system of a graph is such a declaration (\texttt{Decl.kg}): it binds $\Entity$, the
classes with declared subclassing, the resources, each declared at a class below the classes of
which it is a member, as the own class $C_r$ of Definition~\ref{def:kg-theory} is, the formers
of binary joins and of singletons, and the constants of \textsc{Join-E} and \textsc{Sing-E}.
\textsc{Class-F} is Rule~\tttrule{Const} at the declared type $\Setu$ of a class, and
\textsc{Mem} is \textsc{Class-Sub} from the type of the resource. The environment is well formed
when the names of the class system are fresh and distinct and the type of each resource is a
class (Rule~\tttrule{E-KG}). \textsc{Class-Ent} and \textsc{Class-Sub}, along the
reflexive--transitive closure of declared subclassing, are base cases of subsumption in the
declarative system and in $\mathcal{K}_1$, the latter at types definitionally equal to the
classes, as Chapter~\ref{ch:mltt} reads types through their reducts. \textsc{Sing-I} is a typing
rule, a checking rule in $\mathcal{K}_1$, and \textsc{Join-I}, \textsc{Join-L}, and
\textsc{Sing-Sub} are base cases of subsumption, \textsc{Join-I} in $\mathcal{K}_1$ also a
checking rule. The graph types into which \textsc{Join-L} and \textsc{Sing-Sub} lead
(Definition~\ref{def:kg-graph-type}) are the types definitionally equal to a constant of a class
system applied to arguments, as classes, $\Entity$, joins, and singletons are, so that no join
or singleton lies below a universe, a $\Pi$-type, or a $\Sigma$-type; the inversions of
subsumption into these types keep their form. The stratified model of Chapter~\ref{ch:mltt}
extends by a class system: a resource denotes its code, $\Entity$ the set of the codes of the
resources, and a class the set of the codes of its members, the resources whose type lies below
it; the two rules are inclusions. A join denotes the union of the elements of its two types and
a singleton the set of its element, so that \textsc{Sing-I} and \textsc{Join-I} are memberships
and the subsumptions inclusions, and the constants of \textsc{Join-E} and \textsc{Sing-E}, whose
types are propositions, denote~$\emptyset$. The consistency theorems of Chapter~\ref{ch:mltt}
therefore hold for environments with class systems, which is Theorem~\ref{thm:kg-model} for
classes, resources, joins, and singletons. The test case declares a class system after
$\mathsf{Unit}$, $\tttBool$, and $\Nat$ and derives a membership by \textsc{Class-Sub},
memberships in a singleton and a join, and a subsumption by \textsc{Join-L} (\texttt{SClasses}).

\begin{center}
\small
\setlength{\tabcolsep}{4pt}
\begin{longtable}{@{}>{\raggedright\arraybackslash}p{0.22\textwidth}
  >{\raggedright\arraybackslash}p{0.36\textwidth}
  >{\raggedright\arraybackslash}p{0.2\textwidth}
  >{\raggedright\arraybackslash}p{0.12\textwidth}@{}}
\toprule
Book & Lean & Module & Status \\
\midrule
\endhead
Definition~\ref{def:kg-graph} &
  \texttt{Value}, \texttt{FieldType}, \texttt{RecType}, \texttt{KGraph} & \texttt{Graph} &
  definition \\
Definition~\ref{def:kg-req} &
  \texttt{KGraph.DeclSub}, \texttt{Anc}, \texttt{Eqv}, \texttt{Req}, \texttt{RC},
  \texttt{CondReqs} & \texttt{Graph} & definition \\
Theorem~\ref{thm:kg-coherence} &
  \texttt{req\_mono}, \texttt{rsub\_RC\_of\_anc} & \texttt{Graph} & proved \\
Corollary~\ref{cor:kg-equivalence} &
  \texttt{RC\_eq\_of\_eqv} & \texttt{Graph} & proved \\
Definition~\ref{def:kg-value-typing} &
  \texttt{KGraph.VTy}, \texttt{RecTy}, \texttt{CondTy} & \texttt{Graph} & definition \\
Definition~\ref{def:kg-membership} &
  \texttt{Declared}, \texttt{Structural}, \texttt{phi}, \texttt{Member}, \texttt{Mem} &
  \texttt{Graph} & definition \\
Proposition~\ref{prop:kg-membership-decidable} &
  \texttt{vty\_mono}, \texttt{mem\_iff}, \texttt{gfp\_eq\_\allowbreak iterate\_of\_finite},
  \texttt{member\_eq\_iterate} & \texttt{Graph} & proved \\
Lemma~\ref{lem:kg-named-sound} &
  \texttt{mem\_of\_anc} & \texttt{Graph} & proved \\
Lemma~\ref{lem:kg-value-sub} &
  \texttt{KGraph.FSub}, \texttt{FSub.joinIncl}, \texttt{vty\_of\_fsub}, \texttt{RSub},
  \texttt{recTy\_of\_rsub} & \texttt{Graph} & proved \\
Definition~\ref{def:kg-refinement}, Theorem~\ref{thm:kg-refinement-sound} &
  \texttt{RefMem}, \texttt{refMem\_of\_sub} & \texttt{Graph} & proved \\
Definition~\ref{def:kg-admissible} &
  \texttt{Admissible}, \texttt{Valid}, \texttt{ValidGraph}, \texttt{ownRecType} & \texttt{Graph} &
  definition \\
Proposition~\ref{prop:kg-resource-type} &
  \texttt{refMem\_own\_of\_valid}, \texttt{refMem\_RC\_of\_\allowbreak declared} & \texttt{Graph} &
  proved \\
Proposition~\ref{prop:kg-enum-subset} &
  \texttt{En}, \texttt{vty\_en\_iff}, \texttt{fsub\_en\_iri}, \texttt{fsub\_en\_mono} &
  \texttt{Graph} & proved \\
Definition~\ref{def:kg-conditional} &
  \texttt{CondReq}, \texttt{CondTy}, the second conjunct of \texttt{Structural} & \texttt{Graph} &
  definition \\
Definition~\ref{def:kg-canonical} &
  \texttt{FieldSet}, \texttt{Prec}, \texttt{Acyclic}, \texttt{canon}, \texttt{Telescope},
  \texttt{canonTele} & \texttt{Canonical} & definition \\
Theorem~\ref{thm:kg-canonical-order} &
  \texttt{canon\_length\_eq\_iff}, \texttt{telescope\_canon}, \texttt{canon\_congr} &
  \texttt{Canonical} & proved \\
Definition~\ref{def:kg-tfs}, (1)--(3) &
  \texttt{TFSSignature} & \texttt{Feature\-Structures} & definition \\
Lemma~\ref{lem:kg-tfs-inheritance} &
  \texttt{labels\_mono}, \texttt{field\_le}, \texttt{recSub\_of\_le} & \texttt{Feature\-Structures}
  &
  proved \\
Definition~\ref{def:kg-isa}, in the model &
  \texttt{KGraph.IsA}, \texttt{SubclassOf} & \texttt{Propositions} & definition \\
Theorem~\ref{thm:kg-decl-prop} &
  \texttt{isA\_of\_mem}, \texttt{subclassOf\_of\_anc} & \texttt{Propositions} & proved \\
Theorem~\ref{thm:kg-isa-laws} &
  \texttt{isA\_of\_subclassOf}, \texttt{subclassOf\_refl}, \texttt{subclassOf\_trans} &
  \texttt{Propositions} & proved \\
Proposition~\ref{prop:kg-one-way} &
  \texttt{OneWay.\allowbreak isA\_not\_mem} & \texttt{Propositions} & proved \\
Definitions~\ref{def:kg-atomic-valid}, \ref{def:kg-atomic-map} &
  \texttt{AtomicStore}, \texttt{Conforms}, \texttt{AtomicValid}, \texttt{Closed},
  \texttt{fieldType}, \texttt{toGraph} & \texttt{Atomic} & definition \\
Theorem~\ref{thm:kg-atomic-sound} &
  \texttt{declaredRel\_le\_phi}, \texttt{mem\_of\_isA}, \texttt{toGraph\_valid},
  \texttt{refMem\_own} & \texttt{Atomic} & proved \\
The class system of a graph as a declaration, and \textsc{Class-Ent} and \textsc{Class-Sub} in
  the kernel &
  \texttt{KGDecl}, \texttt{Decl.kg}, \texttt{KGDecl.WF}, \texttt{Env.KGClass},
  \texttt{Env.KGAnc}, \texttt{Env.WF} (\texttt{kg}), \texttt{Derivable.\allowbreak
  subClassEnt}, \texttt{subClassSub}, \texttt{Bidir.\allowbreak subClassEnt},
  \texttt{subClassSub} & \texttt{Environment}, \texttt{Wellformed}, \texttt{Typing},
  \texttt{Bidirectional} & definition \\
\textsc{Mem} in the kernel, from the type of a resource &
  \texttt{KGDecl.resType}, \texttt{Chk.\allowbreak alice\_agent},
  \texttt{HasType.\allowbreak alice\_agent} & \texttt{Environment}, \texttt{SClasses} &
  proved \\
Joins and singletons in the kernel: \textsc{Sing-I}, \textsc{Join-I}, \textsc{Join-L}, and
  \textsc{Sing-Sub}, with \textsc{Join-L} and \textsc{Sing-Sub} into graph types &
  \texttt{KGDecl.join}, \texttt{sing}, \texttt{Env.KGJoin}, \texttt{Env.KGSing},
  \texttt{Env.KGConst}, \texttt{Derivable.\allowbreak singI}, \texttt{subJoinI},
  \texttt{subJoinL}, \texttt{subSing}, \texttt{Bidir.\allowbreak chkSing}, \texttt{chkJoin},
  \texttt{subJoinI}, \texttt{subJoinL}, \texttt{subSing} & \texttt{Environment},
  \texttt{Typing}, \texttt{Bidirectional} & definition \\
\textsc{Join-E} and \textsc{Sing-E} as constants, their types propositions &
  \texttt{KGDecl.jcase}, \texttt{singE}, \texttt{KGDecl.\allowbreak jcaseType\_inf},
  \texttt{KGDecl.\allowbreak singEType\_inf} & \texttt{Environment}, \texttt{KTypes} &
  proved \\
The rules of joins and singletons in the test case &
  \texttt{Chk.\allowbreak alice\_sing}, \texttt{HasType.\allowbreak alice\_sing},
  \texttt{Chk.\allowbreak alice\_join}, \texttt{SubK.\allowbreak people\_join} &
  \texttt{SClasses} & proved \\
Theorem~\ref{thm:kg-model} for classes, resources, joins, and singletons, relative to a chain of
  Grothendieck universes &
  \texttt{SEnvModel.kg\_ent}, \texttt{kg\_anc}, \texttt{kg\_join}, \texttt{kg\_sing},
  \texttt{SSet.\allowbreak classCode}, \texttt{joinCode}, \texttt{singCode},
  \texttt{jcaseType\_mem}, \texttt{singEType\_mem}, \texttt{EnvData.kg},
  \texttt{envData\_nonempty}, \texttt{HasType.\allowbreak consistency\_of\_checked},
  \texttt{HasType.\allowbreak consistency\_people} &
  \texttt{SValidity}, \texttt{SEnvKG}, \texttt{SEnvRec}, \texttt{SConsistency},
  \texttt{SClasses} & proved \\
\bottomrule
\end{longtable}
\end{center}

Not formalized are the other rules of Sections~\ref{sec:kg-embedding}
to~\ref{sec:kg-propositions} as extensions of the kernel: record formation, introduction,
projection, and computation, \textsc{Width}, \textsc{Depth}, and \textsc{Rec-Sub} as rules of
the subsumption judgment, \textsc{Class-E}, and with them
Definition~\ref{def:kg-named-type}, Definition~\ref{def:kg-class-type},
Theorem~\ref{thm:kg-subsumption-sound} and Corollary~\ref{cor:kg-ttr-subtyping} as statements
about terms, whose value-level content is \texttt{vty\_of\_fsub} and \texttt{recTy\_of\_rsub},
Definition~\ref{def:kg-tags} with Lemma~\ref{lem:kg-distinct}, Theorem~\ref{thm:kg-exhaustive},
and Theorem~\ref{thm:kg-join-sum}, which eliminate identities in the kernel, and whose
value-level content is \texttt{vty\_en\_iff}, and Theorem~\ref{thm:kg-model} beyond classes,
resources, joins, and singletons. The twelve TTR
definitions of Section~\ref{sec:kg-ttr} are background; the witness clauses of records, meets,
joins, singletons, and lists are the clauses of \texttt{VTy}. Also not formalized are the theory
of a graph and the well-formed ontology (Definitions~\ref{def:kg-theory}
and~\ref{def:kg-wellformed}), the feature structures of Definition~\ref{def:kg-tfs} (4)--(7)
with Proposition~\ref{prop:kg-tfs-membership} and the two conjectures, and the Atomic Data
faithfulness conjecture.

\section{Chapter~\ref{ch:checking}}

Modules are in \texttt{EigeniusTheory/Checking}; the lemmas on $\mathcal{K}_0$ and
$\mathcal{K}_1$ that the chapter uses are in \texttt{EigeniusTheory/TypeTheory}, with
Chapter~\ref{ch:mltt}. Inferred types are types (\texttt{Inf.valid}) and unique up to
definitional equality (\texttt{Inf.unique}, \texttt{Inf.level\_unique}), subsumption into a
universe or a $\Pi$-type inverts (\texttt{SubK.sort\_inv}, \texttt{SubK.pi\_inv\_right}), and
definitional equality is joinability of erasures through annotation-free terms
(\texttt{RuleSet.DefEq.erase\_iff}).

A weak-head normal form is read through the property that
Lemma~\ref{lem:checking-whnf-stable} states of it: every reduct of it keeps its outermost former,
its variable, universe, or constant, and reduces its components (\texttt{Rigid}). The procedures
run over a \emph{weak-head normalization strategy} (\texttt{RuleSet.WHStrategy}), a
deterministic relation from a term to a weak-head normal form among its reducts that is defined
on every strongly normalizing term; strong normalization is accessibility along reduction
(\texttt{SN}). The strategy of Definition~\ref{def:checking-wh} is \texttt{WhStep}, over rule sets
of the shape that it relies on (\texttt{WHShape}): a rule with arguments matches a constructor
application in the last argument of a constant without rules and fires on every application of
the constant to as many arguments of that form. The rules of every environment have that shape
(\texttt{Env.rules\_whShape}), so the strategy is defined over every environment
(\texttt{Env.whStrategy}).

The conversion test is the run of Definition~\ref{def:checking-conv} as a big-step relation with
its answer (\texttt{ConvRun}): a run answers \texttt{true} exactly on definitionally equal terms,
which is soundness and completeness relative to termination, and a run exists on two strongly
normalizing terms (Theorem~\ref{thm:checking-conv}). The measure of the termination argument is
the accessibility of the order ``a reduct or a component'' on strongly normalizing terms
(\texttt{SN.acc\_desc}). Levels are compared as propositions; their decidability is
Proposition~\ref{prop:mltt-levels}.

The algorithmic systems are one inductive family \texttt{Alg} over a strategy, with a flag
selecting $\mathcal{A}_1$, and a premise $C \chkred P$ read as the relation of the strategy from
$C$ to $P$. Soundness
(\texttt{Alg.sound}) holds over an environment that annotates a well-formed environment, with
the strategy reducing over the latter, and states clause~(1) of
Theorem~\ref{thm:checking-sound} up to $\defeq$. Completeness (\texttt{Alg.complete}) is stated
for derivability in $\mathcal{A}_i$, with three hypotheses: the rules of the environment keep
terms annotation-free, as the rules of $|\tttEnv|$ do (Section~\ref{sec:checking-procedures});
the erasures of the terms that infer types in well-formed contexts are strongly normalizing, the
consequence of Hypothesis~\ref{hyp:checking-normalization} after it; and the normal forms of the
strategy are normal for weak-head reduction (\texttt{Whnf}), as those of the strategy of
Definition~\ref{def:checking-wh} are (\texttt{Env.\allowbreak whStrategy\_nf\_whnf}). The second
hypothesis stands for
the termination of the run that Theorem~\ref{thm:checking-complete} assumes: it makes the weak-head
normal forms that the premises name exist, and they are erasures of typed reducts
(\texttt{Alg.whnf\_typed}). Under these hypotheses inference in $\mathcal{A}_i$ is deterministic
(\texttt{Alg.inf\_det}), clause~(1) of Theorem~\ref{thm:checking-sound} holds with $|A| = T$
(\texttt{Alg.sound\_erase}), and inference, checking at the erasure of a type, and subsumption
between the erasures of types agree in $\mathcal{K}_i$ and $\mathcal{A}_i$
(\texttt{Alg.inf\_iff}, \texttt{Alg.chk\_iff}, \texttt{Alg.sub\_iff}).

The procedure for $\mathcal{A}_i$ is a program for each input (\texttt{checker}), in a small
language of continuations (\texttt{Prog}): it answers, calls the procedure on an input, computes a
weak-head normal form, runs the conversion test, or decides a proposition about levels, and
continues with a program for the result, and a combinator tries a second program when the first
rejects. The rules are tried and their premises evaluated in the order of the paragraph after
Definition~\ref{def:checking-alg} and of Section~\ref{sec:checking-kg}, and the checking rules
other than \tttrule{A-Chk-Sub} are an input of their own, so that checking tries them first. A
run (\texttt{Exec}) is the finite tree of the premises evaluated; a terminating run is one that
exists. The propositions about levels of \tttrule{A-Sort}, \tttrule{A-Const}, and
\tttrule{A-Sub-Cum} are decided by their truth, the decision procedure of
Proposition~\ref{prop:mltt-levels}. The answers of runs are correct (\texttt{Run.correct}): an
inferred type and an acceptance are derivations of $\mathcal{A}_i$, and a rejection excludes one,
so the run has one answer (\texttt{Run.det}). Under the hypotheses of completeness, the run
terminates on the inputs of Theorem~\ref{thm:checking-terminates} (\texttt{Run.\allowbreak
terminates\_inf\_chk}, \texttt{Run.\allowbreak terminates\_sub}), and its answer decides the
judgments of $\mathcal{K}_i$ (Corollary~\ref{cor:checking-decidable}).

Normalization by evaluation is one inductive relation \texttt{NbE} on evaluation, elimination,
and readback, read as a procedure whose computations are the finite derivations. An environment
assigns a value to every variable, variables are de Bruijn indices, and the readback goes under a
binder by the variable $0$ and the environment lifted (\texttt{Val.lift}) rather than by a fresh
name. The rules $\delta$ and $\iota$ are those of the rule set: a constant evaluates through its
reduct when a rule fires on it, and an elimination by an argument through the reduct of the
application that the value and the argument stand for (\texttt{Val.quote}), evaluated in the
identity environment, where the text evaluates the right-hand side of the rule at the values
read off the redex. An annotation evaluates as the term it annotates. The terms that values stand
for are simultaneous substitutions of all variables (\texttt{Term.subst}), with their
composition laws. Theorem~\ref{thm:checking-nbe}(1) holds over rules of the shape of weak-head
reduction (\texttt{WHShape}), which the rules of every environment have: on such rules a rule
fires on the readback of an elimination only if it fires on the term the elimination stands for,
which the evaluation has excluded. The relation is deterministic (\texttt{NbE.det}).

$\mathcal{A}_1$ has the rules of Figure~\ref{fig:checking-alg-kg} that correspond to the rules of
Chapter~\ref{ch:knowledge-graphs} in $\mathcal{K}_1$: \tttrule{A-Chk-Sing}, \tttrule{A-Chk-Join},
\tttrule{A-Class-Ent}, \tttrule{A-Class-Sub}, \tttrule{A-Join-I}, \tttrule{A-Join-L}, and
\tttrule{A-Sing-Sub}, and Proposition~\ref{prop:checking-kg} holds for them: soundness,
completeness, termination, and decidability carry over, with the side conditions on the class
system decided by their truth. The components of a join or a singleton that is a type of
$\mathcal{K}_1$ check as Chapter~\ref{ch:knowledge-graphs} types them, read through the
subsumptions of annotations at the head of the spine (\texttt{Inf.join\_inv},
\texttt{Inf.sing\_inv}); a weak-head normal form joinable with a spine of a constant without rules
is a spine of that constant (\texttt{RuleSet.\allowbreak Whnf.\allowbreak eq\_mkApp\_const}). The
subsumption run descends lexicographically, in the left type or, for \tttrule{A-Join-I}, in the
right type with the left one kept, and the checking run for a term descends in the expected type.
Rules \tttrule{A-Class-Var}, \tttrule{A-Rec-Sub}, and \tttrule{A-Ref-Sub} are not formalized,
since the rules of $\mathcal{K}_1$ they decide, for a type of \texttt{Class} below
\texttt{Entity}, records, and refinements, are not rules of \texttt{Bidir}.

For Proposition~\ref{prop:checking-compositional}(1), the program of every input calls the
procedure only on inputs for the immediate subterms of its term, on subsumption, and, for
checking, on the alternatives for the same term (\texttt{checker\_calls}); among those
alternatives, \tttrule{A-Chk-Sing} and \tttrule{A-Chk-Join} check the term itself against a
component of the weak-head normal form of the type. Clause~(2) is
\texttt{ConvRun.\allowbreak true\_of\_lequiv}. Clause~(3) is stated without
Proposition~\ref{thm:inst-cut-redex}, which is not formalized: the redex over closed,
annotation-free propositions is derivable in $\mathcal{A}_i$ from derivations of its body and its
arguments when its annotation is a type, over a strategy that leaves $\Pi$-types unchanged, as
that of Definition~\ref{def:checking-wh} does (\texttt{Alg.\allowbreak cut\_redex}), and every
run on it accepts (\texttt{Run.\allowbreak cut\_redex\_yes}). For
Corollary~\ref{cor:checking-jl}, the procedures decide definitional equality of closed types,
give their weak-head normal forms, and decide checking of instances in $\mathcal{K}_i$
(\texttt{Run.\allowbreak decides\_jl}); the instance of the interface of
Chapter~\ref{ch:justification-logic} over the kernel, \texttt{kernelPS}, decides its relations
classically and types instances in the declarative system, so the corollary is not connected to
it formally.
The library has a normalizer for $\beta$ alone, a function that performs at most a given number
of steps (\texttt{Term.betaNF}), with the soundness of comparing its results
(\texttt{RuleSet.DefEq.of\_betaNF}); the construction of $J$ in
Chapter~\ref{ch:justification-logic} evaluates closed terms with it.

\begin{center}
\small
\setlength{\tabcolsep}{4pt}
\begin{longtable}{@{}>{\raggedright\arraybackslash}p{0.22\textwidth}
  >{\raggedright\arraybackslash}p{0.36\textwidth}
  >{\raggedright\arraybackslash}p{0.2\textwidth}
  >{\raggedright\arraybackslash}p{0.12\textwidth}@{}}
\toprule
Book & Lean & Module & Status \\
\midrule
\endhead
Section~\ref{sec:checking-procedures}, erasures &
  \texttt{RuleSet.\allowbreak DefEq.\allowbreak erase\_iff} &
  \texttt{TypeTheory/\allowbreak Annotated\allowbreak Reduction} & proved \\
Section~\ref{sec:checking-procedures}, strong normalization &
  \texttt{SN}, \texttt{SN.red}, \texttt{SN.\allowbreak of\_mem\_comps},
  \texttt{SN.\allowbreak acc\_desc} & \texttt{Conversion} & definition, proved \\
Definition~\ref{def:checking-wh} &
  \texttt{WhStep}, \texttt{WHShape}, \texttt{whStrategy}, \texttt{Env.\allowbreak rules\_whShape},
  \texttt{Env.\allowbreak whStrategy} & \texttt{WeakHead}, \texttt{EnvShape} &
  definition, proved \\
Lemma~\ref{lem:checking-wh} &
  \texttt{WhStep.toRed}, \texttt{WhStep.det} & \texttt{WeakHead} & proved \\
Lemma~\ref{lem:checking-whnf-stable} &
  \texttt{Rigid}, \texttt{Whnf.rigid} & \texttt{Conversion}, \texttt{WeakHead} & proved \\
Corollary~\ref{cor:checking-whnf-join} &
  \texttt{Rigid.join}, \texttt{Rigid.\allowbreak defEq\_iff} & \texttt{Conversion} & proved \\
Definition~\ref{def:checking-conv} &
  \texttt{RuleSet.\allowbreak WHStrategy}, \texttt{ConvRun} & \texttt{Conversion} & definition \\
Theorem~\ref{thm:checking-conv} &
  \texttt{ConvRun.\allowbreak correct}, \texttt{ConvRun.\allowbreak total} &
  \texttt{Conversion} & proved \\
Theorem~\ref{thm:checking-conv} (1), for $\beta$ &
  \texttt{Term.betaNF}, \texttt{RuleSet.\allowbreak DefEq.\allowbreak of\_betaNF} &
  \texttt{Justification/\allowbreak KernelJ} & proved \\
Lemma~\ref{lem:checking-unique} (1), (2) &
  \texttt{Inf.valid}, \texttt{Inf.unique}, \texttt{Inf.\allowbreak level\_unique} &
  \texttt{TypeTheory/\allowbreak KValidity}, \texttt{TypeTheory/\allowbreak KUnique} & proved \\
Lemma~\ref{lem:checking-sub-inv} (1), (2) &
  \texttt{SubK.sort\_inv}, \texttt{SubK.\allowbreak pi\_inv\_right} &
  \texttt{TypeTheory/\allowbreak KTypes}, \texttt{TypeTheory/\allowbreak Annotated\allowbreak Reduction}
  & proved \\
Lemma~\ref{lem:checking-typed-reducts} &
  \texttt{Inf.\allowbreak typedReduct}, \texttt{KType.\allowbreak typedReduct},
  \texttt{KType.\allowbreak former\_of\_defEq} & \texttt{TypedReducts}, \texttt{Algorithmic} &
  proved \\
Definition~\ref{def:checking-alg} & \texttt{Alg} & \texttt{Algorithmic} & definition \\
Theorem~\ref{thm:checking-sound} &
  \texttt{Alg.sound}, \texttt{Alg.\allowbreak sound\_erase} &
  \texttt{Algorithmic}, \texttt{Alg\allowbreak Completeness} & proved \\
Theorem~\ref{thm:checking-complete} &
  \texttt{Alg.\allowbreak complete}, \texttt{Alg.\allowbreak inf\_det}, \texttt{Alg.\allowbreak inf\_iff},
  \texttt{Alg.\allowbreak chk\_iff}, \texttt{Alg.\allowbreak sub\_iff} &
  \texttt{Alg\allowbreak Completeness} & proved \\
Definition~\ref{def:checking-alg}, the procedure &
  \texttt{Query}, \texttt{Answer}, \texttt{Prog}, \texttt{checker}, \texttt{Exec}, \texttt{Run},
  \texttt{Run.\allowbreak correct}, \texttt{Run.det} & \texttt{Procedure} & definition, proved \\
Theorem~\ref{thm:checking-terminates} &
  \texttt{Run.\allowbreak terminates\_inf\_chk}, \texttt{Run.\allowbreak terminates\_sub} &
  \texttt{Termination} & proved \\
Corollary~\ref{cor:checking-decidable} &
  \texttt{Run.\allowbreak decides\_inf}, \texttt{Run.\allowbreak decides\_chk},
  \texttt{Run.\allowbreak decides\_sub} & \texttt{Termination} & proved \\
Definition~\ref{def:checking-values} &
  \texttt{Val}, \texttt{Elim}, \texttt{Val.lift}, \texttt{Val.quote}, \texttt{Term.subst},
  \texttt{Term.\allowbreak subst\_subst}, \texttt{Term.\allowbreak subst\_up\_inst} &
  \texttt{NbE}, \texttt{Subst} & definition, proved \\
Definition~\ref{def:checking-nbe} &
  \texttt{NJ}, \texttt{NonRedex}, \texttt{NbE}, \texttt{Nbe}, \texttt{NbE.det}, \texttt{Nbe.det} &
  \texttt{NbE} & definition, proved \\
Theorem~\ref{thm:checking-nbe} (1) &
  \texttt{NbE.sound}, \texttt{Val.Good}, \texttt{NbE.good}, \texttt{RuleSet.\allowbreak
  R\_of\_R\_lift}, \texttt{Nbe.sound} & \texttt{NbE} & proved \\
Theorem~\ref{thm:checking-nbe} (2) &
  \texttt{NbE.total}, \texttt{Nbe.total} & \texttt{NbE} & proved \\
Theorem~\ref{thm:checking-nbe} (3) &
  \texttt{Nbe.\allowbreak complete}, \texttt{Nbe.\allowbreak decides}, \texttt{Env.\allowbreak
  nbe\_decides} & \texttt{NbE} & proved \\
Figure~\ref{fig:checking-alg-kg} &
  \texttt{Alg.chkSing}, \texttt{Alg.chkJoin}, \texttt{Alg.\allowbreak subClassEnt},
  \texttt{Alg.\allowbreak subClassSub}, \texttt{Alg.\allowbreak subJoinI},
  \texttt{Alg.\allowbreak subJoinL}, \texttt{Alg.subSing}, \texttt{chkKGProg},
  \texttt{subKGProg} & \texttt{Algorithmic}, \texttt{Procedure} & definition \\
Proposition~\ref{prop:checking-kg} &
  \texttt{Inf.join\_inv}, \texttt{Inf.sing\_inv}, \texttt{KType.\allowbreak join\_of\_defEq},
  \texttt{RuleSet.\allowbreak Whnf.\allowbreak eq\_mkApp\_const}, \texttt{Alg.\allowbreak whnf\_join},
  \texttt{Alg.sound}, \texttt{Alg.\allowbreak complete}, \texttt{Run.\allowbreak terminates\_sub} &
  \texttt{Graph\allowbreak Types}, \texttt{Algorithmic}, \texttt{Alg\allowbreak Completeness},
  \texttt{Termination} & proved, without \tttrule{A-Class-Var}, \tttrule{A-Rec-Sub},
  \tttrule{A-Ref-Sub} \\
Proposition~\ref{prop:checking-compositional} &
  \texttt{Prog.CallsIn}, \texttt{Query.Calls}, \texttt{checker\_calls},
  \texttt{ConvRun.\allowbreak true\_of\_lequiv}, \texttt{Alg.\allowbreak cut\_redex},
  \texttt{Run.\allowbreak cut\_redex\_yes} & \texttt{Properties} & proved \\
Corollary~\ref{cor:checking-jl} & \texttt{Run.\allowbreak decides\_jl} & \texttt{Properties} &
  proved, for $\mathcal{K}_i$ \\
\bottomrule
\end{longtable}
\end{center}

\clearpage

\backmatter
\bibliography{eigenius,eigenius_additional,eigenius_precursors,eigenius_related_work}

\end{document}